\documentclass[11.5pt,a4paper,oneside]{book}
\pdfoutput=1
\usepackage{mathrsfs}
\usepackage{multicol}
\usepackage{amsmath, amssymb}
\usepackage[mathscr]{eucal}
\usepackage{graphicx}
\usepackage{amsfonts}
\usepackage{framed}
\usepackage{hyperref}
\usepackage{makeidx}
\newcommand{\remfigure}[1]{}
\newcommand{\rem}[1]{}
\newcommand{\pa}{{\partial}}
\newcommand{\bG}{{\boldsymbol{G}}}
\newcommand{\bg}{{\boldsymbol{g}}}
\newcommand{\bmu}{\boldsymbol{\mu}}
\newcommand{\bfi}{\bfseries\itshape}
\newcommand{\bom}{\boldsymbol{\omega}}
\newcommand{\boq}{\boldsymbol{q}}
\newcommand{\bhz}{\mathbf{\hat{z}}}
\newcommand{\bx}{\boldsymbol{x}}
\newcommand{\bq}{\boldsymbol{q}}
\newcommand{\bS}{\boldsymbol{S}}
\newcommand{\bu}{\bf u}
\newcommand{\bm}{\boldsymbol{m}}
\newtheorem{theorem}{Theorem}

\newtheorem{corollary}[theorem]{Corollary}

\newtheorem{definition}[theorem]{Definition}

\newtheorem{lemma}[theorem]{Lemma}

\newtheorem{proposition}[theorem]{Proposition}
\newtheorem{remark}[theorem]{Remark}

\newtheorem{assumption}[theorem]{Assumption}
\newenvironment{proof}[1][Proof]{\noindent\textbf{#1.} }{\ \rule{0.5em}{0.5em}}
\newcommand{\comment}[1]{\vspace{1 mm}\par
\marginpar{\large\underline{}}\noindent
\framebox{\begin{minipage}[c]{0.95 \textwidth}
\rm #1 \end{minipage}}\vspace{2 mm}\par}
\def\contract{\makebox[1.2em][c]{\mbox{\rule{.6em}
{.01truein}\rule{.01truein}{.6em}}}}

\begin{document}

\title{
\!\!\!\!\!
\!\!\!\!\!\!\!\!\!\!\!\!
\!\!\!\!\!\!\!\!\!\!\!\!
\!\!\!\!\!\!\!\!\!\!\!\!
\huge\scshape Geometric dynamics of\\ Vlasov kinetic theory\\ and its moments\\\,\bigskip
}

\author{\LARGE\it Cesare Tronci\\\,\\\,\\
\bigskip\,\\ 
\bigskip\,\\
A thesis presented for the degree of \\
\large Doctor of Philosophy of the University of London \\
\large and the Diploma of Imperial College\\
$\,$\\
\includegraphics[angle=0, width=0.1\textwidth]{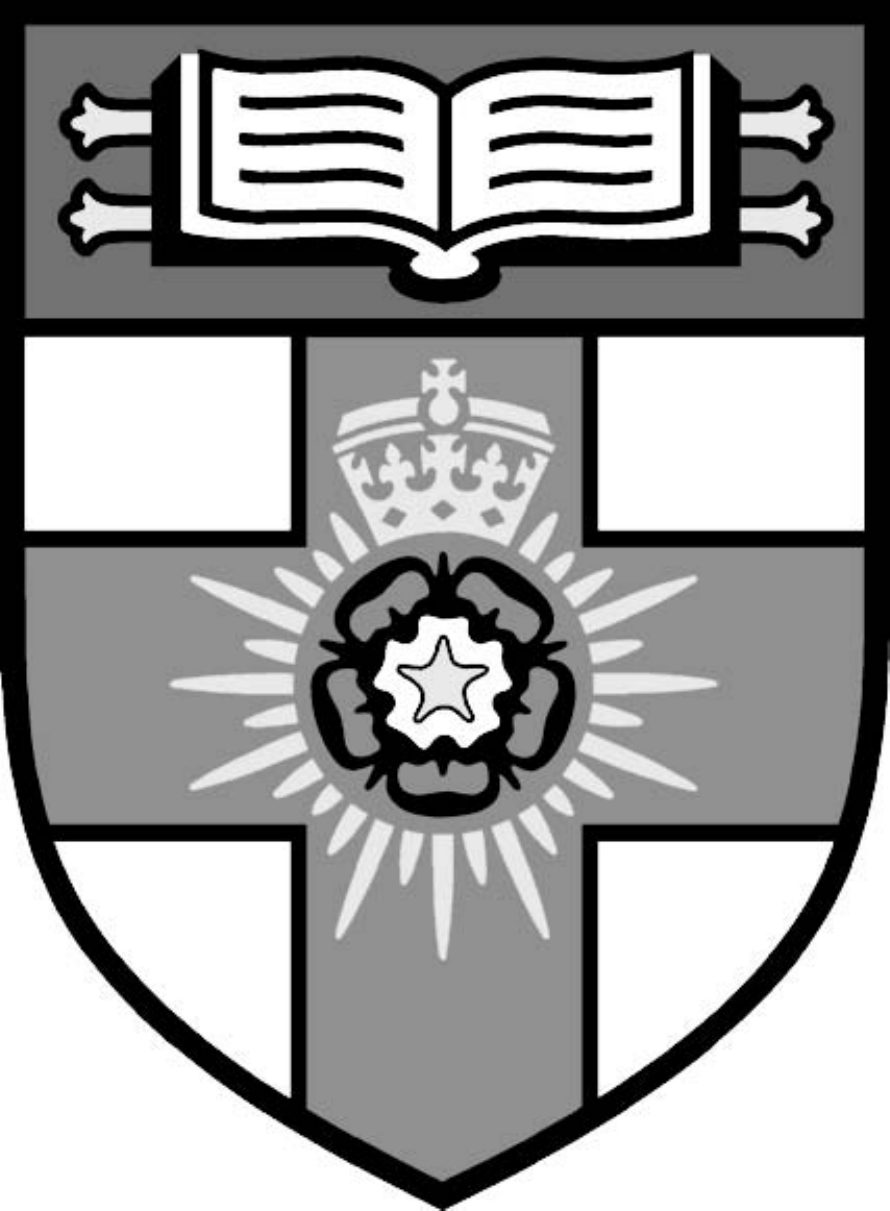}
\quad
\includegraphics[angle=0, width=0.1305\textwidth]{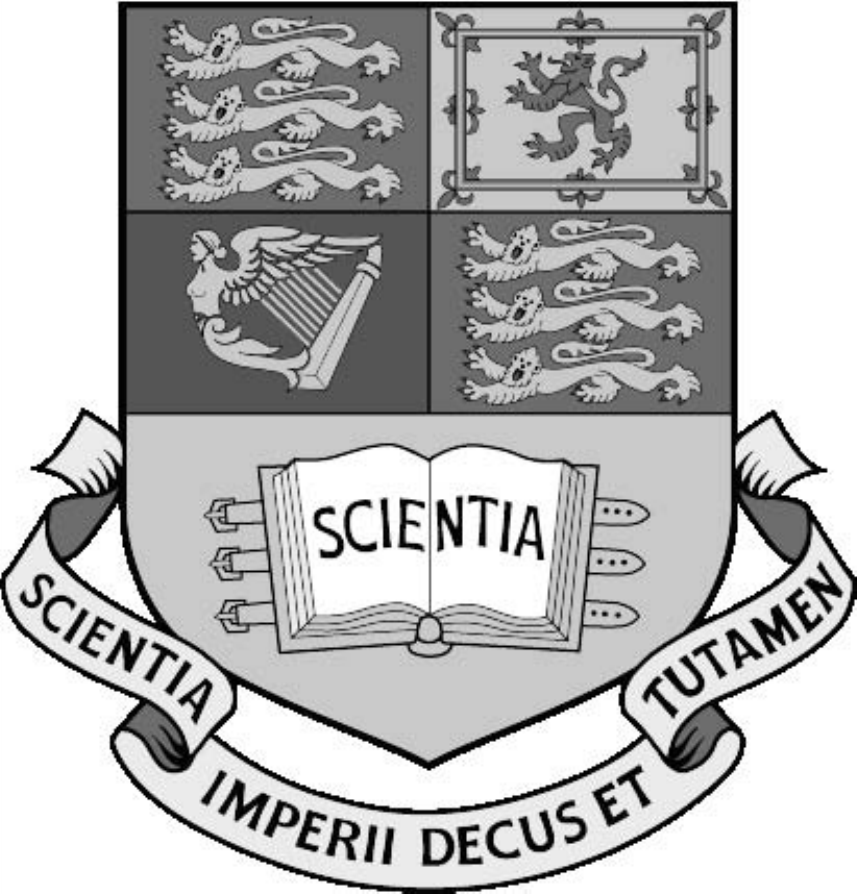}
\medskip
\,\\
}

\date{\normalsize Department of Mathematics\\\normalsize
Imperial College London\\$\,$\\
\bigskip$\,$\\$\,$\\
\bigskip$\,$\\
\medskip$\,$\\
\large April 2008
}
\maketitle


\thispagestyle{plain}
\setcounter{page}{2}
$\,$\\\bigskip$\,$\\$\,$

\bigskip$\,$\\
\begin{flushright}\small\it
``La nature est un temple o\`u  de vivants pilliers$\quad\,\,$\\  
Laissent parfois sortir de confuses paroles;$\qquad\quad\!\!$\\  
L'homme y passe \`a  travers des for\^ets de symboles$\!\!\!$\\  
Qui l'observent avec des regards familiers.\!'' $\qquad$\\ \scriptsize$\,\,$ \\
{\rm\footnotesize (C. Baudelaire, {\it Correspondences}, Les fleurs du mal)\,}

\rem{ 
\vspace{1.2cm}\small\it
``Nature is a temple in which living pillars$\qquad\,\,\qquad$\\  
Sometimes emit confused words;$\qquad\qquad\qquad\qquad$\\  
Man crosses it through forests of symbols$\qquad\,\qquad$\\  
That observe him with familiar glances.'' $\quad\quad\qquad$
} 
\end{flushright}

\newpage
\thispagestyle{plain}
\subsection*{\center Abstract\\}

$\,$\\
The Vlasov equation of kinetic theory is introduced and the Hamiltonian structure of its moments is presented. Then we focus on the geodesic evolution of the Vlasov moments. As a first step, these moment equations generalize the Camassa-Holm equation to its multi-component version. Subsequently, adding electrostatic forces to the geodesic moment equations relates them to the Benney equations and to the equations for beam dynamics in particle accelerators.

Next, we develop a kinetic theory for self assembly in nano-particles. Darcy's law is introduced as a general principle for aggregation dynamics in friction dominated systems (at different scales). Then, a kinetic equation is introduced for the dissipative motion of isotropic nano-particles. The zeroth-moment dynamics of this equation recovers the classical Darcy's law at the macroscopic level. A kinetic-theory description for oriented nano-particles is also presented. At the macroscopic level, the zeroth moments of this kinetic equation recover the magnetization dynamics of the Landau-Lifshitz-Gilbert equation. The moment equations exhibit the spontaneous emergence of singular solutions (clumpons) that finally merge in one singularity. This behaviour represents aggregation and alignment of oriented nano-particles.

Finally, the Smoluchowski description is derived from the dissipative Vlasov equation for anisotropic interactions. Various levels of approximate Smoluchowsky descriptions are proposed as special cases of the general treatment. As a result, the macroscopic momentum emerges as an additional dynamical variable that in general cannot be neglected.

\vspace{4cm}
\begin{framed}{\scriptsize\noindent
I declare that the material presented in this thesis is my own work and any material which is not my own has been acknowledged.

\noindent
\scriptsize Signed: {\it Cesare Tronci}
$\hspace{3cm}$
\scriptsize Date: April 2008}
\end{framed}


\tableofcontents
\listoffigures

\chapter*{Preface}
\addcontentsline{toc}{chapter}{Preface} 
\markright{PREFACE}

This work is the fruit of my research over the last three years, during
my postgraduate studies at Imperial College London. Besides the fundamental
 guide of my supervisor Darryl Holm, the collaboration with John Gibbons and Vakhtang Putkaradze has also been determinant.

\smallskip
The scientific matter of this work is the geometric structure of the Vlasov
equation in kinetic theory and the passage from this microscopic description
to the macroscopic fluid treatment, given by the dynamics of kinetic moments.
Vlasov moments are very well known since the early twentieth century, when
Chapmann and Enskog formulated their closure of the Boltzmann equation \cite{Chapman1960}. The power of the moment approach leaded to the important theory of fluid mechanics and its kinetic justifications in physics. 

\smallskip
In the collisionless Vlasov limit,
the moment hierarchy turns out to conserve a purely geometric structure inherited
by the Vlasov Lie-Poisson bracket. The geometric structure of moment dynamics
is known since the late 70's \cite{KuMa1978,Le1979}
and was found surprisingly in a very different context from kinetic theory, that is the analysis of integrable shallow water equations. The relation
with kinetic theory was found few years later \cite{Gi1981}, but the geometric
properties of moment dynamics were not explored further. Even the fluid closure has always been considered in terms of cold plasma solution of the Vlasov equation, without considering the mathematical property that this solution is equivalent to a {\it truncation} of the moment hierarchy to the first two moments. This property is apparently trivial, although this work
shows that this is crucial in some contexts involving dissipative dynamics,
where the cold plasma solution is not of much use.

\smallskip
This work takes inspiration from the idea that the geometric properties of
moment dynamics deserve further investigation. The topics covered in this
thesis analyze the geometric properties of both Hamiltonian and dissipative
flows. The first part is devoted to exploring the geodesic motion on the
moments and the second part formulates the double bracket equations for dissipative
moment dynamics. The main result is the formulation of a model for the aggregation
of oriented particles, with possible applications in nano-sciences.

\medskip
\subsection*{Plan of the work}
The thesis proceeds in the following order. The first chapter reviews
some background and formulates the motivations by focusing on singular solutions
in continuum theories. 

The second chapter analyzes the geometric structure
of moment dynamics. It contains one main result, that is the identification
of the moment Lie bracket with the symmetric Schouten bracket on symmetric
tensors, which is {\it different} from the Kupershmidt-Manin bracket in multi-index
notation \cite{GiHoTr2008}. 

The third chapter concerns the study of geodesic motion on the
moments: it is explained how this is equivalent to the geodesic motion on
canonical transformations (EPSymp) and this fact determines the existence
of singular solutions, which may reduce to the single-particle dynamics.
At the end of chapter~\ref{EPSymp}, the geodesic motion on the moments is
extended to include anisotropic interactions and this constitutes an introduction
to the topics covered in the last chapter.

The fourth chapter formulates the geometric dissipative dynamics for geometric
order parameters (GOP). This analysis takes inspiration from the geometric structure of Darcy's law \cite{HoPu2005,HoPu2006,HoPu2007} and formulates a geometric dissipation that extends Darcy's law to any tensor quantity, instead of only densities. The behavior of singular solutions is analyzed extensively.
Moreover the application of this framework to the case of the fluid
vorticity leads to the fact that this form of dissipative dynamics embodies
to the double bracket approach, which was established in the early 90's \cite{BlKrMaRa1996}.

The fifth chapter applies the geometric dissipation to the case of the Vlasov
equation and to the Vlasov kinetic moments. The main result of this
section is that Darcy's law follows very naturally as the zero-th moment equation
of the dissipative moment hierarchy. The dissipative moment dynamics is also
applied to formulate appropriate equations such as the dissipative
fluid equations, the $b$-equation and the moment GOP equation, each allowing singular solutions.

The sixth chapter extends the previous dissipative treatment to kinetic theory
for anisotropic interactions. The distribution function now depends on the
orientation of the single (nano)-particle and the moment hierarchy is again
obtained. The analogue of Darcy's law for this case yields two equations, one
for the mass density and the other for the polarization, recovering the Landau-Lifshitz-Gilbert
dissipation term for the magnetization in ferromagnetic media \cite{Gilbert1955}. It is important
to notice that the fluid closure of the dissipative moment hierarchy is
{\it not} obtained through the cold plasma solution of the Vlasov equation,
rather it is obtained by a pure truncation of the moment hierarchy to the
first two moments. This constitutes a good confirmation for the high importance
of moment dynamics in deriving macroscopic continuum models from kinetic
treatments. Further study is devoted to the Smoluchowski approach and it
is shown how this approach presents interesting truncations and specializations,
despite the complicated equations arising from the whole hierarchy.


\medskip
\subsection*{Acknowledgements}

My greatest acknowledgments are addressed to my advisor Darryl Holm for trusting
my capabilities in this field, despite my previous experience in very applied engineering problems. I would also like to thank him for guiding me through the difficult world of scientific research and to let me appreciate more
and more the beauty of geometric models in physics. Thanks, Darryl!

Also I am indebted with Vakhtang Putkaradze at the Colorado State University for his great ideas and for his excellent advices over the last year. My work with him has been a great experience, which I hope will continue in the next years. Moreover I would like to thank John Gibbons at Imperial College London for his expert advice, especially on moment dynamics; this work would have probably never been possible without his expertise in the field. 

Particular acknowledgements also go to Ugo Amaldi at CERN, who first recognized
my mathematical taste and supported my idea of moving to this field. He is the person who helped me most at the beginning of my scientific itinerary and I am enormously indebted to him.  

In addition I feel need to thank Bruce Carlsten, Paul Channell,
Rickey Fahel and Giovanni Lapenta at the Los Alamos National Laboratory for
many helpful discussions and encouragements. 

Finally I would like to thank also my colleagues Matthew Dixon, for his important
advice, and Andrea Raimondo for his keen observations on moment
dynamics.

\chapter{Outline: motivations, results and perspectives}
\label{zeroth}
\markright{OUTLINE: MOTIVATIONS, RESULTS AND PERSPECTIVES}
\section{Mathematical background of kinetic theory}

The importance of kinetic equations in non-equilibrium statistical mechanics
is well known and finds its roots in the pioneering work of Maxwell \cite{Ma1873}
and Boltzmann \cite{Bo95}. The mathematical foundations of kinetic theory reside in {\bfi Liouville's theorem}, stating that no matter how large the number of particles is in a system, they undergo canonical transformations
which preserve the volume element in the global phase space of the system.
More mathematically, one defines a density variable $\rho(q_i,p_i,t)$ (with
$i=1,\dots,N$) for the $N$-particle system. Then one writes the Liouville equation as a characteristic equation on phase space
\[
\frac{d}{dt}\,\rho_t=0
\qquad\text{along}\qquad
\dot{q}_i=\frac{\pa H}{\pa p_i}\,,
\quad
\dot{p}_i=-\frac{\pa H}{\pa q_i}
\]
where $H$ is the $N$-particle Hamiltonian.
\rem{ 
This vector field is related to the $N$-particle Hamiltonian $H$
\[
H(q_i,p_i)=\sum_{i=1}^N \frac12 |p_i|^2 + \sum_{i\neq j} V(q_i-q_j)
\]
where $V$ denotes the potential energy as usual (more general forms of $H$
are also allowed). Once the Hamiltonian is known, one constructs the vector
field $\bf X$ as ${\bf X}=-\mathbb{J}\nabla H=:{\bf X}_H$, which is called
``Hamiltonian vector field'', since it is uniquely defined by $H$. In particular
one has that div$\left(\rho\,{\bf X}_H\right)=\left\{\rho,H\right\}$, for any density
variable $\rho$.
} 
In Eulerian coordinates one has the
\begin{theorem}[Liouville's equation] Given a phase space density $\rho$
for the $N$-particle distribution, its evolution is given by the conservation
equation
\[
\frac{\pa \rho}{\pa t}+\left\{\rho,H\right\}=0
\]
so that the following volume form is preserved
\[
\rho_0(q_i^{(0)},p_i^{(0)})\,{\rm d}\Omega_0=
\rho_t(q_i^{(t)},p_i^{(t)})\,{\rm d}\Omega_t
\]
where ${\rm d}\Omega$ is the infinitesimal volume element in phase space.
\end{theorem}

In the search for approximate descriptions of this system, one may think to deal with global quantities that integrate out the
information on some of the particles. In particular one defines a $n$-particle
distribution as
\[
f_n(z_1,\dots,z_n,t):=\int \rho(z_1,\dots,z_N)\,{\rm d}z_{n+1}\dots{\rm d}z_N
\]
where the notation $z_i=(q_i,p_i)$ has been introduced for compactness of
notation. These quantities are called ``BBGKY moments'' and their equations
constitute an infinite hierarchy of equations known as {\bfi BBGKY hierarchy}
\cite{MaMoWe1984}, or ``Bogoliubov-Born-Green-Kirkwood-Yvon equations''.
This hierarchy is rather complicated, although Marsden, Morrison and Weinstein
\cite{MaMoWe1984} have shown that it possesses a clear geometric
structure in terms of canonical transformations that are symmetric with respect
to their arguments. In particular, this hierarchy is a
Lie-Poisson system, i.e. a Hamiltonian system on a Lie group, as explained
in chapter~\ref{intro}.

Suitable approximations on the equation for the single particle distribution $f:=f_1$ lead to the {\bfi Boltzmann equation}. For the purposes of this work it suffices to write this equation schematically as
\[
\frac{\pa f}{\pa t}+\left\{f,H\right\}=\left(\frac{\pa f}{\pa t}\right)_{coll}
\]
where $H$ is now the $1$-particle Hamiltonian $H=p^2/2+V(q)$. The right hand
side collects the information on pairwise collisions among particles and its explicit expression requires a discussion that is out of the purposes
of this work. Rather it is important to discuss an important approximation
of the Boltzmann equation, the {\bfi Fokker-Planck equation} \cite{Fokker-Plank1931,Ri89}.
Indeed, the hypothesis of stochastic dynamics in terms of Brownian motion
leads to the following fundamental equation
\[
\frac{\pa f}{\pa t}+\left\{f,H\right\}=\gamma\frac{\pa}{\pa p}\left( p f+\beta^{-1}\,\frac{\pa
f}{\pa p}\right)
\] 
where a {\bfi dissipative} drift-diffusion term is evidently substituted to the collision term
of the Boltzmann equation. This term is peculiar of the microscopic stochastic
dynamics expressed by the Langevin equation $\dot{p}=-H_q-\gamma p+\sqrt{2\gamma\beta^{-1}}_{\,}\dot{w}(t)$
for the single particle momentum ($\dot{w}$ is a white noise process). This equation is the most common equation
in kinetic theory and it is probably the most used in physical applications.
 
In many contexts it is possible to neglect the effects of collisions. Such
contexts range from astrophysical topics (cf. e.g. \cite{Ka1991}) to particle beam dynamics (cf. e.g. \cite{Venturini}), which is the very first inspiration for this work, given some previous experience of the author in the field of particle accelerators. In more generality, the hypothesis of negligible
collisions is most commonly used in the physics of plasmas (electrostatic or magnetized). In the case of collisionless dynamics, the resulting
equation
\[
\frac{\pa f}{\pa t}+\left\{f,H\right\}=0
\]
is called {\bfi Vlasov equation} \cite{Vl1961} and its underlying mathematical
structure has been widely investigated over the past decades, especially
in terms of geometric arguments \cite{WeMo,MaWe81,Ma82,MaWeRaScSp,CeHoHoMa1998}.
In particular, Marsden, Weinstein and collaborators \cite{WeMo,MaWe81,Ma82,MaWeRaScSp} have shown that this equation possesses a Lie-Poisson structure on the whole group of canonical transformations. The explicit expression of the Lie-Poisson
bracket is 
\[
\{F,G\}[f]\,=\iint \!f(q,p,t)\left\{\frac{\delta F}{\delta f},\frac{\delta
G}{\delta f}\right\}{\rm d}q\,{\rm d}p
\]
where the Lie bracket $\{\cdot,\cdot\}$ is now the canonical Poisson bracket.
  Even when this equation is coupled with the Maxwell equations and particles are acted on by an electromagnetic field ({\it Maxwell-Vlasov system}), the geometric structure persists \cite{WeMo,MaWe81,MaWeRaScSp}.
This particular result is also due to Cendra and Holm, who showed in their joint work with Hoyle and Marsden \cite{CeHoHoMa1998} how the Maxwell-Vlasov equation has also a Lagrangian formulation. This Lagrangian approach was
first pioneered by Low in the late 50's \cite{Lo58}.

As a Lie-Poisson system, the Vlasov equation possesses the property of being
a kind of {\bfi coadjoint motion} \cite{MaRa99}, so that its evolution map coincides with
the {\it coadjoint group action}
\[
f_t={\rm Ad}^*_{g_t^{-1}}\,f_0
\qquad\text{with}\quad
g_t\in G
\]
as explained in chapter~\ref{intro}. This means that the dynamics is purely
geometric and it is uniquely determined by the canonical nature of particle dynamics. 

A particular kind of Vlasov equation has been proposed by Gibbons, Holm and
Kupershmidt (GHK) \cite{GiHoKu1982,GiHoKu1983} in order to formulate a kinetic
theory for particles immersed in a Yang-Mills field. Without going into the
details, one can refer to it as a collisionless kinetic equation that takes
into account for an extra-degree of freedom of the single particle. In the
case of \cite{GiHoKu1982,GiHoKu1983}, this would be a color charge associated
with chromodynamics. However for the present purposes, this can also be represented
by a spin-like variable which is carried by each particle in the system.
In more generality this equation can be considered as a kinetic equation
for particles with {\it anisotropic interactions}.
The {\bfi GHK-Vlasov equation} considers a distribution function
\[
f=f(q,p,\mu,t)
\quad\text{with}\quad
\mu\in\mathfrak{g}^*
\]
where $\mathfrak{g}^*$ is the dual of some Lie algebra $\mathfrak{g}$. The
equation is written as
\[
\frac{\pa f}{\pa t}+\left\{f,H\right\}+\left\langle\mu,\left[\frac{\pa f}{\pa
\mu},\,\frac{\pa H}{\pa
\mu}\right]\right\rangle=0
\]
where $[\cdot,\,\cdot]$ denotes the Lie bracket and $\langle\cdot,\,\cdot\rangle$
is the pairing. This equation will be determinant for the results presented
in chapter~\ref{orientation}, where a model for oriented nano-particles is
formulated.

Although the Vlasov equation enjoys many geometric properties (Lie-Poisson bracket, coadjoint motion, advection), these are not shared by the Fokker-Planck equation, whose geometric interpretation is far from the theory of symmetry
groups used in this work. Nevertheless, Kandrup \cite{Ka1991} and Bloch and
collaborators \cite{BlKrMaRa1996} have formulated a type of {\bfi dissipative}
Vlasov equation, which preserves the geometric nature of the Hamiltonian
flow while dissipating energy. This theory requires the concept of {\bfi
double bracket dissipation}, i.e. the dissipation is modelled by the subsequent
application of two Poisson brackets and the corresponding equation becomes
\[
\frac{\pa f}{\pa t}+\left\{f,H\right\}=\alpha\left\{f,\left\{f,H\right\}\right\}
\,.
\]
This equation represents an interesting possibility for introducing geometric
dissipation in kinetic equations, but it has never been considered further.
A deeper investigation of this equation is presented in chapter~\ref{DBVlasov}
and extended in chapter~\ref{orientation}.

In the case of isotropic interactions, the Vlasov density $f$ depends on seven variables: six
phase space coordinates plus time. This indicates that the Vlasov equation
is still a rather complicated equation even when numerical efforts are involved.
Thus it is often convenient to find suitable approximations in order to discard
unnecessary information while keeping the main feature of collisionless multi
particle dynamics. To this purpose, one introduces the {\it moments} of the
Vlasov distribution.

\section{Geometry of Vlasov moments: state of the art}

The use of moments in kinetic theory was introduced by Chapmann and Enskog
\cite{Chapman1960}, who formulated their closure of the Boltzmann equation
yielding the equations of fluid mechanics and its kinetic justifications in physics. This result showed how the use of moments is a powerful tool
for obtaining consistent reductions or approximations of the microscopic
kinetic description. Since that time, the mathematical properties of moments have been widely investigated 

The geometric properties of Vlasov moments mainly arose in two very different
contexts, particle beam dynamics and shallow water equations. However it
is important to distinguish between two different classes of moments: {\it
statistical moments} and {\it kinetic moments}. {\bfi Statistical moments}
are defined as
\[
g_{n,\widehat{n}}(t)=\int p^n q^{\widehat{n}} f(q,p,t)\,{\rm d}q\,{\rm d}p
\,.
\]
These quantities first arose in the study of particle beam dynamics \cite{Ch83,Ch90,LyOv88}
from the observation that the beam {\it emittance} $\epsilon:=\left(g_{0,2}\,g_{2,0}-g_{1,1}^{\,2}\right)^{1/2}$
is a laboratory parameter, which is also an invariant function of the statistical
moments. In particular, Channell, Holm, Lysenko and Scovel \cite{Ch90,HoLySc1990,LyPa97}
were the first to consider the Lie-Poisson structure of the moments, whose
explicit expression is given in chapter~\ref{intro} as
\[
\{\,F\,,\,G\,\}
\quad=
\sum_{\widehat{m},m,\widehat{n},n=0}^\infty\,
\frac{\pa F}{\pa g_{\widehat{m},m}}\,
\big(\widehat{m}\,m-\widehat{n}\,n\big)\,
\frac{\pa G}{\pa g_{\widehat{n},n}}\,\,
g_{\,\widehat{m}+\widehat{n}-1,\,m+n-1}
\,.
\]
This
geometric framework allowed the systematic construction of symplectic moment invariants
in \cite{HoLySc1990}, a question that was also pursued by Dragt and collaborators
in \cite{DrNeRa92}. Special truncations and approximations of the equations
for statistical moments have been studied also by Scovel and Weinstein in
\cite{ScWe} in 1994. Besides applications in particle beam physics, the use
of statistical moments has also been proposed in astrophysical problems by
Channell in \cite{Ch95}.

Besides statistical moments, another kind of moments were known to be a powerful
tool in kinetic theory, since they had been used by Chapman and Enskog to recover
fluid dynamics from the Boltzmann equation. These are the {\bfi kinetic moments}
\[
A_n(q,t)=\int p^n f(q,p,t)\,{\rm d}p
\,.
\]
and the following discussion will refer to these quantities as simply ``moments'',
unless otherwise specified. The geometric properties of these moments first arose in 1981 \cite{Gi1981}, when Gibbons recognized that these Vlasov moments are equivalent to the variables introduced by Benney in 1973 \cite{Be1973}, in the context of shallow water
waves. The Hamiltonian structure of these variables was found by Kupershmidt
and Manin \cite{KuMa1978}; later Gibbons recognized how this structure is inherited from the Vlasov Lie-Poisson bracket \cite{Gi1981}. The relation
between moments and the algebra of generating functions was also known to
Lebedev \cite{Le1979}, although he did not recognize the connection with
Vlasov dynamics. The Lie-Poisson
structure for the moments is also called {\bfi Kupershmidt-Manin structure}
and is explicitly written as \cite{KuMa1978}
\[
\{F,G\}=\int\! A_{m+n-1}\left(n\,\frac{\delta F}{\delta A_n}\frac{\pa}{\pa q}\frac{\delta G}{\delta A_m}
-
m\,\frac{\delta G}{\delta A_m}\frac{\pa}{\pa q}\frac{\delta F}{\delta A_n}
\right){\rm d}q
\]
whose derivation will be presented in chapter~\ref{momLPdyn}. The main theorem
regarding moments is thus the following
\begin{theorem}[Gibbons \cite{Gi1981}] The process of taking moments of the
Vlasov distribution is a {\bfi Poisson map}, that is it takes the Vlasov
Lie-Poisson structure to another Lie-Poisson structure, which is given by
the Kupershmidt-Manin bracket.
\end{theorem}
\begin{remark}
It is important to notice that, although the Lie-Poisson moment bracket is well known, the coadjoint group action is not fully understood and this represents
an important open question concerning the geometric dynamics of Vlasov moments.
\end{remark}

Besides their role in the theory of Benney long waves \cite{Be1973}, the geometric structure of the moments has not been considered as a whole so far. Even in that context, the use of the Vlasov equation turns out to be more convenient. Rather the fluid closure of moment dynamics is very well
understood and is given by considering only the first two moments $A_0$ and $A_1$, which coincide with the fluid density and momentum respectively. The
key to understanding the geometric characterization of this closure is to
consider the {\bfi cold plasma solution}, i.e. a singular Vlasov solution of the form
\[
f(q,p,t)=\rho(q,t)\,\delta(p-P(q,t))
\,.
\]
Substituting this expression into the Vlasov equation yields the equations
for $\rho=A_0$ and $P=A_1/\rho$. Marsden, Ratiu and collaborators \cite{MaWeRaScSp}
showed how this solution is a momentum map (cf. e.g. \cite{MaRa99}), which
is called {\bfi plasma-to-fluid map}. This important property has been widely
used to formulate hydrodynamic models from kinetic theory \cite{MaWeRaScSp} and it has been extended to account for Yang-Mills fields in the work of Gibbons, Holm and Kupershmidt \cite{GiHoKu1982,GiHoKu1983}. However these
hydrodynamical models have usually been derived directly from the Vlasov
equation by direct substitution of the cold plasma solution, rather than considering
the moment hierarchy in its own. The two approaches are clearly equivalent
 and this apparently trivial point becomes a key fact in some contexts where the cold plasma is not of much use. An example is provided in chapters~\ref{DBVlasov}
and \ref{orientation}, where the substitution of the cold plasma solution
is evidently avoided as it yields to cumbersome calculations and results
that are not completely clear.

\section{Motivations for the present work}

As mentioned above, the topic of Vlasov moments is first dictated by the
previous scientific experience of the author with particle accelerators.
In particular, beam dynamics issues assume a central role in many questions
of accelerator design, especially for high beam currents ($\sim$ 1--100mA), and the Vlasov approach is a natural step in this matter. The theory of Vlasov statistical moments arose in this environment. However, although the theory of Vlasov statistical moments is completely understood \cite{HoLySc1990,ScWe}, this is not true for kinetic moments. For example, it is not known a priori what geometric nature these moments should have. Is there any chance that their geometric properties could be relevant to beam dynamics and plasma physics? These questions provide the first motivations for approaching the
topic of Vlasov kinetic moments.

Also, it is presented in chapter~\ref{EPSymp} how moment dynamics recovers the integrable {\bfi Camassa-Holm (CH) equation} \cite{CaHo1993} and thus it recovers its singular {\bfi peakon solutions}: one may wonder whether there is an explanation of the CH integrable dynamics in terms of moments. What would be a suitable formulation
of this problem? What is the relation in terms of {\bfi singular solutions}? The fact that the CH equation is recovered by moment dynamics is the main motivation
for seeking possible generalizations of this equation in terms of the moments.
The dynamics of kinetic moments has never been related with singular
solutions and blow--up phenomena in continuum PDE's and this constitutes
another motivation for pursuing this direction.

Moreover, Bloch and collaborators have shown how the {\bfi double bracket dissipation}$\,$ \cite{BlKrMaRa1996,BlBrCr1997} in kinetic theory recovers a form of dissipative Vlasov equation, which has been proposed in astrophysics by Kandrup \cite{Ka1991}. This does not recover the {\bfi single particle solution}. Why? How can this problem be solved? What is the corresponding
interpretation in terms of the moments? The main motivation for pursuing
this direction is that the double bracket dissipation provides an interesting
way of inserting dissipation in collisionless kinetic equations while preserving
the geometric structure of the Vlasov equation.

As it easy to see, there are many open questions that make the geometric
properties of the Vlasov equation and its moments an intriguing field
of research. The next section tries to classify these open questions and
explains what the contribution of this work is.

\section{Some open questions and results in this work}

One can try to classify the open questions in three types: purely geometric
questions, Hamiltonian flows on the moments and dissipative geometric flows.
At this point, one attempts to write a table as follows
\begin{itemize}

\medskip
\item {\bf Purely geometric questions}
\begin{itemize}
\item Is there a {\bfi geometric characterization} of moments? What kind of geometric
quantities are they? Vector fields? differential forms? generic tensors?

\item The BBGKY moments and the statistical moments are well understood as
momentum maps \cite{MaMoWe1984,HoLySc1990}: are kinetic moments momentum maps too? If so, what is the
{\bfi underlying symmetry group}?

\item Statistical moments possess a whole family of invariant functions \cite{HoLySc1990}:
what are the {\bfi moment invariants} for kinetic moments?

\item The {\bfi Euler-Poincar\'e equations} are the Lagrangian counterpart of a Lie-Poisson system \cite{MaRa99}: what are the Euler-Poincar\'e equations for
the moments?
\end{itemize}

\medskip
\item {\bf Hamiltonian flows on the moments}
\begin{itemize}
\item How does the theory of moment dynamics apply to physical problems,
e.g. {\bfi beam dynamics}?

\item Moment dynamics recovers the {\bfi Camassa-Holm equation} \cite{CaHo1993} from the evolution of the first-order moment: why does this happen?

\item Quadratic terms in the moments often appears in applications: what
are the properties of {\bfi purely quadratic Hamiltonians}?

\item Quadratic Hamiltonians define {\bfi geodesic motion} on the moments:
what is its geometric interpretation in terms of Vlasov dynamics?

\item These systems may allow for singular solutions: what kind of solutions
are they? how are they related with the CH peakons?

\item The CH equation is an integrable equation: does geodesic moment dynamics recover other {\bfi integrable cases}?

\end{itemize}

\medskip
\item {\bf Geometric dissipative flows}

\begin{itemize}
\item Is it possible to extend the double bracket dissipation \cite{BlKrMaRa1996} in the Vlasov
equation to allow for the {\bfi single particle solution}?

\item How does the {\bfi double bracket structure} apply to moment dynamics?

\item What kind of macroscopic {\bfi moment equations} arise in this context? what is their meaning?

\item How does the {\bfi GHK-Vlasov equation} \cite{GiHoKu1982,GiHoKu1983} transfer to double bracket dynamics? what is the corresponding moment dynamics?

\item What do {\bfi singular solutions} represent in this case? How do they
interact? what happens in three dimensions?

\item {\bfi Smoluchowski moments} depend on both position and orientation: what are their equations as they arise from double bracket dynamics?

\end{itemize}

\end{itemize}

Analogously, this section illustrates the accomplishments of this work by
following the same scheme.
\begin{itemize}

        \medskip
        \item {\bf Results on the moment bracket}
        \begin{itemize}
        
        \item Chapter~\ref{momLPdyn} shows how the moments have possess a
        deep geometric interpretation in terms of {\bfi symmetric covariant
        tensors} \cite{GiHoTr2007}
        
        \item The moment Lie bracket has been identified with the {\bfi Schouten
        symmetric bracket} on symmetric contravariant tensors \cite{GiHoTr2008},
        as explained
        in chapter~\ref{momLPdyn}.
        
        \item Chapter~\ref{momLPdyn} derives the {\bfi Euler-Poincar\'e equations}
        for the moments and chapter~\ref{EPSymp} illustrates some integrable
        examples \cite{GiHoTr05,GiHoTr2007}
        
        \end{itemize}

        \medskip
 \item {\bf Results on Hamiltonian flows}
        \begin{itemize}
        
        \item Chapter~\ref{EPSymp} shows how the Benney moment equations
        \cite{Be1973} regulate the dynamics of {\bfi coasting beams} in particle
        accelerators \cite{Venturini} and this fact \cite{GiHoTr2007} determines
        the nature of the {\bfi coherent structures} observed in the experiments
        \cite{KoHaLi2001,CoDaHoMa04}.
        
        \item Chapter~\ref{momLPdyn} presents how the Camassa-Holm equation
        \cite{CaHo1993} appears from the restriction of moment dynamics to
        {\bfi cotangent lifts of diffeomorphisms}. This type of flow also
        provides an interpretation of the {\bfi $b$-equation} \cite{HoSt03}
        in terms of moment dynamics \cite{GiHoTr2007}.
        
        \item The {\bfi geodesic flow} on the moments has been formulated
        as a new problem in chapter~\ref{EPSymp}. It has been shown how this
        is equivalent to a {\bfi geodesic Vlasov equation}, that is a {\bfi
        geodesic motion on the symplectic group} of canonical transformations
        \cite{GiHoTr05,GiHoTr2007}.
        
        \item Chapter~\ref{EPSymp} also shows how the CH peakons may be interpreted
        in terms of {\bfi singular solutions for the moments}, i.e. the single
        particle solution \cite{GiHoTr05,GiHoTr2007}
        
        \item The {\bfi two-component CH equation} \cite{ChLiZh2005,Ku2007}
        has
        been shown to emerge as a particular specialization of the geodesic
        moment equations \cite{GiHoTr2007}
        
        \item The geodesic moment equations have been extended to include
        {\bfi anisotropic interactions} \cite{GiHoTr2007}
        
        \end{itemize}

        \medskip
\item {\bf Results on dissipative flows}
        \begin{itemize}
        
        \item Chapter~\ref{GOP} shows how the existence of singular solutions
        can be allowed for a whole class of dissipative equations, called
        {\bfi GOP equations} \cite{HoPu2007}. This is applied 
        to recover the double bracket form of the {\bfi vorticity equation}
        \cite{HoPuTr2007} in chapter~\ref{GOP} and of the {\bfi Vlasov equation}
        \cite{HoPuTr2007-CR} in chapter~\ref{DBVlasov}.
        
        \item Chapter~\ref{DBVlasov} applies the {\bfi double bracket dissipation}
        to formulate {\bfi dissipative equations for the moments} \cite{HoPuTr2007-CR,HoPuTr2007-Poisson},
        whose zero-th order truncation recovers {\bfi Darcy's law} for porous
        media
        
        \item Chapter~\ref{orientation} applies the double bracket dissipation
        to the GHK--Vlasov equation$\,$ \cite{GiHoKu1982,GiHoKu1983} and to moment
        dynamics. The zero-th order truncation constitutes a {\bfi generalization
        of Darcy's law to anisotropic interactions}, recovering Landau-Lifshitz-Gilbert
        dynamics for magnetization in ferromagnetic media \cite{HoPuTr2007-Poisson,HoPuTr08,HoOnTr07}.
        
        \item Chapter~\ref{orientation} explains how this extension of Darcy's
        law admits {\bfi singular solutions} ({\it orientons}) and presents
        analytical results on their behavior \cite{HoPuTr08,HoOnTr07}.
        
        \item {\bfi Smoluchowski moment dynamics} is also derived in chapter~\ref{orientation}
        and particular specializations are presented \cite{HoPuTr2007-Poisson}
        
        \end{itemize}

\end{itemize}
There are two main mathematical ideas behind these results. The first is
that {\bfi taking the moments is a Poisson map} \cite{Gi1981}: this allows to transfer
from the microscopic kinetic side to the macroscopic continuum level. In
particular, this idea is of central importance when deriving fluid--like models
from kinetic equations. The clear example is given by the formulation of
the double bracket for the moments: the dissipative moment dynamics need
not to be determined by direct integration of the Vlasov equation, but rather
they can be constructed by following purely geometric arguments in the theory
of double bracket dissipation.

The second key idea is that continuum models may allow for {\bfi singular solutions}. In the present theory of double bracket, these singular solutions
are not allowed and it is not clear a priori how a smoothing process can
be inserted in order to admit the singularities. The inspiration for the
solution of this problem comes from the {\bfi GOP theory} of Holm and Putkaradze
\cite{HoPu2007}, which derives a class of dissipative equations through a
suitable variational principle. Chapter~\ref{GOP} shows that the way the smoothing process enters in this variational principle determines whether singular solutions exist in the GOP family of equations \cite{HoPu2007,HoPuTr2007}, which also include double bracket equations.

\section{A new model for oriented nano-particles}

The main result of this work is presented in chapter~\ref{orientation}. This result is the formulation of a continuum model that generalizes Darcy's law to oriented nano-particles, starting from first principles
in kinetic theory. The starting point is the double bracket for of the GHK-Vlasov
equation \cite{GiHoKu1982,GiHoKu1983}
\[
\frac{\partial f}{\partial t}=\left\{f,\left\{\mu[f],\frac{\delta E}{\delta f}\right\}_{\!1}\,\right\}_{\!1}
\qquad\text{ with }\qquad
\Big\{f,h\Big\}_{\!1}\,:=\,
\Big\{f,h\Big\}\,+\,\boldsymbol{m}\cdot\,\frac{\partial f}{\partial \boldsymbol{m}}\times\frac{\partial
h }{\partial \boldsymbol{m}}
\]
where $E$ is the energy functional and $\mu[f]$ is a smoothed copy of $f$, i.e. a convolution $\mu[f]=K*f$
with some kernel $K$ \cite{HoPuTr2007-Poisson}.
Once this equation is introduced, one proceeds by considering the leading-order moments
\[
\left(\rho,{\bf M}\right)=\int \left(1,{\boldsymbol{m}}\right)\,f({\bf q,p},\boldsymbol{m},t)\,\,{\rm d}^3{\bf p}\,\,{\rm d}^3\boldsymbol{m}
\]
so that $\rho({\bf q},t)$ is the mass density and ${\bf M}({\bf q},t)$ is the {\it polarization}.
At this point, it suffices to calculate the dissipative equations for $\rho$
and ${\bf M}$, which turn out to be \cite{HoPuTr08}
\begin{align*}
\frac{\partial \rho}{\partial t}
&=
\text{\rm\large div}
\Bigg(
\rho\,
\bigg(
\mu_\rho\, \text{\large$\nabla$}\frac{\delta E}{\delta \rho}
+\,
\boldsymbol\mu_{\bf M}\cdot\text{\large$\nabla$}\frac{\delta E}{\delta \bf M}
\bigg)
\,
\Bigg)\\
\frac{\partial {\bf M}\,}{\partial t}
&=
\text{\rm\large div}\Bigg({\bf M}\,\text{\large$\otimes$}
\left(\mu_\rho\text{\large$\nabla$}\frac{\delta E}{\delta \rho}
+
\boldsymbol\mu_{\bf M}
\cdot\text{\large$\nabla$}\frac{\delta E}{\delta {\bf M}}
\right)
 \Bigg)\!
+
{\bf M}\times
\boldsymbol{\mu}_{\bf M}\times\frac{\delta E}{\delta \bf M}
\end{align*}
where the last term in the second equation is the dissipative term for {\bfi
magnetization
dynamics in ferromagnetics}. Thus the Landau-Lifshitz-Gilbert dissipation
\cite{Gilbert1955} is derived from {\bfi first principles in kinetic theory} and this model can also be applied to systems of ferromagnetic particles.

This model allows for singular solutions of the form \cite{HoPuTr08}
\begin{align*}
\rho({\bf q},t)&=\sum_{i=1}^N\int\delta({\bf q}-{\bf Q}_i(s,t))\,{\rm d}s\\
{\bf M}({\bf q},t)&=\sum_{i=1}^N\int{\boldsymbol w}_{_\text{\!\tiny$\bf M$},i}(s,t)\,\,\delta({\bf
q}-{\bf Q}_i(s,t))\,{\rm d}s
\end{align*}
where $s$ is a coordinate on a submanifold of $\mathbb{R}^3$: if $s$ is a
one-dimensional coordinate, then one gets an {\it orientation filament}, while in two dimensions one has an {\it orientation sheet}.

When the problem is studied in only one spatial dimension, then the singular
solutions take the simpler form \cite{HoPuTr2007-Poisson}
\begin{align*}
\rho(q,t)&=\sum_{i=1}^N \,w_{\rho,i}(t)\,\,\delta(q-Q_i(t))\\
{\bf M}(q,t)&=\sum_{i=1}^N \,{\boldsymbol w}_{_\text{\!\tiny$\bf M$},i}(t)\,\,\delta(q-Q_i(t))
\end{align*}
and $w_\rho$, ${\boldsymbol w}_{_\text{\!\tiny$\bf M$}}$ and $Q$ undergo the following
dynamics
\begin{align*}
\dot{w}_{\rho,i}
&
=0\,,
\qquad
\dot{\boldsymbol w}_{{\bf M},i}={\boldsymbol w}_{{\bf M},i}\times\left(\boldsymbol\mu_{\bf M}\times\frac{\delta E}{\delta \bf M}\right)_{q=Q_i}
\qquad
\dot{Q}_i
=
-\left(\mu_\rho\,\, \frac{\partial}{\partial q}\frac{\delta E}{\delta \rho}
+
\boldsymbol\mu_{\bf
M}\cdot \frac{\partial}{\partial q}\frac{\delta E}{\delta {\bf
M}}\right)_{q=Q_i}
\end{align*}
so that these singular solutions represent the dynamics of $N$ particles.
Numerical simulations show that these solutions may form {\bfi spontaneously from any initial configuration} \cite{HoOnTr07}. The study of pairwise interactions  in chapter~\ref{orientation} shows that there is a wide class of possible situations where these particles exhibit clumping and alignment phenomena \cite{HoOnTr07}.

\section{Perspectives for future work}

Besides its achievements, the present study raises many important
questions concerning various topics, from purely geometric matters to singularities
in double bracket equations.

For example, the result that moment dynamics is determined by the symmetric
Schouten bracket could be used to identify the symmetry group determining
the moment Lie-Poisson structure. This would allow to define moments as momentum
maps \cite{MaRa99}.

\smallskip

The study of the geodesic moment equations generates several open questions.
Chapter~\ref{EPSymp} shows how this hierarchy recovers two important integrable
equations, the CH equation \cite{CaHo1993} and its two-component version
\cite{ChLiZh2005,Ku2007}.
Thus one may wonder if there exist other truncations of the moment hierarchy
with remarkable behavior, such as integrability. The geodesic Vlasov equation
presented in chapter~\ref{EPSymp} is very similar in construction to the
Bloch-Iserles system \cite{BlIsMaRa05} (geodesic flow on Hamiltonian matrices) and it would be interesting to explore this connection further. Also, the dynamics of singular solutions still deserves further investigation, especially in higher dimensions (filaments and sheets). In the case of the CH equation \emph{dual pairs} \cite{MaWe83} emerge in the analysis of singular solutions \cite{HoMa2004}: is this possible for the two-component CH equation? and for other truncations of the moment equations? Similar questions concern the singular solutions of the geodesic moment equations for anisotropic interactions \cite{GiHoTr2007}.

\smallskip

The same questions regarding singular solutions and their behavior can be extended to the double bracket moment equations in chapters~\ref{DBVlasov}
and~\ref{orientation}. In particular, one would wonder how the clumping and alignment phenomena transfer to the case of filaments and sheets. An important question is whether these filaments emerge spontaneously in two or three dimensions. Further development is needed also for the geometric structure of the Smoluchowski moment equations. An analysis of their closures and study of singular solutions is required. Later, one can hope to apply this theory
to real problems involving oriented particles and ferromagnetic materials
in nano-science.


\chapter{Singular solutions in continuum dynamics}
\label{intro}
\section{Introduction}\label{intro1}
The use of geometric concepts in continuum models has highly increased in
the last 40 years and mainly related to physical systems which present some continuous symmetry \cite{MaRa99}. Such an approach has provided an important insight into
the mathematics of fluid models and has been successfully used for physical
modeling and other applications (turbulence \cite{FoHoTi01}, imaging \cite{HoRaTrYo2004}, numerics \cite{BuIs99}, etc.).

\medskip
It has been shown that many important continuum systems in physics (fluid
dynamics \cite{HoMaRa}, plasma physics \cite{HoMaRa}, elasticity \cite{SiMaKr88}, etc.) follow a purely geometric flow, uniquely determined by their total energy and by their symmetry properties. In particular, many geometric
fluid models have been widely studied (LAE-$\alpha$ \cite{HoNiPu06}, LANS-$\alpha$ \cite{FoHoTi01}, etc.) in the last years. One important feature that arises in many
continuum systems is the existence of singular measure-valued solutions.

\medskip
Probably, the most famous example of singular solution in fluids is the point vortex solution for the vorticity equation on the plane. These solutions
are delta-like solutions that follow a multi-particle dynamics. In three
dimensions one extends this concept to vortex filaments or vortex sheets, for which the vorticity is supported on a lower dimensional submanifold (1D
or 2D respectively) of the Euclidean space $\mathbb{R}^3$. The dynamics of these solutions has been widely investigated
and is still a source of important results in both fluid dynamics and geometry.
The existence of these solutions is a result of the nonlocal nature of the
equation describing the dynamics \cite{MaWe83}. Also, these solutions form
an invariant manifold and they are not expected to be created by fluid motion.

\medskip
Another important example of fluid model admitting singular solutions is
the Camassa-Holm (or EPDiff) equation, which is an integrable equation describing shallow water waves (besides its applications in other areas such as turbulence and imaging). However this equation has one more interesting feature, that is the \emph{spontaneous} emergence of singular solutions from any confined
initial configuration. The dynamical variable is the fluid
velocity and the nature of the singular solutions goes back to the trajectory
of the single fluid particle. For this particular case, the singular solutions
also have a soliton behavior. 
\rem{ 
The spontaneous emergence of these solutions in continuum PDE's is a remarkable fact and this behavior is explained through a balance between nonlinearity and nonlocality.
} 

\medskip
Singular solution also arise in plasma physics (magnetic
vortex lines, cf. e.g. \cite{Ga06}), kinetic
theory (phase space particle trajectories, cf. e.g. \cite{GiHoTr05,GiHoTr2007}), and other models for aggregation dynamics in friction dominated systems \cite{HoPu2005,HoPu2006}.
The latter are dissipative continuum models, which involve a fluid velocity
that is proportional to the collective force. In some cases these dissipative
models exhibit the spontaneous formation of singularities that clump
together in a finite time. This behavior is dominated by the dissipation
of energy and describes aggregation of particles.

\medskip
These considerations suggest that the properties of singular solutions in continuum models deserve further investigation. In particular, this work
presents geodesic and dissipative flows that exhibit the spontaneous emergence of singularities. These flows are then related to the kinetic description for multi-particle
systems. The connection from the microscopic kinetic level to the macroscopic level is provided by the kinetic moments. However, before going into the details of kinetic theory, this chapter reviews the mathematical properties of the fluid equations allowing for singular solutions.

\bigskip
\section{Basic concepts in geometric mechanics}
The basic geometric setting for fluid equations is given by Lagrangian (or
Hamiltonian) systems defined on Lie groups and Lie algebras. (This paragraph uses some of the concepts and the notation introduced
in \cite{MaRa99}.) When a system is invariant with respect to the Lie group $G$ over which it is defined, then
it is possible to rewrite its equations on the Lie algebra $\mathfrak{g}=T_e
G$ (or its dual $\mathfrak{g}^*=T^*_e G$) of that group. For example, if one takes the (right) invariant Hamiltonian $H=H(g,p):T^*G\rightarrow\mathbb{R}$,
then one writes
\[
H(gg^{-1},\,pg^{-1})=H(e,\mu)=h(\mu), 
\qquad
\mu\in T^*_e G
\,.
\]
so that the Hamiltonian $h(\mu)$ is defined on the dual Lie algebra $\mathfrak{g}^*$.
Analogously, for a (right) invariant Lagrangian  $L=L(g,\dot{g}):TG\rightarrow\mathbb{R}$
one writes
\[
L(gg^{-1},\,\dot{g}g^{-1})=L(e,\xi)=l(\xi),
\qquad
\xi\in T_e G
\,.
\]
The present work will mainly consider symmetric continuous systems whose equations are already written on the Lie algebra of some Lie group. This theory is called \emph{Euler-Poincar\'e} (or \emph{Lie-Poisson})
reduction and is extensively presented in \cite{MaRa99}.

\medskip
\paragraph{Lie-Poisson and Euler-Poincar\'e equations.} The starting point for the present analysis is the \emph{Lie-Poisson bracket}. 
\begin{definition}
A Hamiltonian system is called {\bfi Lie-Poisson} iff it is defined on the dual of a Lie algebra $\mathfrak{g}^*$ and the Poisson bracket is
given by
\[
\{F,G\}(\mu)=\pm\left\langle\mu,\left[\frac{\delta F}{\delta \mu},\frac{\delta G}{\delta \mu}\right]\right\rangle\qquad\text{with }
\,\,\,\mu\in\mathfrak{g}^*
\]
where $F,G$ are functionals of $\mu$, the notation $\delta F/\delta \mu$ denotes the functional derivative, $[\cdot,\cdot]$ is the Lie bracket and $\langle\cdot,\,\cdot\rangle$
denotes the natural pairing between a vector space and its dual.
\end{definition}
It is important to say that the sign in the bracket depends only on whether
the system is right- or left-invariant (plus and minus respectively). The following sections will explore various examples of Lie-Poisson systems both right and left invariant. However this section keeps the plus sign for right-invariant systems.

The equations arising from this structure are called \emph{Lie-Poisson equations}
and are written as
\begin{equation}\label{liepoisson}
\frac{\partial\mu}{\partial t}+\textrm{\large ad}^*_\text{\normalsize$\frac{\delta H}{\delta \mu}$}\,\, \mu=0
\end{equation}
where the coadjoint operator ad$^*$ is defined as the dual of the Lie bracket
\[
\left\langle \textrm{\large ad}^*_\eta\,\nu,\,\xi\right\rangle:=
\big\langle \nu,\,\textrm{ad}_\eta\,\xi\big\rangle=
\big\langle \nu,\,\left[\eta,\xi\right]\big\rangle
\]
with $\nu\in\mathfrak{g}^*$ and $\eta,\,\xi\in\mathfrak{g}$.

\medskip
If the Hamiltonian is such that the Legendre transform is invertible (\emph{regular
Hamiltonian}), then one can introduce the Lagrangian $L(\xi)$ in terms of the Lie
algebra variable $\xi\in\mathfrak{g}$
\[
\mu=\frac{\delta L}{\delta \xi}
\]
so that the Lagrangian is written as
\[
L(\xi)=\langle\mu,\,\xi\rangle-H(\mu)
\]
and the Euler-Lagrange equations are written in the form
\[
\frac{\partial}{\partial t}\frac{\delta L}{\delta \xi}
+
\textrm{\large ad}^*_\text{\small$\xi$}\, \frac{\delta L}{\delta \xi}=0
\]
which are called {\bfi Euler-Poincar\'e} equations.

\medskip
This work will mainly consider infinite dimensional Lie groups acting
on some manifold $M$. The
most general example is the group of diffeomorphisms Diff($M$), whose Lie algebra $\mathfrak{X}(M)$
consists of all possible vector fields on $M$. The manifold $M$ will be $\mathbb{R}^n$
and the Lie bracket among the vector fields $\bf X$ and $\bf Y$ is given
by the \emph{Jacobi-Lie bracket}
\[
\left[{\bf X,Y}\right]_{J\!L}=\bf (X\cdot\nabla)Y-(Y\cdot\nabla)X
\,.
\]

As it happens in ordinary finite-dimensional classical mechanics, both Lie-Poisson
and Euler-Poincar\'e equations can be derived from the following variational principles
\begin{align*}
\text{Euler-Poincar\'e:}& \qquad \delta\int_{t_1}^{t_2} L(\xi)\,dt=0
\\
\text{Lie-Poisson:}& \qquad \delta\int_{t_1}^{t_2} \big(\langle\mu,\,\xi\rangle-H(\mu)\big)\,dt=0
\end{align*}
for variations of the form $\delta\xi=\dot\eta-\left[\xi,\eta\right]$, where
$\eta(t)$ is a curve that vanishes at the end points $\eta(t_1)=\eta(t_2)=0$.
The second of these variational principles is called \emph{Hamilton-Poincar\'e}
variational principle \cite{CeMaPeRa}.

\medskip
\paragraph{Coadjoint motion.} From above, one can see that Lie-Poisson (or Euler-Poincar\'e) dynamics is
a strongly geometric type of dynamics. This point is even more evident, once one writes the solution of the equations as \cite{MaRa99}
\begin{equation}\label{coadj-mot}
\mu(t)=\textrm{Ad}^*_{g^{-1}(t)}\,\,\mu(0)
\end{equation}
where $g(t)=\exp\left(t\,\delta H/\delta \mu\right)$. The operator 
Ad$^*:G\times\mathfrak{g}^*\mapsto\mathfrak{g}^*$
denotes the {\it coadjoint group action} on the Lie algebra $\mathfrak{g}$
and is defined as the dual of the adjoint group action given by
\[
{\rm Ad}_g\, \xi := \left.\frac{d}{d\tau}\right|_{\tau=0}\,g\circ e^{\tau\,\xi}\circ\,g^{-1}
\qquad
\forall\,\,g\in G,\,\,\xi\in\mathfrak{g}
\,,
\]
so that $\left\langle\mu,\,{\rm Ad}_g\, \xi\right\rangle=\left\langle{\rm Ad}^*_g\,\mu,\, \xi\right\rangle$. Such a motion is called
{\bfi coadjoint motion} and is said to occur on \emph{coadjoint orbits},
where the coadjoint orbit $\mathcal{O}(\mu)$ of $\mu\in\mathfrak{g}^*$ is
the subset of $\mathfrak{g}^*$ defined by
\[
\mathcal{O}(\mu):=G\cdot\mu:=\left\{\textrm{Ad}^*_{g^{-1}}\,\,\mu:g\in G\right\}
\]

In order to see how Lie-Poisson equations (\ref{liepoisson})
are recovered from equation (\ref{coadj-mot}), one takes pairing
of  (\ref{coadj-mot}) with a Lie algebra element $\eta\in\mathfrak{g}$ as
follows
\begin{align}\label{coadj-pairing}
\left\langle
{\mu}(t),\,\eta
\right\rangle
=
\left\langle
\textrm{Ad}^*_{g^{-1}(t)}\,\,\mu(0),\,\eta
\right\rangle
=
\left\langle
\mu(0),\,\textrm{Ad}_{g^{-1}(t)}\,\,\eta
\right\rangle
\end{align}
where
\[
\textrm{Ad}_{g^{-1}(t)}\,\,\eta
=
\left.\frac{d}{d\tau}\right|_{\tau=0}\,e^\text{\small$-\,t\,\frac{\delta H}{\delta \mu}$}\circ\, e^{\text{\small$\tau\,\eta$}}\circ\,e^\text{\small$\,t\,\frac{\delta H}{\delta \mu}$}
\]
Now taking the time derivative of (\ref{coadj-pairing}) and evaluating it
at the initial condition $t=0$ yields
\begin{align*}
\left\langle
\dot{\mu}(0),\,\eta
\right\rangle
&=
\left\langle
\mu(0),\frac{d}{dt}\,\big.\textrm{Ad}_{\,\exp\left(-\,t\,{\delta H}/{\delta \mu}\right)}\,\,\eta\,\Big|_{t=0}
\right\rangle
\\
&=-
\left\langle
\mu(0),\textrm{ad}_\text{\small$\frac{\delta H}{\delta \mu}$}\,\eta
\right\rangle
=-
\left\langle
\textrm{ad}^*_\text{\small$\frac{\delta H}{\delta \mu}$}\,\mu(0),\,\eta
\right\rangle
\end{align*}
where the relation (cf. e.g. \cite{MaRa99})
\[
\textrm{ad}_\xi\,\eta=\frac{d}{dt}\,\big.\textrm{Ad}_{\,\exp\left(t\,\xi\right)}\,\eta\,\Big|_{t=0}
\]
has been used. Consequently, a system undergoing coadjoint orbits is a Lie-Poisson system. In particular, if the trajectory of a Lie-Poisson system starts on $\mathcal{O}$, then it stays in $\mathcal{O}$ \cite{MaRa99}.
This kind of motion explains how the geometry of the Lie group generates the dynamics.

\medskip
\paragraph{Lie derivative of tensor fields.} An important operator which is fundamental for
the purposes of the present work is the Lie derivative. In order to introduce
this operation as an infinitesimal generator, one can focus on the action
of diffeomorphism group on set of tensor fields defined on some manifold $Q$. 
Explicitly, the action $\Phi$ of a group element $g\in{\rm Diff}$ (a smooth invertible change of coordinates $g:q\mapsto g(q)\,$) on a tensor field ${\sf T}(q)$ is given by 
\[
\Phi(g,{\sf T})=g^*\,{\sf T}
\]
where the notation $g^*$ indicates the pull-back operation \cite{MaRa99}. If one considers
a one-parameter subgroup, i.e. a curve $g(t):=g_t$ on the Diff group (such
that $g_0=e$, where $e$ is the identity), then this action transports the tensor {\sf T} along this curve, according to $g^*_t\,\sf
T$. A \emph{Lie algebra action} $\,\xi_M$ on a manifold $M$ is defined by the infinitesimal generator. In particular, if $M$ is the space $\mathcal{T}(Q)$ of tensor fields on $Q$ the infinitesimal generator is evaluated on the tensor
$\sf T$ as follows
\[
\xi_{\,\mathcal{T}(Q)}\,{\sf T}:=
\left.\frac{d}{dt}\right|_{t=0}\Phi(g_t,{\sf T})
=
\left.\frac{d}{dt}\right|_{t=0}g^*_t\,{\sf T}
\]
However, an element of a one-parameter subgroup can be expressed in terms of a Lie algebra element $\xi$ through the exponential map
\[
g_t=e^{t\,\xi}
\]
so that 
\[
\xi_{\,\mathcal{T}(Q)}\,{\sf T}
=
\left.\frac{d}{dt}\right|_{t=0}g^*_t\,{\sf T}
=
\left.\frac{d}{dt}\right|_{t=0}\Phi(e^{t\,\xi},{\sf T})
=
\left.\frac{d}{dt}\right|_{t=0}\left(e^{t\,\xi}\right)^*{\sf T}
\]
Since the Lie algebra of the Diff group is the space of vector fields $\mathfrak{X}$, the one-parameter subgroup $g_t$ is identified with the {\it flow} of the vector field $\xi\in\mathfrak{X}$. Thus $\xi_{\,\mathcal{T}(Q)}\,{\sf T}$ is the $\mathfrak{X}$-Lie algebra action on the space of tensor fields 
$\mathcal{T}(Q)$.
At this point the definition of the Lie derivative is simply
\begin{definition}[Lie derivative of a tensor field]
The Lie derivative of a tensor field ${\sf T}(q)$ on some manifold $Q$ along a vector field $X(q)$ on the same manifold is defined as the infinitesimal
generator of the group of diffeomorphisms acting on $Q$
\[
\pounds_X {\sf T}:=
\xi_{\,\mathcal{T}(Q)}\,{\sf T}
=
\left.\frac{d}{dt}\right|_{t=0} \left(e^{t\,X}\right)^*{\sf T}
\,.
\]
\end{definition}
A particular case is provided by the possibility ${\sf T}=Y\in\mathfrak{X}$, since
now $g^*\,Y=:{\rm Ad}_{g^{-1}}\,Y$ and thus 
\[
\pounds_X\,Y=\left.\frac{d}{dt}\right|_{t=0}{\rm Ad}_{\,\exp(-tX)}\,Y=:{\rm
ad}_{\,-\,X}\,Y=:[X,Y]_{JL}
\]
so that \emph{the Lie derivative of two vector fields is given by the Jacobi-Lie
bracket}. 
\rem{ 
As a general result, the characteristic equation
\[
\frac{d}{dt}\,g^*_t\,{\sf T}(q)=0
\quad\text{\,with\,}\quad
\frac{d}{dt}\,g_t(q)=X(q,t)
\]
is equivalent to
\[
\frac{\pa \hat{\sf T}}{\pa t}+\pounds_{X}\,\hat{\sf T}=0
\]
where and the two formulations are called ``Lagrangian'' and ``Eulerian'' respectively.
} 

\bigskip
\section{Euler equation and vortex filaments}\label{diffvol}

\medskip
An important physical system in the context of geometric dynamics is the Lie-Poisson system for the vorticity of an ideal Euler fluid. As its primary geometric
characteristic, Euler's fluid theory represents fluid flow as
Hamiltonian geodesic motion on the space of smooth invertible maps
acting on the flow domain and possessing smooth inverses. These smooth
maps (diffeomorphisms) act on the fluid reference configuration so
as to move the fluid particles around in their container. And their smooth
inverses recall the initial reference configuration (or label) for the
fluid particle currently occupying any given position in space. Thus, the
motion of all the fluid particles in a container is represented as a
time-dependent curve in the infinite-dimensional group of
diffeomorphisms. Moreover, this curve describing the sequential actions
of the diffeomorphisms on the fluid domain is a special optimal curve that
distills the fluid motion into a single statement. Namely, ``A fluid moves
to get out of its own way as efficiently as possible.'' Put more
mathematically, fluid flow occurs along a curve in the diffeomorphism
group which is a geodesic with respect to the metric on its tangent space
supplied by its kinetic energy. 

For incompressible fluids, one restricts
to diffeomorphisms that preserve the volume element and the fluid is described
by its vorticity, which satisfies a Lie-Poisson equation. This section reviews some of the results presented in \cite{MaWe83}. In order to understand the Lie-Poisson structure, one introduces Euler's vorticity equation as
\begin{equation}
\boldsymbol\omega_t+\text{curl}\big(\boldsymbol\omega\times\mathbf{v}\big)=0
\,.
\end{equation}
where the vorticity is defined in terms of the velocity as $\boldsymbol\omega={\rm
curl}\,\bf v$.
Following \cite{MaWe83}, this equation represents the advection equation for an exact two-form $\omega=\,\bom\cdot d\mathbf{S}$ appearing as the vorticity for incompressible motion along the
fluid velocity $\mathbf{v}$ and thus it can be written in terms of Lie derivative
{\it\pounds\,} along the velocity vector field
\begin{equation}
\omega_t+\pounds_{\bf v}\, \omega
=0
\,.
\end{equation}

The Lie-Poisson bracket for vorticity is written on the dual  
$\mathfrak{X}^*_\textnormal{vol}$ of the Lie-algebra $\mathfrak{X}_\textnormal{vol}$ of volume-preserving diffeomorphisms, which is isomorphic to the set of exact one-forms: $\omega=d\alpha$, where $\alpha$ is a generic one-form. In this case the Jacobi-Lie bracket between two volume-preserving vector fields $\boldsymbol{\xi}_1$ and $\boldsymbol{\xi}_2$ in $\mathbb{R}^3$ may be written as
\[
\left[\,\boldsymbol{\xi}_1\,,\,\boldsymbol{\xi}_2\,\right]_\textit{\footnotesize$
_{J\!L}$}\,
=-\,\text{curl}\left(\,\boldsymbol{\xi}_1\,\times\,\boldsymbol{\xi}_2\,\right)
\,.
\]
In terms of the vector potentials for which $\boldsymbol{\xi}_1=\text{curl }\boldsymbol{\psi}_1$ and $\boldsymbol{\xi}_2=\text{curl }\boldsymbol{\psi}_2$ this bracket becomes
\[
\left[\,\boldsymbol{\xi}_1\,,\,\boldsymbol{\xi}_2\,\right]_\textit{\footnotesize$
_{J\!L}$}\,
=-\,\text{curl }\big(\,\text{curl }\boldsymbol{\psi}_1\,\times\,\text{curl }\boldsymbol{\psi}_2\,\big)\,.
\]
The vector potentials $\boldsymbol{\psi}_1$ and $\boldsymbol{\psi}_2$ are defined up to a gradient of a scalar function
so that one can always choose a gauge in which $\text{div}\,\boldsymbol\psi=0$. Pairing the vector field given by the Lie bracket with a one-form (density) $\boldsymbol{\alpha}$ then yields,  after an integration by parts,
\[
\big\langle
\boldsymbol\alpha\,,\,\left[\,\boldsymbol{\xi}_1\,,\,\boldsymbol{\xi}_2\,\right]_\textit{\footnotesize$
_{J\!L}$}\!
\big\rangle
=
-\,
\big\langle\,
\text{curl }\boldsymbol{\alpha}\,,\,\big(\,\text{curl }\boldsymbol{\psi}_1\,\times\,\text{curl }\boldsymbol{\psi}_2\,\big)
\,\big\rangle
=
-\,
\big\langle\,
\boldsymbol{\omega}\,,\,\big[\,\boldsymbol{\psi}_1\,,\,\boldsymbol{\psi}_2\,\big]
\,\big\rangle
\]
where $\boldsymbol{\alpha}$ is defined up to an exact one-form ${\bf d}f$
and one introduces the notation
\[
\big[\,\boldsymbol{\psi}_1\,,\,\boldsymbol{\psi}_2\,\big]\,
:=\,
\text{curl }\boldsymbol{\psi}_1\,\times\,\text{curl }\boldsymbol{\psi}_2\,.
\]
The bracket $[\,\cdot\,,\,\cdot\,]$ defines a Lie algebra structure on the space of vector potentials whose dual space may be naturally identified with exact two-forms $\omega=\text{curl}\,\alpha$. At this point, the expression for the Lie-Poisson bracket for functionals of vorticity may be introduced as
\[
\{F,\,H\}
\,=\,
\left\langle
\bom\,,\, \left[\,
\frac{\delta F}{\delta\bom}\,,\,\frac{\delta H}{\delta\bom}
\, \right]
\right\rangle
\,=\,
\int\boldsymbol\omega\,\cdot
\left(
\text{curl}\,\frac{\delta F}{\delta\bom}
\times
\text{curl}\,\frac{\delta H}{\delta\bom}
\right)\,\text{d}^3\bf x
\,,
\]
where
\[
H=\frac{1}{2}\int \bom\cdot(-\Delta)^{-1}\bom\,\,{\rm d}^3{\bf x} =
\frac12\int\left|{\bf u}\right|^2 {\rm d}^3{\bf x}=
\frac12\,\|{\bf u}\|^2
\]
is the fluid's kinetic energy expressed in terms of vorticity. 

Now,  vorticity dynamics is an example of {\bf geodesic motion} on a
(infinite-dimensional) Lie group \cite{Ar1966}
\[
\boldsymbol\omega_t
=
-\,{\rm ad}^*_{\,\delta H/\delta \boldsymbol\omega}\,\,\boldsymbol\omega
=
-\,
\text{curl}\left(\boldsymbol\omega\times\text{curl}\,(-\Delta)^{-1}\boldsymbol\omega\right)
=
\text{curl}\,\big(\text{curl}\,\boldsymbol\psi\times\boldsymbol\omega\big)
=
-\pounds_{{\rm curl}\,\boldsymbol\psi}\,\bom
\,.
\]
where the ad$^*$ is now defined as the dual of the $[\![\cdot,\cdot]\!]$ Lie bracket and one has
\[
{\rm ad}^*_{\boldsymbol\psi}\,\bom=\pounds_{{\rm curl}\boldsymbol\psi}\,\bom
\,.
\]

As stated in section~\ref{intro1}, this equation allows singular solutions in the form of vortex filaments, distributions of vorticity supported on a curve ${\bf
Q}(s,t)$. These are represented by the following expression
\[
\bom({\bf x},t)=\int\frac{\partial {\bf Q}(s,t)}{\partial s}\,\delta({\bf x-Q}(s,t))\,{\rm
d}s
\]
where $s$ is a curvilinear coordinate and $\partial_s{\bf Q}$ is the tangent
to the curve. The dynamics of ${\bf Q}(s,t)$ is presented in the work by
Holm and Stechmann \cite{Ho2003,HoSt04}. 
\rem{ 
In general $s$ is a coordinate of dimension $k<3$: if $s$ is one dimensional, then
this solution represents a vortex filament, and if it is two dimensional,
then it represents a vortex sheet. 
} 
These solutions are widely studied in many areas of fluid dynamics as well as in condensed state theory, within the theory of superfluids \cite{RaRe}.

All the arguments above can be projected down onto the plane to
obtain the 2D Euler equation
\begin{equation}
\omega_t
+ 
\big\{ \omega,\,(-\Delta)^{-1}\omega\big\}
=
0
\,,
\end{equation}
where $\{\cdot,\cdot\}$ denotes the canonical Poisson bracket in the plane coordinates $x,y$. This equation also allows for singular solutions, the
point vortex solutions moving
on the plane. The expression for a point vortex is easily written as
\[
\bom(x,y,t)=\delta(x-X(t))\,\delta(y-Y(t))
\]
where the dynamics of $X$ and $Y$ is just ordinary Hamiltonian dynamics for
the two conjugate variables, so that point vortices move around as if they were
particles. This 2D equation is important because it is completely equivalent
to the collisionless Boltzmann equation in kinetic theory and thus provides
a slight introduction to the central topic of this work.

Singular solutions are allowed because of a combination of the form of the equation and the smoothing of the Lie algebra element
$\boldsymbol\psi=\Delta^{-1}\bom$ by the Poisson kernel $\Delta^{-1}$. Indeed,
if one chooses a Hamiltonian that is quadratic with respect to the Euclidean norm ($H=\frac{1}{2}\int \left|\bom\right|^2{\rm d}^3x$), one readily realizes that singular solutions are
forbidden by the dynamics. Consequently the presence of a smooth vector potential is of central importance in the existence of singular solutions. For example,
one could think to modify the equations in order to allow for a different
regularization of the solution, that is changing the norm $\Delta^{-1}$ with
another kernel which, possibly introduces a regularization length-scale.
An example is provided by the \emph{Euler-alpha model} \cite{HoNiPu06} that introduces
a smoothed velocity ${\bf u}=(1-\alpha^2\Delta)^{-1}\, \bf v$, so that upon
defining $\bom={\rm curl}\,\bf u$ (regularized vorticity) and $\boq={\rm curl}\,\bf v$ (singular vorticity), the Hamiltonian is given by 
\begin{align*}
H=1/2\int\!{\bf u\cdot v}\, {\rm d}^3\bf x
&=
1/2\int (1-\alpha^2\Delta)^{-1}\, {\bf v}\cdot {\rm curl}^{-1}{\boq}\,\, {\rm d}^3\bf x
\\
&=
1/2\int (1-\alpha^2\Delta)^{-1}\, {\rm curl}^{-1}{\boq}\cdot {\rm curl}^{-1}{\boq}\,\, {\rm d}^3\bf x
\\
&=
1/2\int {\boq}\cdot {\rm curl}^{-1}(1-\alpha^2\Delta)^{-1}\, {\rm curl}^{-1}{\boq}\,\, {\rm d}^3\bf x
\\
&=
1/2\int {\boq}\cdot (-\Delta)^{-1\,}(1-\alpha^2\Delta)^{-1}{\boq}\,\, {\rm d}^3\bf x
\end{align*}
where the last step is justified by the fact that the integral operators $(1-\alpha^2\Delta)^{-1}$ and ${\rm curl}^{-1}$
commute. In this way, the motion is again geodesic with respect to the singular vorticity $\boq$.
The previous arguments show how the dynamics of the Euler fluid is given by a well know geometric flow,
the {\bfi geodesic flow on the group of volume preserving diffeomorphisms}
\cite{Ar1966,ArKe98}.

The next section shows that this kind of flow plays a central role
in the theory of singular solutions for continuous Hamiltonian dynamics.
This idea relates the formation of singular solutions to geodesic
motion on different infinite-dimensional Lie groups, like the group of diffeomorphisms or the group of canonical transformations (symplectomorphisms). The latter will be a central topic in this work.

\bigskip
\section{The Camassa-Holm and EPDiff equations}

\medskip
The Euler equation admits vortex filaments and these solutions are related to a geodesic flow on an infinite-dimensional Lie group. However for the Euler equations, the singular solutions are an invariant submanifold, that is they do not emerge spontaneously from a smooth initial condition. Now, there are important geometric flows that exhibit a spontaneous emergence of singularities from \emph{any} smooth initial state. One of the most meaningful examples that is also the main inspiration for the present work is the \emph{Camassa-Holm equation} (CH) \cite{CaHo1993}
\[
u_t+2\kappa u_x-u_{xxt}+3uu_x=2u_x u_{xx}+u u_{xxx}
\]
In particular this work focuses on the case when $\kappa=0$ and considers
the case when boundary terms do not contribute to integration by parts (periodic
boundary conditions or fast decay at infinity). It has been shown
\cite{HoMa2004} that this equation is a {\bfi geodesic motion on the group of diffeomorphisms}
(EPDiff). In fact one finds that this equation can be recovered from the
following Euler-Poincar\'e variational principle defined on $\mathfrak{X}(\mathbb{R})$
\[
\delta\!\int_{t_0}^{t_1}\!\!L(u)\,{\rm d}t=0
\qquad\text{ with }\qquad
L(u)=\frac12\int \!u \,(1-\partial_x^2)u\,\,{\rm d}x
\]
In this way the CH equation is the Euler-Poincar\'e equation for a purely
quadratic Lagrangian. Thus the CH equation is again a geodesic flow, which
is given by the geodesic equation on the group of diffeomorphisms. It is easy to
find the Lie-Poisson formulation, via the Legendre transform
\[
m=\frac{\delta L}{\delta u}=u-u_{xx}
\quad
\Rightarrow
\quad
u=(1-\partial_x^2)^{-1} m
\]
where $m=m(x)\,dx\otimes dx\in\mathfrak{X}^*(\mathbb{R})$, the space of one-form densities.
The Hamiltonian becomes
\[
H(m)=\frac12\int\!m\,(1-\partial_x)^{-1}m\,{\rm d}x
\]
and the Lie-Poisson equation is
\[
m_t=-\pounds_u m=-u m_x-2mu_x
\,.
\]
The main result on this equation is its integrability which is guaranteed
by its bi-hamiltonian structure. However there
is another important statement, which is called the \emph{steepening lemma}
\cite{CaHo1993}:

\smallskip
\noindent
{\it Suppose the initial profile of velocity $u(x,t=0)$ has an inflection point at $x = \bar{x}$ to the right of its
maximum, and otherwise it decays to zero in each direction sufficiently rapidly for the Hamiltonian $H(m)$ to be finite. Then the negative slope at the inflection point will become vertical in finite time.}

\begin{figure}[h]\label{peakons_png}\center
\includegraphics[scale=0.169]{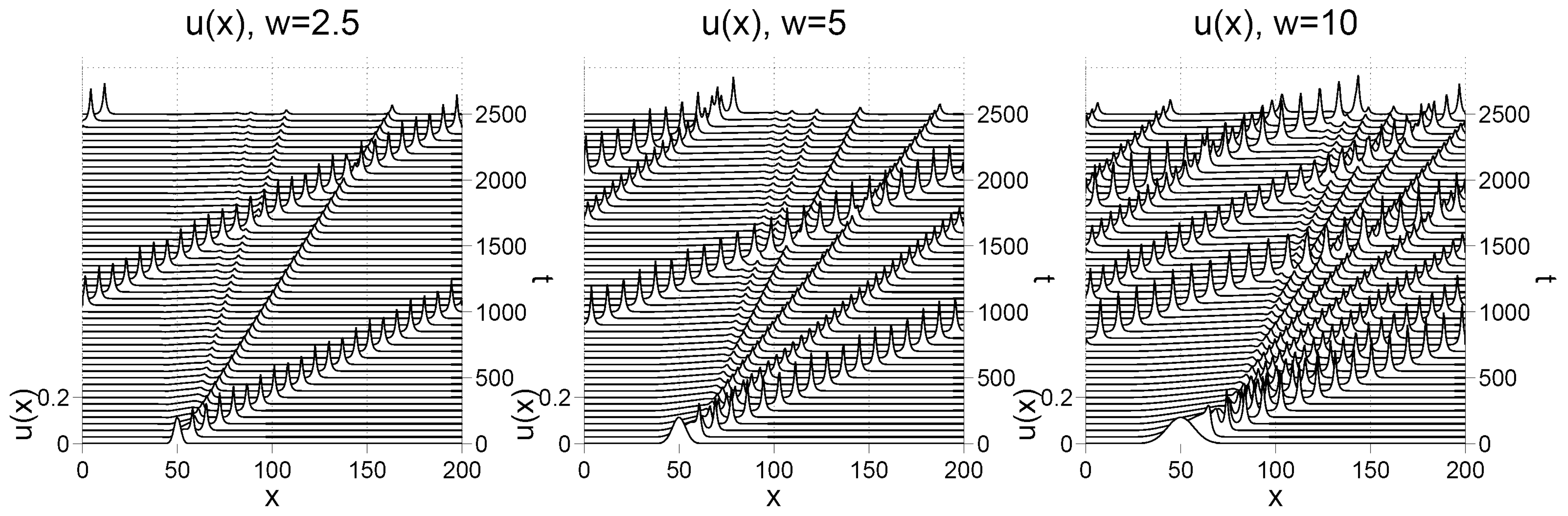}
\caption{
Peakons emerging from Gaussian initial conditions for different
Gaussian widths $w$. Figure from \cite{HoSt03}.}
\end{figure}

\smallskip\noindent
This fact is shown in fig.~\ref{peakons_png} and is particularly relevant when one focuses on the behavior
of singular solutions in PDE's.  Moreover these solutions (called \emph{peakons} in the velocity representation) present soliton behavior and this fact makes their mutual interaction particularly interesting.

\medskip
All the above can be easily generalized to more dimensions, so that the equation
becomes
\begin{equation*}
\mathbf{m}_t
- \mathbf{u}\times{\rm curl\,}\mathbf{m} 
+ \nabla (\mathbf{u}\cdot\mathbf{m}) 
+ \mathbf{m}({\rm div\,}\mathbf{u})
=0
\,.
\end{equation*}
and one takes the following Hamiltonian on $\mathfrak{X}^*(\mathbb{R}^3)$
\[
H({\bf m})=\frac12\int\!{\bf m}\cdot(1-\alpha^2\Delta)^{-1}{\bf m}\,{\rm
d}\bf x
\]
where one has inserted the length-scale $\alpha$, that determines the
smoothing of the velocity ${\bf u}=(1-\alpha^2\Delta)^{-1}{\bf m}$.
The singular solutions (\emph{pulsons}) are written in this representation
as
\begin{align}
\label{singsolepdiff}
{\bf m(x},t)=\sum_i\int{\bf P}_{\!i}(s,t)\,\delta({\bf x-Q}_i(s,t))\,{\rm d}s
\end{align}
where $s$ is a variable of dimension $k<3$. These solutions represent pulse
filaments or sheets, when $s$ has dimension 1 or 2 respectively.

Another generalization is to take another kernel that defines the norm of
$\bf m$ and substitute $(1-\alpha^2\Delta)^{-1\,}\bf m$ with the general convolution  $G\,*\,{\bf m}=\int G({\bf x-x'})\,\bf m(x')\,{\rm d}{\bf x}'$
with some Green's function $G$.
The dynamics of $({\bf Q}_i,{\bf P}_{\!i})$ is given by canonical Hamiltonian dynamics with the Hamiltonian
\[
\mathcal{H}=\frac12\sum_{i,j}\iint {\bf P}_{\!i}(s,t)\cdot{\bf P}_{\!j}(s',t)\,\,G({\bf Q}_i(s,t)-{\bf Q}_j(s',t))\,\,{\rm d}s\,{\rm d}s'
\,.
\]
An important result is the theorem stating that the singular
solution (\ref{singsolepdiff}) is a \emph{momentum map} \cite{HoMa2004}: given a Poisson
manifold (i.e. a manifold $P$ with a Poisson bracket 
$\{\cdot,\cdot\}$ defined on the functions $\mathcal{F}(P)$) 
and a Lie group $G$ acting on it by Poisson maps (so that the Poisson bracket is preserved), a momentum map is defined as a map
\[
\mathbf{J}:P\rightarrow\mathfrak{g}^*
\]
so that
\[
\{F(p),\langle \mathbf{J}(p),\xi\rangle\}=\xi_P[F(p)]
\qquad\quad\forall\, F\in\mathcal{F}(P),\quad\forall\,\xi\in\mathfrak{g}
\]
where $\mathcal{F}(P)$ denotes the functions on $P$ and $\xi_P$ is the vector field given by the infinitesimal generator
\[
\xi_P\,(p)=\left.\frac{d}{dt}\right\vert_{t=0}\!\!\!e^{t\xi}\cdot p \qquad \forall\,p\in
P
\]
Now, fix a $k$-dimensional manifold $S$ immersed in $\mathbb{R}^n$ and consider the embedding ${\mathbf{Q}_i:S\rightarrow\mathbb{R}^n}$. Such embeddings form
a smooth manifold and thus one can consider its cotangent bundle $(\mathbf{Q}_i,\mathbf{P}_i)\in
T^*\text{Emb}(S,\mathbb{R}^n)$. Now consider $\text{Diff}(\mathbb{R}^n)$
acting on $\text{Emb}(S,\mathbb{R}^n)$ on the left by composition $\left(g\mathbf{Q}=g\circ\mathbf{Q}\right)$
and lift this action to $T^*\text{Emb}(S,\mathbb{R}^n)$: this gives the singular
solution momentum map for EPDiff
\[
\mathbf{J}:T^*\text{Emb}(S,\mathbb{R}^n)\rightarrow\mathfrak{X}^*(\mathbb{R}^n)
\qquad\text{with}\qquad
\mathbf{J}(\mathbf{Q},\mathbf{P})=\!\int\mathbf{P}(s,t)\,\delta(\mathbf{x}-\mathbf{Q}(s,t))\,ds
\,.
\]
This result is extensively presented in \cite{HoMa2004}, where different
proofs are given in various cases. A key fact in this regard is that this
momentum map is equivariant, which means it is also a Poisson map. This explains
why the coordinates $(\mathbf{Q},\mathbf{P})$ undergo Hamiltonian dynamics.

The EPDiff equation has been applied in several contexts to turbulence modeling
\cite{FoHoTi01} and imaging techniques \cite{HoRaTrYo2004,HoTrYo2007} and its CH form (also with dispersion) is widely studied in terms of its integrability properties.

\medskip
Again the idea of geodesic flow plays a central role in the behavior of the
pulson solutions. This suggests that a further investigation of geodesic equations on Lie groups is needed with relation with the emergence of singularities  and integrability issues. Chapter \ref{momLPdyn}\, considers the group of canonical transformations (through moment dynamics) and chapter \ref{EPSymp}\, formulates a geodesic flow on it. The results
are encouraging for further investigation, since this flow includes
the EPDiff equation as a special case and provides an extension to its multi-component
versions (some of which are known to be integrable).

\medskip
However, singular solutions do not appear only in Hamiltonian dynamics. There
is another class of systems, which undergo a dissipative dynamics with a
deep geometrical meaning. In fact chapters \ref{GOP},\ref{DBVlasov} and \ref{orientation}\, will show that coadjoint motion does
not necessarily need to be Hamiltonian. This concept is related to the so
called \emph{double bracket dissipation}, which is extensively analyzed in
the second part of this work. In order to introduce how singular solutions
arise in dissipative continuum dynamics, the
next section reviews the main ideas by following the presentation in \cite{HoPu2006}.

\bigskip
\section{Darcy's law for aggregation dynamics}

\bigskip

Many physical processes can be understood in terms of aggregation of individual
components into a ``final product''. This phenomenon is recognizable at different scales: from galaxy clustering \cite{Chandra39,BiTr88} to particles in nano-sensors \cite{MaPuXiBr}. Thus self-aggregation is not necessarily dependent on the particular kind of interaction. 

A related paradigm arises in biosciences, particularly in chemotaxis: the study of the influence of chemical substances in the environment on the
motion of mobile species which secrete these substances. One of the most famous among such models is the Keller–-Segel system of partial differential equations \cite{KellerSegel1970}, which was introduced to explore the effects of nonlinear cross diffusion in the formation of aggregates and patterns by chemotaxis in the aggregation of the slime mold {\it Dictyostelium
discoidium}. The Keller–-Segel (KS) model consists of two strongly coupled reaction-–diffusion equations
\begin{align*}
\rho_t&={\rm div}\big(\rho\,\mu(\rho)\,\nabla\Phi[\rho]+D\,\nabla\!\rho\big)
\\
\epsilon \Phi_t& + \hat{L}\Phi=\gamma\rho
\,.
\end{align*}
expressing the coupled evolution of the concentration of
organism density ($\rho$) and the concentration of chemotactic
agent potential ($\Phi$). The constants $\epsilon,D,\gamma>0$
are assumed to be positive, and the linear operator $\hat{L}$ is taken to
be positive and symmetric. For example, one may choose it
to be the Laplacian $\hat{L} = \Delta$ or the Helmholtz operator $\hat{L} = 1-\alpha^2\Delta$.

Historically, it seems that Debye and H\"uckel in 1923 were
the first to establish this model. They derived the KS evolutionary system  in their article \cite{DeHu1923} on the theory of electrolytes. In particular, they
 present the simplified model with $\epsilon= 0$. Consequently, the simplified evolutionary KS system with $\epsilon= 0$ may also be called the Debye-–H\"uckel
equations.

Later, the same model appeared for modeling aggregation at different scales.
Chandrasekhar formulated the Smoluchowski-Poisson equation for stellar formation
and the ``Nernst-–Planck'' (NP) equations in the same form as KS re-emerged in the biophysics community, for example, in the study of ion transport in biological channels \cite{BaChEi}. The same  system had also surfaced earlier as the drift-diffusion equations in the semiconductor device design literature; see Selberherr (1984) \cite{Se84}. A variant of the KS system re-appeared even more recently as a model of the self-assembly of particles at
nano-scales \cite{MaPuXiBr}. 

\medskip
In order to understand the geometric framework for this kind of equations,
one can consider the Debye-H\"ukel system ($\epsilon=0$) in the limit when
the diffusion is negligible ($D=0$). This system is a conservation equation
\begin{equation}\label{Darcy1}
{\rho}_t=-\,{\rm div}\left(\rho{\bf V}\right)
\qquad\text{ with }\qquad
{\bf V}=-\,\mu(\rho)\nabla\frac{\delta E}{\delta \rho}
\end{equation}
where ${\bf V}$ is called \emph{Darcy's velocity}, $E(\rho)$ is the energy functional,
$\mu(\rho)$ is called
``mobility'' and in general it depends on $\rho$. The physical meaning of
these equations is that when the inertia of the particles is negligible, the \emph{particle velocity is proportional to the force}. This happens in particular for {\bfi friction dominated systems}. Under this
approximation, one can interpret this model as a sort of unifying principle for aggregation and self-assembly of highly dissipative systems at different
scales. The fact that the energy is dissipated is readily seen by calculating
\begin{align*}
\frac{d E}{dt}
&=\left\langle\frac{\pa \rho}{\pa t},\,\frac{\delta E}{\delta \rho}\right\rangle
=\bigg\langle  {\rm\large div}\Big(\rho\, \mu \,\nabla\, \frac{\delta E}{\delta \rho}\Big)
,\, 
\frac{\delta E}{\delta \rho}
\bigg\rangle
\\
&=
-\,
\bigg\langle  
\,\Big(\mu\, \nabla\, \frac{\delta E}{\delta \rho}\Big)^\sharp
,\, 
\Big(\rho \,\nabla\, \frac{\delta E}{\delta \rho}\Big)
\bigg\rangle
=
\nonumber
-\int
\rho\,\mu(\rho)
\left|\nabla\, \frac{\delta E}{\delta \rho}\right|^2
\,d\,^nx
\,.
\end{align*}
so that the energy is monotonically decreasing when the mobility is a positive
definite quantity. The equation (\ref{Darcy1}) is called \emph{Darcy's law}
and in some cases is also known as the porous media equation.

\begin{figure}[h]\label{clumpons_png}
\center
\includegraphics[scale=0.5]{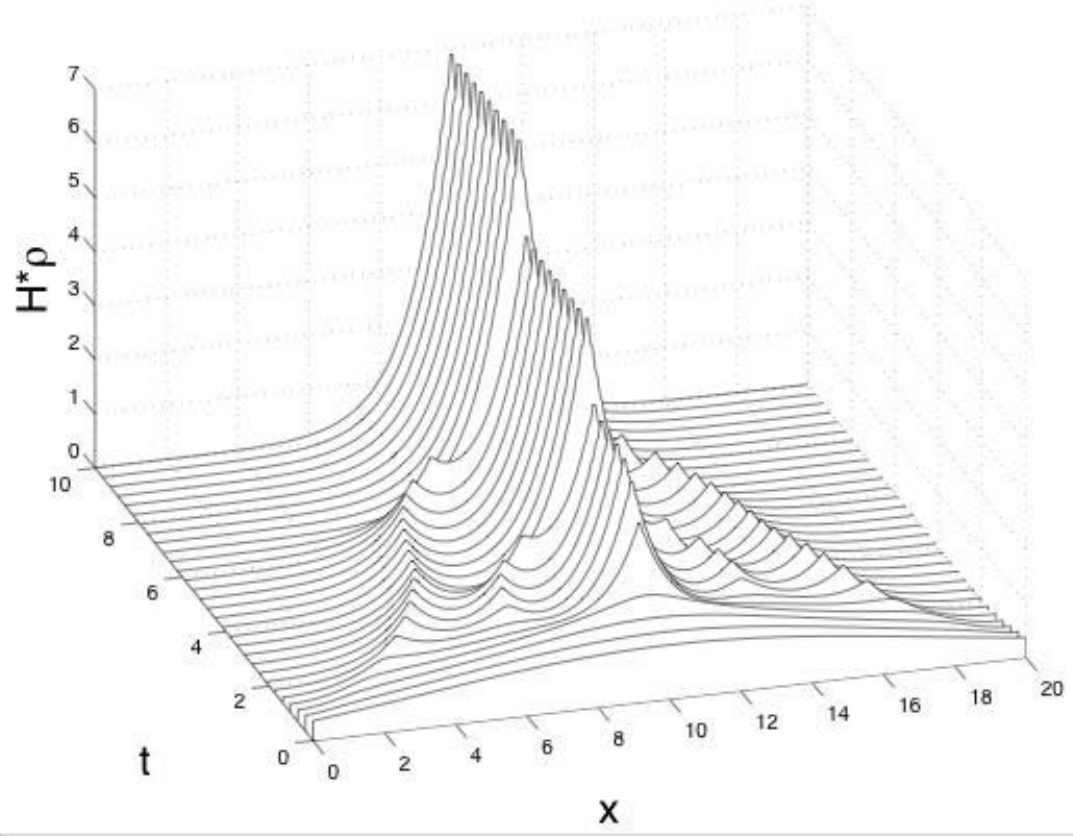}
\caption{Singularities emerging from a Gaussian initial condition. It is
shown how these singularities merge together after their formation. This
figure is taken from \cite{HoPu2005,HoPu2006}.}
\end{figure}

\medskip
At this point one starts discussing the existence of singular solutions.
In fact, Holm and Putkaradze \cite{HoPu2005,HoPu2006} have shown that in 1D this equation
allows for solutions of the form
\[
\rho(x,t)=w_\rho\,\delta(x-q(x,t))
\]
with
\[
\dot{w}_\rho=0,
\qquad
\quad
\dot{q}=-\left.V(x,t)\right|_{x=q}
\]
where $V$ is the Darcy's velocity introduced above. In particular, Holm and
Putkaradze have shown that, for $\mu=(1-\alpha^2\partial_x^2)^{-1}\rho$ and
$E=\frac12\int\!\rho\left(1-\beta^2\partial_x^2\right)^{-1\!}\!\!\rho\,\,{\rm d}x$, this equations possess spontaneously emergent singular solutions from any confined initial distribution, just as it happens for EPDiff in the Hamiltonian case. So, again for a purely quadratic energy functional, this system
possesses singular $\delta$-like solutions, which emerge spontaneously. In particular, a set of singularities emerge from the initial
condition and, after a finite amount of time, these singularities merge in
only one final singular solution, as shown in fig.~\ref{clumpons_png}. This is the reason why these solutions have been named \emph{clumpons}.

This behavior is particularly meaningful for physical applications, since
the merging process is directly related to the concept of aggregation and
self-assembly. Thus the emergence of singular solutions in Darcy's law will control their potential application to self-assembly, especially in
 nano-science.
 
\medskip
In a more general mathematical setting, this equation can be extended
to any geometric quantity as follows.
As a first step, one sees that the equation for $\rho$ is an advection relation
of the type $\rho_t+\pounds_V \rho=0$, so that the Darcy's velocity $V$ acts
on the density $\rho$ as a vector field, as it happens for ordinary fluid
dynamics. Now take the following pairing with a function $\phi$
\[
\left\langle\textit{\large\pounds}_{\!V}\rho,\,\phi\right\rangle:=
\left\langle {\rm div}(\rho V),\,\phi\right\rangle=
-\left\langle \rho\nabla\phi, V\right\rangle=:
\left\langle \rho\diamond\phi,V\right\rangle
\]
where one introduces the diamond operation 
$\diamond:(\rho,\phi)\mapsto\rho\,\nabla\phi\in\mathfrak{X}^*(\mathbb{R})$
which is understood as the ``dual'' of the Lie derivative. If now Darcy's velocity is written as $V=\left(\mu\diamond\delta E/\delta\rho\right)^{\sharp}$, then Darcy's law becomes written in the more abstract way
\[
\rho_t+\pounds_{\left(\mu\,\diamond\frac{\delta E}{\delta \rho}\!\right)^{\!\sharp}}\,\rho=0
\]
This enables one to extend Darcy's law to any geometric order parameter (GOP).
Indeed, given a tensor $\kappa$, one can write the GOP equation
\[
\kappa_t+\pounds_{\left(\mu\,\diamond\frac{\delta E}{\delta \kappa}\!\right)^{\!\sharp}}\,\kappa=0
\]
which is the generalization of the ordinary Darcy's law for the density $\rho$. Holm, Putkaradze and the author \cite{HoPuTr2007} have shown how these equations always admit singular $\delta$-like solutions for any
geometric quantity, when the mobility is taken
as a filtered quantity $\mu=K*\kappa$, through some filter $K$. The reason is that whenever Darcy's velocity is smooth, then the advection
equation admits the single particle solution. The trajectory of the single
particle has an important geometric meaning, since it reflects
the geometry underlying the macroscopic continuum description.

Chapter \ref{GOP}\, will show how this equations are recovered by a symmetric
dissipative bracket and  its geometric properties will be connected with
Riemannian manifolds.

A considerable part of this work is devoted to formulate a microscopic kinetic theory that recovers Darcy's law at the macroscopic fluid level. This process
again involves the theory of kinetic moments (introduced in the next section) as a crucial step in deriving fluid equations. Chapter \ref{orientation}\, extends this treatment to particles with anisotropic
interactions. Again from a suitable kinetic theory, it is possible to derive macroscopic
equations that extend Darcy's law to oriented particles. The next section
presents a slight introduction to kinetic moments and shows how fluid
dynamics is recovered from a truncation of the whole moment hierarchy. 

\bigskip
\section{The Vlasov equation in kinetic theory}

\smallskip
The evolution of $N$ identical particles in phase space $T^*M$ with coordinates  $(q_i, p_i)$ $i=1,2,\dots,N$, may be described by an evolution equation
for their joint probability distribution function. Integrating over all
but one of the particle phase-space coordinates yields an evolution
equation for the single-particle probability distribution function (PDF)
\cite{MaMoWe1984}.
This is the Vlasov equation, which may be expressed as an
advection equation for the phase-space density $f$ along the Hamiltonian vector field ${\bf X}_\textit{\small h}$  corresponding to single-particle motion with Hamiltonian $h(q,p)$:
\begin{eqnarray*}
f_t
=
\big\{f\,,\,h\big\}
=
\,- \,\text{\large div}_{(q,p)}\big(f\, {\bf X}_\textit{\small h}\big)\,
= - \,\textit{\large\pounds}_\text{\small$\mathbf{X}_{h}$}\, f
\end{eqnarray*}
with
\begin{eqnarray*}
{\bf X}_\textit{\small h} (q,p)= \left(\frac{\partial h}{\partial p},-\frac{\partial h}{\partial q}\right)=\frac{\partial h}{\partial p}\frac{\partial}{\partial q}-\frac{\partial h}{\partial q}\frac{\partial}{\partial p}
\end{eqnarray*}

The solutions of the Vlasov equation reflect its heritage in particle
dynamics, which may be reclaimed by writing its many-particle PDF as a
product of delta functions in phase space
\[
f(q,p,t)\,=\,\sum_j \,\delta(q-Q_j(t))\,\delta(p-P_j(t))
\,.
\]
Any number of these delta
functions may be integrated out until all that remains is the dynamics of
a single particle in the collective field of the others. 

In the mean-field approximation of plasma dynamics, this collective field generates the total electromagnetic
properties and the self-consistent equations obeyed by the single
particle PDF are the Vlasov-Maxwell equations. In the electrostatic
approximation, these reduce to the Vlasov-Poisson (VP) equations, which govern
the statistical distributions of particle systems ranging from charged-particle beams \cite{Venturini}, to the distribution of stars in a galaxy \cite{Ka1991}. 
\begin{remark}
A class of singular solutions of the VP equations called the ``cold
plasma'' solutions have a particularly beautiful experimental realization
in the Malmberg-Penning trap. In this experiment, the time average of
the vertical motion closely parallels the Euler fluid equations. In
fact, the cold plasma singular Vlasov-Poisson solution turns out to obey
the equations of point-vortex dynamics in an incompressible ideal flow.
This coincidence allows the discrete arrays of ``vortex crystals''
envisioned by J. J. Thomson for fluid vortices to be realized
experimentally as solutions of the Vlasov-Poisson equations. For a
survey of these experimental cold-plasma results see \cite{DuON1999}. 
\end{remark}

\medskip
The Vlasov equation is a Lie-Poisson system that may be expressed as 
\begin{eqnarray}
\frac{\partial f}{\partial t}
=\,-
\left\{\,f\,,\,\frac{\delta H}{\delta f}\,\right\}
=
\frac{\partial f}{\partial p}\,\frac{\partial}{\partial q}
\frac{\delta H}{\delta f}
-
\frac{\partial f}{\partial q}\,\frac{\partial}{\partial p}
\frac{\delta H}{\delta f}
=: -\,\textrm{\large ad}^*_{\delta H/\delta f}\,f
\label{vlasov-eqn}
\end{eqnarray}
Here the canonical Poisson bracket $\{\,\cdot\,,\,\cdot\,\}$
is defined for smooth functions on phase space with
coordinates $(q,p)$.  
The variational derivative $\delta H/\delta f$ regulates the
particle motion and the quantity ${\rm ad}^*_{\delta h/\delta f}\,f$ is explained as follows.

A functional $G(f)$ of the Vlasov distribution $f$ evolves
according to 
\begin{eqnarray}\nonumber
\frac{dG}{dt}
&=&
\int\hspace{-3mm}\int
\frac{\delta G}{\delta f}\,
 \frac{\partial f}{\partial t} 
\,{\rm d}q\,{\rm d}p \,
= -
\int\hspace{-3mm}\int
\frac{\delta G}{\delta f}
\left\{\,f\,,\,\frac{\delta H}{\delta f}\right\}
\,{\rm d}q\,{\rm d}p 
\\
&=&
\int\hspace{-3mm}\int\! f 
\left\{\frac{\delta G}{\delta f}\,,\,\frac{\delta H}{\delta f}\right\}
\,{\rm d}q\,{\rm d}p 
\,\,=:\,
\left\langle\!\!\!\left\langle f\,,\,
\left\{\frac{\delta G}{\delta f}\,,\,\frac{\delta H}{\delta f}\right\}
\right\rangle\!\!\!\right\rangle
\,=:\,
\big\{\,G\,,\,H\,\big\}
\end{eqnarray}
where one denotes with $\{\cdot,\,\cdot\}$ both the canonical and the
non-canonical Poisson brackets. In this calculation boundary terms were neglected upon integrating by
parts in the third step and the notation
$\langle\!\langle\,\cdot\,,\,\cdot\,\rangle\!\rangle$ is introduced for
the $L^2$ pairing in phase space. The quantity $\{\,G\,,\,H\,\}$ defined
in terms of this pairing is the Lie-Poisson Vlasov (LPV) bracket
\cite{WeMo}. This Hamiltonian evolution equation may also be expressed as
\begin{eqnarray}
\frac{dG}{dt}
=
\big\{\,G\,,\,H\,\big\}
=-
\left\langle\!\!\!\left\langle f\,,\,
\textrm{\large ad}\,_{\delta H/\delta f}
\frac{\delta H}{\delta f}
\right\rangle\!\!\!\right\rangle
=:
-\,
\left\langle\!\!\!\left\langle 
\textrm{\large ad}^*_{\,\delta H/\delta f}\,f\,,\,
\frac{\delta G}{\delta f}
\right\rangle\!\!\!\right\rangle
\end{eqnarray}
which defines the Lie-algebraic operations ad and ad$^*$ in this
case in terms of the $L^2$ pairing on phase space
$\langle\!\langle\,\cdot\,,\,\cdot\,\rangle\!\rangle$:
$\mathfrak{s}^*\times\mathfrak{s}\mapsto\mathbb{R}$. The
notation ${\rm ad}^*_{\delta H/\delta f}\,f$ expresses {\bfi coadjoint action} of $\delta
H/\delta f\in\mathfrak{s}$ on $f\in\mathfrak{s}^*$, where
$\mathfrak{s}$ is the Lie algebra of single particle Hamiltonian
vector fields and $\mathfrak{s}^*$ is its dual under $L^2$
pairing in phase space. This is the sense in which the Vlasov equation
represents coadjoint motion on the {\bfi group of symplectic diffeomorphisms} (symplectomorphisms). 

In order to give an explicit derivation of the LPV
structure from the Jacobi-Lie bracket for Hamiltonian vector fields (here
denoted
by $\mathfrak{X}_\text{can}$), one can follow the same steps as in section~\ref{diffvol} for the volume preserving vector fields $\mathfrak{X}_\text{vol}$ and use the following relation
\begin{align*}
\big[{\bf X}_\textit{\small h},\,
{\bf X}_\textit{\small k}\big]_{_\textit{\scriptsize{\!J\!L}}}
=
-{\bf X}_\textit{\small $\{h,k\}$}
=
-\Omega^\sharp\,{\bf\large d}\{h,k\}
\end{align*}
where $\boldsymbol\Omega=\Omega_{ij}\,{\rm d}q_i\wedge{\rm d}p_j$ is the symplectic form and $\boldsymbol\Omega^\sharp=(\Omega^\sharp)^{ij}\,\partial_{q_i}\wedge\partial_{p_j}$
is its inverse. In what follows we will consider canonical transformations
on the cotangent bundle $T^*\mathbb{R}^n$, so that $\Omega_{ij}=\mathbb{J}_{ij}=(\Omega^\sharp)^{ij}$,
where $\mathbb{J}$ is the symplectic matrix.
Thus, pairing the result with a one-form density ${\bf Y}\in\mathfrak{X}_\text{can}^*$
and integrating by parts yields
\rem{ 
\begin{align*}
\Big\langle{\bf Y},\,\big[{\bf X}_\textit{\small h},\,{\bf X}_\textit{\small k}\big]_{_\textit{\scriptsize{\!J\!L}}}\Big\rangle
=
-\left\langle{\bf Y},\,{\bf X}_\textit{\small$\{h,k\}$}\right\rangle
&=
-\Big\langle{\bf Y},\,\Omega^\sharp\,{\bf\large d}\{h,k\}\Big\rangle\\
&=
\Big\langle {\bf\large d}\{h,k\},\,\Omega^\sharp\,{\bf Y}\Big\rangle
=
\left\langle \delta\big(\Omega^\sharp\,{\bf Y}\big)
,\,\{h,k\}\,\right\rangle
\end{align*}
where the \emph{codifferential} $\delta$ has been introduced here as simply the dual operator of the exterior differential $\bf d$. 
In a vector space the differential is always the gradient and one writes
} 
\begin{align*}
\Big\langle{\bf Y},\,\big[{\bf X}_\textit{\small h},\,{\bf X}_\textit{\small k}\big]_{_\textit{\scriptsize{\!J\!L}}}\Big\rangle
&=
-\left\langle{\bf Y},\,{\bf X}_\textit{\small$\{h,k\}$}\right\rangle
=
-\Big\langle{\bf Y},\,\mathbb{J}\,\nabla\{h,k\}\Big\rangle\\
&=
\Big\langle {\bf\large \nabla}\{h,k\},\,\mathbb{J}\,{\bf Y}\Big\rangle
=
-\left\langle {\rm div}\big(\mathbb{J}\,{\bf Y}\big)
,\,\{h,k\}\right\rangle
=:
-\Big\langle f,\,\big\{h,k\big\}\Big\rangle
\end{align*}
where
\[
f:={\rm div}\big(\mathbb{J}\,{\bf Y}\big)\in\mathcal{F}^*
\]
is evidently a density variable dual to the space $\mathcal{F}$ of functions.
Thus, not only does one identify any Hamiltonian
function $h$ with its associated vector field ${\bf X}_\textit{\small h}$, but
also one associates a density variable $f$ with a one-form density ${\bf
Y}={\bf
Y}_\textit{\!\small f\,}$, which is defined modulo exact one-forms. Finally one checks the isomorphism $\mathfrak{X}_\text{can}\simeq\mathcal{F}$,
so that $\left\langle{\bf
Y}_\textit{\!f\,},{\bf X}_\textit{h}\right\rangle=\left\langle
f,\,h\right\rangle$.
In order to avoid confusion,
one denotes the Lie algebra of the symplectic group simply by {\large$\mathfrak{s}$}.

In higher dimensions, particularly $n = 3$, one may take the direct sum of the Vlasov Lie-Poisson bracket, together with with the Poisson bracket for an electromagnetic field (in the Coulomb gauge) where the electric field $\bf E$ and magnetic vector potential $\bf A$ are canonically conjugate.
For discussions of the Vlasov-Maxwell equations from a geometric viewpoint in the same spirit as the present approach, see \cite{WeMo,MaWe83,Ma82,MaWeRaScSp,CeHoHoMa1998}.
The Vlasov Lie-Poisson structure
was also extended to include Yang-Mills theories in \cite{GiHoKu1982}
and \cite{GiHoKu1983}.

\medskip
In statistical theories such as kinetic theory, the introduction of {\bfi
statistical moments} is a usual tool for extracting useful information from the probability
distribution. It is interesting to see how the dynamics of moments is also
a kind of Lie Poisson dynamics. First, consider moments of the form
\begin{eqnarray}
g_{\,\widehat{m},m}(t)
&=&
\int\hspace{-2.3mm}\int 
q^{\widehat{m}}\,p^m f (q,p,t)\,
\,{\rm d}q\,{\rm d}p 
\,.
\end{eqnarray}
These moments $g_{\,\widehat{m},m}$ are often used in
treating the collisionless dynamics of plasmas and particle beams
\cite{Ch83,Ch90,Dragt,DrNeRa92,Ly95,LyPa97}. This is usually done by considering low-order
truncations of the potentially infinite sum over phase space
moments,  
\begin{eqnarray}
G(t)\,
=\!
\sum_{\widehat{m},m=0}^\infty
a_{\,\widehat{m}m}\,g_{\,\widehat{m},m}
\,,\qquad\,
H(t)\,
=
\sum_{\widehat{n},n=0}^\infty
b_{\,\widehat{n}n}\,g_{\,\widehat{n},n}
\,,
\end{eqnarray}
with $\widehat{m},m,\widehat{n},n=0,1,\dots$. If $H$ is the
Hamiltonian, the sum over moments evolves under the Vlasov
dynamics according to the Lie-Poisson bracket relation
\begin{align}
\frac{dG}{dt}
\,\,=\,\,
\{\,G\,,\,H\,\}
\,\,\,&=\!
\nonumber
\sum_{\widehat{m},m,\widehat{n},n=0}^\infty
\left[
\frac{\pa G}{\pa g_{\widehat{m},m}}\,
\big(\widehat{m}\,m-\widehat{n}\,n\big)\,
\frac{\pa H}{\pa g_{\widehat{n},n}}
\right]
g_{\,\widehat{m}+\widehat{n}-1,\,m+n-1}
\\
&=:\!
\sum_{\widehat{m},m,\widehat{n},n=0}^\infty
\left\langle
g_{\,\widehat{m}+\widehat{n}-1,\,m+n-1}
\,,
\left[
\frac{\pa G}{\pa g_{\widehat{m},m}}\,,
\frac{\pa H}{\pa g_{\widehat{n},n}}
\right]
\right\rangle
\,,
\end{align}
where the Lie bracket 
\[
\big[
a_{\,\widehat{m}m}\,,\,b_{\,\widehat{n}n} 
\big]:=
a_{\,\widehat{m}m}\big(\widehat{m}\,m-\widehat{n}\,n\big)b_{\,\widehat{n}n}
\]
has been defined. Consequently, the Poisson bracket among
the moments is given by \cite{Ch90,Ch95,LyPa97,ScWe}
\[
\left\{
g_{\,\widehat{m},m}\,,\,g_{\,\widehat{n},n}
\right\}
\,\,=\,\,
\big(\widehat{m}\,m-\widehat{n}\,n\big)\,
g_{\,\widehat{m}+\widehat{n}-1,\,m+n-1}
\]
and the moment equations are written as
\begin{align*}
\frac{d g_{\,\widehat{m},m}}{d t}
\,\,=\,\,
\left\{
g_{\,\widehat{m},m}\,,\,H
\right\}
\,\,&=
\sum_{\widehat{n},n=0}^\infty\!
\textrm{\large ad}^*_\text{\normalsize$\frac{\pa H}{\pa g_{\widehat{n},n}}$}\,
\,g_{\,\widehat{m}+\widehat{n}-1,\,m+n-1}
\\
&=
\sum_{\widehat{n},n=0}^\infty\!
\big(\widehat{m}\,m-\widehat{n}\,n\big)\,
\frac{\pa H}{\pa g_{\widehat{n},n}}\,
\,g_{\,\widehat{m}+\widehat{n}-1,\,m+n-1}
\end{align*}
where the infinitesimal coadjoint action ad$^*$ has been defined as usual
\begin{align*}
\sum_{\widehat{m},m,\widehat{n},n=0}^\infty
\big\langle g_{\,\widehat{n},n}\,,\,
\big[a_{\,\widehat{m}m}\,,\,b_{\,\widehat{n}-\widehat{m}+1,\,n-m+1}\big]
\big\rangle
\,\,\,&=\!
\sum_{\widehat{m},m,\widehat{n},n=0}^\infty
\big\langle
{\rm ad}^*_{a_{\widehat{m}m}}\,g_{\,\widehat{n},n}\,,\,
b_{\,\widehat{n}-\widehat{m}+1,\,n-m+1}
\big\rangle
\end{align*}
so that
\begin{align*}
{\rm ad}^*_{a_{\widehat{m}m}}\,g_{\,\widehat{n},n}
&=
\Big(\widehat{m}
\big(\widehat{n}-\widehat{m}+1\big)
-
m
\big(n-m+1\big)
\Big)\,
a_{\widehat{m}m}\,g_{\,\widehat{n},n}
\\
&=
\big(\widehat{m}\,\widehat{n}-m\,n\big)
\,a_{\widehat{m}m}\,g_{\,\widehat{n},n}
-
\big(\widehat{m}^2-m^2\big)
\,a_{\widehat{m}m}\,g_{\,\widehat{n},n}
+
\big(\widehat{m}-m\big)
\,a_{\widehat{m}m}\,g_{\,\widehat{n},n}
\,.
\end{align*}

The symplectic invariants associated with Hamiltonian
flows of these moments \cite{LyOv88,DrNeRa92} were discovered and classified
in \cite{HoLySc1990}. Finite dimensional approximations of the whole moment
hierarchy were discussed in \cite{ScWe,Ch95}. For 
discussions of the Lie-algebraic approach to the control and  steering of
charged particle beams, see 
\cite{Dragt,Ch83,Ch90}.

Other than the statistical moments, also the {\bfi kinetic moments} can be introduced as projection integrals of the PDF over
the momentum coordinates only. In particular, in one dimension one defines the $n$-th moment as
\[
A_n(q,t)=\int_\text{\scriptsize-$\infty$}^\text{\scriptsize+$\infty$}\!\! p^n f(q,p,t)\,\,{\rm d}p
\,.
\]
Kinetic moments arise as important variables not only in kinetic theory, but
also in the theory of integrable shallow water equations \cite{Be1973,Gi1981}. The zero-th kinetic moment is the spatial mass density of particles as a function of space and time. The first kinetic moment is the mean fluid momentum density.

The next chapter shows how kinetic moment equations are also Lie-Poisson
equations and investigates the geometric meaning of these quantities.
Connections are established with well known integrable systems in the context
of shallow water theory and plasma dynamics. Later, a \emph{geodesic}
motion on the moments is constructed which generalizes the Camassa-Holm equation
and its multi-component versions, recovering singular solutions.


\bigskip

\chapter{Dynamics of kinetic moments} 
\label{momLPdyn}

\section{Introduction}

This chapter reviews the moment Lie-Poisson dynamics in the Kupershmidt-Manin
form \cite{KuMa1978,Ku1987,Ku2005} and provides a new geometric interpretation of the moments, which shows how the Lie-Poisson bracket is determined by the Schouten symmetric bracket on contravariant symmetric tensors \cite{GiHoTr2008}. New variational
formulations of moment dynamics are provided and the Euler-Poincar\'e
moment equations are formulated as a result.

This chapter also considers the action of cotangent lifts of diffeomorphisms
on the moments. The resulting geometric dynamics of the Vlasov kinetic moments
possesses singular solutions. These equations turn out to be related to the so called {\it $b$-hierarchy} \cite{HoSt03} exhibiting the spontaneous emergence
of singularities.  Moreover, when the kinetic moment equations are closed at the level of the first-order moment, their singular solutions are found to recover the peaked soliton of the integrable Camassa-Holm (CH) equation for shallow water waves \cite{CaHo1993}. These singular Vlasov moment solutions may also correspond to individual particle motion. The same treatment is extended
to include the dynamics of the zero-th moment, recovering the geometric structures
of fluid dynamics \cite{MaRa99}.

\section{Moment Lie-Poisson dynamics}
\label{Kup-Man}
\subsection{Review of the one dimensional case}
One of the most remarkable features of moment dynamics is that the Lie-Poisson dynamics is inherited from the Vlasov equation \cite{Gi1981}. That is, the evolution of the moments of the Vlasov PDF is also a form of Lie-Poisson dynamics. This fact has been used also in Yang-Mills theories by Gibbons,
Holm and Kupershimdt \cite{GiHoKu1982,GiHoKu1983}. In
order to show why this happens one considers functionals defined by,
\begin{align*}
G
&=
\sum_{m=0}^\infty
\int\hspace{-3mm}\int
 \alpha_m(q)\,p^m\,f\,{\rm d}q\,{\rm d}p
=
\sum_{m=0}^\infty
\int
 \alpha_m(q)\,A_m(q)\,{\rm d}q
=:
\sum_{m=0}^\infty\Big\langle A_m\,,\,\alpha_m\Big\rangle
\,,
\\
H
&=
\sum_{n=0}^\infty
\int\hspace{-3mm}\int
 \beta_n(q)\,p^n\,f\,\,{\rm d}q\,{\rm d}p
\,=\,
\sum_{n=0}^\infty
\int
 \beta_n(q)\,A_n(q)\,\,{\rm d}q
\,=:\,
\sum_{n=0}^\infty\Big\langle A_n\,,\,\beta_n\Big\rangle
\,,
\end{align*}
where $\langle\,\cdot\,,\,\cdot\,\rangle$ is the
$L^2$ pairing on position space.

The functions $\alpha_m$
and $\beta_n$ with $m,n=0,1,\dots$ are assumed to be suitably
smooth and integrable against the Vlasov moments. To
ensure these properties, one may relate the moments to the
previous sums of Vlasov  statistical moments by choosing
\begin{eqnarray}
\alpha_m(q)
&=&
\sum_{\widehat{m}=0}^\infty
a_{\,\widehat{m}m}q^{\,\widehat{m}}
\qquad\text{ and }\qquad
\beta_n(q)
=
\sum_{\widehat{n}=0}^\infty
b_{\,\widehat{n}n}q^{\,\widehat{n}}
\,.
\end{eqnarray}
For these choices of $\alpha_m(q)$ and $\beta_n(q)$, the sums of 
kinetic moments will recover the full set of Vlasov statistical moments.
Thus, as long as the statistical moments of the distribution $f(q,p)$
continue to exist under the Vlasov evolution, one may assume that the
dual variables $\alpha_m(q)$ and $\beta_n(q)$ are smooth functions whose
Taylor series expands the kinetic moments in the statistical moments. These functions are dual to the kinetic moments $A_m(q)$ with $m=0,1,\dots$ under the $L^2$ pairing $\langle\cdot\,,\,\cdot\rangle$ in the spatial variable $q$. In what follows one again assumes boundary conditions giving zero contribution
under integration by parts. This means, for example, that one can ignore boundary terms arising from integrations by parts. In what follows the term ``moment'' means kinetic moment, unless otherwise specified.

The Poisson bracket among the functionals $G=\langle A_m\,,\,\alpha_m\rangle$
and $H=\langle A_n\,,\,\beta_n\rangle$ (summation over $m,n$) is obtained from the Lie-Poisson bracket for the Vlasov equation via the following
explicit calculation, 
\begin{eqnarray*}
\{\,G\,,\,H\,\}
&=&
\sum_{m,n=0}^\infty
\int\hspace{-2.3mm}\int f 
\Big[\alpha_m(q)\,p^m\,,\,\beta_n(q)\,p^n\Big]
\,{\rm d}q\,{\rm d}p
\\
&=&
\sum_{m,n=0}^\infty
\int\hspace{-2.3mm}\int 
\Big[
n\beta_{n}\alpha_{m}^{\prime}\,-m\alpha_{m}\beta_{n}^{\,\prime}\Big]
f\,p^{m+n-1}
\,{\rm d}q\,{\rm d}p 
\\
&=&
\sum_{m,n=0}^\infty
\int
A_{m+n-1}(q)
\Big[
n\beta_{n}\alpha_{m}^{\prime}\,-m\alpha_{m}\beta_{n}^{\,\prime}\Big]
\,{\rm d}q
\\
&=:&
\sum_{m,n=0}^\infty\Big\langle 
A_{m+n-1}
\,,\,
\textrm{\large ad}_{\alpha_m}\,\beta_n
\Big\rangle 
\\
&=&
-\sum_{m,n=0}^\infty
\int
\Big[
n\beta_n A_{m+n-1}'
+(m+n)A_{m+n-1}\beta_n^{\,\prime}\Big]
\alpha_m
\,{\rm d}q
\\
&=:&
-\sum_{m,n=0}^\infty\Big\langle 
\textrm{\large ad}^*_{\beta_n}A_{m+n-1}
\,,\,
\alpha_m
\Big\rangle 
\end{eqnarray*}
where one integrates by parts assuming homogeneous boundary conditions and introduces the 
notation {\rm ad} and {\rm ad}$^*$ for adjoint and coadjoint action,
respectively. Upon recalling the dual relations 
\begin{eqnarray*}
\alpha_m=\frac{\delta G}{\delta A_m}
\quad\hbox{and}\quad
\beta_n=\frac{\delta H}{\delta A_n}
\end{eqnarray*}
the LPV bracket in terms of the moments may be expressed as
\begin{eqnarray}
\{\,G\,,\,H\,\}(\{A\})
&=&
\sum_{m,n=0}^\infty
\int  A_{m+n-1}
\left(\,
n
\frac{\delta H}{\delta A_n} 
\frac{\partial}{\partial q}
\frac{\delta G}{\delta A_m}
-
m
\frac{\delta G}{\delta A_m}
\frac{\partial}{\partial q}
\frac{\delta H}{\delta A_n}
\right)
{\rm d}q
\nonumber \\
&=:&
\sum_{m,n=0}^\infty
\left\langle 
A_{m+n-1}
\,,\,
\left[\frac{\delta G}{\delta A_m}\,,\,
\frac{\delta H}{\delta A_n}\right]
\right\rangle 
\label{KMLP-brkt}
\end{eqnarray}
where one introduces the compact notation $\{A\}:=\{A_n\}$ with $n$ a non-negative integer. 
This is the Kupershmidt-Manin Lie-Poisson (KMLP) bracket
\cite{KuMa1978}, which is defined for
functions on the dual of the Lie algebra with bracket
\begin{eqnarray}\label{LieBracket-mom}
[\,\alpha_m\,,\,\beta_n\,]
=
n\beta_n\partial_q\alpha_m-m\alpha_m\partial_q\beta_n
\,.
\end{eqnarray}
This Lie algebra bracket inherits the Jacobi identity
from its definition in terms of the canonical Hamiltonian
vector fields. 
Also, for $n=m=1$ this Lie bracket reduces to minus the Jacobi-Lie bracket for the vector fields $\alpha_1$ and $\beta_1$.
Thus, one has recovered the following

\noindent
{\bf Theorem (Gibbons \cite{Gi1981})}\\
{\it 
The operation of taking
kinetic moments of Vlasov solutions is a Poisson map. It takes the LPV bracket
describing the evolution of $f(q,p)$ into the KMLP bracket, describing the
evolution of the kinetic moments $A_n(x)$. 
}

A result related to this, for
the Benney hierarchy \cite{Be1973}, was also presented by Lebedev and Manin
\cite{Le1979,LeMa}. Although the moment bracket is a Lie-Poisson bracket,
strictly speaking the solutions for the moments cannot yet be claimed to undergo coadjoint motion, as in the case of the Vlasov PDF solutions, because the group action underlying the Lie-Poisson structure of the moments is not yet understood and thus the Ad$^*$ group operation is not defined.
For example, it is not possible to express the Ad$^*$ operation on the moments by simply starting from the coadjoint motion on the PDF, as shown by the following calculation:
\begin{align*}
\left\langle A_n(t),\,\beta_n\right\rangle&=
\left\langle\!\left\langle f(t),\,p^n\beta_n\right\rangle\!\right\rangle
\\
&=
\left\langle\!\left\langle {\rm Ad}^*_{g^{-1}}f(0),\,p^n\beta_n\right\rangle\!\right\rangle
=\left\langle \int p^n{\rm Ad}^*_{g^{-1}}f(0)\,{\rm d}p,\,\beta_n\right\rangle
\end{align*}
so that
\[
A_n(q,t)=\int p^n{\rm Ad}^*_{g^{-1}}f(0)\,{\rm d}p
=\int p^n\left.\big(f(0)\circ g^{-1}(q,p)\big)\right.{\rm d}p
\]
and the right hand side cannot be expressed as an evolution map for the sequence of moments $\left\{A_n\right\}$.

The evolution of a particular moment $A_m(q,t)$ is obtained
from the KMLP bracket by
\begin{align}\nonumber
\!\!\!\!\!
\frac{\partial  A_m}{\partial t}
=-\,\textrm{\large ad}^*_\text{\normalsize$\frac{\delta H}{\delta A_n}$}\, A_{m+n-1}
&=
\{\,A_m\,,\,H\,\}
\\
&=
-\sum_{n=0}^\infty
\Big(n\frac{\partial}{\partial q} A_{m+n-1}
+
mA_{m+n-1}\frac{\partial}{\partial q}
\Big)\,
\frac{\delta H}{\delta A_n}
\label{KMLPbrkt}
\end{align}
The KMLP bracket among the moments is given by
\begin{eqnarray*}
\{\,A_m\,,\,A_n\,\}
&=&
-n\frac{\partial}{\partial q} A_{m+n-1}
-
mA_{m+n-1}\frac{\partial}{\partial q}
\end{eqnarray*}
expressed as a differential operator acting to the
right. This operation is skew-symmetric under the $L^2$  pairing and the
general KMLP bracket can then be written as \cite{Gi1981}
\[
\{\,G\,,H\,\}\left(  \,\left\{  A\right\}  \right)  =\sum_{m,n=0}^{\infty}
\int\frac{\delta G}{\delta A_{m}}\{\,A_{m}\,,\,A_{n}\,\}\frac{\delta H}{\delta
A_{n}}\,{\rm d}q
\]
so that
\[
\frac{\partial A_{m}}{\partial t}=\sum_{n=0}^{\infty}\{\,A_{m}\,,\,A_{n}
\,\}\frac{\delta H}{\delta A_{n}}.
\]

\medskip
\begin{remark}\label{tensor-mom}
The moments have an important geometric interpretation, which has never appeared
in the literature so far. Indeed one can write the moments as
\begin{equation}
A_n=\int_p \,\otimes^n(p\,dq)\, f(q,p)\, dq\wedge dp=A_n(q)\otimes^{n}\!{\rm d}q\otimes{\rm d}Vol
\end{equation}
where $\otimes^{n}{\rm d}q:={\rm d}q\otimes\dots\otimes {\rm d}q\,$ $n$ times
and \textnormal{d}$Vol$ is the volume element in physical space. Thus,  moments $A_n$ 
belong to the  vector space dual to the contravariant tensors of the type 
$\beta_n=\beta_n(q)\otimes^{n}\!\partial_q$.
These tensors are given a Lie algebra structure by the Lie bracket
\begin{equation}
\left[\alpha_m,\,\beta_n\right]=
\big(\,n\,\beta_n(q)\,\alpha_m'(q)
-
m\,\alpha_m(q)\,\beta_n^{\,\prime}(q)\,\big)\otimes^{n+m-1}\!\partial_q=:\,
\textnormal{\large ad}_{\alpha_m}\, \beta_n
\label{LieStruct-KM}
\end{equation}
so that the ${\,\rm ad}^*$ operator is defined by $\langle\, {\rm ad}^*_{\beta_n} \,A_k,\,\alpha_{k-n+1}\,\rangle:=
\langle\, A_k,\,{\rm ad}_{\beta_n}\,\alpha_{k-n+1}\,\rangle$ 
and is given explicitly
as
\begin{equation}\label{coadjoint-mom}
\textnormal{\large ad}_{\beta_{n}}^{\ast}A_{k}=\left(
n\,\beta_{n}\,\frac{\partial
A_{k}}{\partial q}
+
\left(  k+1\right) \, A_{k}\,\frac{\partial
\beta_{n}}{\partial q}
\right)\otimes^{k-n+1}\!{\rm d}q\otimes{\rm d}Vol\,.
\end{equation}
\end{remark}

\medskip
The equations for the ideal compressible fluid are recovered by the
moment hierarchy by simply truncating at the first order moment. In
fact the moment equations become in this case
\begin{align*}
\frac{\partial A_{0}}{\partial t}&=-\textnormal{\large ad}_{\beta_{1}}^{\ast}\,A_0
=\frac{\partial}{\partial q}\left(A_0\,\frac{\delta H}{\delta A_1}\right)
\\
\frac{\partial A_{1}}{\partial t}&=
-\textnormal{\large ad}_{\beta_{1}}^{\ast}A_{1}
-\textnormal{\large ad}_{\beta_{0}}^{\ast}A_{0}
=
-
\frac{\delta H}{\delta A_1}\frac{\partial A_1}{\!\partial q}
-
2A_{1}\frac{\partial}{\partial q}\frac{\delta H}{\delta A_1}
-A_0\,\frac{\partial}{\partial q}\frac{\delta H}{\delta A_0}
\,.
\end{align*}
which are the equations for ideal compressible fluids when the Hamiltonian
is written as $H=\frac12\int A_1^2/A_0\,{\rm d}x$, so that the fluid velocity
is $u=\delta H/\delta A_1=A_1/A_0$.

Given the beauty and utility of the solution behavior for fluid
equations for the first moments, one is intrigued to know more about
the dynamics of the other moments of Vlasov's equation. Of course, the dynamics of the moments of the Vlasov-Poisson equation is one of the mainstream subjects of plasma physics and space physics, which are the main inspiration and motivation
for the present work.

\subsection{Multidimensional treatment I: background}
\label{3Dmoments}

One can show that the KMLP bracket and the equations of motion may be written in three dimensions in multi-index notation.
By writing $\mathbf{p}^{2n+1}=\,p^{2n}\,  \mathbf{p}$, and checking that:
\begin{align*}
p^{2n} 
& =
\underset{i+j+k=n}{\sum}
\dfrac{n!}{i!j!k!}\,p_{1}^{2i}p_{2}^{2j}p_{3}^{2k}\\
\end{align*}
it is easy to see that the multidimensional treatment can be performed in terms of the quantities
\[
p^{\sigma}
=:
p_{1}^{\sigma_{1}}\,p_{2}^{\sigma_{2}}\,p_{3}^{\sigma_{3}}
\]
\bigskip where $\sigma=\left(\sigma_{1},\sigma_{2},\sigma_{3}\right)\in\mathbb{N}^{3}$. Let $A_{\sigma}$ be defined as \cite{Ku1987,Ku2005}
\[
A_{\sigma}
\left(  
\mathbf{q},t\right)  
=:
{\int}
p^{\sigma}
f\left(\mathbf{q},\mathbf{p},t\right)
{\rm d}^3{\bf p}
\]
and consider functionals of the form
\begin{align*}
G  
& =
{\sum\limits_{\sigma}}
{\iint}
\alpha_{\sigma}\left(  \mathbf{q}\right)  
p^{\sigma}
f\left(  \mathbf{q},\mathbf{p},t\right)
{\rm d}^3{\bf q}\,{\rm d}^3{\bf p}
=:
{\sum\limits_{\sigma\in\mathbb{N}^{3}}}
\left\langle 
A_{\sigma},
\alpha_{\sigma}
\right\rangle \\
H  
& =
{\sum\limits_{\rho}}
{\iint}
\beta_{\rho}
\left(  \mathbf{q}\right)  
p^{\,\rho}
f\left(  \mathbf{q},\mathbf{p},t\right)
{\rm d}^3{\bf q}\,{\rm d}^3{\bf p}
=:
{\sum\limits_{\rho\in\mathbb{N}^{3}}}
\left
\langle A_{\rho},
\beta_{\rho}
\right\rangle
\end{align*}
The ordinary LPV bracket leads to: 

\begin{align*}
\left\{ G,H\right\} 
& =
\sum\limits_{\sigma ,\rho }\iint 
f
\left[ 
\alpha _{\sigma }
\left( 
\mathbf{q}
\right) 
p^{\sigma },
\beta _{\rho}
\left( 
\mathbf{q}
\right)
p^{\,\rho }
\right] 
{\rm d}^3{\bf q}\,{\rm d}^3{\bf p}
=
\\
& =
-
\sum\limits_{\sigma ,\rho }\sum\limits_{j}
\iint
f
\left( 
\alpha _{\sigma }
p^{\,\rho }
\dfrac{\partial p^{\sigma }}{\partial p_{j}}
\dfrac{\partial \beta _{\rho }}{\partial q_{j}}
-
\beta _{\rho }
p^{\sigma }
\dfrac{\partial p^{\,\rho }}{\partial p_{j}}
\dfrac{\partial \alpha _{\sigma }}{\partial q_{j}}
\right) 
{\rm d}^3{\bf q}\,{\rm d}^3{\bf p}
= 
\\
& =
-
\sum\limits_{\sigma ,\rho }\sum\limits_{j}
\iint
f
\left( 
\sigma _{j}
\alpha _{\sigma }\,
p^{\,\rho }\,
p^{\sigma -1_{\hbox{\normalsize\it j}}}\,
\dfrac{\partial \beta _{\rho }}{\partial q_{j}}
-
\rho _{j}
\beta _{\rho }\,
p^{\sigma }\,
p^{\,\rho -1_{\hbox{\normalsize\it j}}}\,
\dfrac{\partial \alpha _{\sigma }}{\partial q_{j}}
\right) 
{\rm d}^3{\bf q}\,{\rm d}^3{\bf p}
= 
\\
& =
-
\sum\limits_{\sigma ,\rho }\sum\limits_{j}
\int
A_{\sigma +\rho -1_{\hbox{\normalsize\it j}}}\,
\left( 
\sigma _{j}
\alpha_{\sigma }
\dfrac{\partial \beta _{\rho }}{\partial q_{j}}
-
\rho _{j}\beta_{\rho }
\dfrac{\partial \alpha _{\sigma }}{\partial q_{j}}
\right) 
{\rm d}^3{\bf q}
=
\\
& =:
-
\sum\limits_{\sigma ,\rho }\sum\limits_{j}
\left\langle
A_{\sigma +\rho -1_{\hbox{\normalsize\it j}}}\,,
\left( 
\mathrm{ad}_{\beta_{\rho }}
\right) 
_{j}
\alpha _{\sigma }
\right\rangle 
= 
\\
& =
-
\sum\limits_{\sigma ,\rho }\sum\limits_{j}
\int 
\left[
\rho _{j}
\beta _{\rho }
\dfrac{\partial }{\partial q_{j}}
A_{\sigma +\rho -1_{\hbox{\normalsize\it j}}}\,
+\left( 
\sigma _{j}+\rho _{j}\right) 
A_{\sigma+\rho -1_{\hbox{\normalsize\it j}}}\,
\dfrac{\partial \beta _{\rho }}{\partial q_{j}}
\right] 
\alpha _{\sigma }\,
{\rm d}^3{\bf q}
= 
\\
& =:
-
\sum\limits_{\sigma ,\rho }\sum\limits_{j}
\left\langle 
\left( 
\mathrm{ad}_{\beta _{\rho }}^{\ast }
\right) _{j}
A_{\sigma +\rho -1_{\hbox{\normalsize\it j}}\,},
\alpha _{\sigma}
\right\rangle 
\end{align*}
where the sum is extended to all $\sigma ,\rho \in \mathbb{N}^{3}$ and
one introduces the notation,
\begin{equation*}
_{1_{\hbox{\normalsize\it j}}\,
=:
\,
(0,...\underset{\underset{\overbrace{
j^{th}\,\text{element}}}{\uparrow }}{,1,}...,0)}
\end{equation*}
so that $\left( 1_{j}\right) _{i}=\delta _{ji}$.

The LPV bracket in terms of the moments may then be written as 
\begin{equation*}
\frac{\partial A_{\sigma }}{\partial t}
=
-
\sum\limits_{\rho \in \mathbb{N}^{3}}
\sum\limits_{j}
\left( 
\mathrm{ad}_{\frac{\delta h}{\delta A_{\rho }}}^{\ast }
\right) 
_\text{\footnotesize$\!\!j$}
A_{\sigma +\rho +1_{\hbox{\normalsize\it j}}}
\end{equation*}
where the Lie bracket is now expressed as 
\begin{equation*}
\left[ 
\frac{\delta G}{\delta A_{\sigma }},
\frac{\delta H}{\delta A_{\rho }}
\right] 
_{\hbox{\normalsize\it j}}
=
\rho _{j}\,
\frac{\delta H}{\delta A_{\rho }}\,
\dfrac{\partial }{\partial q_{j}}
\frac{\delta G}{\delta A_{\sigma }}
-
\sigma _{j}\,
\frac{\delta G}{\delta A_{\sigma }}\,
\dfrac{\partial }{\partial q_{j}}
\frac{\delta H}{\delta A_{\rho }}
\,.
\end{equation*}

\paragraph{Properties of the multidimensional moment bracket.}
The evolution of a particular moment $A_{\sigma }$ is obtained
by 
\begin{align*}
\frac{\partial A_{\sigma }}{\partial t}
& =
\left\{ 
A_{\sigma },h
\right\} 
= 
\\
& =
-
\sum\limits_{\rho}
\sum\limits_{j}
\left[ 
\rho _{j}
\frac{\delta h}{\delta A_{\rho }}
\dfrac{\partial }{\partial q_{j}}
A_{\sigma+\rho -1_{\hbox{\normalsize\it j}}}
+
\left( 
\sigma _{j}+\rho _{j}
\right)
A_{\sigma +\rho -1_{\hbox{\normalsize\it j}}}\,
\dfrac{\partial }{\partial q_{j}}
\frac{\delta h}{\delta A_{\rho }}
\right]
\end{align*}
and the {\bfi KMLP bracket among moments} is given by
\begin{equation*}
\left\{ 
A_{\sigma },A_{\rho }
\right\} 
=
-
\sum\limits_{j}
\left( 
\sigma _{j}
\frac{\partial }{\partial q_{j}}
A_{\sigma +\rho -1_{\hbox{\normalsize\it j}}}\,
+
\rho _{j}
A_{\sigma +\rho -1_{\hbox{\normalsize\it j}}}\,
\frac{\partial }{\partial q_{j}}
\right).
\end{equation*}
Inserting the previous operator in this multi-dimensional KMLP bracket
leads to
\begin{equation*}
\left\{ 
g,h
\right\} 
\left( 
\left\{ 
A
\right\} 
\right) 
=
\sum\limits_{\sigma,\rho}
\int 
\frac{\delta g}{\delta A_{\sigma }}
\left\{
A_{\sigma },
A_{\rho }
\right\} 
\frac{\delta h}{\delta A_{\rho }}\,
{\rm d}^3{\bf q}
\end{equation*}
and the corresponding evolution equation becomes
\begin{equation*}
\frac{\partial A_{\sigma }}{\partial t}
=
\sum\limits_{\rho}
\left\{ 
A_{\sigma },
A_{\rho }
\right\} 
\frac{\delta h}{\delta A_{\rho }}.
\end{equation*}
Thus, in multi-index notation, the form of the Hamiltonian evolution
under the KMLP bracket is essentially {\it unchanged} in going to higher
dimensions. 

\medskip
\subsection{Multidimensional treatment II: a new result}
Besides the multi-index notation, it is also possible to extend the discussion
in remark~\ref{tensor-mom} so to emphasize the tensor interpretation
of the moments. Indeed, one can define the moments as
\begin{equation}\label{tensor-mom-def}
A_n({\bf q}, t)=\int_{T_{\bf q}^*Q} \!\otimes^{n}\left({\bf p}\cdot{\rm
d}{\bf q}\right)\, f({\bf q,
p}, t)\,\,{\rm d}^3{\bf q}\wedge{\rm d}^3{\bf p}
\end{equation}
which can be written in tensor notation as \cite{GiHoTr2008}
\begin{align*}
A_n({\bf q}, t)&=\int_{T_{\bf q}^*Q} \! \left(p_i\,{\rm d}q^i\right)^n f({\bf q,
p}, t)\,{\rm d}^3{\bf q}\wedge{\rm d}^3{\bf p}
\\
&=\int_{T_{\bf q}^*Q}  p_{\,i_1}\dots p_{\,i_n} \,{\rm d}q^{i_1}\dots{\rm d}q^{i_n}\,f({\bf q,
p}, t)\,{\rm d}^3{\bf q}\wedge{\rm d}^3{\bf p}
\\
&=
\left(A_n({\bf q}, t)\right)_{i_1\dots i_n}
\,{\rm d}q^{i_1}\dots{\rm d}q^{i_n}\,{\rm d}^3{\bf q}
\end{align*}
This interpretation is consistent with the moment Lie-Poisson bracket. In fact one can follow the same steps
\begin{align*}
\{\,G\,,\,H\,\}
=&
\int\hspace{-2.3mm}\int f \,
\Big[\alpha_m({\bf q})\contract\otimes^{m}{\bf p},\,\beta_n({\bf
q})\contract\otimes^{n}{\bf p}\Big]
\,{\rm d}^3{\bf q}\wedge{\rm d}^3{\bf p}
\\
=&
\int\hspace{-2.3mm}\int f 
\bigg[\,p_{i_1}\dots p_{i_m}\frac{\pa \left(\alpha_m\right)^{i_1,\dots,i_m}}{\pa q^k}\frac{\pa\, p_{j_1\!}\dots p_{j_n}}{\pa p_k}\,\left(\beta_n\right)^{j_1,\dots,j_n}
\\
&\qquad\qquad\qquad\!-
p_{j_1}\dots p_{j_n}\frac{\pa \left(\beta_n\right)^{j_1,\dots,j_n}}{\pa q^h}\frac{\pa\, p_{i_1\!}\dots p_{i_m}}{\pa p_{h}}\,\left(\alpha_m\right)^{i_1,\dots,i_m}
\bigg]
\,{\rm d}^3{\bf q}\wedge{\rm d}^3{\bf p}
\\
=&
\int\hspace{-2.3mm}\int f 
\bigg[\,n\,p_{i_1}\dots p_{i_m}\,p_{j_1\!}\dots p_{j_{n-1}}
\left(\beta_n\right)^{j_1,\dots,\,j_{n-1\!},\,k}\frac{\pa \left(\alpha_m\right)^{i_1,\dots,i_m}}{\pa q^k}
\\
&\qquad\quad-
m\,p_{j_1}\dots p_{j_n}\,p_{i_1\!}\dots p_{i_{m-1}}
\left(\alpha_m\right)^{i_1,\dots,\,i_{m-1\!},\,h}
\frac{\pa \left(\beta_n\right)^{j_1,\dots,j_n}}{\pa q^h}\,
\bigg]
\,{\rm d}^3{\bf q}\wedge{\rm d}^3{\bf p}
\\
=&
\int\hspace{-2.3mm}\int f \,p_{j_1\!}\dots p_{j_{m+n-1}}
\bigg[\,n
\left(\beta_n\right)^{j_{m+1},\dots,\,j_{m+n-1\!},\,k\,}\frac{\pa \left(\alpha_m\right)^{j_{1},\dots,j_{m}}}{\pa q^k}
\\
&\hspace{3.65cm}-
m
\left(\alpha_m\right)^{j_{n+1},\dots,\,j_{m+n-1\!},\,h\,}
\frac{\pa \left(\beta_n\right)^{j_{1},\dots,j_{n}}}{\pa q^h}\,
\bigg]
\,{\rm d}^3{\bf q}\wedge{\rm d}^3{\bf p}
\\
\rem{ 
=&
\sum_{m,n=0}^\infty
\int\hspace{-2.3mm}\int 
\Big[
n\,(\beta_{n}\contract\nabla)\alpha_{m}\,
-
m\,(\alpha_{m}\contract\nabla)\beta_{n}
\Big]\!
:\otimes^{n+m-1}{\bf p}\,f
\,{\rm d}q\,{\rm d}p 
\\
} 
=&
\sum_{m,n=0}^\infty
\bigg\langle
A_{m+n-1},
\Big[
n\,(\beta_{n}\contract\nabla)\alpha_{m}\,
-
m\,(\alpha_{m}\contract\nabla)\beta_{n}
\Big]
\bigg\rangle
\\
:=&\sum_{m,n=0}^\infty\Big\langle 
A_{m+n-1}
\,,\,
\textrm{\large ad}_{\alpha_m}\,\beta_n
\Big\rangle 
\end{align*}
where the notation $\beta_{n}\contract\nabla$ stands for contraction of indexes
$\beta_{n}\contract\nabla=\left(\beta_{n}\right)^{i_1,\dots,i_n}\contract\partial_{\,i_n}$
and the square bracket in the penultimate step identifies a Lie bracket ${\rm
ad}_{\alpha_m}\,\beta_n$, also known as {\bfi symmetric Schouten bracket} \cite{BlAs79,Ki82,DuMi95} (see remark below). The last expression is the Lie-Poisson bracket on the moments in terms of symmetric tensors.

One also checks that
\begin{align*}
\{G,H\}=&\sum_{n,m=0}^\infty\Big\langle 
A_{m+n-1}
\,,\,
\big[{\alpha_m},\,\beta_n\big]
\Big\rangle\\
=&-
\int\hspace{-2.3mm}\int p_{j_1\!}\dots p_{j_{m+n-1}}
\bigg[\,n\,\left(\alpha_m\right)^{j_{1},\dots,j_{m}}
\frac{\pa}{\pa q^k}\!\left(f \,\left(\beta_n\right)^{j_{m+1},\dots,\,j_{m+n-1\!},\,k\,}\right)
\\
&\hspace{3.25cm}+
m\,f \,
\left(\alpha_m\right)^{j_{n+1},\dots,\,j_{m+n-1\!},\,h\,}
\frac{\pa \left(\beta_n\right)^{j_{1},\dots,j_{n}}}{\pa q^h}\,
\bigg]
\,{\rm d}^3{\bf q}\wedge{\rm d}^3{\bf p}
\\
=&-
\int\hspace{-2.3mm}\int p_{j_1\!}\dots p_{j_{m+n-1}}
\bigg[\,n\,\left(\alpha_m\right)^{j_{1},\dots,j_{m}}
\left(\beta_n\right)^{j_{m+1},\dots,\,j_{m+n-1\!},\,k\,}
\frac{\pa f}{\pa q^k}
\\
&\hspace{3.2cm}
+n\,f\left(\alpha_m\right)^{j_{1},\dots,j_{m}}
\frac{\pa \left(\beta_n\right)^{j_{m+1},\dots,\,j_{m+n-1\!},\,k\,}}{\pa q^k}
\\
&\hspace{3.2cm}+
m\,f \,
\left(\alpha_m\right)^{j_{n+1},\dots,\,j_{m+n-1\!},\,h\,}
\frac{\pa \left(\beta_n\right)^{j_{1},\dots,j_{n}}}{\pa q^h}\,
\bigg]
\,{\rm d}^3{\bf q}\wedge{\rm d}^3{\bf p}
\\
=&
-\sum_{m,n=0}^\infty\bigg(
n\,
\Big\langle 
(\beta_n\contract\nabla)A_{m+n-1},\,\alpha_m
\Big\rangle
+
n\,
\Big\langle 
(\nabla\contract\beta_n)A_{m+n-1},\,\alpha_m
\Big\rangle
\\
&\hspace{7.25cm}
+m\,
\Big\langle 
A_{m+n-1}\nabla\beta_n,\,\alpha_m
\Big\rangle\bigg)
\\
=:&
-\sum_{m,n=0}^\infty\Big\langle 
\textrm{\large ad}^*_{\beta_n}A_{m+n-1}
\,,\,
\alpha_m
\Big\rangle 
\end{align*}
Consequently, we have proven the following
\begin{framed}
\begin{proposition}[\cite{GiHoTr2008}]
The tensor interpretation of the moments (\ref{tensor-mom-def}) leads to a Lie-Poisson
structure, which involves a Lie bracket that generalizes the Jacobi-Lie bracket
to {\sl symmetric contravariant tensors}. This Lie bracket is called {\sl symmetric Schouten bracket} and the corrsponding Lie-Poisson structure is given by
\[
\left\{F,G\right\}=
\sum_{n,m=0}^\infty
\left\langle
A_{m+n-1},\left[ n
\left(\frac{\delta E}{\delta A_n}\contract\,\nabla\right)\frac{\delta F}{\delta A_m}
-
m
\left(\frac{\delta F}{\delta A_m}\contract\,\nabla\right)\frac{\delta E}{\delta A_n}\right]
\right\rangle
\]
\end{proposition}
\end{framed}
In particular, all the considerations made for the one-dimensional
case are valid also in the tensor interpretation of the higher dimensional
treatment.

The tensor equation for the $n$-th moment will then involve $3^n$ components, which are symmetric so that the number of equations for each moment may be appropriately reduced to $1/2\cdot(n+2)!/n!=(n+2)(n+1)/2$. An interesting
example is given by $A_2$, which is the pressure tensor such that Tr$\left(A_2\right)$/2
is the density of kinetic energy. However, given the level of difficulty of this problem, the following discussion will mainly restrict to the one-dimensional case.

\begin{remark}[The symmetric Schouten bracket] The tensor interpretation of the
moments provides a direct identification between the moment Lie bracket and the
so called symmetric Schouten bracket (or concomitant). This bracket was known
to Schouten as an invariant differential operator and its relation with
the polynomial algebra of the phase-space functions is very well known. The
symmetric Schouten bracket has been object of some studies in differential
geometry \cite{BlAs79,DuMi95} and its connection with the Lie-Poisson dynamics for the Vlasov
moments has never been established so far. However it is important to notice
that the Lie-Poisson bracket functional was known to Kirillov \cite{Ki82},
although not in relation with Vlasov dynamics, but rather he studied
such structures in connection with invariant differential operators. In particular, Kirillov
was the only author who noticed how this bracket functional can generate what
 has been here called coadjoint operator (${\rm ad}^*_{\beta_h}\,A_k$),
``which, apparently, has so far not been considered'', he claimed in 1982
(the Kupershmidt-Manin operator was known since 1977). What he claimed
 had been considered were the case $h=1$, which is the Lie derivative, and the case
$h=k$, which is often called ``Lagrangian Schouten concomitant''. It is easy to calculate from eq.~(\ref{coadjoint-mom}) that this operation with $h=k$ is the {\sl diamond operation} ${\rm ad}^*_{\beta_h}\,A_k=\beta_k\diamond A_k$, which will be defined in chapter~\ref{GOP} as the dual of the Lie derivative.
\end{remark}

\section{New variational principles for moment dynamics}
\label{VarPrincHP}
This section shows how the moment dynamics can be derived from
Hamilton's principle both in the Hamilton-Poincar\'e and Euler-Poincar\'e
forms. These variational principles are defined, respectively, on the
dual Lie algebra $\mathfrak{g}^{\ast}$ containing the moments, and on the
Lie algebra $\mathfrak{g}$ itself. For further details about these dual
variational formulations, see \cite{CeMaPeRa} and \cite{HoMaRa}. Summation
over repeated indices is intended in this section

\subsection{Hamilton-Poincar\'e hierarchy} One begins with the Hamilton-Poincar\'{e} principle for the $p-$moments written
as%
\[
\delta
\int_{t_{i}}^{t_{j}}
{\rm d}t
\left.\big(  
\left\langle 
A_{n},
\beta_{n}
\right\rangle 
-
H
\left(  
\left\{
A
\right\}  
\right)  
\right.
\big)  
=
0
\]
(where $\beta_{n}\in\mathfrak{g}$). It is possible to prove that this leads
to the same dynamics as found in the context of the KMLP bracket. To this
purpose, one must define the $n-$th moment in terms of the Vlasov
distribution function. One checks that%
\begin{align*}
0 &  =\delta
\int_{t_{i}}^{t_{j}}
{\rm d}t
\left(  
\left\langle 
A_{n},
\beta_{n}
\right\rangle 
-
H
\left(  
\left\{
A
\right\}  
\right)  
\right)  
=
\\
&  =
\int_{t_{i}}^{t_{j}}
{\rm d}t
\left(  
\delta
\left\langle\! 
\left\langle 
f,
p^{n}
\beta_{n}
\right\rangle\!
\right\rangle 
-
\left\langle\!\!\! 
\left\langle 
\delta f,
\dfrac{\delta H}{\delta f}
\right\rangle\!\!\! 
\right\rangle 
\right)  
=
\\
&  =
\int_{t_{i}}^{t_{j}}
{\rm d}t
\left(  
\left\langle\!\!\! 
\left\langle 
\delta f,
\left(  
p^{n}
\beta_{n}
-
\dfrac{\delta H}{\delta f}
\right)  
\right\rangle\!\!\! 
\right\rangle 
+
\left\langle\!
\left\langle 
f,
\delta
\left(  
p^{n}
\beta_{n}
\right)  
\right\rangle\!
\right\rangle 
\right)
\end{align*}

Now recall that any $g=\delta G/\delta f$ belonging to the Lie algebra
$\mathfrak{s}$ of the symplectomorphisms (whose dual $\mathfrak{s}^*$ contains the
distribution function itself) may be expressed as
\begin{align*}
g  &  =
\frac{\delta G}{\delta f}=
p^{m}
\frac{\delta G}{\delta A_{m}}=
p^{m}\xi_{m}
\end{align*}
by the chain rule. Consequently, one finds the pairing relationship,
\[
\left\langle\!\!\! 
\left\langle 
\delta f,
\left(  
p^{n}
\beta_{n}
-
\dfrac{\delta H}{\delta f}
\right)  
\right\rangle\!\!\! 
\right\rangle 
=
\left\langle 
\delta A_{n},
\left(  
\beta_{n}
-
\dfrac{\delta H}{\delta A_{n}}
\right)  
\right\rangle
\]
Next, recall from the general theory that variations on a Lie group
induce variations on its Lie algebra of the form
\[
\delta w
=
\dot{u}
+
\left[  
g,u
\right]
\]
where $u,w\in\mathfrak{s}$ and $u$ vanishes at the endpoints. Writing
$u=p^{m}\eta_{m}$ then leads to
\begin{align*}
\int_{t_{i}}^{t_{j}}
{\rm d}t
\left\langle\! 
\left\langle 
f,
\delta
\left(  
p^{n}\beta_{n}
\right)
\right\rangle\! 
\right\rangle  
&  =
\int_{t_{i}}^{t_{j}}
{\rm d}t
\left\langle\! 
\left\langle 
f,
\left(  
\dot{u}
+
\left[
p^{n}
\beta_{n},
u
\right]  
\right)  
\right\rangle\! 
\right\rangle 
=\\
&  =
-
\int_{t_{i}}^{t_{j}}
{\rm d}t
\left(  
\left\langle 
\dot{A}_{m},
\eta_{m}
\right\rangle 
-
\left\langle
A_{n+m-1},
\left[\!
\left[  
\beta_{n},
\eta_{m}
\right]\!
\right]  
\right\rangle
\right)  
=\\
&  =-
\int_{t_{i}}^{t_{j}}
{\rm d}t
\left\langle 
\left(  
\dot{A}_{m}
+\mathrm{ad}_{\beta_{n}}^{\ast}
A_{m+n-1}
\right),
\eta_{m}
\right\rangle
\end{align*}
Consequently, the Hamilton-Poincar\'{e} principle may be written entirely
in terms of the moments as
\[
\delta S=
\int_{t_{i}}^{t_{j}}
{\rm d}t
\left\{  
\left\langle 
\delta A_{n},
\left(  
\beta_{n}
-
\dfrac{\delta H}{\delta A_{n}}
\right)  
\right\rangle 
-\left\langle 
\left(  
\dot{A}_{m}
+
\mathrm{ad}_{\beta_{n}}^{\ast}
A_{m+n-1}
\right),
\eta_{m}
\right\rangle 
\right\}  
=
0
\]
This expression produces the inverse Legendre transform%
\[
\beta_{n}=\dfrac{\delta H}{\delta A_{n}}%
\]
(holding for hyperregular Hamiltonians). It also yields the equations of
motion
\[
\frac{\partial A_{m}}{\partial t}=-\mathrm{ad}_{\beta_{n}}^{\ast}A_{m+n-1}
\]
which are valid for arbitrary variations $\delta A_{m}$ and variations
$\delta\beta_{m}$ of the form
\[
\delta\beta_{m}=\dot{\eta}_{m}+\mathrm{ad}_{\beta_{n}}\eta_{m-n+1}
\]
where the variations $\eta_{m}$ satisfy vanishing endpoint conditions,
\[
\left.  \eta_{m}\right\vert _{t=t_{i}}=\left.  
\eta_{m}\right\vert _{t=t_{j}}=0
\]
Thus, the Hamilton-Poincar\'e variational principle recovers the
hierarchy of the evolution equations derived in the previous section
using the KMLP bracket.

\subsection{Euler-Poincar\'e hierarchy} 
The corresponding Lagrangian formulation of the Hamilton's principle now
yields
\begin{align*}
\delta
\int_{t_{i}}^{t_{j}}
L
\left(  
\left\{  
\beta
\right\}  
\right)  
{\rm d}t
&  =
\int_{t_{i}}^{t_{j}}
\left\langle 
\frac{\delta L}{\delta\beta_{m}},
\delta\beta_{m}
\right\rangle
{\rm d}t
=\\
&  =
\int_{t_{i}}^{t_{j}}
\left\langle 
\frac{\delta L}{\delta\beta_{m}},
\left(  
\dot{\eta}_{m}
+\mathrm{ad}_{\beta_{n}}
\eta_{m-n+1}
\right)  
\right\rangle 
{\rm d}t
=\\
&  =
-
\int_{t_{i}}^{t_{j}}
\left(  
\left\langle 
\frac{\partial}{\partial t}
\frac{\delta L}{\delta \beta_{m}},
\eta_{m}
\right\rangle 
+\left\langle 
\mathrm{ad}_{\beta_{n}}^{\ast}
\frac{\delta L}{\delta\beta_{m}}
,\eta_{m-n+1}
\right\rangle 
\right)
{\rm d}t
=\\
&  =
-
\int_{t_{i}}^{t_{j}}
\left(  
\left\langle 
\frac{\partial}{\partial t}
\frac{\delta L}{\delta \beta_{m}},
\eta_{m}
\right\rangle 
+
\left\langle 
\mathrm{ad}_{\beta_{n}}^{\ast}
\frac{\delta L}{\delta\beta_{m+n-1}},
\eta_{m}
\right\rangle 
\right)
{\rm d}t
=\\
&  =
-
\int_{t_{i}}^{t_{j}}
\left\langle 
\left(  
\frac{\partial}{\partial t}
\frac{\delta L}{\delta
\beta_{m}}
+
\mathrm{ad}_{\beta_{n}}^{\ast}
\frac{\delta L}{\delta \beta_{m+n-1}}
\right),
\eta_{m}
\right\rangle 
{\rm d}t
\end{align*}
upon using the expression previously found for the variations
$\delta\beta_{m}$ and relabeling indices appropriately.
The Euler-Poincar\'e equations may then be written as

\[
\frac{\partial}{\partial t}
\frac{\delta L}{\delta \beta_{m}}
+
\mathrm{ad}_{\beta_{n}}^{\ast}
\frac{\delta L}{\delta \beta_{m+n-1}}
=
0
\]
with the same constraints on the variations as in the previous paragraph.
Applying the Legendre transformation
\[
A_{m}=\dfrac{\delta L}{\delta\alpha_{m}}%
\]
yields the Euler-Poincar\'e equations (for hyperregular Lagrangians). This
again leads to the same hierarchy of equations derived earlier using the
KMLP bracket.

\medskip
To summarize, the calculations in this section have proven the following result.
\begin{framed}
\begin{theorem}\label{vartheorem}
With the above notation and hypotheses of hyperregularity the following
statements are equivalent:

\begin{enumerate}
\item {\bfi The Euler--Poincar\'{e} Variational Principle.} The curves $\beta_{n}(t)$ are
critical points of the action
\[
\delta
\int_{t_{i}}^{t_{j}}
L\left(  \left\{  \beta\right\}  \right)  dt=0
\]
for variations of the form
\[
\delta\beta_{m}=\dot{\eta}_{m}+\mathrm{ad}_{\beta_{n}}\eta_{m-n+1}
\]
in which $\eta_{m}$ vanishes at the endpoints
\[
\left.  \eta_{m}\right\vert _{t=t_{i}}=\left.  \eta_{m}\right\vert _{t=t_{j}}=0
\]
and the variations $\delta A_{n}$ are arbitrary.

\item {\bfi The Lie--Poisson Variational Principle.} The curves $(\beta_{n}
,A_{n})\left(  t\right)  $ are critical points of the action
\[
\delta
\int_{t_{i}}^{t_{j}}
\left(  
\left
\langle A_{n},
\beta_{n}
\right\rangle 
-
H
\left(  
\left\{
A
\right\}  
\right)  
\right)  
{\rm d}t
=
0
\]
for variations of the form
\[
\delta\beta_{m}=\dot{\eta}_{m}+\mathrm{ad}_{\beta_{n}}\eta_{m-n+1}
\]
where $\eta_{m}$ satisfies endpoint conditions
\[
\left.  \eta_{m}\right\vert _{t=t_{i}}=\left.  \eta_{m}\right\vert _{t=t_{j}}=0
\]
and where the variations $\delta A_{n}$ are arbitrary.

\item The {\bfi Euler--Poincar\'{e} equations} hold:
\[
\frac{\partial}{\partial t}
\dfrac{\delta L}{\delta\beta_{m}}
+
\mathrm{ad}_{\beta_{n}}^{\ast}
\dfrac{\delta L}{\delta\beta_{m+n-1}}
=
0.
\]

\item The {\bfi Lie--Poisson equations} hold:%
\[
\dot{A}_{m}
=
-
\mathrm{ad}_{\frac{\delta H}{\delta A_{n}}}^{\ast}A_{m+n-1}
\]

\end{enumerate}
\end{theorem}
\end{framed}
\noindent
An analogous result is also valid in the multidimensional
case with slight modifications.

\begin{remark}[Hamilton-Poincar\'e theorems and reduction]
Theorem \ref{vartheorem} belongs to a class of theorems, called {\sl Hamilton-Poincar\'e
theorems} \cite{CeMaPeRa}. These theorems involve a hyperregular Hamiltonian
(or Lagrangian), which is invariant with respect to the action of some
Lie group. The Hamilton-Poincar\'e reduction process \cite{CeMaPeRa} then
allows to write the Hamiltonian (or Lagrangian) on the Lie algebra of that
Lie group, by following the same lines as in the first section of chapter~\ref{intro}. This reduction process is not clear in the case of moment dynamics,
since it would require the explanation of moment Lie-Poisson dynamics as
coadjoint motion on a Lie group. The latter has not been identified yet and
thus it is not possible to write the {\sl unreduced Hamiltonian} \cite{MaRa99}
on the moment Lie group.
\end{remark}

\begin{remark}[Legendre transformation]
In the case of moments, the hypothesis of hyperregularity is a strong assumption. In physical applications, for example, this hypothesis often fails, as it
happens for the Poisson-Vlasov system, whose Hamiltonian is given by $H=\frac12\int\left(A_2+A_0\,\Delta^{-1}\!A_0\right)\, {\rm d}q$. This failure is introduced by the
single particle kinetic energy, which produces the term $\int\! A_2\, {\rm d}q$. This term cannot be Legendre-transformed,
since the quantity $\delta H/\delta A_2$ is not defined as a dynamical variable
(the moment algebra does not include constants). 
Nevertheless, section~\ref{geo-prob} will show that
for the case of {\sl geodesic} motion on the moments, this hypothesis is
always satisfied when the metric is diagonal \cite{GiHoTr2007}. This produces the Euler-Poincar\'e equations on the moment algebra, which extend the CH equation to its multicomponent versions (cf. chapter~\ref{EPSymp}).
\end{remark}

\begin{remark}[Euler-Poincar\'e equations for statistical moments]
By following the same arguments as in the proof of the theorem above, one
sees, that similar results hold also for the {\sl statistical moments} presented in chapter~\ref{intro}. This yields the Euler-Poincar\'e equations arising from a moment Lagrangian $L\left(\left\{a_{\widehat{m}m}\right\}\right)$. Such moment equations on the Lagrangian framework have never been considered so far and it would be interesting to study their dynamics, for example by using simple Lie sub-algebra closures. Such an approach is followed in the next section for kinetic moments.
\end{remark}

\section{Some results on moments and cotangent lifts}\label{colifts}
As explained in the introduction, a first order closure of the moment hierarchy leads to the
equations of ideal fluid dynamics. Such equations represent coadjoint motion with respect to the Lie group of smooth
invertible maps (diffeomorphisms). This coadjoint evolution may be interpreted in terms of Lagrangian variables, which are invariant under the action of diffeomorphisms. This section investigates how the entire moment hierarchy may be expressed in terms of the fluid quantities evolving
under the diffeomorphisms and expresses the conservation laws in this case.

\subsection{Background on Lagrangian variables}
In order to look for Lagrangian variables, one considers the geometric interpretation
of the moments, regarded as \emph{fiber integrals} on the cotangent bundle
$T^*Q$ of some configuration manifold $Q$. A moment is defined as a fiber
integral; that is, an integral on the single
fiber $T^*_q Q$ with base point $q\in Q$ kept fixed
\begin{equation}
A_n(q)=\int_{T^*_q Q}\, p^n \,f(q,p)\, {\rm d}p
\end{equation}
A similar approach is followed for gyrokinetics in \cite{QiTa2004}.
Now, 
\rem{
consider a characteristic curve given by the action of canonical transformations:
\begin{equation}
A_n^{(t)}(q_t)=\int_{T^*_{q_t}\! Q}\, \left[p_t(q_0,p_0)\right]^n \,\,f_{t}\!\left(q_t(q_0,p_0),\,p_t(q_0,p_0)\right)\,\, dp_t(q_0,p_0)
\end{equation}
}
the problem is that in general the integrand does not stay on a single fiber under the action of canonical transformations,
i.e. symplectomorphisms are not \emph{fiber-preserving}
in the general case. However, one may avoid this problem by restricting to a subgroup of these canonical
transformations whose action is fiber preserving.

The transformations in this subgroup (indicated with $T^*\text{Diff}(Q)$) are called \emph{point transformations} or \emph{cotangent lifts} of diffeomorphisms  and they arise from diffeomorphisms on points in configuration space \cite{MaRa99}, such that 
\begin{equation}
q_t=q_t(q_{_0})
\end{equation}
The fiber preserving nature of cotangent lifts is expressed by the preservation
of the canonical one-form:
\begin{equation}
p_t\, dq_t=p_t(q_{_0},p_{_0})\, dq_t(q_{_0})=p_{_0} dq_{_0}
\end{equation}
This fact also reflects in the particular form assumed by the generating
functions of cotangent lifts, which are linear in the momentum coordinate \cite{MaRa99}, i.e.
\begin{equation}
h(q,p)\,=\,\beta(q)\,\frac{\partial}{\partial q}\contract\, p\,dq\,=\,p\,\beta(q)\,.
\end{equation}
where the symbol $\contract$ denotes contraction between the vector field $\beta$ and the momentum one-form $p$.
Restricting to cotangent lifts represents a limitation in comparison with considering the whole symplectic group. However, this is a natural
way of recovering the Lagrangian approach, starting from the full moment dynamics.

\subsection{Characteristic equations and related results}
Once one restricts to cotangent lifts, Lagrangian moment variables may be defined
and conservation laws may be found, as in the context of fluid
dynamics. The key idea is to use the preservation of the canonical one-form
for constructing invariant quantities. Indeed one may take $n$ times
the tensor product of the canonical one-form with itself and write:
\begin{equation}
p_t^{\,n}\left(dq_t\right)^n=p_0^{\,n}\left(dq_0\right)^n
\end{equation}
One then considers the preservation of the Vlasov density
\begin{equation}
f_t(q_t,p_t) \,dq_t\wedge dp_t = f_0(q_0,p_0)\, dq_0\wedge dp_0
\end{equation}
and writes
\begin{equation}
p_t^{\,n} \,f_t(q_t,p_t) \,(dq_t)^n\otimes dq_t\wedge dp_t=
p_0^{\,n}\,f_0(q_0,p_0)\, (dq_0)^n\otimes dq_0\wedge dp_0
\end{equation}
Integration over the canonical particle momenta yields the following characteristic equations
\begin{equation}\label{conslaw}
\frac{d}{dt}\!\left[A^{(t)}_n(q_t)\,(dq_t)^n\otimes dq_t\right] = 0\qquad\text{along}\quad
\dot{q_t}=\frac{\partial h}{\partial p}=\beta(q)
\end{equation}
which recover the well known conservations for fluid density and momentum ($n=0,1$) and can be
equivalently written in terms of the Lie-Poisson equations arising from the KMLP bracket, as shown in the next section.
 Indeed, if the vector field $\beta$ is identified with the Lie algebra variable $\beta=\beta_1=\delta H/\delta A_1$ and $h(A_1)$ is the moment Hamiltonian,
the KMLP form (\ref{KMLPbrkt}) of the moment equations is
\begin{equation}
\frac{\partial  A_n}{\partial t} 
+
\textrm{\large ad}^*_{\beta_1} A_n
=0
\,.
\end{equation}
In this case, the KM $\textrm{ad}^*_{\beta_1}$ operation coincides with the Lie derivative $\pounds_{\!\beta_1}$; so, one may write it equivalently
as
\begin{equation}\label{moment-advection}
\frac{\partial  A_n}{\partial t} 
+{\pounds}_{\!\beta_1} A_n
=0
\,.
\end{equation}
For $n=0,1$, one recovers the advection relations for the  density $A_0$ and the momentum $A_1$ in fluid dynamics. However, unlike fluid dynamics, all the moments are conserved quantities.
This equation is reminiscent of the so called {\it b-equation} introduced in \cite{HoSt03}, for which the vector field $\beta$ is nonlocal and may
be taken as $\beta(q)=G*A_n$ (for any $n$), where $G$ is the Green's function of the Helmholtz operator. When the vector field $\beta$ is sufficiently smooth, this equation is known to possess singular solutions of the form
\begin{equation}
A_n(q,t)=\sum_{i=1}^K P_{n,\,i}(t)\,\delta(q-Q_{i}(t))
\end{equation}
where the $i-$th position $Q_i$ and weight $P_{n,\,i}$ of the singular solution for the $n-$th moment satisfy the following equations
\begin{align}\label{singsol-colifts}
\dot Q_i=\left.\beta(q)\right|_{q=Q_i}
\,,\qquad\quad\,
\dot P_{n,\,i}=-\,n\, P_{n,\,i}\left.\frac{\partial \beta(q)}{\partial q}\right|_{q=Q_i}
\end{align}
One can easily see how these solutions are obtained by pairing
the equation~(\ref{moment-advection}) with the contravariant tensor $\varphi_n$.
One calculates
\begin{align*}
\langle\dot{A}_{n},\varphi_{n}\rangle &  =\sum_i\int dq\,\varphi_{n}(q,t)\,\frac
{d}{dt}\left[  P_{n,\,i}\left(  t\right)  \,\delta(q-Q_i(t))\right]  \\
&  =\sum_i\int dq\,\varphi_{n}\left(  \dot{P}_{n,\,i}\,\delta(q-Q_i)-P_{n,\,i}\left(
t\right)  \,\dot{Q}_i\,\delta^{\prime}(q-Q)\right)  =\\
&  =\sum_i\int dq\,\left(  \widehat{\varphi}_{n}\,\dot{P}_{n,\,i}+P_{n,\,i}\,\dot
{Q_i}\,\widehat{\varphi}_{n}^{\prime}\right)
\end{align*}
where the hat denotes evaluation at the point $q=Q_i(t)$. Analogously one
calculates
\begin{align*}
\left\langle \pounds_{\beta_1\,}A_n,\,\varphi_n\right\rangle
=
\left\langle {\rm ad}^*_{\beta_1\,}A_n,\,\varphi_n\right\rangle
&=
\left\langle A_n,\,{\rm ad}_{\beta_1\,}\varphi_n\right\rangle
=
P_{n,\,i}\left(n\,\widehat{\varphi}_n\,\widehat{\beta}_1'-\widehat{\beta}_1\,\widehat{\varphi}_n'\right)
\end{align*}
and equating the corresponding terms in $\widehat{\varphi}_n$ and $\widehat{\varphi}_n'$
yields the equations for $Q_i$ and $P_{n,\,i}\,$.

Interestingly, for $n=1$ (with $\beta(q)=G*A_1$), these equations recover the peakon solutions of the Camassa-Holm equation \cite{CaHo1993}, which play an important role in the following discussion.
Moreover the particular case $n=1$ represents the single particle solution of the Vlasov equation. However, when $n\neq1$ the interpretation of these
solutions as single-particle motion requires the particular
choice $P_{n,\,i}=(P_i)^{\,n}$. For this choice, the $n-$th weight is identified with the $n-$th power of the particle momentum. 

\paragraph{Higher dimensions and the $b$-equation.} The generalization to higher dimensions when considering the tensorial nature of the moments from section~\ref{3Dmoments},
lead to rather complicated tensor equations. In the one dimensional case, one has $\beta_1=G*A_n$,
so the convolution maps the tensor quantity $A_n$ to the vector field $\beta_1$.
In higher dimensions, one has to be careful in order to let dimensions match
in the expression $\beta_1=G*A_n$. One can think of the kernel $G({\bf q-q}')$
as a contravariant $k$-tensor itself $G^{\,i_1,\dots,i_k}$, so that the convolution operator becomes written as $\int G({\bf q-q}')\contract A_n({\bf q})\,{\rm d}^N{\bf q}$ and a vector field may be constructed by letting $k=n+1$, so that $\beta_1^{\,i_{n+1}}=G^{\,i_1,\dots,i_{n+1}\,}{
*}\left(A_n\right)_{i_n,\dots,i_n}$ (the density ${\rm d}^N{\bf q}$ does
not appear in $\beta_1$ because it has been integrated out in the convolution). For example, the EPDiff equation in
higher dimensions is recovered in the case $n=1$ by writing ${\beta}_1^{\,i}=G^{\,ij} * \left({A}_1\right)_j= G\,\delta^{ij} * \left({A}_1\right)_j=G* ({A}^{\,\sharp}_{1\,})^{i}$. This tensorial interpretation of the kernel will be adopted later in this Chapter, when dealing with quadratic
moment Hamiltonians. 
\begin{framed}
\noindent
At this point, the equation
\begin{equation}\label{b-eq}
\frac{\pa {A_n}}{\pa t}+\pounds_{\,G*A_n}\,A_n=0
\end{equation}
is valid in any number of dimensions and it has the same singular solutions (\ref{singsol-colifts}) as above, provided the variable $P_n$ is now a symmetric covariant tensor
on the configuration manifold $P_n=(P_n)_{i_1,\dots,i_n\,}{\rm d}q^{\,i_1\!}\dots{\rm d}q^{\,i_n}$. However, in more generality the higher dimensional equations allow for solutions of the form
\[
A_n({\bf q},t)=\sum_i\int P_{n,\,i}(s,t)\,\delta({\bf q-Q}_i(s,t))\,{\rm d}s
\]
for which the tensor field $A_n({\bf q},t)$ is supported on a submanifold
of $\mathbb{R}^3$ (a filament if $s$ is a one-dimensional coordinate, a sheet
if $s$ is two-dimensional). One also recovers the single particle trajectory when $P_n=\otimes^n\bf P$.
\end{framed}
 The previous discussion has shown how the one-dimensional version of
this equation coincides with the $b$-equation in \cite{HoSt03} for $b=n+1$. However
in higher dimensions this equation substantially differs from the three-dimensional
version proposed in \cite{HoSt03}, which is a characteristic equation for the tensorial quantity ${\bf m}\cdot{\rm d}{\bf q}\otimes^{b\!}{\rm d}^3{\bf q}=m_i\,{\rm d}q^i\otimes^{b\!}{\rm d}^3{\bf q}$ along the nonlocal vector field $G*\bf m$. This characteristic equation for $\bf m$ has been shown to possess emergent
singular solutions and it would be interesting to check whether this property
is shared with the tensorial $b$-equation (\ref{b-eq}) proposed here.

\paragraph{KMLP bracket and cotangent lifts.}
The previous arguments have shown that restricting to cotangent lifts
leads to a Lagrangian fluid-like formulation of the dynamics of the resulting
$p$-moments. In this case, the moment equations are given by the KMLP bracket when the Hamiltonian depends only
on the first moment ($\beta_1=\delta H/\delta A_1$)
\begin{equation}
\{G,H\}=\sum_n\,
\bigg\langle 
A_{n}
\,,\,
\bigg[\frac{\delta G}{\delta A_n}\,,\,
\frac{\delta H}{\delta A_1}\bigg]
\bigg\rangle 
\end{equation}
If one now restricts the bracket to functionals of only the first moment,
one may check that the KMLP bracket yields the well known Lie-Poisson bracket
on the group of diffeomorphisms
\begin{equation}
\{G,H\}[A_1]=-\left\langle A_1,\,\left(
\frac{\delta G}{\delta A_1}\,
\frac{\partial}{\partial q}
\frac{\delta H}{\delta A_1}
- 
\frac{\delta H}{\delta A_1}\,
\frac{\partial}{\partial q}
\frac{\delta G}{\delta A_1}
\right)\,\right\rangle
\end{equation}
This is a very natural step since diffeomorphisms
and their cotangent lifts are isomorphic. In fact, this is the bracket used for ideal incompressible fluids as well as for the construction of the EPDiff equation, which will be discussed later as an application of moment dynamics.

\subsection{Moments and semidirect products}
Another interesting example of how the Kupershmidt-Manin bracket reduces
to interesting structures is given by considering Hamiltonian functionals
depending only on the first
two moments $A_0$ and $A_1$, instead of only $A_1$. 
In this case one 
modifies equation~(\ref{moment-advection}) as
\begin{equation}\label{moment-advection2}
\frac{\partial  A_n}{\partial t} 
+\textit{\large\pounds}_\text{\!\small$\frac{\delta H}{\delta A_1}$} A_n+n\,A_{n-1}\,\frac{\partial}{\partial
q}\frac{\delta H}{\delta A_0}
=0
\,.
\end{equation}
The last term corresponds to ad$^*_{\delta H/\delta A_0}A_{n-1}$ and is not completely understood as an infinitesimal action, unless one considers the case $n=1$
for which  $\langle\left.{\rm ad}^*_{\delta H/\delta A_0}A_{0},\,\xi\right.\rangle
=\left\langle A_0\diamond{\delta H/\delta A_0},\,\xi\right\rangle$, where
$\xi\in\mathfrak{X}(\mathbb{R})$ is a vector field on the real line and  $\pounds_{\xi\,}A_0=\partial_q\left(A_0\,\xi\,\right)$. It is worth
noticing that this hierarchy of equations also allows for singular solutions
of the form
\begin{equation*}
A_n(q,t)=\sum_{i=1}^K P_{n,\,i}(t)\,\delta(q-Q_{i}(t))
\,.
\end{equation*}
However the dynamics of $Q_i$ and $P_{n,\,i}$ slightly differs from that previously found and is written as
\begin{align}\label{singsol-colfits2}
\dot Q_i=\Big.\beta_1(q)\Big|_{\,q=Q_i}
\,,\qquad\quad\,
\dot P_{n,\,i}=-\,n\left(P_{n,\,i\,}\frac{\partial \beta_1(q)}{\partial q}
+P_{n-1\!,\,i\,}\frac{\partial \beta_0(q)}{\partial q}\right)_{\!q=Q_i}
\end{align}
where the notation $\beta_n=\delta H/\delta A_n$ has been used. 
Again, if $P_{n,\,i}=P^n_i$, then one recovers the single-particle dynamics
undergoing Hamiltonian motion with a Hamiltonian function given by $H_N=\sum_i
P_{i\,}\beta_1(Q_i)+\sum_i\beta_0(Q_i)$. Analogous formulas also hold in
more dimensions.

Also the moment bracket for functionals of only $A_0$ and $A_1$ possesses an interesting
structure, which is a peculiar feature of fluid systems. Indeed one calculates
\begin{align*}
\{G,H\}&=
\bigg\langle 
A_{0}
\,,\,
\bigg[\frac{\delta G}{\delta A_1}\,,\,
\frac{\delta H}{\delta A_0}\bigg]
\bigg\rangle 
+
\bigg\langle 
A_{0}
\,,\,
\bigg[\frac{\delta G}{\delta A_0}\,,\,
\frac{\delta H}{\delta A_1}\bigg]
\bigg\rangle
+
\bigg\langle 
A_{1}
\,,\,
\bigg[\frac{\delta G}{\delta A_1}\,,\,
\frac{\delta H}{\delta A_1}\bigg]
\bigg\rangle
\\
&=
-\left\langle A_1,\,\left(
\frac{\delta G}{\delta A_1}\,
\frac{\partial}{\partial q}
\frac{\delta H}{\delta A_1}
- 
\frac{\delta H}{\delta A_1}\,
\frac{\partial}{\partial q}
\frac{\delta G}{\delta A_1}
\right)\,\right\rangle
-
\left\langle A_0,\,\left(
\frac{\delta G}{\delta A_1}\,
\frac{\partial}{\partial q}
\frac{\delta H}{\delta A_0}
- 
\frac{\delta H}{\delta A_1}\,
\frac{\partial}{\partial q}
\frac{\delta G}{\delta A_0}
\right)\,\right\rangle
\end{align*}
which is the well known Lie-Poisson semidirect product structure \cite{HoMaRa}
on $\left.{\rm
Diff}(\mathbb{R}\right)\circledS\left.{\rm Den}(\mathbb{R}\right)$ where
$\left.{\rm Den}(\mathbb{R}\right):=\mathcal{F}^*(\mathbb{R})$ indicates the
vector space of densities on the real line. 

\paragraph{The moment bracket and continuum models.}
At this point it is easy to see
how the Kupershmidt-Manin bracket is a usual tool for deriving continuum
fluid models from kinetic equations. This machinery is extended in the next
sections to include extra degrees of freedom such as orientation and magnetic
moment for each particle.

\section{Discussion}
After a review of the moment Kupershmidt-Manin bracket, this chapter has
shown how this Lie-Poisson structure can be extended to take into account
of the tensorial nature of the moments. The result is a Lie-Poisson bracket
determined by the {\bfi symmetric Schouten bracket} \cite{BlAs79,Ki82,DuMi95}.
The moments have thus a purely geometric meaning in terms of {\bfi symmetric covariant tensors} \cite{GiHoTr2008}.

The Lie-Poisson structures for the moments have been derived from {\bfi variational
principles}, in terms of {\it Hamilton-Poincar\'e} dynamics \cite{CeMaPeRa}.
As a result, the {\bfi Euler-Poincar\'e equations} \cite{MaRa99,HoMaRa} have been derived from a moment Lagrangian.

In the second part, this chapter has focused on the moment dynamics generated
by {\bfi diffeomorphisms} and their cotangent lifts on the phase space. In
one dimension, the resulting moment equations have the same form as the $b$-equation
\cite{HoSt03}, which thus maybe interpreted as a {\bfi characteristic equation}
for a single Vlasov kinetic moment. This concept extends to {\bfi higher dimensions}, thereby generating a higher dimensional version of the $b$-equation in terms of characteristic equations for symmetric tensors. These higher
dimensional tensor equations are {\it different} from those proposed in \cite{HoSt03}.
The same treatment has been extended to consider semidirect products of diffeomorphisms
and the corresponding equations have been presented. 

\paragraph{Singular solutions and future work.} The moment equations obtained in this chapter have been shown to possess
{\bfi singular solutions}, which reduce to the single particle trajectory
in one dimension. The dynamics of these singularities has been studied and
future work will be focusing on the behavior of singular solutions in some
simple cases of the tensorial $b$-equation in higher dimensions. For example,
one can consider the characteristic equation for $A_2$: in {\it two dimensions},
this equation possesses {\it three independent components}. This problem would represent an interesting opportunity for analyzing the behaviour of
the singular solutions. In particular, one is intrigued to know whether these
solutions emerge spontaneously in a finite time, as it happens in some cases of $b$-equation in one dimension.

\paragraph{Perspectives on momentum maps.} The search for Lagrangian variables in moment dynamics turns out to be a
challenging task and is strictly connected to the geometric nature of
the moments. It is well known they are Poisson maps, but one may wonder whether
they are actually \emph{momentum maps} \cite{Ma82,WeMo,MaWeRaScSp} arising from a group action on the Vlasov
Hamiltonian $H[f]$. This question has never been answered. There are reasons
to believe that the geometric identification of moments with symmetric tensors
is a key step in the construction of momentum maps. In particular, this construction
would require the complete description of the moment Lie-Poisson dynamics
in terms of {\bfi coadjoint motion} after the identification of the symmetry
group acting on the moments. For example, the tensorial description would
involve the symmetric group of permutations, as it happens also in the theory
of statistical moments \cite{HoLySc1990} BBGKY moments \cite{MaMoWe1984}.

\chapter[Geodesic flow on the moments: a new problem]{Geodesic flow on the moments:\\ a new problem}
\label{EPSymp}
\section{Introduction}
This chapter reviews some direct applications of moment dynamics to physical
problems and, as a new result, shows how the one-dimensional system of Benney long wave equations \cite{Be1973} describes the dynamics of coasting beams in particle accelerators \cite{Venturini}. The Benney moment hierarchy is integrable and this explains the nature of the coherent structures observed in the experiments \cite{KoHaLi2001,CoDaHoMa04}.

This chapter also formulates the moment dynamics generated by {\it quadratic} Hamiltonians. This dynamics is a certain type of geodesic motion on the symplectic diffeomorphisms, which are smooth invertible symplectic maps acting on the phase space and possessing smooth inverses. 
In some cases, the  theory of moment dynamics for the Vlasov equation turns out to be related to the theory of shallow water equations. Indeed, the geodesic equations for the first two moments recover both the integrable CH equation \cite{CaHo1993} and its two-component version \cite{Falqui06,ChLiZh2005,Ku2007}, which is again an integrable system of PDE's. The study of such geodesic
moment equations is a new problem, which is here approached for the first
time. Singular solutions are presented as well as an extension of moment
geodesic motion to anisotropic interactions.

\section{Applications of the moments and quadratic terms}\label{Apps} 

\subsection{The Benney equations and particle beams: a new result}
The KMLP bracket (\ref{KMLP-brkt}) was first derived in the context of Benney long waves,
whose Hamiltonian is
\begin{equation}
H=\frac12\int (A_2(q)+gA_0^2(q))\,{\rm d}q.
\end{equation}
The Hamiltonian form $\partial_tA_n=\{A_n,\,H\}$ with the KMLP bracket \ref{KMLPbrkt} leads to the moment equations
\begin{equation}
\frac{\partial A_n}{\partial t}
+
\frac{\partial A_{n+1}}{\partial q}
+g
n A_{n-1}
\frac{\partial A_0}{\partial q}
=0
\end{equation}
derived by Benney \cite{Be1973} as a description of long waves on a
shallow perfect fluid, with a free surface at $y=h(q,t)$. In this
interpretation, the $A_n$ were vertical moments of the horizontal
component of the velocity $p(q,y,t)$:
\begin{equation*}
A_n=\int_{0}^{h} p^n(q,y,t) \,\text{d}y.
\end{equation*}
The corresponding system of evolution equations for $p(q,y,t)$ and $h(q,t)$
is related by hodograph transformation, $y=\int_{-\infty}^p f(q,p',t)\,
\text{d}p'$, to the Vlasov equation
\begin{equation}\label{Vlasov-Benney}
\frac{\partial f}{\partial t}
+
p
\frac{\partial f}{\partial q}
-g
\frac{\partial A_0}{\partial q}
\frac{\partial f}{\partial p}
=0.
\end{equation}
The most important fact about the Benney hierarchy is that it is completely integrable \cite{KuMa1978}. 
\rem{ 
This fact emerges from the following observation. Upon defining a
function $\lambda(q,p,t)$ by the principal value integral, 
\begin{equation*}
\lambda(q,p,t)=
p+P\int_{-\infty}^\infty \frac{f(q,p\,',t)}{p-p\,'} \,\text{d}p\,',
\end{equation*}
it is straightforward to verify \cite{LeMa} that
\begin{equation*}
\frac{\partial \lambda}{\partial t}
+
p
\frac{\partial \lambda}{\partial q}
-
\frac{\partial A_0}{\partial q}
\frac{\partial \lambda}{\partial p}
=0;
\end{equation*}
so that $f$ and $\lambda$ are advected along the same characteristics.
} 
\begin{framed}
\paragraph{Applications to coasting accelerator beams.}
Interestingly, the equation that regulates coasting proton beams in particle accelerators takes exactly the same form as the Vlasov-Benney equation (\ref{Vlasov-Benney}).
(See for example \cite{Venturini} where a linear bunching term is also included.)
The integrability of the \emph{Vlasov-Benney}
equation implies coherent structures. These structures are indeed found experimentally at CERN \cite{KoHaLi2001}, BNL \cite{BlBrGlRaRy03}, LANL \cite{CoDaHoMa04} and FermiLab \cite{MoBaJaLeNgShTa}. (In the last case coherent structures are shown to appear even when a bunching force is present.) 
These structures have attracted the attention of the accelerator community and considerable analytical work has been carried out over the last decade (see for example \cite{ScFe2000}). The existence of coherent structures in coasting proton beams has never been related to the integrability of the governing Vlasov equation via its connection to the Benney hierarchy. However, this connection would explain very naturally why robust coherent structures are seen in these experiments as fully nonlinear excitations. 
\rem{Rather, connections with the well known integrable KdV equation have been proposed \cite{ScFe2000}, but we believe
this is not a natural step since integrability appears directly in the Vlasov-Benney system that governs the collective motion of the beam.}
\end{framed}

\subsection{The wake-field model and some specializations}

Besides integrability of the Vlasov-Benney equation, there are other important applications of the Vlasov equation that have in common the presence of a quadratic term in $A_0$ within the Hamiltonian:
\begin{equation}
H=\frac12\int\! A_2(q)\,{\rm d}q+\frac12\iint\! A_0(q)\,G(q,q')\,A_0(q')\,{\rm d}q\,{\rm d}q'.
\end{equation}
For example, when $G=\left(\partial_q^{\,2}\right)^{-1}$, this Hamiltonian leads to the Vlasov-Poisson
system, which is of fundamental importance in many areas of plasma physics.
Remarkably, the fluid closure of this system has been shown to be integrable in \cite{Pa05}. More generally, this Hamiltonian is widely used for beam dynamics in particle
accelerators: in this case $G$ is related to the electromagnetic interaction of a beam with the vacuum chamber. The {\bfi wake field} is originated by the image charges induced on the walls by the passage of a moving particle: while the beam passes, the charges in the walls are attracted towards the inner surfaces and generate a field that acts back on the beam. This affects the dynamics of the beam, thereby causing several problems such as beam energy spread and instabilities. In the literature, the {\bfi wake function} $W$ is introduced so that \cite{Venturini}
\begin{equation}
G(q,q')=\int_{-\infty}^q \!W(x, q')\,dx
\end{equation}
Wake functions usually depend only on the properties of the accelerator chamber.

An interesting wake-field model has been presented in \cite{ScFe2000}
where $G$ is chosen to be the Green's function of the Helmholtz operator
$\left(1-\alpha^2\partial_q^2\right)$: this generates a \emph{Vlasov-Helmholtz} (VH) equation
\cite{CaMaPu2002} that is particularly interesting for future work. Connections of this equation with the well known integrable KdV equation have been proposed. However this is not a natural step since integrability appears already with no further
approximations in the Vlasov-Benney (VB) system that governs the collective motion of the beam. In particular one would
like to understand the VH equation as a special deformation of the integrable
VB case that allows the existence of {\bfi singular solutions}.  Indeed, the presence of the Green's function $G$ above is a key ingredient for the existence of the single-particle solution, which is \emph{not}
allowed in the VB case. In particular, the single-particle solutions for the Vlasov-Helmholtz equation may be of great interest, since these singular solutions arise from a deformation of an integrable system. In the limit as the deformation parameter $\alpha$ in the Helmholtz Green's function passes to zero ($\alpha\rightarrow0$), one recovers the integrable Vlasov-Benney case. Also, in the limit $\alpha\rightarrow\infty$ the fluid
closure of this system reduces again to an integrable system \cite{Pa05}.
Thus the study of the VH equation and its single particle solution can provide
useful understanding of two integrable limits, the Vlasov-Benney equation ($\alpha\to0$) and the fluid closure of the Vlasov-Poisson system ($\alpha\to\infty$).

\subsection{The Maxwell-Vlasov system}
In higher dimensions, particularly $N=3$, one takes the direct sum of the KMLP bracket, together with with the Poisson bracket for an electromagnetic field (in the Coulomb gauge) where the electric field $\mathbf{E}$ and magnetic vector potential $\boldsymbol{\mathscr{A}}$ are canonically conjugate; then the Hamiltonian (in multi-index notation)
\begin{multline*}
H\left[ \left\{  A\right\}\!,\boldsymbol{\mathscr{A}},\mathbf{E};\varphi\right]
= \dfrac{1}{2}\int\! 
\sum_{j}
\left(  
A_{2_\textit{\scriptsize \!j}}(\mathbf{q})
-
2\,\mathscr{A}_{j}(\mathbf{q})\,A_{1_\textit{\scriptsize \!j}}(\mathbf{q})
\right)\mathrm{d}^3 \mathbf{q}\\
+ \dfrac{1}{2} \int\!
\left(  
\,\vert\boldsymbol{\mathscr{A}}(\mathbf{q})\vert^{2}
+
2\varphi(\mathbf{q})
\right) A_{0}(\mathbf{q})\,
\mathrm{d}^3 \mathbf{q}\\
 \quad +
\frac{1}{2}
\int\!
\left(  
\vert\mathbf{E}(\mathbf{x})\vert^{2}
+
2\,\mathbf{E}(\mathbf{x})\cdot\nabla\varphi(\mathbf{x})
+
\vert\nabla\times\boldsymbol{\mathscr{A}}(\mathbf{x})\vert^{2}\right)\,  \mathrm{d}^3 \mathbf{x}
\end{multline*} 
yields the Maxwell-Vlasov (MV) equations for systems of interacting
charged particles. In the Hamiltonian, $\varphi$ plays the role of a Lagrange multiplier that constraints the variational principle in order to include Gauss' law. For a discussion of the MV equations from a geometric
viewpoint in the same spirit as the present approach, see \cite{Ma82,MaWe81,CeHoHoMa1998}.

\subsection{The EPDiff equation and singular solutions}
Another interesting  moment equation is given by the Euler-Poincar\'e equation on the group of diffeomorphisms (EPDiff) \cite{CaHo1993}. In this case, the Hamiltonian is purely quadratic in the first moments:
\begin{equation}
H=\frac12\iint \!A_1(q)\, G(q,q')\, A_1(q')\, {\rm d}q\,{\rm d}q'
\end{equation}
and the EPDiff equation \cite{HoMa2004}
\begin{equation}
\frac{\partial A_{1}}{\partial t} 
+\frac{\partial A_{1}}{\partial q}\int \!G(q,q') A_{1}(q',t){\rm d}q'
+2A_{1}\,\frac{\partial }{\partial q}\int \!G(q,q') A_{1}(q',t){\rm d}q'=0
\end{equation}
comes from the closure of the KMLP bracket given by cotangent lifts. (Without
this restriction one would obtain again the equations (\ref{conslaw}) with
$\beta=G* A_1$.) Thus this EPDiff equation is a \emph{geodesic} equation on the group
of diffeomorphisms. The Camassa-Holm equation is a particular case in which $G$ is the Green's function of the Helmholtz operator $1-\alpha^2\partial_q^2$. Both the CH and the EPDiff equations are completely
 integrable and have a
large number of applications in fluid dynamics (shallow water theory, averaged
fluid models, etc.) and imaging techniques \cite{HoRaTrYo2004} (medical imaging, contour dynamics,
etc.). 

Besides the complete integrability of the CH equation, the connection between the CH (EPDiff) equation and moment dynamics lies in the fact that  singular solutions appear in both contexts.
The existence of this kind of solution for EPDiff leads to investigate its origin in the context of Vlasov moments. More particularly it is a reasonable
question whether there is a natural extension of the EPDiff equation to all the moments. This would again be a geodesic
(hierarchy of) equation, which would perhaps explain how the singular solutions for EPDiff arise in this larger context.

\begin{remark}
It should be pointed out that the KMLP and VLP formulations are not wholly equivalent; in particular the map from the distribution function $f(q,p)$ to the moments $\{A_n\}$ is explicit, but it is not a trivial problem to reconstruct the distribution from its moments. 
Simple fluid-like closures of the system arise 
very naturally in the KMLP framework, as with the example in Section 3.
\end{remark}

\section{A new geodesic flow and its singular solutions}\label{geo-prob}
\subsection{Formulation of the problem: quadratic Hamiltonians}\label{geodesic-motion}
The previous examples show how quadratic terms in the Hamiltonian produce
interesting behavior in
various contexts. This suggests that a deeper analysis of the role of quadratic
terms may
be worthwhile particularly in connections between Vlasov moment dynamics
and the EPDiff equation, with its singular solutions.
Purely quadratic Hamiltonians are considered  in \cite{GiHoTr05},
leading to the problem of geodesic motion on the space of
moments. 

In this problem the Hamiltonian is the norm on the
moments given by the following metric and inner product,
\begin{align}
H=\frac{1}{2}\|A\|^2
&=
\frac{1}{2}\sum_{n,s=0}^\infty
\int\hspace{-2.3mm}\int
A_n(q)G_{ns}(q,q\,')A_s(q\,')\,{\rm d}q\,{\rm d}q\,'
\label{Ham-metric}
\end{align}
The metric $G_{ns}(q,q\,')$ in (\ref{Ham-metric}) is chosen to be positive definite,
so it defines a norm for $\{A\}\in\mathfrak{g}^*$. The
corresponding geodesic equation with respect to this norm is
found as in the previous section to be,
\begin{eqnarray}
\frac{\partial  A_m}{\partial t}
=
\{\,A_m\,,\,H\,\}
=
-\sum_{n=0}^\infty
\Big(n\beta_n
\frac{\partial}{\partial q} A_{m+n-1}
+
(m+n)A_{m+n-1}\frac{\partial}{\partial q}
\beta_n
\Big)
\label{EPMS-eqn}
\end{eqnarray}
with Lie algebra variables $\beta_n\in\mathfrak{g}$ defined by
\begin{eqnarray}
\beta_n
=
\frac{\delta H}{\delta A_n} 
=
\sum_{s=0}^\infty
\int
G_{ns}(q,q\,')A_s(q\,')\,{\rm d}q\,'
=
\sum_{s=0}^\infty
G_{ns}*A_s
\,.
\label{EPMS-vel}
\end{eqnarray}
Thus, evolution under (\ref{EPMS-eqn}) may be rewritten as
formal coadjoint motion on the dual Lie algebra $\mathfrak{g}^*$
\begin{eqnarray}
\frac{\partial  A_m}{\partial t}
=
\{\,A_m\,,\,H\,\}
=:
-\sum_{n=0}^\infty
{\rm ad}^*_{\beta_n}A_{m+n-1}
\label{A-dot}
\end{eqnarray}
This system comprises an infinite system of nonlinear, nonlocal, coupled
evolutionary equations for the moments. In this system, evolution of
the $m^{th}$ moment is governed by the potentially infinite sum of
contributions of the velocities $\beta_n$ associated with $n^{th}$
moment sweeping the $(m+n-1)^{th}$ moment by a type of coadjoint action.
Moreover, by equation (\ref{EPMS-vel}), each of the $\beta_n$
potentially depends nonlocally on all of the moments.

Equations (\ref{Ham-metric}) and (\ref{EPMS-vel}) may be
written in three dimensions in multi-index notation, as follows:
the Hamiltonian is given by
\[
H
=
\frac{1}{2}
\left\vert 
\left\vert 
A
\right\vert 
\right\vert 
^{2}
=
\frac{1}{2}
\sum\limits_{\mu,\nu}
\iint
A_{\mu}
\left(  
\mathbf{q},
t
\right)  
G_{\mu\nu}
\left(  
\mathbf{q,q}\,^{\prime}
\right)  
A_{\nu}
\left(  
\mathbf{q}\,^{\prime},
t
\right)  
{\rm d}^3
\mathbf{q}\,
{\rm d}^3
\mathbf{q}\,^{\prime}
\]
so the dual variable is written as
\begin{align*}
\beta_{\rho}
=
\frac{\delta H}{\delta A_{\rho}}  
& =
\sum\limits_{\nu}
\iint
G_{\rho\nu}
\left(  
\mathbf{q,q}\,^{\prime}
\right)  
A_{\nu}
\left(  
\mathbf{q}\,^{\prime},
t
\right)  
{\rm d}^3\mathbf{q}\,
{\rm d}^3\mathbf{q}\,^{\prime}
=
\sum\limits_{\nu}
G_{\rho\nu}\ast A_{\nu}.
\end{align*} 

However the equations (\ref{Ham-metric}) and (\ref{EPMS-vel}) are already
valid in higher dimensions if one considers the tensor interpretation of
the moments. This is another case in which the tensor interpretation is helpful.
In this case, the metric is written as
\[
G_{nm}=\left.G_{nm}^{\,i_0,\dots,i_n,\,j_0,\dots,j_m\,}({\bf q,\,q}'\right.)
\]
which takes into account for the tensor nature of the moment equations.

\smallskip
\begin{remark}[Euler-Poincar\'e formulation]
When the metric $G_{nm}$ is diagonal ($G_{nm}=K_{nm}\,\delta^{\,m}_\text{\footnotesize$n$}=:G_n$),
the Hamiltonian becomes hyperregular and one can find the inverse Legendre transform. In order to see this explicitly one can write the Lie algebra
variable $\beta_n$ in one dimension as
\begin{eqnarray*}
\beta_n
=
\frac{\delta H}{\delta A_n} 
=
\int
G_{n}(q,q\,')\,A_n(q\,')\,{\rm d}q\,'
=
G_{n}*A_n
\,.
\end{eqnarray*}
so that, if the $n$-th kernel $G_n$ is the Green's function corresponding
to the inverse of some operator $\widehat{Q}_n$ (so that $G_n=\widehat{Q}_n^{\,-1}$), then one calculates the inverse the Legendre-transform as
\[
A_n=\widehat{Q}_n\,\beta_n
\]
and the problem admits a Lagrangian formulation in terms of the Euler-Poincar\'e variational principle
\begin{equation}
\delta\!\int_{t_1}^{t_2} \!\beta_n\,\widehat{Q}_n\beta_n\,{\rm
d}t\,=\,0
\label{EPSymp-mom}
\end{equation}
and the corresponding Euler-Poincar\'e hierarchy follows.
\end{remark}

The construction of this geodesic motion on the moments is motivated by the
examples provided by Euler and CH equation and is justified by its Lie-Poisson
structure. However the search for singular solutions requires more insight
into the geometric meaning of this infinite hierarchy of equations. In particular,
since the Lie-Poisson dynamics has not been fully interpreted in terms of coadjoint
motion and the underlying Lie group has not been identified, this geodesic
flow needs further investigation.

\subsection{A first result: the geodesic Vlasov equation (EPSymp)}
Importantly, geodesic motion for the moments is equivalent to
geodesic motion for the Euler-Poincar\'e equations on the
symplectomorphisms (EPSymp). 
\begin{framed}
\noindent
This is generated by the following quadratic Hamiltonian
\begin{equation}\label{epsymp}
H\left[  f\right]  =
\frac{1}{2}\iint f\left(q,p\right)  
\mathcal{G}\left(  q,p,q\,^{\prime},p\,^{\prime}\right)  
f\left(q\,^{\prime},p\,^{\prime}\right)  
{\rm d}q\,{\rm d}p\,{\rm d}q\,^{\prime}{\rm d}p\,^{\prime}
\end{equation}
The equivalence with EPSymp emerges when the function $\mathcal{G}$
is written as

\begin{equation}
\mathcal{G}\left(  q,q\,^{\prime},p,p\,^{\prime}\right)  
=
\underset{n,m}{\sum}
\thinspace 
p^{n}\/G_{nm}\left(q,q\,^{\prime}\right) p\,^{\prime\, m}
\,.
\end{equation}
and the corresponding Vlasov equation reads as
\begin{equation}
\frac{\partial f}{\partial t}\,+\,\Big\{f\,,\,\,\mathcal{G}*f\Big\}\,=\,0
\end{equation}
where $\{\cdot,\cdot\}$ denotes the canonical Poisson bracket.
\end{framed}
\noindent
Thus, whenever the metric $\mathcal{G}$ for EPSymp has a Taylor series,
its solutions may be expressed in terms of the geodesic motion for the
moments. More particularly the geodesic Vlasov equation presented here
is nonlocal in both position and momentum. However this equation extends to more dimensions \cite{GiHoTr05} and to any kind of geodesic motion, no matters how the metric is expressed explicitly.
Such an equation reduces to the 2D Euler's equation for $\mathcal{G}=\Delta^{-1}$,
as shown in chapter \ref{intro}, and is surprisingly similar in construction to another important integrable geodesic equation on the linear Hamiltonian
vector fields (Hamiltonian matrices), which has been recently proposed \cite{BlIs,BlIsMaRa05}.

For a more extensive analysis, one can relate the geodesic Vlasov equation
EPSymp with its correspondent equation on Hamiltonian vector fields. To this
purpose one restricts the EPDiff Lagrangian to the symplectic algebra on
$T^*Q$
\[
L[{\bf X}_\textit{\small h}]=\frac12
\big\langle\hat{Q}{\bf X}_\textit{\small h},{\bf X}_\textit{\small h}\big\rangle
\]
where $\hat{Q}:\mathfrak{X}_\text{can}\rightarrow\mathfrak{X}_\text{can}^*$
is an invertible symmetric differential operator.
Upon integration by parts,
this Lagrangian is written on the Hamiltonian functions as
\[
L=\frac12
\Big\langle\hat{Q}{\bf X}_\textit{\small h},\,{\bf X}_\textit{\small h}\Big\rangle
=\frac12
\Big\langle\hat{Q}\,\mathbb{J}\, \nabla h_{\,},\,\mathbb{J} \nabla h\Big\rangle
=\frac12
\Big\langle{\rm div}\big(\mathbb{J}\,\hat{Q}\,\mathbb{J}\, \nabla h\big),\, h\Big\rangle=L[h]
\,.
\]
The Legendre transform 
\[
f=\frac{\delta L}{\delta h}={\rm div}\big(\mathbb{J}\,\hat{Q}\,\mathbb{J}\, \nabla h\big)
\,\Rightarrow\,
h=\left({\rm div}\,
\mathbb{J}\,\hat{Q}\,\mathbb{J}\, \nabla 
\right)^{\!-1}\!f
\]
yields the EPSymp Hamiltonian in the Vlasov form
\[
H[f]=\frac12\,\big\langle f,\hat{O}^{-1}f\big\rangle
\]
with
\[
\hat{O}:={\rm div}\,\mathbb{J}\,\hat{Q}\,\mathbb{J}\, \nabla
\,.
\]
This makes clear the connection between the geodesic Vlasov equation and
the geodesic motion on the Hamiltonian vector fields.
\rem { 
 In this sense the operator
$\hat O$ is a \emph{Beltrami-Laplace operator} $\nabla^2$, which in the case of a finite-dimensional
Riemannian manifold is given by \cite{AbMa} 
\[
\nabla^2=\frac1{\sqrt{{\rm det}\, g}}\frac{\pa}{\pa x^k}\left(g^{ik}\sqrt{{\rm det}\, g}\,\frac{\pa}{\pa x^i}\right)
\]
where $g$ is the metric tensor. Thus in the case of EPSymp the operation
$\Omega^\sharp\hat{Q}\Omega^\sharp$ plays the role of a metric.
} 
 An interesting case occurs when $\hat{Q}$ is the flat operation $\hat{Q}\,{\bf X}_\textit{\small h}=({\bf X}_\textit{\small h})^\flat$, so that
\[
{\rm div}\,
\mathbb{J}\left(\mathbb{J}\,\nabla h\right)^{\flat}=
-\,\Delta h
\,.
\]
Then $\hat{O}$ reduces to minus the Laplacian
\[
\hat{O}=-\,\Delta
\]
and in two dimensions one obtains the Euler Hamiltonian $H[\omega]=1/2\,\big\langle\omega,(-\Delta)^{-1\,}\omega\big\rangle$
with $\omega=f$. 
This analysis explains how
the geodesic motion on the symplectic group is related to the geodesic motion on the volume-preserving diffeomorphisms in the vorticity representation
introduced in chapter~\ref{intro}.
In the more general case when $\hat{Q}$ is a purely differential operator,
one has that $\hat{Q}$ and $\mathbb{J}$ commute and thus $\hat{O}=-\,{\rm div}\,\hat{Q}\, \nabla$. Also if $\hat{Q}$ commutes with the divergence,
then, one has $\hat{O}=-\,\hat{Q}\, \Delta$. However in the most general
case, $\hat{Q}$ is a matrix differential operator that does not commute with $\mathbb{J}$.

\subsection{The nature of singular geodesic solutions}
The geometric meaning of the moment
equations is now explained in terms of coadjoint geodesic motion on the symplectic group and one can therefore characterize singular solutions, since the geodesic Vlasov equation (EPSymp) essentially describes advection in phase
space. Indeed, the geodesic Vlasov equation possesses the single
particle solution
\begin{equation}
f(q,p,t)\,=\,\sum_j \,\delta(q-Q_j(t))\,\delta(p-P_j(t))
\end{equation}
which is a well known singular solution that is admitted whenever the phase-space
density is advected along a smooth Hamiltonian vector field. This happens, for example, in the Vlasov-Poisson system and in the general wake-field model. On the other hand, these singular solutions \emph{do not} appear in the Vlasov-Benney equation.

\begin{framed}
In any number of spatial dimensions, the geodesic
equation (\ref{EPMS-eqn}) possesses exact solutions which are {\it
singular}; that is, they are supported on delta functions in
$q-$space:
equation (\ref{EPMS-eqn}) admits singular
solutions of the form
\begin{eqnarray}
A_n(\mathbf{q},t)
&=&
\sum_{j=1}^N
\int\otimes^{n\,}
{\bf P}_{\!j}(a,t)\,
\delta\big(\mathbf{q}-\mathbf{Q}_j(a,t)\big)\,{\rm d}a
\label{sing-soln}
\end{eqnarray}
\end{framed}
\noindent
in which the integral over coordinate $a$ is performed over
an embedded subspace of the $q-$space and the parameters
$(\mathbf{Q}_{j}\,,\mathbf{P}_{\!j})$ satisfy canonical Hamiltonian equations in
which the Hamiltonian is the norm $H$ in (\ref{Ham-metric})
evaluated on the singular solution Ansatz (\ref{sing-soln}). 

In one dimension, the coordinates $a_j$ are absent and the
equation (\ref{EPMS-eqn}) admits singular solutions of the form
\begin{eqnarray}\label{sing-soln2}
A_n({q},t)
&=&
\sum_{j=1}^N
P_j^n(t)\,
\delta\big({q}-{Q}_j(t)\big)
\end{eqnarray} 

In order to show this is a solution in one dimension, one checks that
these singular solutions satisfy a system of partial differential
equations in Hamiltonian form, whose Hamiltonian couples all the moments
\begin{eqnarray}
\label{H_N}
H_N
=
\frac{1}{2}\sum_{n,s=0}^\infty
\sum_{j,k=1}^N
P^s_j(t)\,P^n_k(t)\,
G_{ns}(Q_j(t),Q_k(t))
\end{eqnarray}
Explicitly, one takes the pairing of the coadjoint equation
\[
\dot{A}_{m}
=
-\sum_{n,s}\textrm{\large ad}_{G_{ns}\ast A_{s}}^{\ast}A_{m+n-1}
\]
with a sequence of smooth functions $\left\{  \varphi_{m}\left(  q\right) 
\right\} 
$, so that:
\[
\langle \dot{A}_{m},\varphi_{m}\rangle 
=
\sum_{n,s}
\left\langle A_{m+n-1},\mathrm{ad}_{G_{ns}\ast A_{s}}\varphi_{m}\right\rangle
\]
One expands each term and denotes $\widetilde{\varphi}_m:=\varphi_m(q,t)\vert_{q=Q_j}$:
\begin{align*}
\langle \dot{A}_{m},\varphi_{m}\rangle
&  =
\sum_{j}
\left(  
\frac{d{P}_{j}^{\,m}}{d t}\,\varphi_{m}  
+
{P}_{j}^{\,m}\,\dot Q_{j}\,
\varphi_{m}^{\,\prime}  
\right)
\end{align*}
Similarly expanding
\begin{align*}
\left.\Big\langle A_{m+n-1},\,\textrm{\large ad}_{G_{ns}\ast A_{s}}\varphi_{m}\Big\rangle\right.
  =
\sum_{j,k}{P}_{k}^{\,s}\,{P}_{j}^{\,m+n-1}
\left(
n\,\widetilde\varphi_{m}^{\,\prime}G_{ns}\left(Q_{j},Q_{k}\right)
-
m\,\widetilde\varphi_{m}
\frac{\partial G_{ns}\left(Q_{j},Q_{k}\right)}{\partial Q_{j}}
\right)
\end{align*}
leads to
\begin{align*}
\frac{d Q_{j}}{d t} 
&  =
\sum_{n,s}\sum_{k}
n\,{P}_{k}^{\,s}\,{P}_{j}^{\,n-1}\,G_{ns}\left(Q_{j},Q_{k}\right)
\\
\frac{d{P}_{j}}{d t} 
&  =
-
\sum_{n,s}\sum_{k}{P}_{k}^{\,s}\,{P}_{j}^{\,n}
\,\frac{\partial G_{ns}\left(Q_{j},Q_{k}\right)}{\partial Q_{j}}
\end{align*}
so that one finally obtains equations for $Q_{j}$ and ${P}_{j}$ in
canonical form, 
\[
\frac{d Q_{j}}{d t}
=
\frac{\partial H_N}{\partial{P}_{j}},
\qquad
\frac{d{P}_{j}}{d t} 
=
-\,\frac{\partial H_N}{\partial Q_{j}}.
\]

\begin{remark}
These singular solutions of EPSymp are also
solutions of the Euler-Poincar\'e equations on the diffeomorphisms
(EPDiff). In the latter case, the single-particle solutions reduce to the pulson solutions for EPDiff \cite{CaHo1993}. Thus, the singular pulson solutions of the EPDiff equation arise naturally from the single-particle dynamics on phase-space.
\end{remark}

\subsection{Some results on the dynamics of singular solutions}
This section presents the problem of the interaction between two singular
solutions. It is easy to show how this system preserves the total momentum
$P=P_1+P_2$. Indeed, one observes that
\begin{align*}
\dot{P}_1&=-\,\sum_{n,m}\,p_1^n\left(
P_1^m\left.\frac{\partial}{\partial Q}\right|_{Q=Q_1}\!\!G_{nm}(Q-Q_1)
+
P_2^m\left.\frac{\partial}{\partial Q}\right|_{Q=Q^1}\!\!G_{nm}(Q-Q_2)\right)
\\
&=
-\,\sum_{n,m}\,P_1^n\,P_2^m\,\partial_{Q_1}G_{nm}(Q_1-Q_2)
\end{align*}
under the assumption that $\left({\partial G_{nm}(Q)}/{\partial Q}\right)_{Q=0}=0$.
and $\partial_{Q_1}G_{nm}(Q_1-Q_2)=-\,\partial_{Q_2}G_{nm}(Q_1-Q_2)$. Thus
$\dot{P}_1+\dot{P}_2=0$ since $\left.G_{nm}(Q_1-Q_2)\right.=\left.G_{mn}(Q_1-Q_2)\right.$.

One can also see this by writing the Hamiltonian
\begin{align*}
H_N=&
\frac12\sum_{n,m}\,p_1^n\,p_1^m
+
\frac12\sum_{n,m}\,p_2^n\,p_2^m\\
&+
\frac12\sum_{n,m}\,p_1^n\,K_{nm}(q^1-q^2)\,p_2^m
+
\frac12\sum_{n,m}\,p_2^n\,K_{nm}(q^1-q^2)\,p_1^m
\\
=&
\frac12\sum_{n,m} \Big(\left(P_1^{n+m}+P_2^{n+m}\right)G_{nm}(0)
+
2\left.G_{nm}(Q^1-Q^2\right)P_1^n\,P_2^m\Big)
\end{align*}
and by checking that
\[
\dot{P}_1=
-\,\sum_{n,m}\,P_1^n\,P_2^m\,\partial_{Q^1}G_{nm}(Q^1-Q^2)
\]
so that $\dot{P}_1+\dot{P}_2=0$.

\paragraph{Convergence of the Hamiltonian.} The problem with the Hamiltonian
$H_N$ is that it evidently diverges in the case when $G_{nm}$ is the Helmholtz kernel $G_{nm}(x)=e^{|x|/\alpha_{nm}}$, which is the case of the Camassa-Holm
equation. However, one can solve this
problem by defining the kernels $G_{nm}$ through the introduction of a sequence
of coefficients $c_{nm}$ such that $c_{nm}\to\infty\quad\!\!\text{with }{n,m\to\infty}$.
For example, one can define
\begin{equation}\label{greens}
G_{nm}(x)=
\frac1{c_{nm}}\left(1-\alpha_{nm\,}\partial^2\right)^{-1}
=
\frac1{c_{nm}}\,e^\text{\normalsize$\,\frac{\left|x\right|}{\alpha_{nm}}$}
\end{equation}
In this case the Hamiltonian $H_N$ becomes
\begin{equation}\label{2Ham}
H_N=
\frac12\sum_{n,m} \left(\frac1{c_{nm}}\left(P_1^{n+m}+P_2^{n+m}\right)
+
2\left.G_{nm}(Q_1-Q_2\right)P_1^n\,P_2^m\right)
\end{equation}
and if $c_{nm}\to\infty$ sufficiently rapidly, then the Hamiltonian converges.
A particular choice inspired by Taylor series could be $c_{nm}=(n+m)!$. For
example one evaluates the sum
\begin{align*}
\frac12\sum_{n,m}\frac{P^{n+m}}{(n+m)!}
=
\frac12&\bigg(1+(P^{1+0}+P^{0+1})+\frac12(P^{1+1}+P^{2+0}+P^{0+2})
\\
&
+\frac1{3!}(P^{3+0}+P^{0+3}+P^{1+2}+P^{2+1})
\\
&+\frac1{4!}(P^{4+0}+P^{0+4}+P^{1+3}+P^{3+1}+P^{2+2})
+\dots+\frac{n+1}{n!}\,P^n\bigg)
\end{align*}
which evidently diverges. Consequently, the right choice for $c_{nm}$ becomes
$c_{nm}=(n+m+1)!$ so that
\begin{align*}
\frac12\sum_{n,m}\frac{P^{n+m}}{(n+m)!}
=
\frac12\sum_{n}\frac{n+1}{(n+1)!}\,P^n
=\frac12\sum_{n}\frac{1}{n!}\,P^n
=
\frac12\,e^P
\end{align*} 
Thus, upon redefining $c_{nm}=(n+m+1)!/2$ for convenience, the Hamiltonian becomes
\begin{equation}\label{2Ham-conv}
H_N=
e^{\,P_1}+e^{\,P_2}
+
\sum_{n,m}\left.G_{nm}(Q_1-Q_2\right)P_1^n\,P_2^m
\end{equation}
which yields a particle velocity of the form
\[
\dot{Q}_{\,1}=e^{\,P_1}+
2\sum_{n,m}\,n\left.G_{nm}(Q_1-Q_2\right)P_1^{n-1}P_2^m
\]
Thus one has the following
\begin{proposition}
With the choice of metric (\ref{greens}) and for $c_{nm}=(n+m+1)!/2$, the two particle Hamiltonian (\ref{2Ham}) converges to the expression (\ref{2Ham-conv}).
\end{proposition}

Now one can specializes to the case when the metric $G_{nm}$ is diagonal
($G_{nm}=G_n\,\delta_{nm}$),
so that the Hamiltonian becomes
\begin{equation}\label{2Ham-diag}
H_N
=\frac12\sum_{n}\sum_{i,j}\,P_i^{n}\,G_{n}(Q_i-Q_j)\,P_j^n
\end{equation}
that is, for $i,j=1,2$
\[
H_N=
\frac12\sum_{n} \left(\frac1{c_{n}}\left(P_1^{2n}+P_2^{2n}\right)
+
2\left.G_{n}(Q_1-Q_2\right)P_1^n\,P_2^n\right)
\]
and if one chooses $c_n=(2n)!/2$ (the factor 2 is just a convenient choice), then one can write
\begin{equation}\label{2Ham-diag-conv}
H_N=
\cosh(P_1)+\cosh(P_2)
+
\sum_n\left.G_{n}(Q_1-Q_2\right)P_1^n\,P_2^n
\end{equation}
This result can be summarized as
\begin{proposition}
The two particle Hamiltonian (\ref{2Ham-diag}) with the metric
\[
G_n(x)=\frac2{(2n)!}\,e^{\frac{|x|}{\alpha_{n}}}
\]
converges to the expression in (\ref{2Ham-diag-conv}).
\end{proposition}

The quadrature formulas for these systems are left for further study as well
as the expressions for phase shifts in the collisions. However it is interesting
to notice the particular forms assumed by the Hamiltonian $H_N$ which are very different from the usual expression used in physics $H=T+V=\left.1/2\,g^{kh}(Q)\,P_k\,P_h+V(Q)\right.$.

\begin{remark}[Remark about higher dimensions]
The singular solutions (\ref{sing-soln}) with the
integral over coordinate $a$ exist in higher
dimensions. The higher dimensional singular solutions satisfy a system of
canonical Hamiltonian integral-partial differential equations, instead of
ordinary differential equations.
\end{remark}

\begin{remark}[Connections with EPDiff]
The singular solutions of EPSymp are also
solutions of the Euler-Poincar\'e equations on the diffeomorphisms
(EPDiff), provided one considers only the first order moment
\cite{HoMa2004}. In this case, the singular solutions reduce in one dimension to the pulson
solutions for EPDiff \cite{CaHo1993}. 

Thus the pulson solution for EPDiff has been shown to arise very naturally as the closure of single-particle dynamics given by cotangent lifted diffeomorphisms on phase-space.
\end{remark}

\subsection{Connections with the cold plasma solution}
A more general kind of singular solution for the moments
may be obtained by considering the {\bfi cold-plasma solution} of the Vlasov
equation
\begin{equation}
f(q,p,t)=\sum_j\,\rho_j(q,t)\,\delta(p-P_j(q,t))
\end{equation} 
For example, the single particle solution is recovered by putting $\rho_j(q,t)=\delta\left(q-Q_j(t)\right)$.
Moreover exchanging the variables $q\leftrightarrow p$ in the
single particle PDF leads to the following expression
\begin{equation}
f(q,p,t)=\sum_j \psi_j(p,t)\delta(q-\lambda_j(p,t))
\end{equation}
which is always a solution of the Vlasov equation because of the symmetry
in $q$ and $p$. This leads to the following singular solutions for the moments:
\begin{equation}
A_n(q,t)=\sum_j \int\!{\rm d}p\, p^n\,\psi_{j}(p,t)\,\delta(q-\lambda_j(p,t))
\end{equation}
At this point, if one considers a Hamiltonian depending only on $A_1$ (i.e.
one considers the action of cotangent lifts of diffeomorphism), then it is
possible to drop the $p$-dependence in the $\lambda$'s and thereby recover to the
singular solutions previously found for eq. (\ref{conslaw}). In order to
understand this point, one can proceed as follows. Let $\lambda_j$ be independent
of $p$ and define $\lambda_j=:Q_j(t)$; thus one writes the moments as
\begin{align*}
A_n(q,t)=&\sum_j \int\! p^n\,\psi_{j}(p,t)\,{\rm d}p\,\,\delta(q-Q_j(t))
=:\sum_j P_{n,\,j}(t)\,\delta(q-Q_j(t))
\end{align*}
where one defines $P_{n,\,j}(t):=\int\! p^n\,\psi_{j}(p,t)\,{\rm d}p$.
In order to calculate the dynamics of $P_n$ and $Q$, it suffices to substitute
the expression above in the moment equations (\ref{A-dot}) and to calculate the pairing with contravariant $n$-tensors $\varphi_n$. This procedure leads
to
\begin{align*}
\dot{P}_{n}  & =-\,
n\sum_m
\,P_{m+n-1}\,
\widehat{\beta}_{m}^{\,\prime}\\
P_{n}\,\dot{Q}  & =
\sum_{m}
m\,P_{m+n-1}\,
\widehat{\beta}_{m}
\end{align*}
which hold for all non-negative integers $n$. In particular, fixing $n=0$ yields
$\dot{P}_0=0=\int\dot{\psi}(p,t)\,{\rm d}p$, consistently with the hypothesis
$\int f\,{\rm d}q\,{\rm d}p=1=P_0\,$. More importantly, fixing $n=0$ yields
the dynamics for the coordinate $Q$
\[
\dot{Q}=\sum_m m\,P_{m-1}\,\widehat{\beta}_m
\,.
\] 
Multiplying by $P_n$ again leads to the powers $P_n=P^n$. In fact,
since $Q$ is independent on $n$ one obtains
\[
P_n\sum_m m\,P_{m-1}\,\widehat{\beta}_m
=
\sum_m m\,P_{n+m-1}\,\widehat{\beta}_m
\]
which means that
\[
P_k P_h=P_{h+k}
\quad\Rightarrow\quad
P_n=P^n
\quad\!
\forall n\geq0
\]
and thus obtaining the singular solutions (\ref{sing-soln2}), corresponding
to single particle dynamics, that is $\psi(p,t)=\delta(p-P(t))$. It is worth
noticing that the single particle solution arises only when considering moment
dynamics, while it is always possible to allow for a Vlasov solution of the
form $f=\psi(p,t)\,\delta(q-Q(t))$. This happens because the power $P^n$ in the singular solutions restricts the moments to be necessarily symmetric,
when they are considered as covariant tensors. This result differs from that
obtained for eq. (\ref{conslaw}), which always allows the solution 
$A_n(q,t)=P_n(t)\,\delta(q-Q(t))$. The reason is that eq. (\ref{conslaw})
is generated by the action of diffeomorphisms, which always preserves the symmetric
nature of the tensor $P_n$ in the dynamics. This is not true for all the canonical transformations,
whose general action does not keep $P_n$ symmetric during
its evolution; rather the tensor $P_n$ becomes a tensor power $P_n=P^n$, which is symmetric
by definition. 
In this spirit, the solution
\begin{equation*}
A_n(q,t)=\sum_j \int p^n\,\psi_{j}(p,t)\,\delta(q-\lambda_j(p,t))\,{\rm d}p
\end{equation*}
represents a more general singular solution than the solutions (\ref{singsol-colifts}),
since it embodies the action of more general canonical transformations, which are not
cotangent lifts of diffeomorphisms on the configuration manifold.

\subsection{A result on truncations: the CH-2 equation} 
The problem presented by the coadjoint motion equation  (\ref{A-dot}) for geodesic evolution of moments under EPDiff may be
simplified, by truncating the Poisson bracket to a finite set. 
Such truncations
are not in general consistent with the full dynamics; in the rarer cases
where they are consistent, they will be referred to as ``reductions'' \cite{GiTs1996}. 
These moment dynamics may be truncated to a Hamiltonian system, at any stage by simply modifying the Lie algebra in the KMLP bracket to vanish for weights $m+n-1$ greater than a chosen cut-off value.
\begin{framed}
For example, if one truncates the sums to $m,n=0,1,2$ only, then
equation (\ref{A-dot}) produces the coupled system of partial
differential equations,
\begin{align}\nonumber
\frac{\partial A_{0}}{\partial t} 
&  =
-\partial_{q}\left(  A_{0}\beta_{1}\right)  
-2A_{1}\partial_{q}\beta_{2}
-2\beta_{2}\partial_{q}A_{1}
\\
\frac{\partial A_{1}}{\partial t} 
&  =
-A_{0}\partial_{q}\beta_{0}-2A_{1}\partial_{q}\beta_{1}
-\beta_{1}\partial_{q}A_{1}-3A_{2}\partial_{q}\beta_{2}
-2\beta_{2}\partial_{q}A_{2}
\label{A2-Truncation}\\
\frac{\partial A_{2}}{\partial t} 
&  =
-2A_{1}\partial_{q}\beta_{0}
-3A_{2}\partial_{q}\beta_{1}-\beta_{1}
\partial_{q}A_{2}
\nonumber
\end{align}
The fluid closure of system (\ref{A2-Truncation}), which may be called \emph{EPSymp fluid}, neglects $A_2$ and may be written as
\begin{align}\nonumber
\frac{\partial A_{0}}{\partial t} &  =-\partial_{q}\left(  A_{0}\beta_{1}\right)  \\
\frac{\partial A_{1}}{\partial t} &  =-A_{0}\partial_{q}\beta_{0}-2A_{1}\partial_{q}\beta_{1}-\beta_{1}%
\partial_{q}A_{1}
\label{FluidClosureSystem}
\end{align}
When $A_1=(1-\alpha^2\partial_q^2)\beta_1$ and $\beta_0=A_0$, this system becomes the two-component Camassa-Holm system (CH-2) studied in \cite{ChLiZh2005,Falqui06,Ku2007}.
\end{framed}
\noindent
For this case, the fluid closure system (\ref{FluidClosureSystem}) is equivalent to the compatibility for $d\lambda/dt=0$ of a system of two linear equations,
\begin{eqnarray}
\partial_x^2\psi 
&+& \Big( -\,\frac{1}{4} 
+ A_1\lambda + A_0^2\lambda^2\Big )\psi = 0
\label{LaxPair-eigen}
\\
\partial_t\psi
 &=& 
-\,\Big( \frac{1}{2\lambda} + \beta_0\Big)\partial_x\psi
+ \frac{1}{2}\psi\partial_x \beta_1
\label{LaxPair-evol}
\end{eqnarray}
The first of these (\ref{LaxPair-eigen}) is an eigenvalue problem known as the {\bfi Schr\"odinger equation with energy dependent potential}. Because the eigenvalue $\lambda$ is time independent, the evolution of the nonlinear fluid closure system (\ref{FluidClosureSystem}) is said to be {\bfi isospectral}. The second equation (\ref{LaxPair-evol}) is the evolution equation for the eigenfunction $\psi$. 

The fluid closure system for geodesic flow of the first two Vlasov moments also has a semidirect product structure on Diff$(\mathbb{R}^3)\circledS$Den$(\mathbb{R}^3)$ \cite{HoMaRa} which allows for singular solutions for both $A_0$ and $A_1$ in the case that $\beta_s=G*A_s$, $s=0,1$. 
The behavior of these singular solutions will be investigated in future work. In particular one would like to understand whether these singularities may emerge spontaneously as for the EPDiff equation.
\begin{remark}[CH-2 equation for imaging]
Remarkably, a similar system of equations also arises in the study of imaging using a process of template matching with active templates, known as metamorphosis \cite{HoTrYo2007}. In this context these equations are called {\rm EP}G$\circledS$H,
which emphasizes the semidirect product structure.
\end{remark}
\begin{remark}[Euler-Poincar\'e equations for the EPSymp fluid]
As mentioned in section~\ref{geodesic-motion}, the moment equations for EPSymp have an Euler-Poincar\'e
formulation, which is given by the hierarchy of equations (\ref{EPSymp-mom}).
This hierarchy can be truncated to obtain the Euler-Poicar\'e equations for
the fluid closure (\ref{FluidClosureSystem}). In order to keep close to the
formulation of the Camassa-Holm equation, one can choose $\hat{Q}_n=1-\alpha^2_n\partial_q^2$
in the equations (\ref{EPSymp-mom}). If $\alpha_1=1$, then one obtains 
\begin{align*}
\lambda_t-\alpha_0^2\lambda_{qqt}&=-\left(u\lambda-\alpha_0^2 u \lambda_{qq}\right)_q
\\
u_t-u_{qqt}&=-3uu_q+2u_q u_{qq}+u u_{qqq}
-\lambda_q\left(\lambda-\alpha_0^2\lambda_{qq}\right)
\end{align*}
with $A_1=\left(1-\partial_q^2\right)\beta_1$ and one introduces the notation $(\beta_{0},\,\beta_1)=(\lambda,u)$. This yields an extension
of the two component Camassa-Holm equation, which is {\sl nonlocal in both
density and momentum}. Again, for $\alpha_0\to0$, one
recovers the results in \cite{ChLiZh2005,Falqui06,Ku2007}. Of course, integrability issues for this system remain to be pursued elsewhere.
\end{remark}

\begin{remark}[Singular solutions]
The interaction of two singular solutions of the EPSymp fluid may be easily
analyzed by truncating the Hamiltonian~(\ref{H_N}) to consider only $n=0,1$. This yields
\[
H_2=
\frac12 \Big(\left.P_1^{2}+P_2^{2}
+
2\left.G_{1}(Q_1-Q_2\right)P_1\,P_2
+
2\left.G_{0}(Q_1-Q_2\right)
\Big)\right.
\]
By proceeding in the same way as in \cite{HoSt03}, one defines
\begin{align*}
P&=P_1+P_2
\,,\quad
Q=Q_1+Q_2
\,,\quad
p=P_1-P_2
\,,\quad\,\,
q=Q_1-Q_2
\end{align*}
so that, the Hamiltonian can be written as
\[
\mathcal{H}=\frac12P^2-\frac14(P^2-p^2)\left(1-G_1(q)\right)+G_0(q)
\]
At this point one writes the equations
\begin{align*}
\frac{dP}{dt}&=-2\frac{\partial \mathcal{H}}{\partial Q}=0
\,,
\hspace{5.65cm}
\frac{dQ}{dt}=2
\frac{\partial \mathcal{H}}{\partial P}=P\left(1+G_1(q)\right)
\\
\frac{dp}{dt}&=
-2\frac{\partial \mathcal{H}}{\partial q}=
-\frac12\left(P^2-p^2\right)G_1^{\,\prime}(q)-2G_0^{\,\prime}(q)
\,,
\hspace{0.72cm}
\frac{dq}{dt}=
2\frac{\partial \mathcal{H}}{\partial p}=
-p\left(1-G_1(q)\right)
\end{align*}
that yield
\[
\left(\frac{dq}{dt}\right)^2=
P^2\left(1-G_1(q)\right)^2
-4
\left(H-G_0(q)\right)
\left(1-G_1(q)\right)
\]
and finally lead to the quadrature
\[
dt=\frac{dG_1}{G_1^{\,\prime}\sqrt{P^2\left(1-G_1(q)\right)^2
-4\left(H-G_0(q)\right)\left(1-G_1(q)\right)}}
\,.
\]
\end{remark}

\medskip
Setting $A_0$ and $A_2$ both initially to zero in (\ref{A2-Truncation}) reduces these three equations to the single equation 
\begin{eqnarray}
\frac{\partial  A_1}{\partial t}
&=&
-\,\beta_1\,{\partial_q}A_1
-\,2A_1\,{\partial_q}\beta_1
\,.
\end{eqnarray}
Finally, if one assumes that $G$ in the convolution $\beta_1=G*A_1$ is the
Green's function for the operator relation 
\begin{equation}
A_1=(1-\alpha^2\partial_q^2)\beta_1
\end{equation}
for a constant lengthscale $\alpha$, then the evolution equation
for $A_1$ reduces to the integrable Camassa-Holm (CH) equation
\cite{CaHo1993} in the absence of linear dispersion. This is the one-dimensional EPDiff equation, which has singular (peakon) solutions. 

Thus, even very drastic restrictions of the moment system still lead to interesting special cases, some of which are integrable and possess emergent coherent structures among their solutions. That such strong restrictions of the moment system leads to such interesting special cases bodes well for future investigations of the EPSymp moment equations. Before closing, it is useful to mention
other open questions about the solution behavior of the moments of EPSymp.

\subsection{Extending EPSymp to anisotropic interactions}
An example of how the geodesic motion on the moments can be extended to include
extra degrees of freedom is provided by the work of Gibbons, Holm and Kupershmidt
\cite{GiHoKu1982,GiHoKu1983}, where the authors consider a Vlasov distribution depending also
on a dual Lie algebra variable $g\in\mathfrak{g}^*$ undergoing Lie-Poisson dynamics in finite
dimensions. Following
the treatment in \cite{GiHoKu1983}, take the purely quadratic Hamiltonian on $\mathfrak{s}^*\left(T^*\mathbb{R}\oplus\mathfrak{g}^*\right)$ (with $\mathfrak{s}:=T_e\rm Symp$) defined by
\[
H[f]=\int\hspace{-0.25cm}\int\hspace{-0.25cm}\int 
f(q,p,g)\,(\mathcal{G}*f)(q,p,g) \,\,{\rm d}q\,{\rm d}p\,{\rm d}^3\!g
\]
with notation $g=g_ae^a$, pairing $\langle e^a,\,e_b\rangle=\delta^a_b$ and Lie bracket $[g_b,\,g_c]=\epsilon^a_{bc}g_a$
\[
(\mathcal{G}*f)(q,p,g)=\int\hspace{-0.25cm}\int\hspace{-0.25cm}\int 
\mathcal{G}(q,q',p,p',g,g') f(q',p',g') \,\,{\rm d}q'\,{\rm d}p'\,{\rm d}^3\!g'
\]
The geodesic Vlasov equation is given in \cite{GiHoKu1982} as
\[
\frac{\partial f}{\partial t}
\,=\,-\,\Big\{\,f,\,
\mathcal{G}*f\,
\Big\}_1
\,,
\]
where $\{\,\cdot\,,\,\cdot\}_1$ is the sum of the canonical bracket on $T^*\mathbb{R}$ and the Lie-Poisson bracket on $\mathfrak{g}^*$,
\[
\Big\{\,f\,,\,k\Big\}_1
=
\Big\{\,f\,,\,k\Big\}
+
\bigg\langle g\,,\, \left[\frac{\pa f}{\pa g}\,,\,\frac{\pa k}{\pa g}\,\right] \bigg\rangle
\,,
\]
in vector notation for elements of $so(3)^*$.
Now, assume that the kernel $\cal G$ can be expanded as
\[
\mathcal{G}(q,q',p,p',g,g')\,=\,
K_0(q,q')\,+\,p\,K_1(q,q')\,p'\,+\,g_a\, {\bar K}^{ab}(q,q')\,g'_b
\]
so that the quadratic Hamiltonian becomes
\[
H=
\int\rho(q)\,(K_0*\rho)(q)\,{\rm d}q
\,\,+
\int M(q)\,(K_1*M)(q)\,{\rm d}q
\,\,+
\int \left\langle G(q),\,(\bar{K}\bullet G)(q)\right\rangle\,{\rm d}q
\]
where one defines
\[
\bar{K}\bullet G(q):=\int {\bar K}^{ab}(q,q')\,G_b(q')\,{\rm d}q'\,\, e_a
\in so(3)
\,.
\]
The moment equations for 
mass density
$\rho(q,t)=\int f\,{\rm d}p\,{\rm d}^3\!g$, 
momentum density $M(q,t)=\int pf\,{\rm d}p\,{\rm d}^3\!g$ and 
orientation density $G(q,t)=\int gf\,{\rm d}p\,{\rm d}^3\!g$ 
are presented in  \cite{GiHoKu1982}. 
\begin{framed}
\noindent
For the quadratic Hamiltonian above these become
\begin{align}
\frac{\partial \rho}{\partial t}
&=
-\frac{\partial}{\partial q}
\left(\rho\,u\right)
\\
\frac{\partial G}{\partial t}
&=
-\frac{\partial}{\partial q}
\left(G\,u\right)
+
{\rm ad}^*_{\,\bar K\bullet\, G}\,\,G
\\
\frac{\partial M}{\partial t}
&=-\pounds_u M 
-\rho\,\frac{\partial}{\partial q}\!\left(K_0*\rho\right)
-\left\langle G,\,\frac{\partial}{\partial q}\!\left(\bar K\bullet 
G\right)\right\rangle
\end{align}
where $u=K_1*M$.
When $G\in\mathcal{F}^*(\mathbb{R})\otimes\mathfrak{so}(3)^*$, then ${\rm ad}^*_{\,\bar K\bullet\, G}\,\,G=-\,(\bar K \bullet\mathbf{G})\times\mathbf{G}$
and one recognizes the Hamiltonian part of the Landau-Lifschitz equation on the right hand side in the second equation
\[
\frac{\partial \bf G}{\partial t}
=
-\frac{\partial}{\partial q}
\left({\bf G}\,u\right)+\,\mathbf{G}\times\frac{\delta H}{\delta \mathbf{G}}
\,.
\]
\end{framed}
\noindent
For $K_1=(1-\partial_q^2)^{-1}$ and $K_0=\delta(q-q')$, this extends the Camassa-Holm system to several components. Such an approach will be also
followed in Chapter \ref{orientation}\, for aggregation and self-assembly of oriented nano-particles in the context of double bracket dissipation.

\paragraph{Singular solutions.}
This section presents the interaction of two singular solutions of the equations
presented in this section in the particular case of $\mathfrak{g}=\mathfrak{so}(3)$
in the simple case when the density variable $\rho$ is neglected. The result generalizes the pulson solutions to the possibility of \emph{oriented} pulsons,
which may be called ``orientons''.

One starts with the Hamiltonian
\[
\mathcal{H}=\frac12\left\langle M, K*M\right\rangle+\frac12\left\langle \bG, H*\bG\right\rangle
\]
and by inserting the singular solution ansatz 
\[
M(q,t)=\sum_i P_i(t)\,\delta(q-Q_i(t))
\,,\qquad
\bG(q,t)=\sum_i \bmu_i(t)\,\delta(q-Q_i(t))
\]
the Hamiltonian becomes
\[
\mathcal{H}=\frac12\sum_{i,j} p_i\, p_j \,K^{ij}
+
\frac12\sum_{ij}\,\left\langle \bmu_i,\, H^{ij\,}\bmu_j\right\rangle
\]
with
\[
K^{ij}=K(Q_i-Q_j)
\quad\text{and }\quad
H^{ij}=H(Q_i-Q_j)
\]
equations of motions
\begin{align*}
\dot{Q}_{\,i}&=\frac{\partial \mathcal{H}}{\partial p_i}
=
\sum_j \left.K(Q_i-Q_j\right)p_j
\\
\dot{p}_i&=-\frac{\partial \mathcal{H}}{\partial Q_i}
=
-\,p_i\sum_j
\left.K^\prime(Q_i-Q_j\right) p_j
-\sum_j
\left\langle
\bmu_i,\left.H^\prime(Q_i-Q_j\right) \bmu_j
\right\rangle
\\
\dot{\bmu}_i&=\textrm{\large ad}^*_\text{\normalsize$\frac{\partial \mathcal{H}}{\partial \bmu_i}$}\,
\bmu_i
=
\sum_j\textrm{\large ad}^*_{_\text{\small$\!\left.H(Q_i-Q_j\right)\bmu_j$}}
\bmu_i
\end{align*}
For simplicity, we restrict to the case $\bmu\in\mathfrak{so}(3)$. This
does not affect the validity of the following result, which is true for any
finite-dimensional Lie-algebra.

It is straightforward to verify that the orienton--orienton system has the following eight constants of
motion
\[
H
\,,\quad
P=p_1+p_2
\,,\quad
\bmu=\bmu_1+\bmu_2
\,,\quad
\theta_{ij}=\bmu_i\cdot\bmu_j
\quad
\forall\,i,j=1,2
\]

\noindent
In order to prove the conservation of $P$, take the equation for $p_1$:
\begin{align*}
\dot{p}_1=&-\,p_1\left(
p_1\left.\frac{\partial}{\partial q}\right|_{q=Q_1}\!\!K(Q_1-q)
+
p_2\left.\frac{\partial}{\partial q}\right|_{q=Q_1}\!\!K(Q_2-q)\right)
\\
&-
\left\langle
\bmu_1,\left(
\left.\frac{\partial}{\partial q}\right|_{q=Q_1}\!\!H(Q_1-q)\, \bmu_1
+
\left.\frac{\partial}{\partial q}\right|_{q=Q_1}\!\!H(Q_2-q)\,\bmu_2
\right)
\right\rangle
\\
=&
-\,p_1\,p_2\,\partial_{Q_1}K(Q_2-Q_1)
-\,\left\langle\bmu_1,\partial_{Q_1}H(Q_2-Q_1)\,\bmu_2\right\rangle
\end{align*}
so that $\dot{p}_1+\dot{p}_2=0$, since $\partial_{Q_1}K(Q_2-Q_1)=-\partial_{Q_2}K(Q_2-Q_1)$
(analogously for $H$).

\noindent
Also one proves
\begin{align*}
\dot{\bmu_1}+\dot{\bmu_2}&=
\textrm{\large ad}^*_{_\text{\small$H^{12}\bmu_2$}}\bmu_1
+
\textrm{\large ad}^*_{_\text{\small$H^{21}\bmu_1$}}\bmu_2
\\
&=
\textrm{\large ad}^*_{_\text{\small$H^{12}\bmu_2$}}\bmu_1
-
\textrm{\large ad}^*_{_\text{\small$H^{12}\bmu_2$}}\bmu_1
=0\,.
\end{align*}
The conservation of $\theta$ is another simple result. This conclusion is not affected by the insertion of the density variable $\rho=\int f\, {\rm d}p\,{\rm d}\bmu$ in the dynamics.

\section{Open questions for future work}

\paragraph{Singular solutions for EPSymp.} 
Several open questions remain for future work. The first of these is whether the singular solutions found here will emerge
spontaneously in EPSymp dynamics from a smooth initial Vlasov PDF. This
spontaneous emergence of the  singular solutions does occur for EPDiff.
Namely, one sees the singular solutions of EPDiff emerging from {\it any}
confined initial distribution of the dual variable. What happens with the
singular solutions for EPSymp? Will they emerge from a confined smooth initial
distribution, or will they only exist as an invariant manifold for
special initial conditions? Of course, the interactions of these singular
solutions in higher dimensions and for various metrics and the properties of their collective dynamics is a question for future work. The same questions
apply to the case of anisotropic interactions. For example, the interaction
of two filaments carrying an extra degree of freedom in two or three dimensions
would be a very interesting problem, which could also shed light on the questions
arising in chapter~\ref{orientation}.


\paragraph{Similarities with the Bloch-Iserles equation.}
A finite dimensional integrable equation has been recently proposed by Bloch and Iserles, which
may be written in the even-dimensional case as the geodesic equation on the group of the linear canonical transformations Sp($\mathbb{R},2n$) \cite{BlIsMaRa05}. Given an antisymmetric matrix $N$, this equation is usually written on the space of symmetric matrices as
\[
\dot{X}=[X^2,N]
\] 
where the bracket is the usual matrix commutator. On the other hand, it is
well known that
\[
\widehat{X}=NX\in\mathfrak{sp}(\mathbb{R},2n)
\]
is a Hamiltonian matrix associated to the symplectic form $N^{-1}$ (if $N$
is not invertible, this system is still integrable). At this point a Lie algebra isomorphism can
be constructed between symmetric and Hamiltonian matrices \cite{BlIsMaRa05}, through the Lie
bracket relation
\[
N[X,Y]_N=[\widehat{X},\widehat{Y}]
\qquad\quad\text{ with }\qquad
[X,Y]_N:=XNY-YNX
\]
The Bloch-Iserles equation arises now as the Euler-Poincar\'e equation on the
Lie algebra $[\cdot,\cdot]_N$ of symmetric matrices, where the Lagrangian
$l(X)$ is given by
\[
l(X)=\frac12\,{\rm Tr}(X^2)
\]
By the isomorphism above, this equation is then equivalent
to the Euler-Poincar\'e equation on the Hamiltonian matrices $\mathfrak{sp}(\mathbb{R},2n)$:
thus one wonders what connections there may be between this equation and the geodesic Vlasov equation (EPSymp) which has been proposed in this paper, given the surprisingly similar nature of these two equations. In particular
one wonders whether integrability properties may arise also for EPSymp, with a certain choice of metric. In finite dimensions, a certain class of geodesic flows on Lie groups is well known to be integrable from the work of Mi\u s\u cenko and Fomenko \cite{MiFo1978}. Nevertheless, the Bloch-Iserles system
does not belong to the Mi\u s\u cenko-Fomenko class \cite{BlIsMaRa05}. In infinite dimensions, some important examples of geodesic flows on Diff$_\text{vol}$ (Euler's equation) and Diff (CH equation) are also integrable. Thus it is a reasonable question whether the geodesic moment hierarchy corresponding to EPSymp may exhibit integrable dynamics. A positive answer is already available for the fluid closure, recovering the CH and CH-2 equations. An investigation of the relations between the EPSymp equation and the Bloch-Iserles system would be fundamental to answer such questions.

\chapter{GOP theory and geometric dissipation}
\label{GOP}
\section{Introduction} 
The approach to a critical point in free energy of a continuum material may produce pattern formation and self-organization.   Diverse examples of such processes include the formation of stars and galaxies at large scales, growth of colonies of organisms at mesoscales and self-assembly of proteins or micro/nanodevices at micro- and nanoscales  \cite{Whitesides2002}. Some of these processes, such as  nano-scale self-assembly of molecules, are of great technological interest.   
Due to the large number of particles involved in nano-scale self-assembly, the development of continuum descriptions for aggregation or self-assembly is a natural approach
toward its theoretical understanding and modeling. This chapter shows how
such continuum descriptions may be formulated in  order to allow the existence
of singular solutions. \smallskip

A useful concept for deriving a continuum description of macroscopic  pattern formation (e.g., aggregation) due to microscopic  processes is the notion of {\bfi order parameter}. Order parameters are continuum variables that describe macroscopic effects due to microscopic variations of the internal structure \cite{Ho2002}. They take values in a vector space called the order parameter space that respects the underlying geometric structure of the microscopic variables. The canonical example is the description of the local directional asymmetries of nematic liquid crystal molecules by a spatially and temporally varying macroscopic continuum field of unsigned unit vectors called ``directors'',
see, e.g. Chandrasekhar \cite{Ch1992}  and de Gennes and Prost \cite{deGePr1993}.
\smallskip

The classic examples of continuous equations for aggregation are those  of Debye-H\"uckel \cite{DeHu1923} and  Keller-Segel (KS) \cite{KellerSegel1970} for which the order parameter is the density of particles. The physics of these models consists of Darcy's law,
introduced in chapter \ref{intro}: 
$\partial_t\rho =\mbox{div} \left(\rho\,\mathbf{u}\right)$, coupled with an evolution equation for velocity $\mathbf{u}$ which depends on the density $\rho$ through a free energy $E$ as  $\mathbf{u}\simeq \mu \nabla \delta E/\delta \rho$ (velocity proportional to force), in which `mobility' $\mu$ may also depend on the density. 

The idea of a velocity proportional to the
force has its roots in the work of George Gabriel Stokes, who formulated his famous drag law for the resistance of spherical particles moving in a viscous fluid at low Reynolds numbers (dominance of viscous forces). It is commonly assumed that all processes in fluids at micro- and nano-scales are dominated by viscous forces and the Stokes approximation applies. The Stokes result states that a round particle moving through ambient fluid will experience a resistance force that is proportional to the velocity of the particle. Conversely, in the absence of inertia, the velocity of a particle will be proportional to the force applied to it since resistance force and applied force must balance.
This law, that ``force is proportional to velocity'' is also known as {\it
Darcy's law}. At this point it is clear how  dissipation and friction are key
concepts in the development of this theory. As a result, {\bfi friction dominated
systems} described by Darcy's law exhibit aggregation and self-assembly phenomena
that can be recognized mathematically through the formation of singularities
clumping together in a finite time \cite{HoPu2005, HoPu2006}.

  Previous investigation by Holm and Putkaradze \cite{HoPu2007} extended Darcy's Law to incorporate nonlocal, nonlinear and anisotropic effects in self-organization of aggregating particles of finite size. This theory produces a whole family
of geometric flows that describe certain dissipative
dynamics. In particular, the Holm-Putkaradze theory formulates a form of {\bfi geometric dissipation} for continuum systems describing the evolution of order parameters
(geometric order parameter (GOP) equations). \smallskip

The main goals of this chapter are 
\begin{itemize}
\item to present the author's contribution to the Holm-Putkaradze theory;
\item to present a particular application to vorticity dynamics.
\end{itemize}

The first allows for the existence of singular solutions in order to capture coherent structures. This is done by introducing a spatial averaging that follows these coherent structures in a Lagrangian sense. In principle, the averaging process can be inserted in two different ways, although only one of them allows the formation of singularities.

The second scope is to present an application that prepares for the developments in the next chapters. It is shown how the Euler vorticity equation (for an exact two-form) can be extended to include a Darcy-like dissipation term.  This formalism generalizes earlier modified fluid equations of this type in Bloch et al. \cite{BlKrMaRa1996, BlBrCr1997} and Vallis et al. \cite{VaCaYo1989} so that
the theory now allows for point vortex solutions or vortex filaments (and
sheets) in three dimensions.

\rem{ 
In Sec.~\ref{Meth-MSV} this Lagrangian averaging approach yields nonlocal equations expressed in fixed spatial coordinates. 
Rather general mathematical and physical requirements (correct geometry of the measured variable and independence of the method of measurement) in Sec.~\ref{sec:thermodynamics} yield a surprisingly narrow class of allowable equations.
Remarkably, these general considerations yield the same evolution equation (\ref{GOP-eqn-brkt}) for an arbitrary geometric quantity as found by using the Lagrangian averaging method, except for an ambiguity in the roles of the order parameter and its mobility, which live in the same vector space. One can consider aggregation of geometric order parameters in continua that are either stationary, or flowing. \smallskip
} 

Applications of this general framework also include the dynamics of geometric
quantities such as scalars, densities (Darcy's law), one- and two-forms. Each
flow recovers the singular solutions. Depending on the geometric type of the order parameter, the space of singular solutions may either form an invariant manifold, or these solutions may emerge from smooth confined initial conditions. In the latter case, the singular solutions dominate the long-term aggregation dynamics. From the physical point of view, such localized, or quenched solutions would form the core of the processes of self-assembly and are therefore of great practical interest. The formation of these localized solutions is driven by a combination of  nonlinearity and nonlocality. Their evolution admits a reduced description, expressed completely in terms of coordinates on their singular embedded subspaces.


\section{Theory of geometric order parameter equations}

\subsection{Background: geometric structure of Darcy's law} 
Darcy's law for the geometric order parameter $\rho$
(density) \cite{HoPu2005,HoPu2006} is written in terms of an energy functional
$E=E[\rho]$ and a {\it mobility} $\mu$ which takes into account of the typical
size of the particles in the system (in general it depends on $\rho$). In
formulas, one has the equation
\begin{equation}\label{DLaw}
\frac{\pa \rho}{\pa t}=
{\rm div}\!\left(\rho\,\mu[\rho]\,\nabla\frac{\delta E}{\delta
\rho}\right)
\,.
\end{equation}
This may be stated in terms of Lie derivatives
in two possible ways as
\begin{equation}\label{exchange-darcy}
\frac{\pa \rho}{\pa t}=
\textit{\large\pounds}_\text{\!\footnotesize$\left(\rho\nabla\frac{\delta E}{\delta
\rho}\right)^{\!\sharp}$}\,\,\mu[\rho]
\qquad\text{or}\qquad
\frac{\pa \rho}{\pa t}=
\textit{\large\pounds}_\text{\!\footnotesize$\left(\mu[\rho]\nabla\frac{\delta E}{\delta \rho}\right)^{\!\sharp}$}\,\rho
\end{equation}
\rem{ 
\begin{quote}
``The local value of $\rho$ remains invariant along the characteristic curves of a flow, whose velocity $\bf u[\rho]$ depends on $\rho$.''
\end{quote}
This principle may be formulated in symbols as, 
\begin{eqnarray}
\frac{d\rho}{dt}(\mathbf{x}(t),t)=0
\quad\hbox{along}\quad
\frac{d\mathbf{x}}{dt}=\mathbf{u}[\rho]
\,.
\label{GOPprincip}
\end{eqnarray}
where the flow velocity $\mathbf{u}[\rho]$ is to be expressed .
In the case of particle density $\rho$ in $n$-dimensional space, for example, the number of particles
$\kappa=\rho\,\mbox{d}^n\mathbf{x}$ has physical meaning.
} 
where sharp $(\,\cdot\,)^\sharp$ denotes raising the vector index
from covariant to contravariant, so its divergence may be taken (the sign
in the right hand side is taken in agreement with the dissipative nature of the dynamics, as it is shown in Sec.~\ref{sec:geom-phys}).
The evident difference between these two forms is that, unlike the first
form, the second equation can be written as the characteristic equation 
\begin{eqnarray}
\frac{d\rho}{dt}(\mathbf{x}(t),t)=0
\quad\hbox{along}\quad
\frac{d\mathbf{x}}{dt}=\mathbf{u}[\rho]=
\left(\mu[\rho]\nabla\frac{\delta E}{\delta \rho}\right)^{\!\sharp}
\label{GOPprincip}
\end{eqnarray}
so that velocity $\mathbf{u}$ depends on density $\rho$ through the
gradient of the variation of the free energy
$E$ (velocity proportional to thermodynamic force with mobility $\mu[\rho]$)
\cite{HoPu2005,HoPu2006,HoPu2007}.
\rem{ 
Here, the time
derivative of $\rho$ invokes the fundamental chain rule for the
product of the density function times the volume element,
$\rho(\mathbf{x}(t),t)\,\mbox{d}^n\mathbf{x}(t)$. Preservation of this
product along
$d\mathbf{x}/dt=\mathbf{u}[\rho]$ yields 
\begin{eqnarray}
(\partial_t \rho
+
\mathbf{u}[\rho]\cdot \nabla \rho
+ \rho\,{\rm div}\,\mathbf{u}[\rho] )\,\mbox{d}^n\mathbf{x}(t)
= 0
\,,
\end{eqnarray} 
As mentioned above, Darcy's Law approach assumes that velocity $\mathbf{u}$ depends on density $\rho$ through the
gradient of the variation of free energy
$E$ (velocity proportional to thermodynamic force with mobility $\mu[\rho]$)
\cite{HoPu2005,HoPu2006,HoPu2007}. This assumption leads
to the expected continuity equation for density,
\begin{equation} 
\label{cont-eqn}
\partial_t \rho
=-\, {\rm div}\, 
\rho\,\mathbf{u}[\rho]
\quad\hbox{where}\quad
\mathbf{u}[\rho] = - (\mu\, \nabla \delta E/\delta \rho)^\sharp
\,,
\end{equation} 
and sharp $(\,\cdot\,)^\sharp$ denotes raising the vector index
from covariant to contravariant, so its divergence may be taken (the sign
in the expression of $\bf u$ is taken to be negative in agreement with the
dissipative nature of the dynamics, as it is shown in Sec.~\ref{sec:geom-phys}).
} 
\smallskip

The Holm-Putkaradze (HP) theory \cite{HoPu2007} generalizes this type of
geometric flow method underlying Darcy's Law approach to apply to other order parameters (denoted by $\kappa$) with different geometrical meaning (not just densities).
\rem{ 
 Geometrical considerations might lead one to expect  that such a generalization would take the mathematical form,
\begin{eqnarray}
\partial\kappa/\partial t + \pounds_{u[\kappa]}\kappa=0
\,,
\label{GOPprincipmath}
\end{eqnarray}
where $\pounds_u\kappa$ denotes the Lie derivative with respect to the vector field $u=\mathbf{u}\cdot\nabla$ of any geometrical quantity $\kappa$ \cite{MaRa99}.
} 
The key question for understanding the physical modeling that would be needed in making such a generalization is, ``What is the corresponding Darcy's law
for an order parameter $\kappa$?'' Namely, how does one determine the corresponding geometric flow for an arbitrary geometrical quantity $\kappa$? The first problem is that there is no reason to consider only one of the two geometric formulations in (\ref{exchange-darcy}). Although a characteristic form would be preferable because of its reacher geometric  meaning, no choice can be performed a priori.

As a further step in the investigation of the geometric structure in Darcy's
law (\ref{DLaw}), one seeks a variational formulation of the equations (\ref{exchange-darcy}),
that could shed more light on how these formulations arise. Thus one takes the $L^2$ pairing of (\ref{DLaw}) with a test function $\phi$ and sets it equal to the variation $\delta E$ of the free energy \cite{HoPu2006,HoPu2007}
\begin{eqnarray*}
\left\langle \frac{\pa \rho}{\pa t} \,,\, {\phi} \right\rangle 
=
\left\langle 
\delta\rho \,,\, 
\frac{\delta E}{\delta
\rho} 
\right\rangle 
\end{eqnarray*}
where the variation $\delta \rho$ satisfies
\[
\delta \rho 
=
-\,
{\rm div}\big(\rho\,\mu\, \nabla \phi\big)
\]
in order to recover equation (\ref{DLaw}) (the calculation proceeds by integration by
parts, \cite{HoPu2005,HoPu2006}). In order to analyze the geometric
structure, one needs to express the variational principle in terms of geometric
covariant quantities and this leads to the same ambiguity as in (\ref{exchange-darcy}).
Two possibilities are available:
\begin{equation}\label{variations}
\delta \rho 
=
- \,
\textit{\large$\pounds$}_\text{\!\footnotesize$\left(\mu \,\nabla \phi\right)^{\sharp}$}\,\rho
\qquad\text{or}\qquad
\delta \rho 
=
- \,
\textit{\large$\pounds$}_\text{\!\footnotesize$\left(\rho \,\nabla \phi\right)^{\sharp}$}\,\mu
\end{equation}
which are determined by the relative position of $\rho$ and $\mu$ in the
formulas.
\rem{ 
A surprising clue leading to a geometrical analog of Darcy's Law emerges when considering the spontaneous appearance of singularities in solutions of (\ref{cont-eqn}) for which $\mu[\rho]$ and $\delta E/\delta \rho$
depend on the {\em average density}, rather than its pointwise value
\cite{HoPu2005,HoPu2006}. Those singular solutions obey a weak form of
the continuity equation (\ref{cont-eqn}), expressed by pairing it in $L^2$ with an arbitrary smooth test function $\phi$ and integrating twice by parts, as
\begin{eqnarray}
\label{HP-rho}
\langle \partial_t \rho \,,\, {\phi} \rangle 
&=&
\langle 
\delta\rho \,,\, 
\delta E/\delta \rho 
\rangle 
\quad \mbox{to find} 
\\ 
\delta \rho 
=
-\,
{\rm div}\left(\rho\mu \nabla \phi\right)
&=&
- \,
\pounds_{\mathbf{v}({\phi})}\,\mu
\quad\mbox{so}\quad
\mathbf{v}(\phi)
=
(\rho \nabla \phi)^\sharp
\nonumber
\end{eqnarray}
where spatial integration defines the real-valued pairing
$\langle\,\cdot\,,\,\cdot\,\rangle$ between densities and their dual space of
scalar functions. Expression (\ref{HP-rho}) for the weak solutions of the
continuity equation (\ref{cont-eqn}) provides the clue for expressing
Darcy's law for the evolution of an arbitrary geometric quantity
$\kappa$, not just a density. 
} 

Now, in order to express (\ref{DLaw}) and (\ref{exchange-darcy}) in a completely
geometric covariant form, one writes out the integrations by parts explicitly
 and makes use of the {\it diamond operation} introduced in chapter~\ref{intro}
(see below). Upon performing the second choice in (\ref{variations}), one
obtains \cite{HoPu2007,HoPuTr2007}
\begin{align}
\Big\langle \frac{\partial \rho}{\partial t} \,,\, {\phi} \Big\rangle 
=
\bigg\langle 
\delta \rho\,,\, 
\frac{\delta E}{\delta \rho} 
\bigg\rangle 
=
-\,
\bigg\langle 
\pounds_{\mathbf{v}({\phi})}\,\mu \,,\, 
\frac{\delta E}{\delta \rho} 
\bigg\rangle 
&=:
-\,
\bigg\langle 
\mathbf{v}({\phi}) \,,\,\mu \diamond
\frac{\delta E}{\delta \rho} 
\bigg\rangle 
\nonumber
\\ 
=
\bigg\langle 
{\rm div}\, \mu(\rho \nabla \phi)^\sharp \,,\, 
\frac{\delta E}{\delta \rho} 
\bigg\rangle 
&=
-\,
\bigg\langle 
(\rho\nabla\phi)^\sharp
\,,\,
\mu\nabla\frac{\delta E}{\delta \rho} 
\bigg\rangle 
\nonumber
\\ 
&=:\!
-\,
\bigg\langle 
(\rho\diamond\phi)^\sharp
\,,\,
\mu\diamond\frac{\delta E}{\delta \rho} 
\bigg\rangle 
\label{HP-rho-calc}
\,,
\end{align}
while performing the first choice in (\ref{variations}) switches $\rho\leftrightarrow\mu$
in the last two lines, so that (\ref{DLaw}) may be written in the following geometric forms
\begin{equation}\label{exchange-diamond}
\frac{\pa \rho}{\pa t}=-\,
\textit{\large\pounds}_\text{\!\footnotesize$\left(\rho\diamond\frac{\delta E}{\delta
\rho}\right)^{\!\sharp}$}\,\,\mu[\rho]
\qquad\text{or}\qquad
\frac{\pa \rho}{\pa t}=-\,
\textit{\large\pounds}_\text{\!\footnotesize$\left(\mu[\rho]\diamond\frac{\delta E}{\delta \rho}\right)^{\!\sharp}$}\,\rho
\end{equation}
corresponding to the two different cases in (\ref{exchange-darcy}).
As in chapter~\ref{intro}, the third equality on the first line defines the diamond ($\diamond$) operation as the \emph{dual} of the Lie derivative under integration by parts for any pair $(\kappa,b)$ of dual variables and any vector field $\mathbf{v}$ \cite{HoMaRa}. That is 
\begin{eqnarray}
\langle \kappa \diamond b, \mathbf{v} \rangle=\langle \kappa, 
-\pounds_\mathbf{v} b \rangle
\,.
\label{diamond-def}
\end{eqnarray}
\rem{ 
Consequently, in the calculation above one may identify $\mathbf{v}({\phi})=
(\rho \nabla \phi)^\sharp
=-\,(\rho\diamond\phi)^\sharp$. Using
the definition of diamond once more gives
\begin{equation}
\bigg\langle \frac{\partial \rho}{\partial t} \,,\, {\phi} \bigg\rangle 
=
-\,\bigg\langle 
(\rho\diamond\phi)^\sharp
\,,\,
\mu\diamond\frac{\delta E}{\delta \rho} 
\bigg\rangle
=
-\,\bigg\langle 
\pounds_{\!\left(\mu\diamond \frac{\delta E}{\delta \rho}\right)^{\!\sharp} }\,\rho
\,,\,
\phi
\bigg\rangle 
\label{HP-kappa}
\end{equation}
This evolution equation is expressed geometrically in the same form as (\ref{cont-eqn}) and (\ref{GOPprincipmath}), where the velocity vector field $\mathbf{u}[\rho]=(\mu\diamond {\delta E}/{\delta \rho})^\sharp$.  
} 
\rem{
Replacing $\rho\to\kappa$ here generalizes (\ref{cont-eqn}) to any
quantity $\kappa$ in an arbitrary vector space.  
}

It is readily seen how the geometric properties of the result in (\ref{HP-rho-calc})
are unchanged by switching $\rho\leftrightarrow\mu$ and the only difference
is that the second choice in (\ref{variations}) yields a characteristic equation
for $\rho$. However, at this stage there is no particular reason to choose
between the two possibilities.

\subsection{GOP equations: a result on singular solutions}

The arguments in the previous section show that Darcy's law can be applied
to any tensor quantity $\kappa$, since equations (\ref{exchange-diamond})
do not depend on the particular nature of $\rho$ as a density variable. The
Lie derivative is defined for any tensor along a generic vector field and
thus the substitution $\rho\to\kappa$ is completely justified in geometric
terms. Thus one obtains
\begin{equation}\label{kappa-exchange}
\frac{\pa \kappa}{\pa t}=-\,
\textit{\large\pounds}_\text{\!\footnotesize$\left(\kappa\diamond\frac{\delta E}{\delta
\kappa}\right)^{\!\sharp}$}\,\,\mu[\kappa]
\qquad\text{or}\qquad
\frac{\pa \kappa}{\pa t}=-\,
\textit{\large\pounds}_\text{\!\footnotesize$\left(\mu[\kappa]\diamond\frac{\delta E}{\delta \kappa}\right)^{\!\sharp}$}\,\kappa
\end{equation}
It is interesting to notice that the two possibilities are identical when
$\mu\propto\kappa$, say for simplicity $\mu=\kappa$. In this case one obtains a type
of {\bfi geometric order parameter equation} (GOP) which is written as
\begin{equation}\label{mu=kappa}
\frac{\pa \kappa}{\pa t}=-\, \textit{\large\pounds}_\text{\!\footnotesize$\left(\kappa\diamond\frac{\delta E}{\delta \kappa}\right)^{\!\sharp}$}\,\kappa
\end{equation}

This equation indeed identifies the type of flow for the order parameter
$\kappa$ arising from the geometric extension of Darcy's law. However, equation (\ref{DLaw}) with generic mobility $\mu=\mu[\rho]$ has one more feature, besides its purely geometric character. This feature is the emergence of {\bfi singular solutions}. For example, in one dimension the equation (\ref{DLaw}) admits particle-like solutions of the form
\cite{HoPu2005,HoPu2006}
\[
\rho(x,t)=\sum_{n=1}^N w_n(t)\,\delta(x-Q_n(t))
\]
corresponding to the trajectories of $N$ particles in the system (one has
$\dot{w}_n=0$). The spontaneous emergence of this kind of solution \cite{HoPu2005,HoPu2006} is a remarkable result on its own, within the context of blow-up phenomena in nonlinear PDE's. However, the behaviour of these solutions exhibits one more interesting feature: these particle-like structures merge together in finite time \cite{HoPu2005,HoPu2006},
thereby recovering aggregation and self-assembly phenomena. This point leads
to the question: is it possible to generalize the existence of singular solutions
to GOP theory? For example, in the one dimensional case, one would expect
solutions of the GOP equation for $\kappa$ of the form
\[
\kappa(x,t)=\sum_{n=1}^N p_n(t)\,\delta(x-Q_n(t))
\,.
\]
It is a direct verification that this type of solution never exists for
any equation of the form (\ref{mu=kappa}). Thus one is motivated to look
at one of the forms in (\ref{kappa-exchange}). Upon pairing the first equation in (\ref{kappa-exchange}) with a dual element $\phi$, direct substitution of the singular solution ansatz yields
\begin{multline*}
\bigg\langle
\frac{\pa\kappa}{\pa t},\,\phi
\bigg\rangle 
=
\sum_n\,\frac{\partial p_n}{\partial t}\cdot \phi \left({Q}_n(t) \right)
+
\sum_n\,
\frac{\partial Q_n}{\partial t}\cdot \phi^{\,\prime}\left(Q_n(t) \right) 
\\
=-\,
\bigg\langle 
\textit{\large\pounds}_\text{\!\footnotesize$\left(\kappa\diamond\frac{\delta E}{\delta \kappa}\right)^{\!\sharp}$}\,\mu\,,\,\phi
\bigg\rangle
=-\,
\bigg\langle 
\mu\diamond\phi,\,
\left(\kappa\diamond\frac{\delta E}{\delta \kappa}\right)^{\!\sharp}
\bigg\rangle
=-\,
\bigg\langle 
\kappa,\,
\textit{\large\pounds}_\text{\!\small$\left(\mu\diamond\phi\right)^{\sharp}$}\,\text{\small$\frac{\delta E}{\delta \kappa}$}
\bigg\rangle
\\
=-\,
\sum_{n=1}^N\,
p_n(t)\contract
\left.
\textit{\large\pounds}_\text{\!\small$\left(\mu\diamond\phi\right)^{\sharp}$}\,\text{\small$\frac{\delta E}{\delta \kappa}$}\,
\right|_{x=Q(t)}
\end{multline*}
where the dot symbol $\bf\contract$ denotes contraction of indexes.
In order for the singular solutions to exist, one would match terms in $\phi$
and $\phi^{\,\prime}$ and obtain the evolution equations for $p_n$ and $Q_n$,
as it happens for the density variable $\rho$ in Darcy's law \cite{HoPu2005,HoPu2006}. However, in general the term in the last line may involve higher derivatives, not just first order (for example, if $\kappa$ is a one-form density, then
diamond is again a Lie derivative, which generates second order derivatives
in $\phi$). Therefore, the first choice in (\ref{kappa-exchange}) is not suitable to recover the singular solutions in the general case of an order parameter $\kappa$. Instead, by following the same procedure for
the second equation in (\ref{kappa-exchange}), one obtains
\begin{multline}
\label{HP-kappa2}
\bigg\langle
\frac{\pa\kappa}{\pa t},\,\phi
\bigg\rangle 
=
\sum_n\,\frac{\partial p_n}{\partial t}\cdot \phi \left({Q}_n(t) \right)
+
\sum_n\,
\frac{\partial Q_n}{\partial t}\cdot \phi^{\,\prime}\left(Q_n(t) \right)
\\
=-\,
\bigg\langle 
\textit{\large\pounds}_\text{\!\footnotesize$\left(\mu\diamond\frac{\delta E}{\delta \kappa}\right)^{\!\sharp}$}\,\kappa\,,\,\phi
\bigg\rangle
=-\,
\sum_{n=1}^N\,
p_n(t)\contract
\left.
\textit{\large\pounds}_\text{\!\footnotesize$\left(\mu\diamond\frac{\delta E}{\delta
\kappa}\right)^{\sharp}$}\,\phi\,
\right|_{x=Q(t)}
\end{multline}
Now, from the general theory of Lie differentiation \cite{AbMaRa} one recognizes
that the last term on the second line contains only terms that are {\it linear} in $\phi$ and its first order derivatives and does not involve any higher order derivatives of $\phi$. Thus, in higher dimensions one finds the following conclusion \cite{HoPuTr2007}
\begin{framed}
\begin{theorem}
The second equation of (\ref{kappa-exchange}) always allows for singular
solutions of the form
\begin{equation}
\kappa({\bf x},t)=\sum_{n=1}^N \int p_n(s,t)\,\delta({\bf x-q}_n(s,t))\,{\rm d}s
\label{GOPsing}
\end{equation}
for any tensor field $\kappa$, provided $\mu$ and $\delta E/\delta \rho$ are sufficiently smooth.
\end{theorem}
\end{framed}
As in earlier chapters, the variable $s$ is a coordinate on a submanifold
of $\mathbb{R}^3$: if $s$ is a one-dimensional coordinate, then $\kappa$
is supported on a curve (filament), if $s$ is two dimensional, then $\kappa$
is supported on a surface (sheet) immersed in physical space. The proof proceeds
by direct substitution.
\begin{framed}
At this point, one defines GOP equations as characteristic equations of the
type
\begin{eqnarray}
\frac{d\kappa}{dt}(\mathbf{x}(t),t)=0
\quad\hbox{along}\quad
\frac{d\mathbf{x}}{dt}=
\left(\mu[\kappa]\nabla\frac{\delta E}{\delta \kappa}\right)^{\!\sharp}
\label{GOP-char}
\end{eqnarray}
or, in Eulerian coordinates,
\begin{equation}
\frac{\pa \kappa}{\pa t}=-\,
\textit{\large\pounds}_\text{\!\footnotesize$\left(\mu[\kappa]\diamond\frac{\delta E}{\delta \kappa}\right)^{\!\sharp}$}\,\kappa
\label{GOPprincipmath}
\end{equation}
\end{framed}
\noindent
The fact that Darcy's law (\ref{DLaw}) is symmetric in $\rho$ and $\mu$ is
the reason why singular solutions are recovered for both possibilities in
(\ref{exchange-diamond}). This property is peculiar of Darcy's law and does
not hold in general. The distinction between the two cases identifies the
geometric structure of the GOP family of equations.

Moreover, it is important to notice that the geometry underlying this dynamics
is uniquely determined by the group of diffeomorphisms, whose infinitesimal
generator coincides with the Lie derivative, as explained in chapter~\ref{intro}.
However, these equations can be further generalized to consider {\it different} Lie group actions, such as the rotations $SO(3)$ \cite{HoPu2007}. Indeed,
if $\kappa$ belongs to a generic $\mathfrak{g}$-module $V$ (i.e. a vector space acted on by the Lie algebra $\mathfrak{g}$), then the GOP equation
becomes
\[
\frac{\pa \kappa}{\pa t}=-
\left(\mu[\kappa]\diamond\frac{\delta E}{\delta \kappa}\right)^{\!\sharp}\kappa
\]
where $\xi\,\kappa\in V$ denotes the action of the Lie algebra element $\xi\in\mathfrak{g}$
on the order parameter $\kappa\in V$ and the diamond is now defined as
$
\langle \kappa \diamond b,\, \xi \rangle:=\langle \kappa, \,
\xi\,\kappa \rangle
$
.
In order to distinguish between the various Lie groups, the next chapters
will use different symbols for the diamond operation.

The next question in the formulation of GOP theory is the particular meaning
assumed by the {\it generalized mobility} $\mu[\kappa]$. This quantity
has been related to the typical particle size in Darcy dynamics, but the
physical meaning of this quantity is not yet clear in the case of
a generic GOP equation for the order parameter $\kappa$. The next section
presents the mobility as a smoothed quantity that keeps into account the dynamics of jammed states in the system, by introducing a typical length-scale
\cite{HoPu2005,HoPu2006,HoPu2007}.

\subsection{More background: multi-scale variations}\label{Meth-MSV}
One seeks a variational principle for a continuum description of coherent structures. This includes the evolution of particles of finite size that may clump together under crowded conditions. In crowded conditions, finite-sized particles typically reach jammed states, sometimes called  rafts, that may be locally locked together over a coherence length of several particle-size scales. Thus, a variational principle for the evolution of coherent structures such as jammed states in particle aggregation must accommodate more than one length scale. A  multi-scale variational principle may be derived by considering the variations as being applied to rafts, or patches, of jammed states of a certain size (the coherence length). However, the approach of
Holm and Putkaradze \cite{HoPu2007} is based on applying a Lagrangian coordinate average that moves with the clumps of particles. In this approach, the variation ($\delta\kappa$) of the order parameter ($\kappa$) at a given fixed point in space is determined by a family of smooth maps $\varphi(s)$ depending continuously on a parameter $s$ and acting on the average value $\bar{\kappa}$ defined by
\begin{eqnarray}
\bar{\kappa} = H*\kappa = \int H(y-y')\kappa(y') dy'
\end{eqnarray}
in the frame of motion of the jammed state. Thus, $y$ is a Lagrangian coordinate, defined in that frame. This motion itself is to be determined by the variational principle. The kernel $H$ represents the typical size and shape of the coherent structures, which in this example would be rafts of close-packed finite-size particles. The mobility $\mu$ of the rafts depends on $\bar{\kappa}$, and so the corresponding variation of the local quantity $\kappa$ at a fixed point in space may be modeled as
\begin{eqnarray}
\delta\kappa 
=\frac{d}{ds}\Big|_{s=0}\Big(\mu(\bar{\kappa})\varphi^{-1}(s)\Big)
\,.
\end{eqnarray}
Here $\varphi(s)y=x(s)$ is a point in space, which $\varphi^{-1}(s)$ returns  to its Lagrangian label $y$ and $\varphi(0)$ is the identity operation.
The average $\bar{\kappa}=H*\kappa$ is applied in a \emph{Lagrangian} sense, following a locally locked raft of particles along a curve parameterized by time $t$  in the family of smooth maps. The latter represents the motion of the raft as  $\varphi(t)y=x(t)$, whose velocity tangent vector is still to be determined. When composed from the right the derivative at the identity of the action of $\varphi(s)$ results in a variation $\delta\kappa$ at a fixed point in space given by
\begin{eqnarray}
\delta\kappa =-\pounds_{\mathbf{v}(\varphi)}\mu[\kappa]
\quad\hbox{with}\quad
\mathbf{v}(\varphi) = \varphi'\varphi^{-1}\big|_{s=0}
\,,
\end{eqnarray}
thereby recovering the proper expression for the variation, i.e. the second
equation of (\ref{variations}).
In the GOP theory of Holm and Putkaradze \cite{HoPu2007} $\mu[\kappa]$ is a general functional of $\kappa$, not just a function of $\bar{\kappa}$. \smallskip

\rem{ 
Taking the steps for $\kappa$ corresponding to those for $\rho$ in (\ref{HP-rho-calc}) but with the Lie derivative appropriate for the geometric nature of $\kappa$ allows one to close the GOP equation (\ref{GOPprincipmath}) explicitly as
\begin{eqnarray}
\label{GOP-eqn-brkt}
\frac{\partial \kappa}{\partial t}
=
-\,\pounds_{{u}[\kappa]}
\kappa
\,,
\quad\hbox{in which}\quad
{u}[\kappa]
=
\Big(\mu[\kappa]\, {\diamond}\frac{\delta E}{\delta \kappa} \Big)^\sharp
\,.
\end{eqnarray}
Here the co-vector $(\mu[\kappa]\,\diamond\delta E/\delta \kappa)$ resulting from the diamond operation is defined as in equation (\ref{diamond-def}) and sharp $(\,\cdot\,)^\sharp$ raises its vector index to make it contravariant. Although Lagrangian averaging was invoked in defining the mobility as a coherence property, equation 
(\ref{GOP-eqn-brkt}) is expressed in fixed spatial coordinates. Its Lagrangian heritage is recognized upon rewriting it equivalently in the characteristic form (\ref{GOPprincip}).
} 

\subsection{Properties of the diamond operation}

The {\bfi diamond operation} $\diamond$  is defined  in (\ref{diamond-def}) for Lie
derivative $\pounds_\eta$ acting on dual variables $a\in V$ and $b\in V^*$
($V$ being a vector space) by
\begin{equation}
\langle b \diamond a \,,\, \eta \rangle
\equiv
-\,
\langle b \,,\, \pounds_\eta\, a \rangle
=:
-\,
\langle b \,,\, a\, \eta \rangle
\,,
\end{equation}
where Lie derivative with respect to right action of the diffeomorphisms on elements of $V$ is also denoted by concatenation on the right.  
The diamond operator takes two dual quantities $a$ and $b$ and produces a quantity  dual to a vector field, \emph{i.e.}, a $a\diamond b$ is a one-form density. In abstract notation $\diamond:V\times V^*\to\mathfrak{X}^*$.  The $\diamond$ operation is also known as the ``dual representation'' of this right action of the Lie algebra of vector fields  on the representation space $V$ \cite{HoMaRa}. When paired with a vector field $\eta$, the diamond operation has the following three useful properties \cite{HoMaRa,HoPu2007}:
\begin{enumerate}
\item
It is antisymmetric
\[
\langle \, b \diamond a
+
a \diamond b\,,\, \eta \rangle
=
0
\,.
\]
\item
It satisfies the product rule for Lie derivative
\[
\langle \, \pounds_\xi \,(b \diamond a)\,,\, \eta \, \rangle
=
\langle\, (\pounds_\xi\,b) \diamond a
+
b \diamond (\pounds_\xi\,a)\,,\, \eta \, \rangle.
\]
\item
It is antisymmetric under integration by parts
\[
\langle \,
db \diamond a
+
b \diamond da\,,\, \eta 
\, \rangle
=
0
\,.
\]
\end{enumerate}
These three properties of $\diamond$ are useful in computing the
explicit forms of the various geometric flows for order
parameters (\ref{HP-rho-calc}). Of course, when the order parameter is a
density undergoing a gradient flow, then one recovers Darcy's law (\ref{DLaw}) from (\ref{GOPprincipmath}).

\subsection{Energy dissipation in GOP theory}
\label{sec:geom-phys}
As mentioned in the first section of this chapter, the physical nature of Darcy's law resides in energy dissipation and friction \cite{HoPu2005,HoPu2006,HoPu2007}. Thus, a faithful generalization
in the context of GOP theory needs to accommodate energy dissipation. This allows the introduction of the \emph{dissipation bracket} \cite{HoPu2007}, so that the equations can be written in an alternative bracket form. 
The corresponding energy equation follows from (\ref{GOPprincipmath}) as
\begin{align}
\frac{d E}{dt}
=
\Big\langle \frac{\partial \kappa}{\partial t} \,,\, 
\frac{\delta E}{\delta \kappa} \Big\rangle
&=
\left\langle  -\,
\pounds_{(\mu[\kappa]\, {\diamond}\frac{\delta E}{\delta \kappa})^\sharp}
\kappa,\,\frac{\delta E}{\delta \kappa}
\right\rangle
=
-
\left\langle  
\Big(\mu[\kappa] \,\diamond\, \frac{\delta E}{\delta \kappa}\Big)
,\, 
\Big(\kappa\,\diamond\,\frac{\delta E}{\delta \kappa}\,\Big)^\sharp
\right\rangle
.
\label{GOP-erg}
\end{align}
Holm and Putkaradze \cite{HoPu2007} observed that equation (\ref{GOP-erg}) defines the following bracket notation for the time derivative of a functional $F[\kappa]$,
\begin{align}
\frac{d F[\kappa]}{dt}
=
\Big\langle \frac{\partial \kappa}{\partial t} \,,\, 
\frac{\delta F}{\delta \kappa} \Big\rangle
&=
\left\langle  
-\,\pounds_{(\mu[\kappa]\, {\diamond}\frac{\delta E}{\delta \kappa})^\sharp}
\kappa
\,,\,
\frac{\delta F}{\delta \kappa}
\right\rangle
\nonumber \\
&=
-\,
\left\langle  
\Big(\mu[\kappa] \,\diamond\, \frac{\delta E}{\delta \kappa}\Big)
,\, 
\Big(\kappa\,\diamond\,\frac{\delta F}{\delta \kappa}\,\Big)^\sharp
\right\rangle
=:
\{\!\{\,E\,,\,F\,\}\!\} [\kappa]
\label{bracketdef}
\end{align}
The properties of the GOP brackets $ \{\!\{\,E\,,\,F\,\}\!\}$ defined in equation (\ref{bracketdef}) are determined by the diamond operation and the choice of the mobility $\mu[\kappa]$. For physical applications, one should choose a mobility that satisfies strict dissipation of energy, \emph{i.e.}
$
  \{\!\{\,E\,,\,E\,\}\!\} \leq 0. 
$
A particular example of mobility that satisfies the energy dissipation requirement  is $\mu[\kappa]=\kappa M[\kappa]$, where $M[\kappa] \geq 0$ is a non-negative scalar functional of $\kappa$ \cite{HoPu2007}. (That is, $M[\kappa]$ is a number.) Requiring the mobility to produce energy dissipation does not limit the mathematical properties of the GOP bracket.  For example, Holm and Putkaradze
showed that the dissipative bracket possesses the Leibnitz property with any choice of mobility \cite{HoPu2007}. That is, it satisfies the Leibnitz rule for the derivative of a product of functionals
\begin{proposition}[Leibnitz property \cite{HoPu2007}]
The GOP bracket (\ref{bracketdef}) satisfies the Leibnitz property. That is, it satisfies 
\[
\{ \{ \,EF\,,\,G\,\} \}[\kappa] =F \{ \{ \,E\,,\,G\,\} \}[\kappa] +E\{ \{ \,F\,,\,G\,\} \}[\kappa] 
\] 
for any functionals $E,F$ and $G$ of $\kappa$. 
\end{proposition}
\begin{proof}
For arbitrary scalar functionals  $E$ and $F$ of $\kappa$ and any smooth vector field $\eta$, the Leibnitz property for the functional derivative and for the Lie derivative  together imply
\begin{eqnarray*} 
\bigg\langle  
\mu \diamond 
\left( 
\frac{\delta (EF )}{\delta \kappa} 
\right) \, , \,  
\eta
\bigg\rangle 
&=&
\bigg\langle  
\mu \diamond 
\left( 
E \frac{\delta F}{\delta \kappa}
+
F \frac{\delta E}{\delta \kappa} 
\right) 
\, , \,  
\eta
\bigg\rangle 
\\ 
\\
&=&
\bigg\langle  \mu \, , \,  
-\pounds_\eta 
\left( 
E \frac{\delta F}{\delta \kappa}
+
F \frac{\delta E}{\delta \kappa} 
\right)
\bigg\rangle  
\\
\\
&=&
E \bigg\langle  \mu \, , \,  
-\pounds_\eta 
\frac{\delta F}{\delta \kappa}
\bigg\rangle  
+
F
\bigg\langle  \mu \, , \,   
-\pounds_\eta 
\frac{\delta E}{\delta \kappa} 
\bigg\rangle  
\\
&=&
E\bigg\langle  
\mu \diamond 
\frac{\delta F}{\delta \kappa}
\, , \,  
\eta
\bigg\rangle
+
F\bigg\langle  
\mu \diamond 
\frac{\delta E}{\delta \kappa}
\, , \,  
\eta
\bigg\rangle
\end{eqnarray*}
Choosing $\eta= \Big(\kappa\diamond \frac{\delta G}{\delta \kappa} \Big)^\sharp$ then proves the proposition that the bracket (\ref{bracketdef}) is Leibnitz.
\end{proof}

In addition, the dissipative bracket formulation (\ref{bracketdef}) allows one to reformulate the GOP equation (\ref{GOPprincipmath}) in terms of flow on a Riemannian manifold with a metric defined through the dissipation bracket. The following
section reviews the results in \cite{HoPu2007}.

\paragraph{Connection to Riemannian geometry.} 
Following \cite{Ot2001}, Holm and Putkaradze \cite{HoPu2007} used their GOP bracket to introduce a metric tensor on the manifold connecting a ``vector'' $\partial_t \kappa$ and ``co-vector'' $\delta E/\delta \kappa$. 
That is, one expresses the evolution equation (\ref{GOPprincipmath}) in the weak form  as 
\begin{equation} 
\bigg \langle 
\frac{\partial \kappa}{\partial t} \, , \, \psi
\bigg \rangle
=
-\,
g_\kappa \left(  
\frac{\delta E}{\delta \kappa} \, , \, \psi 
\right) 
\label{Otto-def} 
\end{equation} 
for an arbitrary element $\psi$ of the space dual to the $\kappa$ space, and where 
$g_\kappa(\cdot \, , \, \cdot)$ is a symmetric positive definite function -- metric tensor -- defined on vectors from the dual space. 

First one notice that 
for any choice of mobility producing a symmetric bracket (in particular, $\mu[\kappa]=\kappa M[\kappa]$) 
 \[
 \{\!\{\,E\,,\,F\,\}\!\} = \{\!\{\,F\,,\,E\,\}\!\}
 \,,
 \]
so that that symmetric bracket defines an inner product between the functional derivatives \cite{HoPu2007},
 \begin{equation} 
 \{\!\{\,E\,,\,F\,\}\!\} =: 
 -\,
 g_\kappa \Big(
  \frac{\delta E}{\delta \kappa} \,,\,
\frac{\delta F}{\delta \kappa}
\Big) =
-\,
\bigg \langle 
\mu \diamond \frac{\delta E}{\delta \kappa} 
\, , \, 
\left( \kappa \diamond \frac{\delta F}{\delta \kappa} \right)^\sharp 
\bigg \rangle 
 \label{gtensor} 
 \end{equation} 
 Alternatively, (\ref{gtensor}) can be understood as a symmetric positive definite function of two 
 elements of dual space $\phi,\psi$: 
 \begin{equation} 
 g_\kappa \Big(
\phi \,,\,
\psi
\Big)
= 
\bigg \langle 
\mu \diamond \phi 
\, , \, 
\left( \kappa \diamond \psi \right)^\sharp 
\bigg \rangle . 
 \label{gtensor2} 
 \end{equation} 
 Notice that $g(\phi,\phi) \geq 0$, since $\{E,E\}\leq0$ and these arguments
 maybe summarized in the following 
 
 \begin{proposition}[Metric property \cite{HoPu2007}]
For the choice of metric tensor (\ref{gtensor2}), the GOP equation (\ref{GOPprincipmath}) may be expressed as the metric relation (\ref{Otto-def}). 
 \end{proposition}

This approach harnesses the powerful machinery of Riemannian geometry to the mathematical analysis of the GOP equation (\ref{GOPprincipmath}). This opens a wealth of possibilities, but it also limits the analysis to mobilities $\mu$ for which the GOP bracket (\ref{bracketdef}) is symmetric and positive definite, as in the modeling choice  $\mu[\kappa]=\kappa M[\kappa]$.

\paragraph{Previous dissipative brackets.} Historically, the use of symmetric brackets for introducing dissipation into Hamiltonian systems seems to have originated with works of Grmela \cite{Gr1984}, Kaufman \cite{Ka1984} and Morrison \cite{Mo1984}. See \cite{Ot05} for references and further engineering developments. This approach introduces a sum of two brackets, one describing the Hamiltonian part of the motion and the other obtained by representing the dissipation with a symmetric bracket operation involving an entropy defined for that purpose. 
Being expressed in terms of the diamond operation $(\,\diamond\,)$ for an arbitrary geometric order parameter $\kappa$, the dissipative bracket in equation (\ref{bracketdef}) differs from symmetric brackets proposed in earlier work.  The  geometric advection law (\ref{GOPprincipmath}) for the order parameter will be shown below to arise from thermodynamic principles that naturally yield the dissipative bracket (\ref{bracketdef}).  Moreover, being written as a Lie derivative, the equation of motion (\ref{GOPprincipmath}) respects the geometry of the transported quantity. The dissipative brackets from the earlier literature do not appear to be expressible as a geometric transport equation in Lie derivative form. 

\subsection{A general principle for geometric dissipation}
\label{sec:thermodynamics} 
Equations (\ref{GOPprincipmath}) may be justified by more general principles. Consider using an arbitrary functional $F$  in (\ref{bracketdef}) as a basis for the derivation of an equation for $\kappa$. 
Suppose $\kappa$ is an observable quantity for a physical system, and that system evolves due to the inherent free energy $E[\kappa]$ in the absence of external forces. This is the physical picture one envisions, for example, when thinking about processes of self-assembly in nanotechnology. Suppose one would like to measure the time evolution of a functional $F[\kappa]$, which may for example represent as total energy or total momentum. 
\begin{framed}\noindent
For an \emph{arbitrary} functional $F[\kappa]$ and for a  \emph{given} free energy $E[\kappa]$, the GOP bracket yields 
\begin{equation} 
\frac{d F}{d t}
=
 \bigg\langle 
 \partial_t \kappa \, , \, \frac{\delta F}{\delta \kappa}
 \bigg \rangle 
=
 \bigg\langle 
 \delta \kappa \, , \, \frac{\delta E}{\delta \kappa}
 \bigg \rangle 
 =
 \delta E
 \,.
 \label{arbitraryF}
\end{equation}
\end{framed}
\noindent
The main postulate here is that, in principle, one can determine the evolution of the system indirectly by probing many different \emph{global} quantities $F[\kappa]$ (for example, the moments of a probability distribution). It is only natural to assume that the law for the evolution of $\kappa$ should be independent of the choice of which quantities $F$ are used to determine it.

Surprisingly, this rather general sounding assumption sets severe restrictions on the nature of the variation $\delta \kappa$. In particular, 
\vspace{-3mm}
\begin{enumerate} \itemsep -2mm
\item The variation $\delta \kappa$ must be linear in $\delta F/\delta \kappa$, since the left hand side of (\ref{arbitraryF}) is also linear in $\delta F/\delta \kappa$. 
\item The variation $\delta \kappa$ must transform the same way as $\kappa$, as it must be dual to $\delta F/\delta \kappa$. This introduces the \emph{mobility}  $\mu$ that must be of the same type as $\kappa$. 
\item  The variation $\delta \kappa$ must specify a quantity at the tangent space to the space of all possible $ \kappa$. The proper geometric way to specify this quantity is through the Lie derivative $\pounds_\mathbf{v}$ with respect to some vector field $\mathbf{v}$.  
\end{enumerate} \vspace{-3mm}

There are only two ways to specify $\delta \kappa$ so that it  obeys these three thermodynamic and geometric constraints, when one insists that only a single new physical quantity $\mu[\kappa]$ is introduced. Namely, 
\begin{eqnarray} 
\label{choice1}
 \mbox{either} \quad 
\delta \kappa = - \pounds_{\mathbf{v}\,} \kappa  
& \quad  \mbox{with} \quad 
& 
\mathbf{v}=-\left( \mu[\kappa] \diamond \frac{\delta F}{\delta \kappa} \right)^\sharp
\,, 
 \\  \mbox{or} \quad
\delta \kappa = - \pounds_{\mathbf{v}\,} \mu[\kappa] 
& \quad \mbox
{with} \quad 
& 
\mathbf{v}=-\left( \kappa \diamond \frac{\delta F}{\delta \kappa} \right)^\sharp
\,.
\label{choice2} 
\end{eqnarray} 
Both of (\ref{choice1}) and (\ref{choice2}) are consistent with all three geometric and thermodynamics requirements. However, the first possibility (\ref{choice1}) prevents formation of measure-valued solutions in $\kappa$, when $\kappa$ is chosen to be a 
1-form, a 2-form or a vector field.  In contrast, the second possibility (\ref{choice2}) yields the \emph{conservation
law} (\ref{GOPprincipmath}), which is a characteristic equation admitting measure-valued solutions for an arbitrary geometric quantity $\kappa$. The remainder of this chapter deals with (\ref{choice2}) and investigates the corresponding evolution equation (\ref{GOP-char}). The alternative choice (\ref{choice1}) would have reversed the roles of $\kappa$ and $ \mu[\kappa] $ in the Lie derivative.


\section{Review of scalar GOP equations}
\label{sec:scalars} 

The fundamental example is an {\it active} scalar, for which $\kappa=f$ is a function. For this particular example, the exposition follows the work
by Holm and Putkaradze \cite{HoPu2007}. The evolution of a scalar
by equation (\ref{GOPprincipmath}) obeys
\begin{equation}
\partial_t\,f =
 -\,\pounds_{(\mu[f]\diamond\frac{\delta E}{\delta f})^\sharp}f
= -\,\Big(\frac{\delta E}{\delta f}\nabla \mu[f] \Big)^\sharp\cdot\nabla f
\,.\label{scalareq} 
\end{equation}
Equation (\ref{scalareq}) can be rewritten in characteristic form as 
\begin{equation}
df/dt=0
\quad\hbox{along}\quad
d\mathbf{x}/dt
=
\Big(\frac{\delta E}{\delta f}\nabla \mu[f] \Big)^\sharp
\,.
\end{equation}
The characteristic speeds of this equation are {\em nonlocal}
when $\delta E/\delta f$ and $\mu$  are chosen to
depend on the {\em average value}, $\bar{f}$. It is interesting that such problems arise commonly in the theory of quasi-geostrophic convection and may lead to the development of  singularities in finite time \cite{Geostrophic, Chae2005, Cordoba2005}. 

\rem{ 
One may verify that the scalar equation (\ref{scalareq}) admits weak solutions
(\ref{clumpons}). Figure~\ref{fig:scalarevolution} 
shows the spatio-temporal
numerical evolution of $H*f$ given by (\ref{scalareq}) with initial conditions of
the type (\ref{clumpons}) with  $\delta$-functions whose strengths are random
numbers between $\pm 1/8$. We have taken $\delta E/\delta f=H*f:=\bar{f}$ where
$H$ is the inverse Helmholtz operator $H(x)=e^{-|x|/\alpha}$ with $\alpha=1$ and mobility $\mu[f]=\exp(1-\bar{f}^2)$. We
see the evolution of sharp ridges in $\bar{f}=H*f$, which corresponds to
$\delta$-functions in the solutions $f(x,t)$. \smallskip
} 

Explicit equations for the evolution of strengths $p_n$ and
coordinates $\mathbf{q}_n$ for a sum of $\delta$-functions in (\ref{GOPsing}) may be derived using (\ref{HP-kappa2}) when $\mu[f]=H*f=\bar{f}$. The singular solution parameters satisfy \cite{HoPu2007} 
\begin{eqnarray}
\frac{\partial p_n(t,s)}{\partial t} 
&=& 
{p}_n(t,s) 
\,{\rm div}\,
\Big(\frac{\delta E}{\delta f}
\nabla \mu[f]\Big)^\sharp
\bigg|_{\bx = \bq_n(t,{s})}
\label{weak-fsoln-peqn}\\
{p}_n(t,s) \, \frac{\partial \mathbf{q}_n(t,s)}{\partial t} 
&=&
{p}_n(t,s)\, 
\Big(\frac{\delta E}{\delta f}
\nabla \mu[f]\Big)^\sharp
\bigg|_{\bx = \bq_n(t,{s})}
\label{weak-fsoln-qeqn}
\end{eqnarray}
for $n=1,2,\dots,N$. For the choice $\mu[f]=\bar{f}$, a solution containing a {\it single} $\delta$-function satisfies
$\dot{p}=-A p^3$, so an initial condition $p(0)=p_0$, evolves
according to $1/p(t)^2=1/p_0^2+4\alpha^2 t$ \cite{HoPu2007}. 
\remfigure{ 
\begin{figure} [h]
\centering 
\includegraphics[width=6in]{Figure2.eps} 
 \caption{ 
Evolution of the $\delta$-function strength $1/p(t)^2$ versus time (circles) from \cite{HP-GOP-2007}.
The theoretical prediction $1/p_0^2+4 t$ is shown as a solid line obtained
without any fitting parameters. 
\label{fig:ampevolution} 
 } 
 \end{figure} 
}

\section{New GOP equations for one-forms and two-forms}
\label{sec:1forms} 
\subsection{Results on singular solutions}
As particular examples, this section develops nonlocal characteristic
equations for the evolution of one- and two-forms. So one specializes equation
(\ref{GOPprincipmath}) to consider the differential 1-form
$\kappa=\mathbf{A} \cdot \mbox{d} \mathbf{x}$ and the 2-form $\kappa=\mathbf{B} \cdot \mbox{d} \mathbf{S}$  in  three-dimensional space. For this, one begins by computing the the Lie derivative and the diamond operation for these cases. In Euclidean coordinates, the Lie derivatives for these two choices of $\kappa$ are:
\begin{eqnarray}
%
-\pounds_\mathbf{v}\,(\mathbf{A}\cdot d\bx)
&=&
-\left((\mathbf{v}\cdot\nabla)\mathbf{A}+A_j\nabla v^j\right)
\cdot d\bx
\nonumber \\
&=& \left(\mathbf{v}\times{\rm curl}\,\mathbf{A}
-\nabla(\mathbf{v}\cdot\mathbf{A})\right)\cdot d\bx\,,
\nonumber \\
-\pounds_\mathbf{v}\,(\mathbf{B}\cdot d\bS)
&=&
-\,d \big(v\contract (\mathbf{B}\cdot {d\bS})\big)
-\,v\contract d (\mathbf{B}\cdot {d\bS})
\nonumber \\
&=& 
-\,d \big((\mathbf{v}\times\mathbf{B})\cdot d\bx\big)
-\,v\contract ({\rm div}\,\mathbf{B} d\,^3\,x)
\nonumber \\
&=& \left({\rm curl}\,(\mathbf{v}\times\mathbf{B})
- \mathbf{v}\,{\rm div}\,\mathbf{B}
\right)\cdot {d\bS}
\nonumber
\label{w-eqn}
\end{eqnarray}
Both of these expressions are familiar from fluid dynamics, particularly magnetohydrodynamics (MHD).

From these formulas for Lie derivative in vector form and the definition of diamond in equation (\ref{diamond-def}), one computes explicit expressions for the diamond operation with 1-forms and 2-forms,
\begin{eqnarray}
\left\langle
\boldsymbol\mu[\mathbf{A}]\diamond \frac{\delta E}{\delta \mathbf{A}}
\,,\,\mathbf{u}\right\rangle 
&=&
\left\langle
\frac{\delta E}{\delta \mathbf{A}} \times\,{\rm curl}\,\boldsymbol\mu[\mathbf{A}]
-\, \boldsymbol\mu[\mathbf{A}]\,{\rm div}\,\frac{\delta E}{\delta \mathbf{A}} \,,\,\mathbf{u}\right\rangle
\nonumber \\
\left\langle
\boldsymbol\mu[\mathbf{B}]\diamond \frac{\delta E}{\delta \mathbf{B}}
\,,\,\mathbf{u}\right\rangle 
&=&
\left\langle
\boldsymbol\mu[\mathbf{B}] \times
\,{\rm curl}\,\frac{\delta E}{\delta \mathbf{B}} 
- \, \frac{\delta E}{\delta \mathbf{B}} \, {\rm div}\,\boldsymbol\mu[\mathbf{B}]
\,,\,\mathbf{u}\right\rangle
\nonumber
\end{eqnarray}
for any vector field $\mathbf{u}$. 
\rem{ 
\begin{remark}
These formulas result from computing the diamond operation using the standard action of Lie derivatives on differential forms, instead of the action used earlier to calculate equation (\ref{diamond-vectot-pot}) which employed vector potentials for divergenceless vector fields.
\end{remark}
} 

The explicit forms of the GOP equations (\ref{GOPprincipmath}) are 
\begin{align}\label{GOP1-new}
\frac{\partial \mathbf{A}}{\partial t}
&=-\,
\nabla\left({\bf v}_1\cdot\,\mathbf{A}\right)
+
{\bf v}_1\,{\rm curl}{\bf A}
\,,\qquad
{\bf v}_1:=\left(\frac{\delta E}{\delta \mathbf{A}} \times\,{\rm curl}\,\boldsymbol\mu[\mathbf{A}]
-\, \boldsymbol\mu[\mathbf{A}]\,{\rm div}\,\frac{\delta E}{\delta \mathbf{A}}\right)^\sharp
\\
\label{GOP2-new}
\frac{\partial \mathbf{B} }{\partial t}
&=
\textrm{curl}\left({\bf v}_2
\boldsymbol{\times}\mathbf{B}\right)
-
{\bf v}_2 \,{\rm div}{\bf B}
\,,\qquad
{\bf v}_2:=
\left(
\boldsymbol\mu[\mathbf{B}] \times
\,{\rm curl}\,\frac{\delta E}{\delta \mathbf{B}} 
- \, \frac{\delta E}{\delta \mathbf{B}} \, {\rm div}\,\boldsymbol\mu[\mathbf{B}]
\right)^\sharp
\end{align}
and in vector notation one has the following result
\begin{proposition}
The geometric order parameter equations (\ref{GOP1-new}) and (\ref{GOP2-new}) for any one-forms $\bf A$  and any two-form $\bf B$ have singular solutions of the form (\ref{GOPsing}), where
\begin{align}
\dot{\bf q}_a(t,s)
&=\left.\mathbf{v}_1(\bx)\right\vert_{{\bf \bx = q}_a}
\nonumber\\
\dot{\bf p}_a(t,s)
&=
{\bf 
\mathbf{p}}_a(t,s)\,\left.\left({\nabla\!\!\cdot\!
\mathbf{v}_1(\bx)}\right)\right\vert_{{\bf \bx = q}_a}
-
\left.{\nabla
\mathbf{v}_1(\bx)}
\right\vert_{{\bf \bx = q}_a}
\!\!
\cdot
\,
\mathbf{p}_a(t,s)
\label{pq-eqns-1form}
\end{align}
for closed one-forms $\bf A$ and
\begin{align*}
\dot{\bf q}_a(t,s)
&={\bf v}_2(\bx)
\vert_{{\bf \bx = q}_a}\\
\dot{\bf p}_a(t,s)
&=
\mathbf{p}_a\text{$\!\!^T$}(t,s)\cdot\nabla
{\bf v}_2(\bx)\vert_{{\bf \bx = q}_a}
\end{align*}
for closed two-forms $\bf B$.
\end{proposition}
\begin{proof}
Consider equation (\ref{GOP1-new}) for {\bf A}.
Pairing this equation with a smooth vector field $\boldsymbol\phi$, 
substituting the singular solution ansatz (\ref{GOPsing}), integrating by parts where necessary and matching all terms in $\boldsymbol\phi$
on the two sides yields equations (\ref{pq-eqns-1form}) for the $\bf q$'s and $\bf p$'s.\smallskip

The result for closed 2-forms is proven by noticing that
\[
{\rm curl}\left( {\bf v}_2\times {\bf B}\right)={\bf B}\text{$^T$}\!\cdot\nabla {\bf v}_2-{\bf v}_2\text{$^T$}\!\cdot\nabla {\bf B}-{\bf B}\, {\rm div}\,{\bf v}_2+ {\bf v}_2\,{\rm div}\,{\bf B}
\]
then following the same steps as for the case of exact 1-forms.
\end{proof}

\smallskip
\subsection{Exact differential forms}
When considering the GOP equations (\ref{GOPprincipmath}) for exact differential forms (${\rm curl}\,\mathbf{A}=0={\rm div}\,\mathbf{B}$), the singular solutions also exist satisfy analogous relations. One may see this, by following the same procedure. An important simplification in this case is to take ${\rm curl}\,\boldsymbol\mu[\mathbf{A}]=0={\rm div}\,\boldsymbol\mu[\mathbf{B}]$.
In this case the GOP equations for {\bf A} and {\bf B} take the simpler form
\begin{align}\label{GOP1}
\frac{\partial \mathbf{A}}{\partial t}
&=
\text{\large$\nabla$}\left(\mathbf{A}\cdot\left(\boldsymbol\mu[\mathbf{A}]\,{\rm div}\,\frac{\delta E}{\delta \mathbf{A}}\right)^\sharp\right)
\\
\label{GOP2}
\frac{\partial \mathbf{B} }{\partial t}
&=-\,
\textrm{curl}\left(
\mathbf{B}\times\left(
\boldsymbol\mu[\mathbf{B}] \times
\,{\rm curl}\,\frac{\delta E}{\delta \mathbf{B}} 
\right)^\sharp\right)
\end{align}
where the expressions for ${\bf v}_1$ and ${\bf v}_2$ have been inserted
explicitly.

Moreover, for exact one- and two-forms, the vector equations above can be reduced to nonlocal
nonlinear scalar characteristic equations (\ref{scalareq}) for
the potentials \cite{HoPu2007}. Note that in $\mathbb{R}^3$ (which is of interest to us here) every closed form is exact since $\mbox{curl } \mathbf{A}=\mathbf{0}$ gives $\mathbf{A}= \nabla \psi$ 
for some scalar $\psi$ and $\mbox{div } \mathbf{B}=0$ necessitates $\mathbf B=\mbox{curl } \mathbf{C}$ for some vector $\mathbf{C}$. The characteristic equations for the potentials are derived in the following 
\begin{proposition}[GOP equations for scalar potentials \cite{HoPu2007}] The vector equations (\ref{GOP1}) and (\ref{GOP2}) for exact 1-forms ${\bf A}=\nabla\psi$
and exact 2-forms ${\bf B}={\rm curl}\left(\Psi\,\bf\hat z\right)$ are equivalent to
scalar GOP equations of the type (\ref{scalareq}), in terms of the potentials $\psi$ and $\Psi$. 
Specifically, one finds
\begin{equation}
\frac{\partial \psi }{\partial t}= \Big( \frac{\delta E}{\delta\psi}\,
\nabla \vartheta[\psi] \Big)^\sharp
\boldsymbol{\cdot}\nabla \psi
\,,
\label{APsieq}
\end{equation}
and
\begin{equation} 
\label{psieqB2} 
\frac{\partial \Psi}{\partial t}
=  \Big(\frac{\delta E}{\delta \Psi} \nabla \Phi[\Psi] \Big)^\sharp \cdot \nabla \Psi
\,, 
\end{equation}
where one defines $\boldsymbol\mu[\bf A]:=\nabla \vartheta[\psi]$ and $\boldsymbol\mu[{\bf B}]:={\rm curl}\left(\Phi[\Psi]\,\bf\hat z\right)$.
\end{proposition}
\begin{proof}
Inserting the expression ${\bf A}=\nabla\psi$ in eq. (\ref{GOP1}) yields
\begin{align*}
\frac{\partial \psi }{\partial t}&= 
\Big( \boldsymbol\mu[{\bf A]}\,\,{\rm div}\,\frac{\delta E}{\delta\bf
A}
\Big)^\sharp
\boldsymbol{\cdot}\nabla \psi\\
&=
\Big( \nabla \vartheta[\psi]\,\,\frac{\delta E}{\delta\psi}
\Big)^\sharp
\boldsymbol{\cdot}\nabla \psi
\end{align*}
\rem{
a 1-form allows a simplification in the case
$\mathbf{A}=\nabla \psi$: the equation for potential $\psi$ reads,}
with nonlocal ${\delta E/\delta\psi}$ and $\mu[\psi]$.
\rem{
\begin{equation}
\frac{\partial \psi }{\partial t}= \Big( \frac{\delta E}{\delta\psi}
\nabla \mu[\psi] \Big)^\sharp
\boldsymbol{\cdot}\nabla \psi
\,.
\label{APsieq}
\end{equation} 
This equation for the potential $\psi$ has the same nonlocal
characteristic structure as the scalar equation (\ref{scalareq}).
} 

Similarly, the evolution of 2-form fluxes 
$\mathbf{B} \cdot \mbox{d}\bS= B_x \,\mbox{d}y \wedge \mbox{d}z
+B_y\, \mbox{d}z 
\wedge \mbox{d}x+ B_z\, \mbox{d}x \wedge \mbox{d}y$ also simplifies,
\rem{Again,
the evolution for a general flux $\mathbf{B} \cdot d\bS$ is
quite complicated, but it can be simplified  for the case  ${\rm
div}\,\mathbf{B}=0$ when 
}
when $\mathbf{B}=\nabla\Psi\times\bf\hat{z}$ where $\Psi$ only depends on two spatial coordinates $(x,y)$. Then, 
\[ 
\mbox{curl}\, \frac{\delta E}{\delta \mathbf{B}}
=\mathbf{\hat{z}} \frac{\delta E}{\delta \Psi}. 
\] 
and 
\[ 
\mathbf{\boldsymbol\mu[B]} \times \mbox{curl}\, \frac{\delta E}{\delta \mathbf{B}}
=
(\nabla \Phi \times \mathbf{\hat{z}})
\times \mathbf{\hat{z}} 
\,\frac{\delta E}{\delta \Psi} 
= - \, \frac{\delta E}{\delta \Psi} \,\,\nabla \Phi 
\,.
\] 
Equation (\ref{scalareq}) may be written for the stream function
$\Psi$ (removing the curl from both sides of (\ref{GOP2}))  
\begin{equation} 
\label{psieqB} 
\mathbf{\hat{z}}\frac{\partial \Psi}{\partial t}
=
- \, \Big(\frac{\delta E}{\delta \Psi} \nabla \Phi \Big)^\sharp
\times \mathbf{B}
\rem{ \hskip1cm 
\mbox{with} \hskip1cm {\rm div}\, {\boldsymbol{\mu}}[\mathbf{B}] = 0}
\end{equation}
Then, simplification of two
cross products leads to 
\begin{equation} 
\label{psieqB3} 
\frac{\partial \Psi}{\partial t}
=  \Big(\frac{\delta E}{\delta \Psi} \nabla \Phi \Big)^\sharp \cdot
\nabla \Psi. 
\end{equation}

\rem{
In this case, equation (\ref{scalareq}) may be written for the stream function
$\Psi$ in  $\mathbf{B}=\rem{\mbox{curl}\,\Psi\mathbf{\hat{z}} 
=}\nabla \Psi \times \mathbf{\hat{z}}$ as 
\begin{equation} 
\label{psieqB2} 
\frac{\partial \Psi}{\partial t}
=  \Big(\frac{\delta E}{\delta \Psi} \nabla \Phi \Big)^\sharp \cdot \nabla \Psi
\,, 
\end{equation}
when we choose mobility to be $\mu=\mbox{curl}\,(\mathbf{\hat{z}} \Phi) 
=\nabla \Phi\times\mathbf{\hat{z}} $.
}
Hence, choosing ${\delta E/\delta \Psi}$ and $\Phi$ to depend on the average value $\bar{\Psi}$ again yields a nonlocal characteristic equation.
\end{proof}


\subsection{Singular solutions for exact forms and their potentials}
Equations (\ref{APsieq}) and (\ref{psieqB2}) do allow singular $\delta$-like solutions of the form (\ref{GOPsing})
for $\psi$ and $\Psi$. These solutions, however, 
lead to $\delta'$-like singularities in the forms $\mathbf{A}$ and $\mathbf{B}$. One may understand this point by deriving the expressions for  $\psi$ and $\Psi$ corresponding to the  \emph{clumpon} solutions of the form  (\ref{GOPsing}) for  $\mathbf{A}$ and $\mathbf{B}$. 

For example, taking the divergence of an exact one-form $\mathbf{A}=\nabla\psi$ yields 
$
\nabla\cdot\mathbf{A}=\Delta\psi.
$
Upon using the Green's function of the Laplace operator $G(\mathbf{\bx},\text{\bfi y})=-\left|\mathbf{\bx}-\text{\bfi y}\,\right|^{-1}$, an expression for $\psi$ emerges in terms of $\mathbf{A}$:
\begin{align*}
\psi(\mathbf{\bx},t)&
=-\!\int  \nabla_{\!\mathbf{\bx}'\,}G(\mathbf{\bx},\mathbf{\bx}')\cdot\!\mathbf{A}(\mathbf{\bx}',t)\,d\mathbf{\bx}'.
\end{align*}
Inserting the singular solution (\ref{GOPsing}) for $\bf A$ then yields
\begin{align*}
\psi(\mathbf{\bx},t)&=-\sum_i \int\! ds\,  \mathbf{P}_i(s,t)\cdot\nabla_{\!\mathbf{Q}_i}G(\mathbf{\bx},\mathbf{Q}_i(s,t)).
\end{align*}
However, this singular solution for the potential is \emph{not} in the same form as (\ref{GOPsing}), since the singularities for $\psi$ do not manifest themselves as  $\delta$-functions. 
\smallskip

A similar procedure applies to the case of exact two-forms $\mathbf{B}(x,y)=\text{\rm
curl}\!\left(\Psi(x,y)\,\bf\hat{z}\right)$,
so that $\text{\rm curl}\,\mathbf{B}=\Delta\left(\Psi\, \bf\hat{z}\right)$. One has 
\[
\Psi(\bx)\,=\,{\bf\hat{z}}\,\cdot\,
\sum_i\int\!ds\, \mathbf{P}_i(s,t)\times \nabla_{\mathbf{Q}_i}G(\mathbf{\bx},\mathbf{Q}_i(s,t)), \]
where $\mathbf{Q}$ is in the plane $(x,y)$.
Thus, the equations (\ref{APsieq}) and (\ref{psieqB2}) for $\psi$ and $\Psi$
allow for two species of singular solutions. One of them takes the form
(\ref{GOPsing}), while the other corresponds to a $\delta$-like solution
of the same form (\ref{GOPsing}) for $\mathbf{A}$ and $\mathbf{B}$.

A deeper explanation of this fact can be given in a general context as follows. Consider the advection
equation for an exact form $\kappa=d\lambda$, with potential $\lambda$
\[
\left( \partial_t  +
\,\pounds_{u}\right)
d\lambda=0.
\]
At this point, one remembers
that the exterior differential commutes with the Lie derivative so that the
equation for the potential $\lambda$ is again an advection equation with
the same characteristic velocity
\[
\left( \partial_t  +
\,\pounds_{u}\right)
\lambda=0
\]
At this point, one obtains singular $\delta$-like solutions of the form
(\ref{GOPsing}) for both $\kappa$ and $\lambda$ (provided the characteristic velocity $u$ is sufficiently smooth).

\rem{
Thus, the singular solutions of (\ref{APsieq}) and (\ref{psieqB2}) for $\psi$ and $\Psi$ differ from the singular solutions of the form  (\ref{clumpons}) for  $\mathbf{A}$ and $\mathbf{B}$.
}

\section{Applications to vortex dynamics}
\label{sec:euler}
\subsection{A new GOP equation for fluid vorticity}
The developments above produce an interesting opportunity for the addition of dissipation to ideal fluid equations. This opportunity arises from noticing that the dissipative diamond flows that were just derived could just as well be used with any type of evolution operator, not just the Eulerian partial time derivative. 
For example, if one chooses the geometric order parameter $\kappa$ to be the exact two-form $\omega=\bom\cdot d\mathbf{S}$ appearing as the vorticity in Euler's equations for incompressible motion with fluid velocity $\mathbf{u}$, then the GOP  equation (\ref{GOPprincipmath}) with Lagrangian time derivative may be introduced as a modification of Euler's vorticity equation as follows,
\begin{equation}
\underbrace{\
\partial_t\,\omega+\pounds_u \,\omega\
}
_{\hbox{Euler}}
=
\underbrace{\
\pounds_{\left(\mu[\omega]\, {\diamond}\frac{\delta E}{\delta \omega}\right)^{\!\sharp}}\,
\omega\
}
_{\hbox{Dissipation}}
\,.
\label{vortdiss}
\end{equation}
Euler's vorticity equation is recovered when the left hand side of this equation is set equal to zero. This modified geometric form of vorticity dynamics supports point vortex solutions, requires no additional boundary conditions, and dissipates kinetic energy for the appropriate choices of $\mu$ and $E$.  Equation (\ref{vortdiss}) will be derived after making a few remarks about the geometry of the vorticity governed by Euler's equation. \smallskip

The Lie-Poisson structure of the vorticity equation as been presented in chapter~\ref{intro} and it is written as
\[
\partial_t\,\boldsymbol\omega
\,=\,
-\,{\rm ad}^*_{\,\delta H/\delta \boldsymbol\omega}\,\,\boldsymbol\omega\,=
\text{curl}\big(\boldsymbol\omega\times\text{curl}\,\Delta^{-1}\boldsymbol\omega\big)=
\text{curl}\big(\boldsymbol\omega\times\text{curl}\,\boldsymbol\psi\big)
\]

In order to write the GOP evolution equation (\ref{GOPprincipmath}) for $\boldsymbol\omega$ one
must compute the diamond operation $\diamond$ for the ${\rm ad}^*$ action, which is defined in terms of Lie derivative by
\begin{equation}
{\rm ad}^*_{\boldsymbol\psi}\bom
=
\pounds_{{\rm curl}\,\boldsymbol\psi}\,\bom
\,.
\end{equation}
The computation of the $\diamond$ operation follows from its definition in equation (\ref{diamond-def}). For any two velocity vector potentials $\boldsymbol\phi$ and $\boldsymbol\psi$, and an exact two form $\bom$ one finds
\begin{align}
\left\langle
\boldsymbol\phi\diamond\bom,\boldsymbol\psi
\right\rangle
&=
-\,
\left\langle
\boldsymbol\phi,\pounds_{{\rm curl}\,\boldsymbol\psi}\,\bom
\right\rangle
=
\left\langle
\boldsymbol\phi,\text{curl}\big(\bom\times\text{curl}\,\boldsymbol\psi\big)
\right\rangle\\
&=
\left\langle\,
\text{curl}\,\boldsymbol\phi\times\bom,\text{curl}\,\boldsymbol\psi\,
\right\rangle
=
\left\langle\,
\text{curl}\,\big(\text{curl}\,\boldsymbol\phi\times\bom\big),\boldsymbol\psi\,
\right\rangle
\,.
\end{align}
Consequently, up to addition of a gradient, the diamond operation is given in vector form as
\begin{equation}
\boldsymbol\phi\diamond\bom=
\text{curl}\,\big(\text{curl}\,\boldsymbol\phi\times\bom\big)
\,.
\label{diamond-vectot-pot}
\end{equation}
The insertion of this expression in the bracket (\ref{bracketdef}) gives the GOP equation for $\bom$,
\begin{equation}
\partial_t\,\boldsymbol\omega
=
\text{curl}\left(\boldsymbol\omega\times\text{curl curl}\left(\boldsymbol\mu[\boldsymbol\omega]\times
\text{curl}\,\frac{\delta E}{\delta\boldsymbol\omega}\,\right)\right)
\,.
\label{GOP-eqn-omega}
\end{equation}
\begin{framed}
\noindent
Consequently, equation (\ref{vortdiss}) emerges in the equivalent forms, \begin{align}
\partial_t\,\boldsymbol\omega
&=
-\,{\rm ad}^*_{\boldsymbol\psi}\, \boldsymbol\omega\,
+\,{\rm ad}^*_{\,\,({\rm ad}^*_{\boldsymbol\psi}\,\, \boldsymbol\mu[\boldsymbol\omega])^{\,\sharp}}\,\, \boldsymbol\omega
\nonumber
\\
&=
\text{curl}\left(\boldsymbol\omega
\times\text{curl}\left(\,-\,\frac{\delta H}{\delta\boldsymbol\omega}
+ \text{curl}\left(\boldsymbol\mu[\boldsymbol\omega]
\times
\text{curl}\,\frac{\delta E}{\delta\boldsymbol\omega}\right)
\,\right)\right)
\,.
\label{vortdiss1}
\end{align}
\end{framed}
\noindent
The full dynamics for the vorticity in equation (\ref{vortdiss}) is specified up to the choices of the mobility $\boldsymbol\mu[\bom]$ and the energy in the dissipative bracket $E[\bom]$. By definition, the mobility belongs to the dual space of volume-preserving
vector fields which is here identified with exact two-forms, thus one can
write the mobility in terms of its vector potential as $\boldsymbol\mu=\text{curl}\,\boldsymbol\lambda$ and rewrite the GOP equation (\ref{GOP-eqn-omega}) as 
\begin{equation}
\partial_t\,\bom
=
\text{curl}\left(\bom\times\text{curl}\,\text{curl}
\left[\,\boldsymbol\lambda\,,\,\frac{\delta E}{\delta\bom}\, \right] \right)
\end{equation}
This equation raises questions concerning the dynamics
of vortex filaments with nonlocal dissipation, following the ideas in \cite{Ho03}, where connections were established between the Marsden-Weinstein bracket \cite{MaWe83} and the Rasetti-Regge bracket for vortex dynamics \cite{RaRe}. Ideas for dissipative bracket descriptions in fluids have been introduced previously, see Bloch et al. \cite{BlKrMaRa1996, BlBrCr1997} and references therein. In particular, equation (\ref{vortdiss1}) recovers equations (2.2-2.3) of Vallis et al. \cite{VaCaYo1989} when $E=H$ and $\boldsymbol\mu=\alpha\,\bom$ for a constant $\alpha$. 
\begin{remark}[Coadjoint dissipative dynamics] From the first line in equation
(\ref{vortdiss1}), one sees that the vorticity dynamics is a form of {\it
coadjoint motion} (cf. chapter~\ref{intro}). Indeed, the equation can also
be written as
\begin{align}
\frac{\partial\boldsymbol\omega}{\partial t}
&=
-\,\textnormal{\large ad}^*_{\big[\text{\,\footnotesize$\boldsymbol\psi-\!\left({\rm ad}^*_{\boldsymbol\psi}\,\, \boldsymbol\mu[\boldsymbol\omega]\right)^{\sharp}\,$}\big]}
\, \boldsymbol\omega\,
\nonumber
\end{align}
which shows how the vorticity evolves on coadjoint orbits of the group {\rm
Diff}$_\textnormal{vol}$,
generated by the Lie algebra element $\boldsymbol\psi-\!\left({\rm ad}^*_{\,\boldsymbol\psi}\,\, \boldsymbol\mu[\boldsymbol\omega]\right)^{\sharp}$. In particular, the Casimir functionals corresponding to Hamiltonian dynamics are also preserved by the geometric dissipation. This observation is a key step in the theory of geometric dissipation and it leads to the preservation of entropy when this theory is applied to kinetic equations (cf. chapter~\ref{DBVlasov}).
\end{remark}

The GOP equation (\ref{vortdiss}) may be expressed as 
the Lie-derivative relation for conservation of vorticity flux, 
\begin{equation}
\partial_t\,(\bom\cdot d\mathbf{S})
\,=\,-\,\pounds_{\mathbf{u}-\mathbf{v}}\,
(\bom\cdot d\mathbf{S})
\,,
\label{Lie-der-GOP-vortex}
\end{equation}
in which the velocities $\mathbf{u}$ and $\mathbf{v}$ may be written in terms of the commutator $[\,\cdot\,,\,\cdot\,]$ of divergenceless vector fields as,
\begin{equation}
\mathbf{u}
=
{\rm curl}\,\frac{\delta H}{\delta\boldsymbol\omega}
\,,\quad
\mathbf{v}
=
{\rm curl}\,{\rm curl}\left(\boldsymbol\mu[\boldsymbol\omega]
\times
\mathbf{\tilde{u}} \right)
=
{\rm curl}\,\big[\,\boldsymbol\mu[\bom]
\,,\,\mathbf{\tilde{u}}\,\big]
\quad\hbox{where}\quad
\mathbf{\tilde{u}}
=
{\rm curl}\,\frac{\delta E}{\delta\boldsymbol\omega}
\,,
\label{uandv-3D}
\end{equation}
The compact form (\ref{Lie-der-GOP-vortex}) clearly underlines the dissipative
nature of the dynamics, for which the transport velocity {\bf u} is appropriately
decreased by the nonlocal dissipative velocity {\bf v}.

Since both $\mathbf{u}$ and $\mathbf{v}$ are divergenceless, the vorticity equation (\ref{Lie-der-GOP-vortex}) may also be expressed as a commutator of divergenceless vector fields, denoted as $[\,\cdot\,,\,\cdot\,]$,
\begin{equation}
\partial_t\,\bom + (\mathbf{u}-\mathbf{v})\cdot\nabla\bom
-
\bom\cdot\nabla(\mathbf{u}-\mathbf{v})
=
\partial_t\,\bom
+
[\mathbf{u}-\mathbf{v},\,\bom]
=
0
\,.
\label{vorticity-commute}
\end{equation}
Thus, the vorticity is advected by the total velocity $(\mathbf{u}-\mathbf{v})$ and is stretched by the total velocity gradient. In this form one recognizes that the singular vortex filament solutions of (\ref{vorticity-commute}) will move with the total velocity $(\mathbf{u}-\mathbf{v})$, instead of the Biot-Savart velocity $(\mathbf{u}={\rm curl}^{-1}\bom)$ alone.  

\rem{ 
Note that adding geometric dissipation to the vorticity equation as in equation (\ref{vorticity-commute}) does not require an additional boundary condition for fixed boundaries. 
} 

\subsection{Results in two dimensions: point vortices and steady flows} The GOP  equation (\ref{vortdiss}) for vorticity including both inertia and dissipation takes the same form as the Euler vorticity equation in {\it two dimensions}, but with a modified stream function. Indeed, by a standard calculation with stream functions in two dimensions, equations (\ref{uandv-3D}) and (\ref{Lie-der-GOP-vortex}) imply the following dynamics, expressed in terms of $\omega:=\bhz\cdot\bom$ and $\mu:=\bhz\cdot\boldsymbol\mu$
\begin{equation}
\partial_t\,\omega
+ \big[ \omega,\,
\psi - [\mu[\omega],\tilde{\psi}]\,\big]
=
0
\,,
\label{vortdiss-2D}
\end{equation}
where $\psi=\delta H/\delta\omega$, 
$\tilde{\psi}=\delta E/\delta\omega$ and $[ f,\,g]$ is the symplectic bracket, given for motion in the $(x,y)$ plane by the two-dimensional Jacobian determinant, 
\rem{
\begin{equation}
\big[ f,\,g\big]
=
\frac{\partial f}{\partial x} 
\,\frac{\partial g}{\partial y} 
-
\frac{\partial g}{\partial x} 
\,\frac{\partial f}{\partial y} 
=
\bhz\cdot\nabla f \times \nabla g
\,.
\label{bracket-2D}
\end{equation}
}
\begin{equation}
\big[ f,\,g\big]dx\wedge dy
=
df\wedge dg
\,.
\label{bracket-2D}
\end{equation}
Equation (\ref{vortdiss-2D}) takes the same form as Euler's equation for vorticity, but with a modified stream function, now given by the sum $\psi - [\,\mu,\,\tilde{\psi}\,]$. 

\begin{remark}
The GOP equation for vorticity in two dimensions (\ref{vortdiss-2D}) recovers equation (4.3) of Vallis et al. \cite{VaCaYo1989} when one chooses $\mu=\alpha\omega$ for a constant $\alpha$ and $E= \frac{1}{2}\int \omega\,\psi \,dxdy$. However, for this choice of mobility, $\mu$,  point vortex solutions are excluded. 
\end{remark}

\begin{proposition}[Point vortices]
The GOP  equation for vorticity in two dimensions (\ref{vortdiss-2D}) possesses point vortex solutions, with any choices of $\mu[\omega]$ and $\tilde{\psi}$ for which 
$K=\psi -[\mu\left[\omega\right],\tilde{\psi}]$ is sufficiently smooth.
\end{proposition}

\begin{proof}
Pairing equation (\ref{vortdiss-2D}) with a stream function
$\eta$ yields
\begin{equation}
\langle\, \eta,\,\partial_t{\omega}\,\rangle
=\left\langle\, 
\big[\,\eta,\,K\left[\omega\,\right] \big],\,
\omega\,\right\rangle
\quad\text{ where }\quad
K\left[\omega\right]
=
\psi -[
\mu\left[\omega\right],\tilde{\psi}
]
\label{omega-psi-eqn}
\end{equation}
Inserting the expression
\[
\omega(x,y,t)=\Gamma(t)\,\delta(x-X(t))\,\delta(y-Y(t))
\]
into the previous equation and integrating against a smooth test function yields
\[
\dot{\Gamma}\,\eta  + \Gamma\, \dot{X}\,\frac{\partial \eta}{\partial X}
 + \Gamma\, \dot{Y}\,\frac{\partial \eta}{\partial Y}
 =
 \Gamma\,
 \frac{\partial \eta}{\partial X}\,
 \frac{\partial {K}}{\partial Y}
 -
 \Gamma\,
 \frac{\partial {K}}{\partial X}\,
 \frac{\partial \eta}{\partial Y}
 \,,
\]
where $\eta$ and $K$ are evaluated at the point $(x,y)=(X(t),Y(t))$.
Thus, the point vortex solutions for equation (\ref{vortdiss-2D}) on the $(X,Y)$ plane satisfy
\begin{equation}
\dot{\Gamma} =0\,,
\qquad\quad
\dot{X} = \frac{\partial {K}}{\partial Y}\,,
\qquad\quad
\dot{Y} = -\,\frac{\partial {K}}{\partial X}
\,,
\label{Ham-2D}
\end{equation}
whose solutions exist provided the function $K$  is sufficiently smooth.
\end{proof}

\begin{remark}
Solutions of the symplectic Hamiltonian system (\ref{Ham-2D}) extend for the case of evolution of arbitrary many point vortices  for the GOP vorticity equation (\ref{vortdiss-2D}) in two dimensions. These solutions represent a set of $N$ vortices at positions $(X_k(t),Y_k(t))$ ($k=1,\ldots,N$) moving in the plane. Properties of the corresponding point vortex solutions of Euler's equations in the plane are discussed for example in \cite{Sa1992}.
\end{remark}

\paragraph{Steady states of the dissipative vorticity equation in 2D.}
The two dimensional version of the vorticity equation provides a simple opportunity
for investigating the stationary solutions. For example, it is obvious that
the equation 
\begin{equation*}
\big[ \omega,\,
\psi - [\mu[\omega],\tilde{\psi}]\,\big]
=
0
\,,
\end{equation*}
is always satisfied when  $\psi - [\mu[\omega],\tilde{\psi}]=\Phi(\omega)$,
where $\Phi$ is a function of the vorticity $\omega$. In fact, the chain
rule yields
\[
\big[\omega,\,\Phi(\omega)\big]\,=\,\omega_x\,\Phi'(\omega)\,\omega_y\,-\,\omega_y\,\Phi'(\omega)\,\omega_x\,=\,0
\,.
\]
This is an evident
consequence of the fact that geometric dissipation preserves the coadjoint
nature of the Hamiltonian flow thereby recovering the Casimir functionals
$C[\omega]=\int\Phi(\omega)\,{\rm d}x\,{\rm d}y$. However, more can be said about the relation
occurring between the steady flows of the Hamiltonian dynamics and those
corresponding to geometric dissipation. Indeed the observation above means
that 
\begin{proposition}
If $[\mu[\omega],\tilde{\psi}]$ depends only on $\omega$ 
(say $\widetilde\Phi(\omega)=[\mu[\omega],\tilde{\psi}]$), and if $\bar\omega$
is a stationary state of the Hamiltonian vorticity equation (so that $[\bar\omega,\psi]=0$),
then {\sl the equilibria of the dissipative flow will coincide with those of the
Hamiltonian flow} and the level sets of $\omega$ and $\psi$ will evolve until
they coincide. The same holds if $[\mu[\omega],\tilde{\psi}]=M[\omega]\,[\omega,\psi]$,
where $M[\omega]$ is a pure functional of $\omega$ and if $[\mu[\omega],\tilde{\psi}]=\alpha\,[\omega,\psi]$.
\end{proposition}

\begin{remark}[Extending to kinetic equations]
The validity of this statement can be extended to other systems undergoing
geometric dissipation. The necessary condition is that the Hamiltonian
flow corresponding to the dissipative system under consideration undergoes
coadjoint dynamics, so that Casimir functionals are known. An example
is provided in chapter~\ref{DBVlasov}, where the geometric dissipation is
applied to kinetic equations. The equilibria of the resulting dissipative
Vlasov equation are {\sl explained exactly by the proposition above}, which applies
in this case upon substituting the vorticity $\omega$ with the distribution
function $f$ on phase space.
\end{remark}

\subsection{More results in three dimensions}
As a consequence of the modified vorticity equation (\ref{vorticity-commute}) in commutator form, one easily checks the following properties.
\begin{itemize}
\item
{\bfi Ertel's theorem} is satisfied by the vector field $\,\,\bom\cdot\nabla$ associated to vorticity. By using the commutator notation and the material
derivative $D/Dt$, one can write
\begin{equation}
\frac{D\alpha}{Dt}
:=
\frac{\partial\alpha}{\partial t}+\left({\bf u-v}\right)\cdot\nabla\alpha
=
\bom\cdot\nabla\alpha
\,,\quad\text{ so that } \quad
\left[\frac{D}{Dt}\,,\,\bom\cdot\nabla\,\right] \alpha
=0
\,,
\end{equation}
for any scalar function $\alpha({\bf x},t)$. 
\item
An analogue of the Kelvin's circulation theorem holds for equation (\ref{vorticity-commute}). Upon expressing the vorticity as $\bom={\rm
curl }\,\bf u$, one writes the following dissipative form of the Euler equation for the velocity $\bf u$
\begin{equation}
\partial_t {\bf u+(u-v)\cdot\nabla u}
-u_j\nabla v^j
=-\nabla p
\,,\qquad
\nabla \cdot \mathbf{u} = 0 \, ,
\end{equation}
where $\mathbf{v}$ is given in (\ref{uandv-3D}). This equation may also be expressed as 
\begin{equation}
\partial_t {\bf u+u\cdot\nabla u}
+ \nabla \Big(p 
+ \mathbf{u}\cdot\mathbf{v} \Big)
= -
\underbrace{\
{\bf v \times {\rm curl}\, u }\
}_{\hbox{\rm Vortex force}}
\,,\qquad
\nabla \cdot \mathbf{u} = 0 \,,
\label{GOP-vortexforce}
\end{equation}
by using a vector identity.
An equivalent alternative is the Lie derivative form,
\begin{equation}
\left(\partial_t+\pounds_{\bf u-v}\right)\left(\bf u\cdot \textnormal{d}x\right)=-\,\textnormal{d}p
\,.
\end{equation}
Hence, one finds that a {\bfi modified circulation theorem} is satisfied,
\begin{equation}\label{euler-dissip}
\frac{d}{dt}\,\,\oint_{\mathcal{C}\left(\bf u-v\right)} \!\!\!\!\!\bf u\cdot \textnormal{d}x\,\,=\,\,0
\end{equation}
for a loop $\mathcal{C}(\bf u-v)$ moving with the ``total'' velocity $\,\bf u-v$. That is, two velocities appear in the modified circulation theorem. One is the ``transport velocity'' $\,\bf u-v$ and the other is the ``transported velocity'' $\,\bf u$.
\item
From equations (\ref{euler-dissip}) and (\ref{Lie-der-GOP-vortex}) one
checks that
\begin{equation}
\left(\partial_t+\pounds_{\bf u-v}\right){\bf\left(\bom\cdot\textnormal{d}S\wedge
u\cdot\textnormal{d}x\right)
=
-\,\bom\cdot\textnormal{d}S\wedge\textnormal{d}}p
=
-\textnormal{div}\left(p\,\bom\right)\textnormal{d}^3\bf x
\end{equation}
so that the {\bfi helicity of the vorticity $\bom$ is conserved}
\begin{equation}
\frac{d}{dt}\,\int\!\!\!\!\int\!\!\!\!\int_\textnormal{Vol}\! \bom\cdot\bf u \,\,\textnormal{d}^3 x\,=\,0
\end{equation}
\end{itemize}
One may summarize these remarks as follows:
\begin{quote}
\begin{framed}\textsl{%
All of these classical geometric results for ideal incompressible fluid mechanics follow for the modified Euler equation. These results all persist (including preservation of helicity) when transport velocity is replaced as $(\mathbf{u}-\mathbf{v})\to-\,\mathbf{v}$.
}
\end{framed}
\end{quote}


This completes the present investigation of GOP  vortex dynamics. An obvious extension would be to consider GOP vortex patches in two dimensions. Instead of pursuing such GOP vorticity considerations further, the next section
applies GOP theory to different well known cases in continuum Hamiltonian
mechanics. 

\section{Two more examples}
The developments discussed above produce an interesting opportunity for the addition of dissipation to various other continuum equations. Following the introduction of the dissipative Euler equation above,  one could extend the dissipative diamond flows with any type of evolution operator.  This section sketches how one might develop this idea further, by illustrating its application in three more physically relevant examples.

\subsection{Dissipative EPDiff equation}
Consider adding geometric dissipation to the Euler-Poincar\'e equation on the diffeomorphisms (EPDiff)  \cite{HoMa2004} for the evolution of a one-form density $m$ defined by
\begin{equation}
 m=\mathbf{m}\cdot d\bx\otimes d^3x
 \,.
\end{equation}
This addition gives the {\it dissipative EP equation},
\begin{equation}
 \partial_t\, m 
 + {\rm ad}^*_{\delta H/\delta m}\, m 
 = 
-\, \pounds_{\!\big( \mu[m]\,\diamond\, {\delta E/\delta m} \big)^\sharp }\,\, m
\,=\,
-\, {\rm ad}^*_{\!\big( \mu[m]\,\diamond\, {\delta E/\delta m} \big)^\sharp }\,\,m
 \,.
 \label{EPDiff-diss}
\end{equation}
When $H[m]$ is the Hamiltonian for the Lie-Poisson theory corresponding to EPDiff, then the vector field $\delta H/\delta m=u$ is the characteristic velocity for the Euler-Poincar\'e equation. For a one-form density $m$, the diamond operation is given by ${\rm ad}^*$, which is equivalent to Lie derivative. That is,
\begin{equation}
 \mu[m]\,\diamond\, {\delta E/\delta m} 
 =
 {\rm ad}^*_{\delta E/\delta m}\,  \mu[m]
 =
 \pounds_{\delta E/\delta m} \, \mu[m]
 \,.
\end{equation}
The further choice $\mu[m]=\alpha m$ for a positive constant $\alpha$ recovers equation (6.10) of Bloch et al.  \cite{BlKrMaRa1996}. When in addition, $\mu[m]=K*m$ for a smoothing kernel $K$, then equation (\ref{EPDiff-diss}) supports singular solutions of the type discussed in Holm and Marsden \cite{HoMa2004}. 


\paragraph{Peakon dynamics for the dissipative Camassa-Holm equation.} 
In one dimension, the GOP version of the EPDiff equation (\ref{EPDiff-diss}) reduces to,
\begin{equation}
\partial_t m +(u-v)m_x
+ 2m(u-v)_x
 =
0
 \,,
 \label{GOP-CH}
\end{equation}
where $u=\delta H / \delta m$ for a specified Hamiltonian $H[m]$.
The other velocity $v$ is given in one dimension by
\begin{equation}
v = \left(\,{\rm ad}^*_{\delta E / \delta m}\,\mu[m]\,\right)^\sharp
= \frac{\delta E }{ \delta m}\,\partial_x\mu[m]
+ 2 \mu[m]\,\partial_x\,\frac{\delta E }{ \delta m}
\,,
\end{equation}
for arbitrary (smooth) choices of $\mu[m]$ and $E[m]$.
Now consider the singular solution form for $m$ given by a sum of $N$ delta functions,
\begin{equation}\label{m-delta1}
m(x,t) = \sum_{i=1}^N p_i(t)\,\delta(x-q_i(t))
\,,
\end{equation}
and take quadratic functionals $H[m]=1/2\,\langle m,G*m\rangle$ and $E[m]=1/2\,\langle m,W\!*m\rangle$ so that
\[
u(x)=G*m=\sum_{j=1}^N \,p_j\,G(x-q_j)
\]
and, since $\mu[m]=K*m$,
\begin{align*}
v(x)&=W\!*m\,\,\partial_x K*m
+ 2\, K*m\,\,\partial_x W\!*m\\
&=
\sum_{j,k=1}^N p_j\,p_k\,\bigg(K(x-q_j)\,\partial_x \!W(x-q_k)+2\,W(x-q_k)\,\partial_x\!
K(x-q_j)\bigg)\\
&=
\sum_{j,k=1}^N p_j\,p_k\,\mathcal{R}(x-q_j,x-q_{\,k})
\end{align*}
where one defines $\mathcal{R}$ for compactness of notation.
Substituting the above expressions into the GOP EPDiff equation (\ref{GOP-CH}) and integrating against a smooth test function yields the following relations for time derivatives of $p_i(t)$ and $q_i(t)$: 
\begin{eqnarray} \label{q-eqn} 
\dot{q}_i & = & 
\left.
\Big(u(x) - v(x) \Big)\right\vert_{x=q_i(t)}\\ \nonumber
& = &
\sum_{j=1}^{N}\,\, p_j\,G(q_i-q_j)\,-\sum_{j,k=1}^{N}\,p_j\,p_k\,\mathcal{R}(q_i-q_j,q_i-q_k)
\,, 
\\
\dot{p}_i & = & 
-\,p_i
\left.
\Big(u'(x) + v'(x) \Big)\right\vert_{x=q_i(t)}\\ \nonumber
& = &
\, -\,p_i\sum_{j=1}^{N}\,\,p_j\,\partial_{q_i}\!G(q_i-q_j)\,+p_i\sum_{j,k=1}^{N}\,p_j\,p_k\,\partial_{q_i}\!\mathcal{R}(q_i-q_j,q_i-q_k)
\,,
\label{p-eqn}
\end{eqnarray} 
\rem{ 
where the Hamiltonian $H_N$ is given by
\begin{equation}\label{Ham-pulson}
H_N =  \frac{1}{2}\sum_{i,j=1}^N p_i\,p_j
\left(G(q_i-q_j) + \sum_{k=1}^N p_k\mathcal{R}(q_i-q_j,q_i-q_k) \right)
\,.
\end{equation}
}      
The choices of $H$, $E$ and $\mu$ as functionals of $m$ determine the ensuing dynamics of the singular solutions. 
In particular,  in the case when the velocity $u$ is given by
\begin{equation}
u[m] = \frac{\delta H }{ \delta m} 
= (1-\alpha^2\partial_x^2)^{-1}m
= \int e^{-\frac{|x-x'|}{\alpha}}\, m(x'){\rm d}x'
\,,
\quad\hbox{for}\quad
H = \frac{1}{2}\int m\,u[m] \, {\rm d}x
\,.
\end{equation}
Equation (\ref{GOP-CH}) is a GOP version of the integrable Camassa-Holm equation with peaked soliton solutions \cite{CaHo1993}. Nonlinear interactions of $N$ traveling waves of this system may be investigated by following the approach of Fringer and Holm \cite{FrHo2001}. 

\smallskip
\paragraph{Remarks on dissipative semidirect product dynamics.}\label{DarcyFluid}

The equations derived above consolidate the idea that any continuum equation in characteristic form,
\[
\left(\partial_t+\pounds_u\right)\kappa=0
\,,
\]
may be modified to include dissipation via the substitution $u\rightarrow u+v$, in which $v$ is the dissipative velocity term expressed in equation (\ref{GOPprincipmath}).
This idea may also be extended to the semidirect product framework 
presented in \cite{HoMaRa}, in order to include compressible fluid flows and plasma fluid models such as the barotropic fluid model or MHD. Instead
of constructing GOP equations for such structures, chapter~\ref{DBVlasov}
derives the barotropic fluid equations from moment dynamics in the kinetic
approach. Once the structure of these equations is identified, the same can
be applied to other examples such as MHD.
\rem{ 
An immediate application of the semidirect product framework would be an investigation of ideas of  {\it selective decay} in the approach to {\it topological equilibria} for example in MHD, as first suggested by Taylor \cite{Ta1974} based on work of Woltjer \cite{Wo1958} and later elaborated by Moffat \cite{Mo1985} and others. A possible counterpoint would be to treat the additional helicity-conserving geometric force as a means of driving a magnetic dynamo, rather than relaxing to an equilibrium.
} 

\smallskip
\subsection{Dissipative Vlasov dynamics} 
One may also extend the diamond dissipation framework to systems such as the Vlasov equation in the symplectic framework of coordinates and momenta as independent variables. This extension requires the introduction of the Vlasov Lie-Poisson bracket,
defined for phase space densities $f(q,p)\,\textnormal{d}q\wedge \textnormal{d}p$ on $T^*\mathbb{R}^N$ as
\begin{equation}
\{F,H\}(f)=\iint f\,\left\{\frac{\delta F}{\delta f}\,,\,\frac{\delta H}{\delta f}\right\}dq\wedge dp \, . 
\end{equation}
where the bracket $\{\,\cdot\,,\,\cdot\,\}$ in the integral is the canonical Poisson bracket. At this point a new definition of the diamond operator is required. This is found by the well known identification of Hamiltonian
vector fields and their generating functions,
\rem{ 
that gives the relation
\begin{equation}
-\pounds_{X_h}\,f=[f,h]_{qp}
\end{equation}
for any phase space density $f(q,p)$ and any function $h(q,p)$. 
This relation
} 
 which  identifies
the symplectic Lie algebra action on the Vlasov distribution and the new kind of \emph{symplectic} diamond.
\rem{ 
 which may be  computed by applying the general definition as
\begin{equation}
\left\langle g\star f\,,\,h\right\rangle
=
\left\langle g\,,\,-\pounds_{X_h\,} f\right\rangle
=
\Big\langle \big [g\,,\,f\big]_{qp}\,,h \Big\rangle
\end{equation}
for any two functions $g$ and $h$. 
} 
This treatment is extensively presented in chapter~\ref{DBVlasov} and introduces
the idea of a microscopic description for Darcy's law. This section reports the final result. 
Extending the previous discussions to the symplectic case, one
can write the following form of GOP dissipative Vlasov equation,
\begin{equation}
\frac{\partial f}{\partial t}
+
\left\{\,f\,,\,\frac{\delta H}{\delta f}\,\right\}
\,=\,
\left\{\,f\,,\,\left\{\,\mu[f]\,,\,\frac{\delta E}{\delta f}\,\right\}
\,\right\}
\label{Vlasov-diss-chp4}
\end{equation}
where, in general, the functionals $H$ and $E$ are independent. 
This equation has the same form as the equations for a dissipative class of Vlasov plasmas in astrophysics, proposed by Kandrup \cite{Ka1991} to model gravitational radiation reaction. Kandrup's formulation for an azimuthally symmetric particle distribution is recovered by choosing a linear phase space mobility $\mu=\alpha f$ with positive constant $\alpha$ and taking $E$ to be $J_z[f]$ the total azimuthal angular momentum for the Vlasov distribution $f$. 
\rem{ 
Equation (\ref{Vlasov-diss-chp4}) recovers the dissipative bracket formulations of both Kaufman \cite{Ka1984} and Morrison \cite{Mo1984} when $E=H_f-S$, where $S$ is the Vlasov  entropy functional $S=\int f\log{f}$, the quantity $H_f$ is the single-particle Hamiltonian and $\mu[f]=\alpha f$.  For these choices
} 
More generally, if one chooses $\mu[f]=\alpha f$ and $E$ to be the Vlasov
Hamiltonian $H[f]$, the dissipative Vlasov equation (\ref{Vlasov-diss-chp4})  assumes the {\bfi double bracket} form,
\begin{equation}
\frac{\partial f}{\partial t}
+
\left\{\,f\,,\,\frac{\delta H}{\delta f}\,\right\}
\,=\,\alpha\,
\left\{\,f\,,\,\left\{\,f\,,\,\frac{\delta H}{\delta f}\,\right\}
\,\right\}
.
\label{Vlasov-diss-BlKrMaRa}
\end{equation}
This is also the Vlasov-Poisson equation in Bloch et al. \cite{BlKrMaRa1996}. However, in contrast to the choices in \cite{BlKrMaRa1996, Ka1984, Mo1984, Ka1991}, the GOP form of the Vlasov equation (\ref{Vlasov-diss-chp4}) allows more general mobilities such as $\mu[f]=K*f$ (which denotes convolution of $f$ with a smoothing kernel $K$). The GOP choice has the advantage of recovering the one-particle solution as its singular solution. 
The investigation of this equation and the consequent kinetic theory is the
subject of chapter~\ref{DBVlasov}, which will present important connections
with the theory of \emph{double bracket dissipation} and will show how a
geometric form of dissipation can be introduced for kinetic moments.

\section{Discussion} 
This chapter has provided a contribution to the GOP theory of Holm and Putkaradze
(HP) \cite{HoPu2007} for the construction of dissipative evolutionary equations in the form (\ref{GOPprincipmath}) for a variety of different types of geometric order parameters $\kappa$. As a result, the HP method now produces a plethora of fascinating singular solutions for these evolutionary GOP equations. Each GOP equation is expressed as a characteristic equations in a certain geometric sense. However, the characteristic velocities in these equations may be {\bfi nonlocal}. That is, the characteristic velocities may depend on the solution in the entire domain. The equations may possess either or both of the following structures: (i) a conservative Lie-Poisson Hamiltonian structure; (ii) a dissipative Riemannian metric structure. The two types of evolution are combined by simply adding the characteristic velocities in their Lie derivatives. 
Similar types of equations were discussed by Bloch et al. \cite{BlKrMaRa1996} who studied the effects on the stability of equilibrium solutions of continuum Lie-Poisson Hamiltonian systems of adding a type of geometric dissipation that preserves the coadjoint orbits of the Hamiltonian systems. Such equations have the form
\[
\frac{dF}{dt}= \{F,H\} - \{\{F,H\}\} 
\]
for two bracket operations, one antisymmetric and Poisson ($\{F,H\}$) and the other symmetric and Leibnitz ($\{\{F,H\}\}$).  

The GOP theory has been shown to apply in a number of continuum flows with geometric order parameters, each allowing singular solutions. The various types of singular solutions include point vortices, vortex filaments and sheets, solitons and single particle solutions for Vlasov dynamics. 

In some cases, the singular solutions emerge from smooth confined initial conditions
\cite{HoPu2005,HoPu2006,HoPu2007}. In other cases, such emergent behavior does not occur.  It remains an open question to determine whether the singular solutions of a given geometric type will emerge from smooth initial conditions.
In particular, the existence of a ``steepening lemma'' \cite{CaHo1993,HoPu2005,HoPu2006} (cf. chapter~\ref{intro}) for a certain class of order parameters, the GOP equations would guarantee the emergence of singularities {\it in finite time}
for some choices of energy $E$ and mobility $\mu[\kappa]$. For example, one
may conjecture that this property is actually valid in the case of the dissipative
EPDiff equation (one-form densities) and the formulation of a ``steepening lemma'' would be necessary to prove this conjecture. After they are created, the singular solutions evolve with their own dynamics. Investigations of the interactions of these singular solutions and the types of motions available to them will be discussed in the remainder of this work. The present chapter has derived the dynamical equations for these singular solutions in various cases. 

\chapter{Geometric dissipation for kinetic equations}
\label{DBVlasov}
\section{Introduction}

Non-linear dissipation in physical systems can modeled by the sequential application of  two Poisson brackets, just as in magnetization dynamics \cite{Gilbert1955}.
A similar double bracket operation for modeling dissipation has been proposed for the Vlasov equation. Namely,
\begin{equation}
\frac{\partial f}{\partial t}+ \left\{\,f\,,\,\frac{\delta H}{\delta f}\right\}
=
\alpha \left\{\,f\,,\,\left\{\,f\,,\,\frac{\delta H}{\delta f}\,\right\} \right\}
\,,
\label{Kandrup-dbrkt}
\end{equation}
where $\alpha>0$ is a positive constant, $H$ is the Vlasov Hamiltonian  and 
$\{ \cdot \, , \, \cdot \}$ is the canonical Poisson bracket. When $\alpha\to0$, this equation reduces the Vlasov equation for collisionless plasmas. For $\alpha>0$, this is the {\bfi double bracket dissipation} approach for the Vlasov-Poisson equation introduced in Kandrup \cite{Ka1991} and developed in Bloch {\it et al.} \cite{BlKrMaRa1996}.
This double bracket approach for introducing dissipation into the Vlasov equation differs from the standard Fokker-Planck linear diffusive approach \cite{Fokker-Plank1931}, represented by the equation
\[
\frac{\partial f}{\partial t}+\left\{\,f\,,\,\frac{\delta H}{\delta f}\right\}
=
\frac{\partial}{\partial p}\Bigg(\!\left(\gamma\, p+D\,\frac{\partial}{\partial p}\right)f\Bigg)
\]
which adds dissipation on the right hand side as the Laplace operator in the momentum coordinate $\Delta_p f$. 

 An interesting feature of the double bracket approach is that the resulting symmetric bracket gives rise to a metric tensor and an associated Riemannian (rather than symplectic) geometry for the solutions, as explained in chapter~\ref{GOP}.  The variational approach also preserves the {\bfi advective} nature of the evolution of Vlasov phase space density, by coadjoint motion under the action of the canonical transformations on phase space densities.

 As Otto \cite{Ot2001} explained, the geometry
of dissipation may be understood as emerging from a variation principle. This chapter follows the variational approach to derive the following generalization of the double bracket structure in equation (\ref{Kandrup-dbrkt}) that recovers previous cases for particular
choices of modeling quantities \cite{HoPuTr2007-CR}, 
\begin{framed}
\begin{equation}
\frac{\partial f}{\partial t}
+
\left\{\,f\,,\,\frac{\delta H}{\delta f}\,\right\}
\,=\,
\left\{\,f\,,\,\left\{\,\mu(f)\,,\,\frac{\delta E}{\delta f}\,\right\}
\,\right\}
\,.
\label{Vlasov-diss}
\end{equation}
\end{framed}
\noindent
Eq. (\ref{Vlasov-diss}) extends the double bracket operation in (\ref{Kandrup-dbrkt}) and reduces to it when $H$ is identical to $E$ and $\mu(f)=\alpha\,f$.
The form (\ref{Vlasov-diss}) of the Vlasov equation with dissipation allows for more general mobilities than those in \cite{BlKrMaRa1996,Ka1991,Ka1984,Mo1984}.
For example, one may choose $\mu[f]=K*f$ (in which $*$ denotes convolution in phase space). As in \cite{HoPuTr2007} the smoothing operation in the definition of $\mu(f)$ introduces a fundamental length scale (the filter width) into the dissipation mechanism.  Smoothing also has the advantage of endowing (\ref{Vlasov-diss}) with the one-particle solution as its singular solution. 
The generalization Eq. (\ref{Vlasov-diss}) may also be justified by using a thermodynamic and geometric arguments \cite{HoPuTr2007}. 
In particular, this generalization extends the classic Darcy's law (velocity being proportional to force) to allow the corresponding modeling at the microscopic statistical  level.

\subsection{History of double-bracket dissipation}
Bloch, Krishnaprasad, Marsden and Ratiu (\cite{BlKrMaRa1996} abbreviated BKMR) observed that linear dissipative terms of the standard Rayleigh dissipation type are inappropriate for dynamical systems undergoing coadjoint motion. Such systems are expressed on the duals of Lie algebras and they commonly arise from variational principles defined on tangent spaces of Lie groups. A well known example of coadjoint motion is provided by Euler's equations for an ideal incompressible fluid \cite{Ar1966}. Not unexpectedly, adding linear viscous dissipation to create the Navier-Stokes equations breaks the coadjoint nature of the ideal flow.  Of course, ordinary viscosity does not suffice to describe dissipation in the presence of orientation-dependent particle interactions. 

Restriction to coadjoint orbits requires nonlinear dissipation, whose gradient structure differs from the Rayleigh dissipation approach leading to Navier-Stokes viscosity.  As a familiar example on which to build their paradigm, BKMR emphasized a form of energy dissipation (Gilbert dissipation \cite{Gilbert1955}) arising in models of ferromagnetic spin systems that preserves the magnitude of angular momentum. In the context of Euler-Poincar\'e or Lie-Poisson systems, this means that coadjoint orbits remain invariant, but the energy decreases along the orbits. BKMR discovered that their geometric construction of the nonlinear dissipative terms summoned  the double bracket equation of Brockett \cite{Br1988, Br1993}. In fact, the double bracket form is well adapted to the study of dissipative motion on Lie groups since it was originally constructed as a gradient system \cite{Br1994}. 

While a single Poisson bracket operation is bilinear and antisymmetric, a double bracket operation is a symmetric operation. Symmetric brackets for dissipative systems, particularly for fluids and plasmas, were considered previously by Kaufman \cite{Ka1984, Ka1985}, Grmela \cite{Gr1984, Gr1993a, Gr1993b}, Morrison \cite{Mo1984, Mo1986}, and Turski and Kaufman \cite{TuKa1987}. The dissipative brackets introduced in BKMR were particularly motivated by the double bracket operations introduced in Vallis, Carnevale, and Young \cite{VaCaYo1989} for incompressible fluid flows. 

\subsection{The origins: selective decay hypothesis}
One of the motivations for Vallis et al.  \cite{VaCaYo1989} was the {\bfi selective decay hypothesis}, which arose in turbulence research \cite{MaMo1980} and is consistent with the preservation of coadjoint orbits. 
According to the selective decay hypothesis, energy in strongly nonequilibrium statistical systems tends to decay much faster than certain other ideally conserved properties. In particular, energy decays much faster in such systems than those ``kinematic'' or ``geometric'' properties that would have been preserved in the ideal nondissipative limit {\it independently of the choice of the Hamiltonian}. Examples are the Casimir functions for the Lie-Poisson formulations of various ideal fluid models \cite{HoMaRaWe1985}. 

The selective decay hypothesis was inspired by a famous example; namely, that enstrophy decays much more slowly than kinetic energy in 2D incompressible fluid turbulence \cite{Kr1967}.
\rem{ 
Kraichnan \cite{Kr1967} showed that the decay of kinetic energy under the preservation of enstrophy causes dramatic effects in 2D turbulence. Namely, it causes the well known ``inverse cascade'' of kinetic  energy to {\it larger} scales, rather than the usual ``forward cascade'' of energy to smaller scales that is observed in 3D turbulence! 
} 
In 2D ideal incompressible fluid flow the enstrophy (the $L^2$ norm of the vorticity) is preserved on coadjoint orbits. That is, enstrophy is a Casimir of the Lie-Poisson bracket in the Hamiltonian formulation of the 2D Euler fluid equations. Vallis et al. \cite{VaCaYo1989} chose a form of dissipation that was expressible as a double Lie-Poisson bracket. This choice of dissipation preserved the enstrophy and thereby enforced the selective decay hypothesis for all 2D incompressible fluid solutions, laminar as well as turbulent. 

Once its dramatic effects were  recognized in 2D turbulence,  selective decay was posited as a governing mechanism in other systems, particularly in statistical behavior of fluid systems with high variability. For example, the slow decay of magnetic helicity was popularly invoked as a possible means of obtaining magnetically confined plasmas \cite{Ta86}.  Likewise, in geophysical fluid flows, the slow decay of potential vorticity (PV) relative to kinetic energy strongly influences the dynamics of weather and climate patterns much as in the inverse cascade tendency in 2D turbulence. The use of selective decay ideas for PV thinking in meteorology and atmospheric science has become standard practice since the fundamental work in \cite{HoMcRo1985, Yo1987}.

A form of selective decay based on double-bracket dissipation is also the
basis of equation (\ref{Kandrup-dbrkt}), proposed in astrophysics by Kandrup \cite{Ka1991} for the purpose of modeling gravitational radiation of energy in stars. In this case, the double-bracket dissipation produced rapidly growing instabilities that again had dramatic effects on the solution. The form of double-bracket dissipation proposed in Kandrup \cite{Ka1991} is a strong motivation for the present work and it also played a central role in the study of instabilities in BKMR.

\section{Double bracket structure for kinetic equations}
\subsection{Background review}
This section starts by reviewing the ideas on geometric dissipative
terms for conservation laws formulated in chapter~\ref{GOP}.
Suppose that on physical grounds one knows that a certain quantity $\kappa$
is conserved, \emph{i.e.,} $d \kappa(\bx, t) / d t=0$ on $d \bx / d t=\bu$, where
$\bu$ is the velocity of particle constituting the continuum at the given point $\bx$. The nature of the conservation law depends on the geometry of
the conserved quantity $\kappa$ and the conservation law may be alternatively written in the Lie Derivative form $\partial_t \kappa +\pounds_{\bu} \kappa=0$. The physics
of the problem dictates the nature of the quantity $\kappa$. 
   In order to close the system, an expression for
$\bu$ must be established. In the treatment for geometric order parameters, one takes the inspiration from self-organization
phenomena and pattern formation of spherical particles. In this case one relates the velocity to density using the Darcy's
law that establishes a linear dependence of the local particle velocity $\bu$
and force acting on the particle $\nabla \delta E/\delta \rho$ as 
$\bu = \mu[\rho] \nabla \delta E/\delta \rho$. Here, $E[\rho]$ is the total
energy of the system in a given configuration and $ \delta E/\delta \rho$
is the potential at a given point. 

In mathematical
terms, the generalization to any geometric order parameter arises from the
action of a Lie algebra $\mathfrak{g}$ on some vector space $V$. A frequent
example of such an action is the Lie derivative, that is the basis for any order parameter equation on configuration space. Given a tensor $\kappa$ on
the configuration space $Q$ and an element $\xi\in\mathfrak{X}$ of the Lie
algebra $\mathfrak{X}$ of vector fields, the action of $\xi$
on $\kappa$ is defined
as 
\[
\xi\,\kappa:= \pounds_{\xi}\,\kappa
\,.
\]
The importance of the Lie derivative
in configuration space is given by the fact that any geometric quantity evolves
along the integral curves of some velocity vector field whose explicit expression
depends only on the physics of the problem. At this point the diamond operation
is defined as the dual operator to Lie derivative. More precisely, given a tensor $\zeta$ dual to $\kappa$, one defines $\langle \,\kappa \diamond \zeta, \,\xi \,\rangle:=\langle\, \kappa\,, -\pounds_\xi\, \zeta \,\rangle$, so that $\kappa \diamond \zeta\in\mathfrak{X}^*$. Once this operation has been defined, the general equation
for an order parameter $\kappa$ is written as $\partial_t \kappa+\pounds_\mathbf{u}\,\kappa$,
where $\bf u$ is called {\it Darcy's velocity} and is given by $\mathbf{u}=\left(\mu \diamond \delta E/\delta \rho\right)^\sharp$.


This chapter aims to model dissipation in Vlasov kinetic systems through a suitable generalization of Darcy's law. 
Indeed, it is reasonable to believe that the
basic ideas of Darcy's Law in configuration space can be transferred to a phase space treatment giving rise to the kinetic description of self-organizing
collisionless multiparticle systems. 
%
The main issue here  is to accurately
consider  not only the geometry of particle distribution,
but also the structure of the phase space itself. As is well known, the properties
of the phase space (momentum and position) are completely different from the configuration space (position only)
because of the symplectic relation between the momentum and position. This
structure of the phase space warrants a suitable modification of the diamond
operator.
The following sections
will construct kinetic equations for geometric order parameters that respect the symplectic nature of the phase space by considering the Lie algebra of
generating functions of canonical transformations (symplectomorphisms).

\subsection{A new multiscale dissipative kinetic equation}
The first step is to establish how a geometric quantity evolves on phase
space, so that the symplectic nature is preserved.  For this,  one regards the action
of the symplectic algebra as an action of the generating functions $h$ on
$\kappa$, rather then an action of vector fields. Here $\kappa$ is a tensor
field over the phase space. The action is formally expressed
as
\[
h\,\kappa=\pounds_{X_\textit{\!\scriptsize h}}\,\kappa
\,.
\]
The dual operation of the action
(here
denoted by $\star$) is then defined as 
\[
\langle \,\kappa\star \zeta,\, h \,\rangle = \langle\, \kappa,-\, \pounds_{{X_\textit{\!\scriptsize h}}}\,\zeta \,\rangle
\,.
\]
Here $X_h(q,p)$ is the Hamiltonian
vector field  generated by a Hamiltonian function $h(q,p)$ through
the definition $X_\textit{\footnotesize h} \contract \,\omega:=dh$. Notice that the star operation takes values in the space $\mathcal{F}^*$ of phase space densities  $\kappa\star \zeta\in\mathcal{F}^*$. In the particular case of interest here, $\kappa$ is the phase space density $\kappa=f\, dq\wedge dp$ and $\zeta=g$,  a function on phase space. In this case, the star operation is simply the canonical Poisson bracket, $\kappa\star g =\{f,g\}\,dq\wedge dp$.
\begin{framed}
It it possible to employ these considerations to 
find the \emph{purely dissipative} part of the kinetic equation for a particle
density on
phase space. To this purpose, one chooses variations of the form 
\[
\delta f=-\pounds_{{\,X_{h}(\phi)}} \,\,\mu(f)=-\,\{\mu(f),h(\phi)\}
\qquad\text{with}\qquad
h(\phi)=(f \star \phi)^\sharp =\{f \, , \, \phi\}
\]
where $(\,\cdot\,)^\sharp$ in $(f \star \phi)^\sharp$ transforms a phase space density to a scalar function. The operation $(\,\cdot\,)^\sharp$ will be understood in the pairing below. One then follows the steps:
\begin{align*}
\left\langle \phi,\frac{\partial f}{\partial t} \right\rangle
= 
\left\langle \frac{\delta E}{\delta f},\delta f \right\rangle
&=
 \left\langle \frac{\delta E}{\delta f}, -\,\bigg\{\mu(f),\,h(\phi)\bigg\} \right\rangle
 \\
&=
 \Bigg\langle \left\{\mu(f) , \frac{\delta E}{\delta f}\right\}, \bigg\{f ,
\phi \bigg\}
\Bigg\rangle
= -\,\Bigg\langle \phi , \left\{f,\left\{\mu(f),\frac{\delta E}{\delta f}\right\}\right\} \Bigg\rangle .
\end{align*}
Therefore, a functional $F(f)$ satisfies the following evolution equation in bracket notation \cite{HoPuTr2007-CR},
\begin{eqnarray}
\frac{d F}{dt}
=
\left\langle \frac{\partial f}{\partial t} \,,\, 
\frac{\delta F}{\delta f} \right\rangle
=
-\,\Bigg\langle \left\{\,\mu(f)\,, \frac{\delta E}{\delta f}\right\},\, \left\{\,f\,,\frac{\delta F}{\delta f}\right\} \Bigg\rangle
=:
\{\!\{\,E\,,\,F\,\}\!\} 
\,.
\label{diss-bracket}
\end{eqnarray}
\end{framed}
The mobility $\mu$ and dissipation energy functional $E$ appearing in (\ref{diss-bracket})
are modeling choices and must be selected based on the additional input from physics. The bracket (\ref{diss-bracket}) reduces to
the dissipative bracket in Bloch et al. \cite{BlKrMaRa1996} for the modeling choice of $\mu(f)=\alpha f$ with some $\alpha>0$. In this case the dissipation energy $E$ was taken to be the Vlasov Hamiltonian (see below), but in the present approach it also can be taken as a modeling choice. This extra freedom allows for more physical interpretation and treatment of the dissipation. 

\begin{proposition}\label{energy-dissip}
There exist choices of mobility $\mu[f]$ for which the bracket (\ref{diss-bracket}) dissipates energy $E$.
\end{proposition}
\noindent
{\bf Proof.} The dissipative bracket in equation (\ref{diss-bracket}) yields $\dot{E}=\{\!\{\,E\,,\,E\,\}\!\}$ which is negative when $\mu[f]$ is chosen appropriately.  For example, $\mu[f]=f M[f]$, where $M[f] \geq 0$ is a non-negative scalar functional of $f$. (That is, $M[f]$ is a number.) 

\begin{remark}
The dissipative bracket (\ref{diss-bracket}) satisfies the Leibnitz rule for the derivative of a product of functionals. In addition, it allows one to reformulate the equation (\ref{Vlasov-diss}) in terms of flow on a Riemannian manifold with a metric defined through the dissipation bracket, as discussed in more detail in chapter~\ref{GOP}.
\end{remark}\smallskip

\section{Properties and consequences of the model}

\subsection{GOP theory and double bracket dissipation: background}
The previous section has shown how the GOP theory can be applied to kinetic
equations if one considers the symplectic structure of Vlasov dynamics. The
result is a kinetic equation in double bracket form. At this point one may
wonder what is meant by ``double bracket'' in rigorous mathematical terms.
The preceding discussion has presented the double bracket as simply the composition
of two Poisson brackets and the reason is that this composition always yields
a quantity of definite sign, so that the energy functional can be taken to
decrease monotonically in time. However, the double bracket structure has
deep geometric roots, in particular for dissipative systems whose ideal limit
can be written in Lie-Poisson form. Chapter~\ref{GOP} presented the GOP bracket
and presented its application to several cases, but some of them turn out
to be more special then others. Indeed, the ideal limit of the GOP equations
for vorticity and one-form densities reduce to well known Lie-Poisson systems:
the Euler and EPDiff equations. On the other hand, for densities and differential
forms, the GOP equations cannot be reduced to non-dissipative cases without
obtaining trivial dynamics $\kappa_t=0$.

In order to better understand the geometric structure of a Lie-Poisson double
bracket equation, one starts with GOP theory and observes that the Lie algebra
action on the Lie algebra $\mathfrak{g}$ itself is always given by
\[
\xi\,\eta={\rm ad}_\xi\,\eta
\]
so that the correspondent diamond operation is given by
\[
\left\langle \mu\diamond\eta,\,\xi\right\rangle=
\left\langle \mu,-\,{\rm ad}_\xi\,\eta\right\rangle=
\left\langle \mu,\,{\rm ad}_\eta\,\xi\right\rangle=
\left\langle {\rm ad}_\eta^*\,\mu,\,\,\xi\right\rangle
\]
that is, the diamond operation is given by the infinitesimal coadjoint operator
$\rm ad^*$. Inserting this result in the GOP bracket yields for the geometric
order parameter $\kappa\in\mathfrak{g}^*$
\begin{align*}
\frac{d F}{dt}
=
\{\!\{\,E\,,\,F\,\}\!\}
:\!&=
-\,\Bigg\langle \left(\mu(\kappa)\diamond \frac{\delta E}{\delta \kappa}\right)^{\!\sharp},
\, \kappa\,\diamond\,\frac{\delta F}{\delta \kappa} \Bigg\rangle
\\
&=
-\,\Bigg\langle \left({\rm ad}^*_\text{\normalsize$\frac{\delta E}{\delta \kappa}$}\,\mu(\kappa)\right)^{\!\sharp} ,
\, {\rm ad}^*_\text{\normalsize$\frac{\delta F}{\delta \kappa}$}\,\kappa \Bigg\rangle
\\
&=
-\,\Bigg\langle \kappa,\left[\frac{\delta F}{\delta \kappa},\left({\rm ad}^*_\text{\normalsize$\frac{\delta E}{\delta \kappa}$}\,\mu(\kappa)\right)^{\!\sharp}\right] \Bigg\rangle
\,.
\end{align*}
This is the {\bfi Lie-Poisson double bracket structure} and one easily recognizes
the Lie-Poisson form, which becomes evident by taking a Hamiltonian functional 
\[
\mathscr{H}[\kappa]
\quad\text{such that}\quad
\frac{\delta \mathscr{H}}{\delta \kappa}=\left({\rm ad}^*_\text{\normalsize$\frac{\delta E}{\delta \kappa}$}\,\mu(\kappa)\right)^{\!\sharp}
\,.
\]
Of course, the Hamiltonian $\mathscr{H}$ is not the energy of the dissipative
system under consideration, which is instead given by the energy functional $E$. Rather it is the conserved energy of another system that has physically
nothing to do with the original one. (It should be noticed that the existence
of such Hamiltonian is not certain: it is possible that this does not even
exist.) Also, the new Lie-Poisson system is always left-invariant. In fact, the sign in the bracket does not depend on
whether the Lie algebra action is left or right, since the signs cancel because
of the product of two ad$^*$ terms in the bracket. The sign depends only
on the requirement that the original system dissipates energy: inverting
the sign yields a monotonic increase of the functional $E$.

\begin{remark}[Double bracket formulation of Toda lattice] It is worth noticing
that the Toda lattice also has a double bracket formulation on the special
linear algebra $\mathfrak{sl}(\mathbb{R},n)$ of real matrices \cite{BlBrRa92}. In this case,
the double bracket is explicitly given by
\[
\{\{E,F\}\}=-\,{\rm Tr}\left( \left[A,\frac{\delta E}{\delta A}\right]^{\!T}
\left[A,\frac{\delta F}{\delta A}\right]\right)
\]
where the operator $[\cdot,\cdot]$ denotes the commutator of two matrices.

This structure has been extended at the continuum level in \cite{BlFlRa95}. In this
case the double bracket is formally the same as (\ref{diss-bracket}), although
the canonical Poisson bracket is calculated on new coordinates $(z,\theta)\in[0,1]\times[0,2\pi[$
that represent the coordinates on the annulus. In this sense, this structure
becomes related to the area preserving diffeomorphisms of the annulus. 
\end{remark}

\rem{ 
\begin{remark}[Double bracket dissipation for quantum mechanics] It is well
known how the Heisenberg equation of quantum mechanics for a self-adjoint operator $A$ is a Lie-Poisson system \cite{Ma82} on the group \textnormal{U($\mathscr{H}$)} of unitary transformations
of a Hilbert space $\mathscr{H}$. Indeed, the Lie algebra $\mathfrak{u}(\mathscr{H})$
consists of skew adjoint operators $A^\dagger=-A$ under the commutator operation; via the inner product
$\langle B,\,A\rangle=\textnormal{Tr}(BA^\dagger)$ one identifies $\mathfrak{u}(\mathscr{H})$
with its dual $\mathfrak{u}^*(\mathscr{H})$. Upon writing the Hamiltonian
as $H(A)=\langle iH_{\rm op},\,A\rangle\in\mathbb{R}$, the Lie-Poisson structure
is given by 
\[
\{F,H\}(A)=-\left\langle A, \left[\frac{\delta F}{\delta A},iH_{\rm op}\right]\right\rangle
\]
which yields the Heisenberg equation
\[
\dot{A}=i\left[H_{\rm op},A\right]
\,.
\]
Now this structure suggests the possibility of inserting the double bracket
dissipation in the equation to obtain the coadjoint dissipative evolution
\[
\dot{A}=-\,i\left[A,H_{\rm op}\right]+i\left[A,\left[A,E_{\rm op}\right]\right]
\]
This form of dissipative Heisenberg equation could be used for the density
matrix in applications involving quantum computation theory.
\end{remark}
} 

\begin{remark}[Double bracket and complex maps]
The double bracket structure also appears in the study of complex maps.
Indeed, let $f:M\to \mathbb{C}$, with $M$ a symplectic manifold. Then the following equation
appears \cite{Donaldson1999} in minimizing the norm $E=||f||^2=\int |f|^2\, {\rm d}x\,{\rm d}y$
\[
\frac{\partial f}{\partial t}=-\frac12\left\{f,\left\{f,f^*\right\}\right\}
\]
where $\{\cdot,\,\cdot\}$ is now the canonical Poisson bracket in $(x,y)$.
\end{remark}

\subsection{A first consequence: conservation of entropy}
From the arguments in the previous section is now clear that any Double bracket
Lie-Poisson system can be written as
\[
\frac{\pa \kappa}{\pa t}+{\rm ad}^*_\text{\normalsize$\frac{\delta H}{\delta \kappa}$}\,\kappa={\rm ad}^*_\text{\!\footnotesize$\bigg(\!{\rm ad}^*_\text{\footnotesize$\frac{\delta E}{\delta \kappa}$}\,\mu(\kappa)\!\bigg)^{\!\sharp}\,$}\,\kappa
\]
or, in more compact form
\[
\frac{\pa \kappa}{\pa t}+{\rm ad}^*_{\,\Gamma[\kappa]}\,\kappa=0
\qquad\text{with}\qquad
\Gamma[\kappa]:=\frac{\delta H}{\delta \kappa}-\left({\rm ad}^*_\text{\normalsize$\frac{\delta E}{\delta \kappa}$}\,\mu(\kappa)\!\right)^{\!\sharp}\in\,\mathfrak{g}
\,.
\]
For the Vlasov equation, this becomes
\[
\frac{\pa f}{\pa t}+\Big\{f,\,\Gamma[f]\Big\}=0
\qquad\text{with}\qquad
\Gamma[f]:=\frac{\delta H}{\delta f}-\left\{\mu(f),\frac{\delta E}{\delta f}\right\}
\,.
\]
Now, from the Lie-Poisson theory of the Vlasov equation such an equation is known to possess the following property
\begin{framed}
\begin{proposition}[Casimir functionals]
\label{Casimir-conserv}
For an arbitrary smooth function $\Phi$ the functional $C_\Phi=\int\! \Phi(f)$ is preserved for any energy functional $E$. 
\end{proposition}
\noindent
\begin{proof}
It suffices to calculate the bracket 
\begin{align}
\frac{dC_\Phi}{dt}=\{\{C_\Phi,E\}\}
:\!&=-\,\Bigg\langle 
\left\{\,\mu(f)\,, \frac{\delta E}{\delta f}\right\},\, \left\{\,f\,,\frac{\delta C_\Phi}{\delta f}\right\} \Bigg\rangle
\\
&=
-\, \Bigg\langle \left\{\,\mu(f)\,, \frac{\delta E}{\delta f}\right\},\, \bigg\{\,f\,,\,\Phi'(f)\,\bigg\} \Bigg\rangle=0
. 
\end{align}
\end{proof}\\
An important corollary follows, concerning the entropy functional \cite{HoPuTr2007-CR}:
\begin{corollary}\label{entropy-conserv}
The entropy functional $S=\int\!f\,\log f$ is preserved by the dynamics in
equation (\ref{Vlasov-diss}) for any energy functional $E$.
\end{corollary}
\end{framed}
\rem{ 
\begin{remark}
The existence of Casimirs and the corresponding preservation of any entropy defined solely in terms of $f$ arises because the dissipative bracket (\ref{diss-bracket}) generates coadjoint motion, which is  {\bfi reversible}. This property is shared with Kandrup's bracket, which is recovered for $\mu(f)=\alpha f$ for constant $\alpha>0$.
\end{remark}
} 

This result can appear surprising because the major part of dissipative continuum
systems involve an increase of entropy, basically connected with the Brownian
motion of the particles that constitute the system. This Brownian motion
yields diffusion processes and continuous particle trajectories,
 which are far from being differentiable. This is the reason why
the single particle trajectory cannot be a solution of the continuum description.
Moreover, in the mathematical description, Brownian motion is related via the Langevin stochastic equation to a source of noise that represents a loss of information
in the system. Basically, one introduces a Langevin force in the single particle
trajectory that finally leads to the Laplace operator. The microscopic noise
is the reason
why the entropy functional is monotonically increasing in time and therefore
the information on particle paths is definitely lost.

However, the double bracket Vlasov equation is \emph{not} related with Brownian
motion and it is constructed in a completely deterministic fashion, so that
no diffusion process is involved in the kinetic description. To see this,
it suffices to write the double bracket Vlasov flow as coadjoint motion in
the form
\[
f(t)={\rm Ad}^*_{g^{-1}(t)}\,f(0)
\qquad\text{ with }\qquad
g(t)=e^{t\,\Gamma[f]}
\,.
\]
This relation well enlightens the geometric nature of the motion, which is
purely given by the group action of the symplectic group on its (dual) Lie
algebra. However, not only is this of mathematical importance, but it also
has important physical implications. In fact, this form of dissipation yields
a completely {\bfi reversible dynamics} and it is clear how inverting the
group element at each time gives the reversed time evolution. In this sense,
the reversibility of dynamics yields the conservation of the entropy functional.

Importantly, this fact is \emph{not} related with the single particle paths, which may or may not be a solution of the equation. For example, the preservation
of entropy is shared by Kandrup's dynamics ($\mu(f)=\alpha\,f$). However, the evolution under Kandrup's equation does not allow single particle solutions. The absence of the single particle
solution might appear as a common element between Kandrup's equation and the
usual diffusive Fokker-Plank approach. However it is not possible to establish
such a relation, since diffusive processes destroy
the geometric nature of the dynamical variable and the microscopic physics
underlying the two approaches is very different.
Also, the existence of single particle paths
as a solution of the equation may always be allowed in the double bracket equation by introducing the mobility
on phase space, which is nothing but a smoothed version of the Vlasov distribution.
This smoothing process yields the singular $\delta$-like solutions representing
the single particle trajectories, as it is shown in the next section.

\begin{proposition}[cf. \cite{HoPuTr2007-CR}]
Variations of the form $\delta f=-\pounds_{{\,X_{h}(\phi)}} \,\,f=-\,[f,h(\phi)]$ with $h(\phi)=\mu(f) \star \phi =[\mu(f) \, , \, \phi]$ in (\ref{diss-bracket}) yield the dissipative double bracket
\begin{eqnarray*}
\frac{d F}{dt}
=
-\,\Bigg\langle \left\{\,f\,, \frac{\delta E}{\delta f}\right\},\, \left\{\,\mu(f)\,,\frac{\delta F}{\delta f}\right\} \Bigg\rangle
=:
\{\!\{\,E\,,\,F\,\}\!\} 
\,.
\end{eqnarray*}
with $\mu(f)\leftrightarrow f$ switched in the corresponding entries with
respect to (\ref{diss-bracket}). This bracket yelds entropy dynamics of the form
\begin{align*}
\frac{dS}{dt}=\{\{S,E\}\}
&=
-\, \Bigg\langle \frac{\mu(f)}{f}
\,,
\Bigg\{\,f\,, \left\{\,f\,, \frac{\delta E}{\delta f}\right\} \Bigg\} \Bigg\rangle
\neq0
. 
\end{align*}
\end{proposition}
\begin{proof}
One repeats the calculation for deriving (\ref{diss-bracket}) and insert
the new variation $\delta f=-\pounds_{{\,X_{h}(\phi)}} \,\,f=-\,[f,h(\phi)]$
to obtain
\begin{align*}
\left\langle \phi,\frac{\partial f}{\partial t} \right\rangle
= 
\left\langle \frac{\delta E}{\delta f},\delta f \right\rangle
&=
 \left\langle \frac{\delta E}{\delta f}, -\,\bigg\{f,\,h(\phi)\bigg\} \right\rangle
 \\
&=
 \Bigg\langle \left\{f , \frac{\delta E}{\delta f}\right\}, \bigg\{\mu(f) ,
\phi \bigg\}
\Bigg\rangle
= -\,\Bigg\langle \phi , \left\{\mu(f),\left\{f,\frac{\delta E}{\delta f}\right\}\right\} \Bigg\rangle .
\end{align*}
The evolution of the entropy functional is obtained by direct substitution
of its expression $S=\int f\log f$ as follows
\begin{align*}
\frac{dS}{dt}=\{\{S,E\}\}
&=
-\, \Bigg\langle \left\{\,f\,, \frac{\delta E}{\delta f}\right\},\, \bigg\{\,\mu(f)\,,\,\log f\,\bigg\} \Bigg\rangle
\\
&=
-\, \Bigg\langle \frac{\mu(f)}{f}
\,,
\Bigg\{\,f\,, \left\{\,f\,, \frac{\delta E}{\delta f}\right\} \Bigg\} \Bigg\rangle
. 
\end{align*}
\end{proof}
\begin{remark}
For entropy increase, this alternative variational approach would require $\mu(f)$ and $E(f)$ to satisfy an additional  condition (e.g., $\mu(f)/f$ and $\delta E/\delta f$ functionally related). However, the Vlasov dissipation induced in this case would not allow the reversible single-particle solutions, consistently with the loss of information associated with entropy increase. 
\end{remark}

\subsection{A result on the single-particle solution}

The discussion from the previous sections produces an interesting
opportunity for the addition of dissipation to kinetic equations. This opportunity arises from noticing that the dissipative bracket derived here could just as well be used with any type of evolution
operator. 
In particular, one may consider introducing a double bracket to modify Hamiltonian dynamics as in the approach by Kaufman \cite{Ka1984} and Morrison \cite{Mo1984}.
 In particular, the dissipated energy may
naturally be associated with the Hamiltonian arising from the corresponding Lie-Poisson theory for the evolution of a particle distribution function
$f$. Therefore, it is possible to write the total dynamics generated by any functional 
$F(f)$ as $\dot{F}=\left\{F,H\right\}+\left\{\left\{F,E\right\}\right\}$ where  
$\left\{\cdot \, , \, \cdot \right\}$ represents the Hamiltonian part of the dynamics. 
This gives the {\it dissipative Vlasov equation} of the form (\ref{Vlasov-diss}) with $E=H$, where $H(f)$ is the Vlasov Hamiltonian. To illustrate these ideas it is worthwhile to compute the singular (measure-valued) solution of equation (\ref{Vlasov-diss}), which represents the reversible motion of a single particle
\cite{HoPuTr2007-CR}.

\begin{theorem}\label{singparticles}
Taking $\mu(f)$ to be an arbitrary function of the smoothed distribution
$\bar{f}=K*f$ for some kernel $K$  allows for single particle solutions $f=\sum_{i=1}^Nw_i
\delta(q - {Q}_i(t) ) \delta(p- {P}_i(t))$. 
The single particle dynamics is governed by canonical equations with Hamiltonian given by
\[
\mathcal{H}=\left(\frac{\delta H}{\delta f}-
\left\{\mu\left(f\right),\frac{\delta H}{\delta f}\right\}\right)_{(q,p)=(Q_i(t),P_i(t))}
\]
\end{theorem}
\noindent
{\bf Proof.}
One writes the equation of motion (\ref{Vlasov-diss}) in the following compact form
\[
\frac{\partial f}{\partial t}=-\,\left\{\,f,\,\mathcal{H}\,\right\}
\,,\qquad\text{ with }\quad
\mathcal{H}
:=
\left(\frac{\delta H}{\delta f}-
\left\{\mu\left(f\right),\frac{\delta H}{\delta f}\right\}\right)
\]
and substitute the single particle solution ansatz
$
f(q,p,t)\,=\,\sum_i w_i\,\delta(q-Q_i(t))\,\delta(p-P_i(t))
$.
Now take the pairing with a phase space function $\phi$ and write
$
\langle\, \phi,\,\dot{f}\,\rangle=-\left\langle\, 
\left\{\,\phi,\,\mathcal{H}\,\right\},\,
f\,\right\rangle
$.
Evaluating on the delta functions proves the theorem.
\begin{remark}
The quantity $-
\{\mu\left(f\right),{\delta H}/{\delta f}\}$
plays the role of a Hamiltonian for the advective dissipation process by coadjoint motion. This Hamiltonian is constructed from the momentum map $J$
defined by the $\star$ operation (Poisson bracket). That is, 
\[
J_h(f,g) 
= \langle g, -\pounds_{X_h}f\rangle 
= \langle g, \{h,f\}\rangle 
= \langle h, \{f,g\}\rangle 
= \langle h, f \star g\rangle
\,.
\]
\end{remark}
\rem{
\begin{equation*}
\dot{w}_i=0
\,,\quad
\dot{Q}_i=\frac{\partial\mathcal{H}}{\partial P_i}
\,,\quad
\dot{P}_i=
-\,\frac{\partial\mathcal{H}}{\partial Q_i}
\end{equation*}
where one evaluates $\mathcal{H}$ at the phase point $(q,p)=(Q_i(t),P_i(t))$.
}

\section{Geometric dissipation for kinetic moments}
This section shows how Eq. (\ref{Vlasov-diss}) leads very naturally to a nonlocal form of Darcy's law. In order to show how this equation is recovered, one first reviews the Kupershmidt-Manin structure for kinetic moments.
The discussion proceeds by considering a one-dimensional configuration space; an extension to higher dimensions would also be possible by considering the treatment in chapter~\ref{momLPdyn}.

\subsection{Review of the moment bracket}
Chapter~\ref{momLPdyn} has shown how the equations for the moments of the Vlasov equation are a Lie-Poisson system \cite{Gi1981,GiHoTr05,GiHoTr2007}. The $n$-th moment is defined as
\[
A_n(q):=\int p^n\, f(q,p)\, dp\,.
\]
and the dynamics of these quantities is regulated by the {Kupershmidt-Manin structure}
\[
\{F,G\}=
\left\langle A_{m+n-1},\,
\left[\frac{\delta F}{\delta A_n},\frac{\delta G}{\delta A_m}\right]
\right\rangle
\,,
\]
where summation over repeated indices is omitted and the Lie bracket $\left[\cdot,\cdot\right]$ is defined as
\[
\left[\alpha_m,\,\beta_n\right]\,=\,
n\,\beta_n(q)\,\alpha_m'(q)
-
m\,\alpha_m(q)\,\beta_n^{\,\prime}(q)
\,=:\,
\textrm{\large ad}_{\alpha_m}\, \beta_n
\]
The moment equations are
\[
\dot{A}_n=-\,\textrm{\large ad}^*_{\beta_n}\,A_{m+n-1}=
-\left(  n+m\right) \, A_{n+m-1}\,\frac{\partial
\beta_{n}}{\partial q}
-
n\,\beta_{n}\,\frac{\partial
A_{n+m-1}}{\partial q}
\,,
\]
where $\beta_n=\delta H/\delta A_n$ and the ${\sf ad}^*$ operator is defined by $\langle\, {\sf ad}^*_{\beta_n} \,A_k,\,\alpha_{k-n+1}\,\rangle:=
\langle\, A_k,\,{\sf ad}_{\beta_n}\,\alpha_{k-n+1}\,\rangle$.

\subsection{A multiscale dissipative moment hierarchy}
At this point one can consider the following Lie algebra action on Vlasov densities \cite{HoPuTr2007-CR}
\[
\beta_n\,f:=\text{\it\large\pounds}_{X_{p^n\beta_n}}f=\big\{\,f,\,p^n\beta_n\big\}
\qquad\text{(no sum)}
\]
which is obviously given by the action of the Hamiltonian function $
h(q,p)=p^n\beta_n(q)$. Now, the dual action is given by
\begin{align}
\Big\langle f\,\text{\large$\star$}_{n}\, g,\,\beta_n\Big\rangle:=
\Big\langle f,\, \beta_n\, g\Big\rangle &=
\Big\langle f\!\star g\,,\, p^n \beta_n(q) \Big\rangle =
 \left\langle \int\{f, g\}\,p^n\,dp\,,\,\beta_n \right\rangle
\label{stardef}
\end{align}
and the dissipative bracket for the moments (\ref{diss-bracket}) is written in this notation as
\begin{framed}
\begin{align*}
\{\!\{\,E\,,\,F\,\}\!\}
&= 
-\,\Bigg\langle \int\! p^n\left\{\,\mu[f]\,, \frac{\delta E}{\delta f}\right\}dp,\,\int\!
p^n \left\{f\,,\frac{\delta F}{\delta f}\right\}dp \Bigg\rangle
\\&=
-\left\langle\textrm{\large ad}^*_{\beta_k}\, \widetilde{\mu}_{\,k+n-1},\,
\left(\textrm{\large ad}^*_{\alpha_m}A_{m+n-1}\right)^{\sharp}\,\right\rangle
\end{align*}
\end{framed}
where one substitutes
\[
\frac{\delta E}{\delta f}=p^k\beta_k\,,
\qquad
\frac{\delta F}{\delta f}=p^m\alpha_m\,,
\qquad
\widetilde{\mu}_s(q):=\!\int\! p^s \mu[f]\,dp
\,.
\]

Thus the purely dissipative moment equations are \cite{HoPuTr2007-CR}
\begin{framed}
\begin{equation}\label{mom-diss-dyn}
\frac{\pa {A}_n}{\pa t}=\textrm{\large ad}^*_{\gamma_m}A_{m+n-1}
\qquad\text{with}\qquad
\gamma_m:=\left(\textrm{\large ad}^*_\text{\normalsize$\frac{\delta E}{\delta A_k\!}$}\, \widetilde{\mu}_{\,k+m-1}\right)^\sharp
\end{equation}
\end{framed}
which arise from the subsequent application of two moment Lie-Poisson
brackets, as it can be seen by the nested ad$^*$ operator. Thus the dissipative
moment equations reflect the double-bracket construction of the flow.
\begin{remark}
The explicit expression
of $\widetilde{\mu}_n$ may involve all the moments. In order to see this,
it is necessary to consider the smoothed distribution $\mu[f]$, whose $\widetilde{\mu}_n$
is the $n$-th moment. One can write its functional derivative as
\[
\frac{\delta \mu}{\delta f}=\sum_s p^s\nu_s(q)
\]
so that
\[
\mu[f]=\iint H(q,p,q',p')\,f(q',p')\,{\rm d}q'\,{\rm d}p'
\quad\Rightarrow\quad
H(q,p,q',p')=\sum_s p\text{\small$^\prime$}^s\widetilde{H}_s(q,p,q')
\]
and thus
\[
\mu[f]=\sum_s\int \widetilde{H}_s(q,p,q')\,\left(\int p\text{\small$^\prime$}^s
f(q',p'){\rm d}p'\right){\rm d}q'
=
\sum_s\int \widetilde{H}_s(q,p,q')\,A_s(q')\,{\rm d}q'
\]
At this point the moment $\widetilde{\mu}_i$ is written as
\[
\int\!p^i\mu(f)\,{\rm d}p=\sum_s\int
\left(\int\!\widetilde{H}_s(q,p,q')\,p^i\,{\rm d}p\right)
A_s(q')\,{\rm d}q'=\sum_s G_{si}\,*\,A_s:=\widetilde{\mu}_i
\]
where one defines
\[
G_{si}(q,q'):=\int\!\widetilde{H}_s(q,p,q')\,p^i\,{\rm d}p
\]
Consequently, the smoothed moments $\widetilde{\mu}_n$ can depend on all
the moments, although one can choose $\widetilde{\mu}_n=\widetilde{\mu}_n[A_n]$
for simplicity.
\end{remark}

\section{Properties of the dissipative moment hierarchy}

\subsection{A first result: recovering Darcy's law}
If one now writes the equation for $\rho:=A_0$ and
consider only $\gamma_0$ and $\gamma_1$, then one recovers the following form of Darcy's law \cite{HoPu2005,HoPu2006,HoPuTr2007-CR}
\begin{align}
\dot\rho=\,\textrm{\large ad}^*_{\gamma_1}\rho=\,
\frac{\partial}{\partial q}\!\left(\rho\,\mu[\rho]\,\frac{\partial}{\partial q}\frac{\delta E}{\delta
\rho}\right)
\label{Darcy-rho}
\end{align}
where one chooses $E=E[\rho]$ and $\widetilde{\mu}_0=\mu[\rho]$, so that $\,\gamma_1=\widetilde{\mu}_0 \,\partial_q\beta_0$.

\paragraph{Special cases.}
Two interesting cases may be considered at this point. In the first case one makes Kandrup's choice in (\ref{Kandrup-dbrkt}) for the mobility at the kinetic level $\mu[f]=f$, so that Darcy's law is written as
\[
\dot\rho=\frac{\partial}{\partial q}\!\left(\rho^2\,\frac{\partial}{\partial q}\frac{\delta E}{\delta
\rho}\right)\,.
\]
Kandrup's case applies to the dissipatively induced instability of galactic dynamics \cite{Ka1991}. The previous equation is Darcy's law description of this type of instability. 
In the second case, one considers the mobility $\mu[\rho]$ as a functional of $\rho$ (a number), leading to the equation
\[
\dot\rho=\mu[\rho]\,\frac{\partial}{\partial q}\!\left(\rho\,\frac{\partial}{\partial q}\frac{\delta E}{\delta
\rho}\right)\,,
\]
which leads to the classic energy dissipation equation, 
\[
\frac{dE}{dt}=-\int\!\rho\,\mu[\rho]\,\left|\frac{\partial}{\partial q}\frac{\delta E}{\delta
\rho}\right|^2 {\rm d}q.
\]

\subsection{A new dissipative fluid model and its properties}

The dissipative bracket on the moments provides an answer to the question
formulated in section~\ref{DarcyFluid}. In particular, it formulates the
dissipative equation of a fluid undergoing Darcy dissipative dynamics. As
already mentioned in section~\ref{DarcyFluid}, one would expect that these
fluid equations require the substitution $u \to u+v$, where $v$ is Darcy's
velocity. However, the whole discussion in chapter~\ref{GOP} considers an
energy functional $E$ depending only on the density $\rho$, so that $v=\mu[\rho]\,\partial_q
\delta E/\delta \rho$. In general, one
can consider an energy functional also depending on the fluid momentum $m$,
for example in the case $E=H$, where $H$ is the fluid Hamiltonian. This section
formulates this model, by taking into account this dependence on $m$.

One starts with the moment equations
\[
\dot{A}_n=\textrm{\large ad}^*_{\gamma_m}A_{m+n-1}
\qquad\text{with}\qquad
\gamma_m:=\left(\textrm{\large ad}^*_{\beta_k}\, \widetilde{\mu}_{\,k+m-1}\right)^\sharp
\]
and expand
\begin{align*}
\gamma_1=
\left(\textrm{\large ad}^*_{\beta_k}\, \widetilde{\mu}_{\,k}\right)^\sharp
&=
\left(\textrm{\large ad}^*_{\beta_0}\, \widetilde{\mu}_{\,0}\right)^\sharp
+
\left(\textrm{\large ad}^*_{\beta_1}\, \widetilde{\mu}_{\,1}\right)^\sharp
\\
&=\mu_0\frac{\pa \beta_0}{\pa q}+2\,\mu_1\frac{\pa \beta_1}{\pa q}+\beta_1\frac{\pa \mu_1}{\pa q}
\\
\gamma_0=\left({\rm ad}^*_{\beta_1}\,\mu_0\right)^\sharp
&=
\frac{\pa}{\pa q}\left(\mu_0\,\beta_1\right)
\,.
\end{align*}
By changing notation
\[
\beta_1=\frac{\delta E}{\delta m}\,,
\qquad
\beta_0=\frac{\delta E}{\delta \rho}\,,
\qquad
\gamma_1=-\,v\,,
\qquad
\mu_0=\mu_\rho[\rho]\,,
\qquad
\mu_1=\mu_m[m]
\]
one writes the expression of Darcy's velocity
\begin{align*}
v&=-\,
\mu_\rho\,\frac{\pa}{\pa q}\frac{\delta E}{\delta \rho}
-\,
2\,\mu_m\,\frac{\pa}{\pa q}\frac{\delta E}{\delta m}
-\,
\frac{\delta E}{\delta m}\frac{\pa \mu_m}{\pa q}
\\
&=
\left(\mu_\rho\diamond\frac{\delta E}{\delta \rho}\right)^\sharp
+
\left(\mu_m\diamond\frac{\delta E}{\delta m}\right)^\sharp
\end{align*}
Also one finds
\begin{align}
\gamma_0=\left(\pounds_\text{\!\normalsize$\frac{\delta E}{\delta m}$}\,\mu_\rho \right)^\sharp
\end{align}
and the fluid equations are
\begin{align*}
\frac{\pa \rho}{\pa t}+\pounds_v\,\rho&=0
\\
\frac{\pa m}{\pa t}+\pounds_v\,m&=-\,\rho\,\diamond\left(\pounds_\text{\!\normalsize$\frac{\delta E}{\delta m}$}\,\mu_\rho \right)^\sharp
\end{align*}
Now, if one wants to incorporate the Hamiltonian part with velocity $u=\delta
H/\delta m$, then this yields
\begin{framed}
\begin{align}\nonumber
\frac{\pa \rho}{\pa t}+\pounds_{u+v}\,\rho&=0
\\
\frac{\pa m}{\pa t}+\pounds_{u+v}\,m&=
\rho\,\diamond\left(\frac{\delta H}{\delta \rho}
-
\left(\pounds_\text{\!\normalsize$\frac{\delta E}{\delta m}$}\,\mu_\rho \right)^\sharp\right)
\label{darcyfluid}
\end{align}
\end{framed}
\noindent
Thus, these equations show that the total fluid velocity is indeed $u+v$. However
now Darcy's velocity $v$ also depends on the fluid momentum $m$ and its smoothed
version $\mu_m$. Moreover the diamond term on the right hand side is also
modified by a dissipative term, so that the contribution of pressure (right
hand side in the second equation) is itself
``dissipated'', consistently with the double bracket structure.

This is a particular example of how the kinetic moments are powerful in deriving
macroscopic continuum models from microscopic kinetic treatments. This section has
derived the dissipative moment equations by simply implementing the moment
double bracket without worrying about the semidirect product structure of
the equations with no dissipation. And still, the semidirect product structure
evidently appears in the dissipative moment equations, that have the same
form as the non-dissipative case.

The simplest case of Darcy fluid is the one dimensional case. For simplicity,
the Hamiltonian part may be omitted and one can consider the purely dissipative fluid equations, which are written as
\begin{align*}
\dot\rho&+
\frac{\pa}{\pa q}\!
\left(
\rho\,\mu_\rho\,\frac{\pa}{\pa q}\frac{\delta E}{\delta \rho}
+
\rho\,\mu_m\,\frac{\pa}{\pa q}\frac{\delta E}{\delta m}
+
2\,\rho\,\mu_m^{\,\prime}\,\frac{\delta E}{\delta m}
\right)=0
\\
\dot{m}
&+
m\,\frac{\pa}{\pa q}\!
\left(
\mu_\rho\,\frac{\pa}{\pa q}\frac{\delta E}{\delta \rho}
+
\mu_m\,\frac{\pa}{\pa q}\frac{\delta E}{\delta m}
+
2\,\mu_m^{\,\prime}\,\frac{\delta E}{\delta m}\right)
+2\,m'
\left(
\mu_\rho\,\frac{\pa}{\pa q}\frac{\delta E}{\delta \rho}
+
\mu_m\,\frac{\pa}{\pa q}\frac{\delta E}{\delta m}
+
2\,\mu_m^{\,\prime}\,\frac{\delta E}{\delta m}\right)
\\
&\hspace{11cm}=
\rho\,\frac{\pa^2}{\pa q^2}\!\left(\mu_\rho\,\frac{\delta E}{\delta m}\right)
\end{align*}
where the prime stands for derivation. It is interesting to notice that the
right hand side of the second equation does not prevent the existence of
singular solutions. Indeed, the substitution of the single particle solution ansatz
\[
(\rho,m)(q,t)=(w,P)(t)\,\delta(q-Q(t))
\]
into the equation does not generate second-order derivatives of delta functions,
provided the vector field $\delta E/\delta \rho$ is sufficiently smooth (it is useful to recall that the energy functional $E$ does not need to coincide
with the Hamiltonian of the non-dissipative case). Thus the existence of
singular solutions is allowed also for the Darcy fluid and one can address
the question whether these solutions appear spontaneously, for example, in
the case of a purely quadratic energy functional of the form
\[
E[\rho,m]=\frac12\iint\rho(q)\,G_\rho(q-q')\rho(q')\,{\rm d}q\,{\rm d}q'
+
\frac12\iint m(q)\,G_m(q-q')m(q')\,{\rm d}q\,{\rm d}q'
\]
which coincides with the truncation of the quadratic moment Hamiltonian (\ref{Ham-metric})
in Chapter~\ref{EPSymp} to only $A_0$ and $A_1$. In this sense, these equations
represent a dissipative version of the EPSymp fluid equations (\ref{FluidClosureSystem}),
which preserve their geometric nature and allow for singular solutions. Future
research will study the behavior of singularities under
competition of length-scales involved in the smoothed quantities $\mu_\rho$
and $\mu_m$, in the same spirit of \cite{HoOnTr07}. For example, since this
system contains Darcy's law for aggregation and self-assembly (the first two
terms in the equation for $\rho$) and it has the same geometric structure, one may seek conditions for the same aggregation phenomena in the dynamics of singular solutions.

\section{Further generalizations}

\subsection{A double bracket structure for the $b$-equation.} 

The dissipative Kupershmidt-Manin bracket provides a hint to the possibility
of inserting the double bracket moment structure into the $b$-equation (developed in \cite{HoSt03} and treated
in Chapter~\ref{momLPdyn}). In general, the $b$-equation is a characteristic
equation for a covariant symmetric tensor (density) along a smooth nonlocal vector field. The analogy between this equation and the moment hierarchy arises because of the important property that ${\rm ad}^*_{\beta_n}=\pounds_{\beta_n}$
iff $n=1$. Thus the Kupersmhmidt-Manin operator enters naturally in this
problem, since it establishes a Lie algebra structure in the space of symmetric
contravariant tensors (dual to the symmetric covariant tensor-densities). One could be tempted to call ``dissipative'' the double bracket term in the equation; however, since the $b$-equation is {\it not} Hamiltonian (at least under the Kupershmidt-Manin structure), one should be careful when talking about dissipation in this context: it is not clear a priori what should be dissipated. 
Rather, the moment double bracket is a way of preserving the geometric nature of the equations and in the case of the $b$-equation it can be interesting
to see how the action of diffeomorphisms behaves under this construction.

In order to formulate a double bracket version of the $b$-equation, one may proceed by writing the characteristic equation for the moment $A_n$ and separating
the simple advection term from the double bracket term. 
\[
\frac{\pa A_n}{\pa t}+\pounds_{\beta_1}\,A_n=\pounds_{\gamma_1}\,A_n
\]
Upon recalling that ${\rm ad}^*_{\beta_n}\,A_k$ is a symmetric covariant $(k-n+1)$-tensor-density,
one notes that ${\rm ad}^*_{\beta_n}\,A_n$ is a one form-density, so that
$\left({\rm ad}^*_{\beta_k}\,A_k\right)^\sharp$ is a vector field for any
integer $k$. In particular, upon substituting $A_k\to\mu_k[A_k]$ and summing over $k$ one obtains the dissipative vector field $\gamma_1=\sum_k\left({\rm ad}^*_{\beta_k}\,\mu_k\right)^\sharp$ from the last section, which is needed in the right hand side of the equation above in order to construct the double bracket term. The resulting equation is
\begin{equation}\label{mom-GOP1}
\frac{\pa A_n}{\pa t}+\pounds_{\beta_1}\,A_n=\pounds_{\left(\sum_k{\rm ad}^*_{\beta_k}\,\mu_k\right)^\sharp}\,A_n
\end{equation}
Since one wants $\beta_1$ to regulate the moment dynamics, it is possible to fix $k=1$, so that one writes
\[
\frac{\pa A_n}{\pa t}+\pounds_{\beta_1}\,A_n=
\pounds_{\left({\rm ad}^*_{\beta_1}\,\mu_1\right)^\sharp}\,A_n
\]
\rem{ 
At this point, one recalls that ${\rm ad}^*_{\beta_n}\,\mu_n=n\,\beta_n\,\mu_n'+(n+1)\,\mu_n\,\beta_n'$
and let $\beta_n$ be given by
\[
\beta_n=\int G(q-q')\,A_n(q')\,{\rm d}q'
\]
In higher dimensions one should take into account of the dimensionality of $\beta_n$. For example, in order to make $\beta_n$ a contravariant $n$-tensor, the quantity $G$ should be a contravariant $1+n$-tensor. However this problem
is not present in one dimension.

It is interesting to notice that
\[
\left\langle{\rm ad}^*_{\beta_n}\,A_n,\,\alpha_1\right\rangle
=
-\left\langle A_n,\,{\rm ad}_{\alpha_1}\,{\beta_n}\right\rangle
=
-\left\langle A_n,\,\pounds_{\alpha_1}\,{\beta_n}\right\rangle
:=
\left\langle A_n\diamond \beta_n,\,{\alpha_1}\right\rangle
\]
which allows to write the equation in the GOP form as
\[
\frac{\pa A_n}{\pa t}+\pounds_{\beta_1}\,A_n=\pounds_\text{\small$\left(\mu_n\diamond{\beta_n}\right)^\sharp$}\,A_n
\]
} 
If $k=1$, one recalls that the ad$^*_{\beta_1}$ coincides with Lie derivative and the equation reduces to
\[
\frac{\pa A_n}{\pa t}+\pounds_{\beta_1}\,A_n=\pounds_\text{\!\small$\Big(\pounds_{\beta_1}\,\mu_1\Big)^\sharp$}\,A_n
\]
At this point one performs the choice $\beta_1=G*A_n$ (as in the $b$-equation) and lets the smoothed moment $\mu_1=\mu[A_n]$ depend only on $A_n$ (instead of $A_1$), since one recalls that in the general case $\mu_n$ can depend on any sequence of moments. 
\begin{framed}
\noindent
The result is the
equation
\begin{equation}\label{DB-beq}
\frac{\pa A_n}{\pa t}
+
\pounds_{\beta_1}\,A_n=
\pounds_\text{\!\small$\Big(\pounds_{\beta_1}\,\mu[A_n]\Big)^\sharp$}\,A_n
\quad\text{ with }\quad
\beta_1=G*A_n
\end{equation}
where $\mu[A_n]$ is some filtered version of the $n$-th moment $\mu[A_n]=H*A_n$.
\end{framed}
\noindent
In the particular case $n=1$, one recovers the dissipative EPDiff equation introduced in Chapter~\ref{GOP}, exactly as it happens  in the ordinary case with simple advection.
In this sense, the double bracket bracket can be understood as ``double advection'',
since it involves a sequential application of two Lie derivatives.

\subsection{A GOP equation for the moments}
The development of the double bracket form of the $b$-equation provides an
interesting hint to formulate a GOP form of the moment equations. In order
to see this, one can consider again the equation (\ref{mom-GOP1}) and discard
simple advection to obtain
\[
\frac{\pa A_n}{\pa t}
=
\pounds_{\left(\sum_k{\rm ad}^*_{\beta_k}\,\mu_k\right)^\sharp}\,A_n
\]
Since one may want to avoid the presence of moments different from $A_n$,
it is possible to fix $k=n$. Also, one recalls that 
${\rm ad}^*_{\beta_n}\,\mu_n=n\,\beta_n\,\mu_n^{\,\prime}+(n+1)\,\mu_n\,\beta_n^{\,\prime}$.
\begin{framed}
\noindent
It is interesting to notice that
\[
\left\langle{\rm ad}^*_{\beta_n}\,A_n,\,\alpha_1\right\rangle
=
-\left\langle A_n,\,{\rm ad}_{\alpha_1}\,{\beta_n}\right\rangle
=
-\left\langle A_n,\,\pounds_{\alpha_1}\,{\beta_n}\right\rangle
:=
\left\langle A_n\diamond \beta_n,\,{\alpha_1}\right\rangle
\]
which allows to write the equation in the GOP form as
\begin{equation}\label{GOPmom}
\frac{\pa A_n}{\pa t}
=
\textit{\large\pounds}_\text{\small$\!\left(\mu_n\diamond\frac{\delta E}{\delta A_n}\right)^\sharp$}\,A_n
\end{equation}
\end{framed}
\noindent
In this case the quantity $\mu_n$ is chosen to depend only on $A_n$ as $\mu_n[A_n]=H*A_n$
and the energy functional $E=E[A_n]$ is left as a modeling choice. Two particular
cases are $n=0$ and $n=1$. In the first case, the GOP equation reduces to
Darcy's law, while in the second case, the equation reduces to the \emph{purely}
dissipative EPDiff equation. Thus the fact that this equation reduces to
such interesting cases promises well for future research.

\begin{framed}
\begin{remark}
The property ${\rm ad}^*_{\beta_n}\,A_n=A_n\diamond\beta_n$ leads to another
way of writing the equation for the first-order moment $A_1$ in the Hamiltonian
hierarchy $\dot{A}_n=-\sum_m{\rm ad}^*_{\beta_m}\,A_{m+n-1}$. Indeed, it is evident
how the equation for $A_1$ may be written in the form \makebox{$\dot{A}_1=-\sum_m A_m\diamond\beta_m$}.
This particular form of the diamond operator between symmetric tensors was known to Schouten, since it arises naturally from his symmetric bracket \cite{BlAs79,DuMi95}
(also cf. chapter~\ref{momLPdyn}), and it is called ``Lagrangian Schouten concomitant'' \cite{Ki82}.
\end{remark}
\end{framed}

\smallskip
\section{Discussion and open questions}
\rem{
In this paper, we have developed a geometric approach to the derivation of
the kinetic equations and considered as an example a dissipative Vlasov
equation.
We have achieved this result through geometric generalization of the Darcy's law  for the symplectic $(\bq,\bp)$ phase space. Continuing our geometric
approach, it is possible to derive a dissipative Vlasov equation for the
particles with internal degrees of freedom, for example, energy $E$  dependent
on the mutual position and orientation of particles in the three dimensional
space.  
}
This chapter has developed a new symplectic variational approach for modeling dissipation in kinetic equations that yielded a {\bfi double bracket structure in phase space}. This approach has been focused on the Vlasov example and
it yielded the existence of single-particle solutions. In general, it is possible to extend the present theory to the evolution of an arbitrary geometric quantity defined on any  smooth manifold \cite{HoPuTr2007}. For example, the restriction of the geometric formalism for symplectic motion considered here to cotangent lifts of diffeomorphisms recovers the corresponding results for fluid momentum. 

The last section has provided a consistent {\bfi derivation of Darcy's law} from first principles in kinetic theory, obtained by inserting dissipative terms into the Vlasov equation which respect the geometric nature of the system. This form of the Darcy's law has been studied and analyzed
in \cite{HoPu2005,HoPu2006}, where it has been shown to possess emergent singular solutions ({\it clumpons}), which form spontaneously and collapse together in a finite time, from any smooth confined initial condition.

Also, the last section has formulated the dissipative version of compressible
fluids by following the {\bfi moment double bracket approach}. These fluid equations (\ref{darcyfluid})
can be called ``Darcy fluid'' and it is an interesting question whether these
equations possess emergent {\bfi singularities}, like in the case of EPDiff. \rem{ 
Indeed, these equations represent the dissipative
extension of the EPSymp fluid introduced in chapter~\ref{EPSymp}. Since the
double bracket dissipation preserves the spontaneous emergence of singularities
in the EDiff equation, one may conjecture that the same phenomenon occurs
for the Darcy fluid. 
} 
In order to establish whether the singular solutions
appear spontaneously in finite time, one needs to prove a ``steepening Lemma'' \cite{CaHo1993} (cf. chapter~\ref{intro}) for the equations of the EPSymp fluid. For example, a similar result has been found for Darcy's law by Holm and Putkaradze \cite{HoPu2005,HoPu2006}.

Further speculations have involved the {\bfi double bracket form of the $b$-equation}
(\ref{DB-beq}). This has
two relevant properties, one of which is that it reduces to the dissipative
EPDiff equation for the first order moment. The second and more interesting property is that this equation
allows for the existence of {\bfi singular solutions} for any integer $n$. It would
be interesting to check whether these solutions emerge from any confined
initial distribution for some values of $n$, as it happens for the ordinary
$b$-family. This is a possible road for further analysis. Similar considerations
also apply to the {\bfi GOP equations for the moments} (\ref{GOPmom}).

One may also extend the present phase space treatment and the corresponding
moment bracket to include an additional set of dimensions corresponding to internal degrees of freedom (order parameters, or orientation dependence) carried by the microscopic particles, rather than requiring them to be simple point particles. This is a standard approach in condensed matter theory, for example in liquid crystals, see, e.g., \cite{Ch1992,deGePr1993}. These questions are pursued in the next chapter, which is the main chapter and
contains only new results.

\chapter{Anisotropic interactions: a new model}
\label{orientation}

\section{Introduction and background}
\subsection{Geometric models of dissipation in physical systems}
This chapter explains how the geometry of double-bracket dissipation makes its way from the microscopic (kinetic theory) level to the macroscopic (continuum) level, when the particles in the microscopic description carry an internal variable that is orientation dependent.  Without orientation dependence, the moment equations derived previously yield a nonlocal variant of the famous Darcy's law \cite{Darcy1856}. When orientation is included, the resulting Lie-Darcy moment equations identify the macroscopic parameters of the continuum description and govern their evolution.

In previous work, Gibbons, Holm and Kupershmidt \cite{GiHoKu1982,GiHoKu1983} (abbreviated GHK) showed that the process of taking moments of the Vlasov equation for such particles is a Poisson map.  GHK used this property to derive the equations of {\bfi chromohydrodynamics}. These are the equations of a fluid plasma consisting of particles carrying Yang-Mills charges and interacting self-consistently via a Yang-Mills field. The GHK Poisson map for chromohydrodynamics provides the guidelines for an extension of the Kupershmidt-Manin (KM) bracket \cite{KuMa1978} for the moments. GHK considered only Hamiltonian motion and did not consider the corresponding double-bracket Poisson structure of dissipation. That is the subject of the present chapter. 


\subsection{Goal and present approach}
The goal of the present work is to determine the macroscopic implications of introducing nonlinear double-bracket dissipation at the microscopic level, so as to respect the coadjoint orbits of canonical transformations for dynamics that depends upon particle orientation. The present approach introduces this orientation dependence into the microscopic description by augmenting the canonical Poisson bracket in position $q$ and momentum $p$ so as to include the Lie-Poisson part for orientation $g$ taking values in the dual $\mathfrak{g}^*$ of the Lie algebra $\mathfrak{g}$, with eventually $\mathfrak{g}=\mathfrak{so}(3)$ for physical orientation. Thus this chapter makes use of the {\bfi total Poisson bracket from GHK}, 
\begin{equation} 
\Big\{f,h\Big\}_{\!1}:=\,
\Big\{f,h\Big\}+\left\langle g,\,\left[\frac{\partial f}{\partial g},\frac{\partial
h }{\partial g}\right]\right\rangle 
\,,
\label{PB1}
\end{equation} 
where $[\,\cdot\,,\,\cdot\, ]:\, \mathfrak{g}\times\mathfrak{g}\to \mathfrak{g}$ is the Lie algebra bracket and $\langle\,\cdot\,,\,\cdot\, \rangle:\, \mathfrak{g}^*\times\mathfrak{g}\to \mathbb{R} $ is the pairing between the Lie algebra $\mathfrak{g}$ and its dual $\mathfrak{g}^*$. For rotations, $\mathfrak{g}=\mathfrak{so}(3)$ and the bracket $[\,\cdot\,,\,\cdot\, ]$ becomes the cross product of vectors in $\mathbb{R}^3$. Correspondingly, the pairing $\langle\,\cdot\,,\,\cdot\, \rangle$ becomes the dot product of vectors in $\mathbb{R}^3$. This chapter considers the double-bracket dynamics of $f(q,p,g,t)$ resulting from replacing the canonical Poisson brackets in Eq. (\ref{Vlasov-diss}) by the direct sum of canonical and Lie-Poisson brackets $\{\,\cdot\,,\,\cdot\,\}_1$ in Eq. (\ref{PB1}). One then takes moments of the resulting dynamics of $f(q,p,g,t)$ with respect to momentum $p$ and orientation $g$, to obtain the dynamics of the macroscopic description. The moments with respect to momentum $p$ alone provide an intermediate set of dynamical equations for the $p-$moments, 
\[
A_n(q,g,t):=\!\int p^n\,f(q,p,g,t)\,dp \,. 
\]
These intermediate dynamics are reminiscent of the Smoluchowsi equation for the probability $A_0(q,g,t)$. However, the intermediate dynamics of the $p-$moments cannot be identical to the Smoluchowsi equation even for the probability $A_0(q,g,t)$, because {\it the kinetic double-bracket dissipation is deterministic, not stochastic}. This chapter presents closed sets of equations for the intermediate dynamics of $A_0(q,g,t)$ and $A_1(q,g,t)$. A closed set of continuum equations for the $(p,g)$ moments is also found. The final closure provides the macroscopic continuum dynamics for the set of moments of the double-bracket  kinetic equations (\ref{Vlasov-diss}) under the replacement  $\{\,\cdot\,,\,\cdot\,\}\to \{\,\cdot\,,\,\cdot\,\}_1$ with respect to $\{1,p,g,p^2,pg,g^2\}$. This macroscopic continuum closure inherits the geometric properties of the double bracket, because the process of taking these moments is a Poisson map, as observed in GHK. 


\section{Geometric dissipation for anisotropic interactions \label{sec:Anisotropic}}

\subsection{A dissipative version of the GHK-Vlasov equation}
Following GHK, one introduces a particle distribution which depends not
only on the position and momentum coordinates $q$ and $p$, but also on an extra coordinate $g$ associated with \emph{orientation}. The coordinate $g$ belongs to the dual of a certain Lie algebra $\mathfrak{g}$, which for anisotropic interactions would be $\mathfrak{g}=\mathfrak{so}(3)$. However, this chapter will formulate the problem in a more general context and analyze the case of rotations separately. In the non-dissipative case, the Vlasov equation is written in terms of a Poisson bracket, which is the direct sum of the canonical $(pq)$-bracket and the Lie-Poisson bracket on the Lie algebra $\mathfrak{g}$. Explicitly, this Poisson  bracket is written as
\begin{equation} 
\Big\{f,h\Big\}_{\!1}:=\,
\Big\{f,h\Big\}+\left\langle g,\,\left[\frac{\partial f}{\partial g},\frac{\partial
h }{\partial g}\right]\right\rangle 
\, . 
\label{bracket1} 
\end{equation} 
The non-dissipative Vlasov equation now becomes
\[
\frac{\partial f}{\partial t}
=
-
\left\{f,\frac{\delta H}{\partial f}\right\}_{\!1}
=
-\,
\widehat{X}_{\!\frac{\delta H}{\delta f}}(f)
\,,
\]
where one defines the vector field $\widehat{X}_h$ associated with the Hamiltonian function $h$ as
\[
\widehat{X}_h
:=
\frac{\partial h}{\partial p}\frac{\partial }{\partial q}
-
\frac{\partial h}{\partial q}\frac{\partial }{\partial p}
+\left\langle{\rm ad}^*_{\frac{\partial h}{\partial g}}\,g,\,\frac{\partial }{\partial g}\right\rangle
=
X_h+
\left\langle{\rm ad}^*_{\frac{\partial h}{\partial g}}\,g,\,\frac{\partial }{\partial g}\right\rangle 
\, . 
\]
The Vlasov equation is thus a {\bfi characteristic equation} for the evolution governed by the flow of the vector field
$\widehat{X}_{{\delta H}/{\delta f}}$,  determined by the action
of this vector field on the density $f$. 

One can identify  $\widehat{X}_h$
with $h$ and define an action $h\cdot f:=\widehat{X}_h(f)$, so that its dual operation denoted by $(\star)$ is defined by
\begin{align*}
\Big(f\star k, h \Big)
&=
\Big(k,-\,h\cdot f \Big)
=
\Big(\,k, \,\big\{h,f \big\}_{1} \,\Big)
\\
&=
\Big(\,k, \,\big\{h,f \big\} \,\Big)
-
\left(\,k, \,\left\langle g,\,\left[\frac{\partial f}{\partial g},\frac{\partial h}{\partial g}\right]\right\rangle\,\right)
\\
&=
-\Big(\,k, \,\big\{f,h \big\} \,\Big)
-
\int \!\left\langle k\,\,\text{\rm\large ad}^*_\frac{\partial f}{\partial g}\, g,\,\frac{\partial h}{\partial g}\right\rangle {\rm d}q\,{\rm d}p\,{\rm d}g
\\
&=
\Big(\,\big\{f,k \big\}, \,h\,\Big)
+
\int \!h\,\,\frac{\partial}{\partial g}\cdot\!\Big(k\,\,\text{\rm\large ad}^*_\frac{\partial f}{\partial g}\, g\Big)\, {\rm d}q\,{\rm d}p\,{\rm d}g
\\
&=
\Big(\,\big\{f,k \big\}, \,h\,\Big)
+
\int \!h\,\left\langle\text{\rm\large ad}^*_\frac{\partial f}{\partial g}\, g,\,\frac{\partial k}{\partial g}\right\rangle \,{\rm d}q\,{\rm d}p\,{\rm d}g
\\
&=
\Big(\,\big\{f,k \big\}, \,h\,\Big)
+
\int \!h\,\left\langle g,\,\left[\frac{\partial f}{\partial g},\frac{\partial k}{\partial g}\right]\right\rangle \,{\rm d}q\,{\rm d}p\,{\rm d}g
=
\Big(\,\big\{f,k \big\}_1,\,h\,\Big)
\,.
\end{align*}
where in the 5th line one uses the following argument
\[
\frac{\partial}{\partial g}\cdot\,\textrm{\large
ad}^*_{\frac{\partial f}{\partial g}}\, g
=
\frac{\partial}{\partial g_c}\left(g_a\,C^a_{bc}\,\frac{\partial f}{\partial g_b}\right)
=
\widehat{g}_{\,bc}\,\,\frac{\partial^2 f}{\partial g_c \,\partial g_b}
=
0
\,,
\]
with $\widehat{g}_{\,bc}:=g_a\,C^a_{bc}=-\widehat{g}_{\,cb}$. This is justified by the antisymmetry of $C^a_{bc}$ and by the symmetry
of $\partial_{g_c}\partial_{g_b}$. Thus, $f\star k=\left\{f,k\right\}_{1}$.
\begin{framed}
Upon applying the same arguments as in the previous chapter and making use
of the general theory of the double bracket dissipation, one
finds the purely dissipative Vlasov equation in double-bracket form, 
\begin{equation}
\frac{\partial f}{\partial t}
=
\left\{f,\left\{\mu[f],\frac{\delta E}{\partial f}\right\}_{\!1}\right\}_{\!1}
\, . 
\label{Vlasov-diss1}
\end{equation} 
where the derivative $\delta E/\delta f$ is the energy of the single particle
(the following discussion treats the energy $E[f]$ and the Hamiltonian $H[f]$
indifferently).
\end{framed}
\noindent
This equation  has exactly the same form as in (\ref{Vlasov-diss}), but now one substitutes the direct sum Poisson bracket 
$\left\{\cdot\,,\cdot\right\}_1$ in  (\ref{bracket1}) instead of the 
canonical Poisson bracket $\left\{\cdot\,,\cdot\right\}$. This formulation  can now be used to derive the double-bracket dissipative version
of the Vlasov equation for particles undergoing anisotropic interaction.

\subsection{Dissipative moment dynamics: a new anisotropic model}
To derive the moment dynamics with orientation dependence, 
one  follows the same steps as in the previous section, beginning by introducing  the quantities
\[
A_n(q,g):=\!\int p^n\,f(q,p,g)\,dp
\,\,\,\quad\text{with}\quad
g\in\mathfrak{g}^*
\,.
\]
One may find the entire hierarchy of equations for these moment quantities and then integrate over $g$ in order to find the equations for the mass density $\rho(q):=\int \!A_0(q,g)\,dg$ and the continuum charge density $G(q)=\int\! g\,A_0(q,g)\,dg$. Without the integration over $g$, such an approach would yield the Smoluchowski approximation for the density $A_0(q,g)$, usually denoted by $\rho(q,g)$. This approach is followed in the Sec.~\ref{sec:Smoluchowski}, where the dynamics of $\rho(q,g)$
is presented explicitly. 

This section  extends the Kupershidt-Manin approach as in GHK to generate the dynamics of moments with respect to both momentum $p$ and charge $g$. The main complication is that the Lie algebras of physical interest (such as $\mathfrak{so}(3)$) are not one-dimensional and in general are not Abelian. Thus, in the general case one needs to use a multi-index notation as in \cite{Ku1987,Ku2005,GiHoTr05}. 
One can introduce multi-indices $\sigma:=(\sigma_1, \sigma_2,\,\dots,\sigma_N)$,
with $\sigma_i\geq0$, and define $g^\sigma:=g_1^{\sigma_1}\!\!\dots g_N^{\sigma_N}$, where $N={\rm dim}(\mathfrak{g})$. Then, the moments are expressed as
\[
A_{n\,\sigma}(q)\,:=\int p^n\,g^\sigma\,f(q,p,g)\,dp\,dg
\,.
\]
This multi-dimensional treatment leads to cumbersome calculations. For the purposes of this section, one is primarily interested in the equations for $\rho$ and $G$, so one restricts to consider only moments of the form
\[
A_{n,\nu}=\int p^n\,g_\nu\,f(q,p,g)\,dp\,dg
\qquad \nu=0,1,\dots,N \, . 
\]
Here one defines $g\,_0=1$ and $\,g_a=\langle\, g,\,\mathbf{e}_a\rangle$ where $\mathbf{e}_a$ is a basis of the Lie algebra and $\langle\, g_b\mathbf{e}^b,\,\mathbf{e}_a\rangle=g_a\in\mathbb{R}$ represents the result of the pairing $\langle\, \cdot\,,\,\cdot\,\rangle$ between an element of the Lie algebra basis and an element of the dual Lie algebra.
One writes the single particle Hamiltonian as $h=\delta H/\delta
f=p^n g_\nu\,\delta H/\delta A_{n,\nu}=:p^n g_\nu \,\beta_n^\nu(q)$, which means  that one employs the following assumption. 

\medskip

\begin{assumption} \label{linassump}
The single-particle Hamiltonian $h=\delta H/\delta f$ is linear in $g$ and can be expressed as 
\[ h(q,p,g)= p^n\,\psi_n(q)+p^n\!\left\langle g,\,\overline\psi_n(q) \right\rangle \, ,
\] 
where $\psi_n(q)\in \mathbb{R}$ is a real scalar function and $\overline\psi_n(q)\in \mathbb{R}\otimes\mathfrak{g}$ is a real Lie-algebra-valued function. This assumption will be used throughout the present chapter, except in Section \ref{sec:Smoluchowski}. 
\end{assumption}


\medskip

\paragraph{Dual Lie algebra action.}
The action of $\beta_n^\nu$ on $f$ is defined as
\[
\beta_n^{\,\nu}\cdot f=\big\{p^n g_\nu\,\beta_n^{\,\nu},\,f\big\}_{\!1}
\qquad\text{ (no sum)} \, . 
\]
The dual of this action is denoted by $(\star_{n, \nu})$. It may be  computed analogously to the equation (\ref{stardef}) in the previous chapter and  found to be  
\begin{align*}
f\,\,\text{\Large$\star$}_{n,\nu}\,k
\,=&\,
\iint  p^n g_\nu\,\big\{f,\,k\big\}_{\!1}\,{\rm d}p \,{\rm d}g\\
\,=&\,
\int g_\nu \, g_\sigma \,{\sf ad}_{\alpha_m^{\sigma}}^*A_{m+n-1} \,{\rm d}g
+
\int g_\nu
\left\langle g,\left[\frac{\partial A_{m+n}}{\partial g},\frac{\partial (g_\sigma\,\alpha_m^\sigma)}{\partial g}\right]\right\rangle{\rm d}g
\\
\,=&\,\,
{\sf ad}_{\alpha_m^{\sigma}}^* \int\! g_\nu \, g_\sigma \,A_{m+n-1} \,{\rm d}g
+
\int g_\nu
\left\langle g,\left[\frac{\partial A_{m+n}}{\partial g},\frac{\partial (g_a\,\alpha_m^a)}{\partial g}\right]\right\rangle{\rm d}g 
\, . 
\end{align*}
Here, $k=p^m\,g_\sigma\,\alpha_m^\sigma(q)$ and one uses the definition of the moment 
\[
A_n(q,g)=\int p^n\,f(q,p,g)\,dp
\,.\]

\paragraph{Evolution equation.}
Having characterized the dual Lie algebra action, one
may write the evolution equation for an arbitrary functional $F$ in terms of the dissipative bracket as follows: 
\begin{equation} 
\dot{F}= \{\!\{\,F\,,\,E\,\}\!\}=-\left\langle\!\!\!\left\langle \left(\mu[f]\,\,\text{\Large$\star$}_{n,\nu}\,\,\frac{\delta E}{\partial f}\right)^\sharp\!,\,f\,\,\text{\Large$\star$}_{n,\nu}\,\,\frac{\delta F}{\partial f} \right\rangle\!\!\!\right\rangle
\label{dissbracket1} 
\end{equation}
where the pairing $\left\langle\!\!\!\left\langle\,\cdot\,,\cdot\,\right\rangle\!\!\!\right\rangle$
is given by integration over the spatial coordinate $q$.
Now we fix $m=0$, $n=1$.  The equation for the evolution of $F=A_{0,\lambda}:=\int g_\lambda
\,A_0 \,{\rm d}g{\rm d}p$ is found from (\ref{dissbracket1}) to be
\begin{align}\nonumber
\frac{\partial A_{\,0,\lambda}}{\partial t}=
&\,
{\sf ad}_{\gamma_{1,\nu}^\sharp}^*\int \!g_\nu \, g_\lambda \,A_0 \,{\rm d}g\,
\\
&+
\int g_\lambda \left\langle g,\left(\,
\left[\frac{\partial A_{1}}{\partial g},\frac{\partial (g_\sigma\,{\gamma}_{1,\sigma}^{\,\,\sharp})}{\partial g}\right]+\left[\frac{\partial A_{0}}{\partial g},\frac{\partial (g_\sigma\,{\gamma}_{0,\sigma}^{\,\,\sharp})}{\partial g}\right]\,\right)\right\rangle{\rm d}g
\nonumber
\\
=&
\frac{\partial}{\partial q}\left({\gamma_{1,\nu}^\sharp}\int \!g_\nu \, g_\lambda \,A_0 \,{\rm d}g\right)\nonumber
\\
&+
\int g_\lambda \left\langle g,\left(\,
\left[\frac{\partial A_{1}}{\partial g},\frac{\partial (g_a\,{\gamma}_{1,a}^{\,\,\sharp})}{\partial g}\right]+\left[\frac{\partial A_{0}}{\partial g},\frac{\partial (g_a\,{\gamma}_{0,a}^{\,\,\sharp})}{\partial g}\right]\,\right)\right\rangle{\rm d}g \, , 
\label{moments1}
\end{align}
where one defines the analogues of Darcy's velocities: 
\begin{align*}
\gamma_{0,\nu}:=\,\mu[f]\,\,\text{\Large$\star$}_{0,\nu}\,\,\frac{\delta E}{\delta f}
&=
\int \!g_\nu
\left\langle g,\left[\frac{\partial \widetilde\mu_{k}}{\partial g},\frac{\partial (g_a\,\beta_k^a)}{\partial g}\right]\right\rangle{\rm d}g
\\
&=\int \!g_\nu
\left\langle g,\left[\frac{\partial \widetilde\mu_{0}}{\partial g},\frac{\partial (g_a\,\beta_0^a)}{\partial g}\right]\right\rangle{\rm d}g
\end{align*}
and
\begin{align*}
\gamma_{1,\nu}:=\,\mu[f]\,\,\text{\Large$\star$}_{1,\nu}\,\,\frac{\delta E}{\delta f}
&=\,
{\sf ad}_{\beta_k^{\,\sigma}}^*
\int g_\nu \,g_\sigma \,\widetilde\mu_{k} \,{\rm d}g
+
\int \!g_\nu
\left\langle g,\left[\frac{\partial \widetilde\mu_{k+1}}{\partial g},\frac{\partial (g_a\,\beta_k^a)}{\partial g}\right]\right\rangle{\rm d}g
\\
&=\,
\frac{\,\,\partial \beta_0^{\,\sigma}}{\partial q}
\int \!g_\nu \, g_\sigma \,\widetilde\mu_{0} \,{\rm d}g
+
\int g_\nu
\left\langle g,\left[\frac{\partial \widetilde\mu_{1}}{\partial g},\frac{\partial (g_a\,\beta_0^a)}{\partial g}\right]\right\rangle{\rm d}g
\,.
\end{align*}
Here one assumes that the energy functional $E$ depends only on $A_{0,\lambda}$
(recall that $\beta_n^\lambda:=\delta E/\delta A_{n,\lambda}$), so that it
is possible to fix $k=0$ in the second line. These equations above will be treated as
a higher level of approximation in section~\ref{higher}. Now, one further simplifies the treatment
by discarding all terms in $\gamma_{1,a}$, that is one truncates the summations
in equation (\ref{moments1}) to consider only terms in $\gamma_{0,0}$, $\gamma_{0,a}$ and $\gamma_{1,0}$. This is equivalent to consider a single-particle Hamiltonian
of the form
\[
h(q,p,g)=\psi_0(q)+\langle g,\bar{\psi}_0(q)\rangle + p\,\psi_1(q).
\]
 With this simplification the equation (\ref{moments1})
becomes
\begin{align}
\frac{\partial A_{\,0,\lambda}}{\partial t}=
&\,
{\sf ad}_{\gamma_{1,0}}^*\int g_\lambda \,A_0 \,{\rm d}g\,
+
\int g_\lambda \left\langle g,\left(\,
\left[\frac{\partial A_{0}}{\partial g},\frac{\partial (g_\sigma\,{\gamma}_{0,\sigma}^{\,\,\sharp})}{\partial g}\right]\,\right)\right\rangle{\rm d}g
\nonumber
\\
=&
\frac{\partial}{\partial q}\left({\gamma_{1,0}}\int  g_\lambda \,A_0 \,{\rm d}g\right)
+
\int g_\lambda \left\langle g,\left(\,
\left[\frac{\partial A_{0}}{\partial g},\frac{\partial (g_a\,{\gamma}_{0,a}^{\,\,\sharp})}{\partial g}\right]\,\right)\right\rangle{\rm d}g \, , 
\label{moments-simplified}
\end{align}
and the expression for $\gamma_{1,0}$ is
\begin{align*}
\gamma_{1,0}:\!&=\,\mu[f]\,\,\text{\Large$\star$}_{1,0}\,\,\frac{\delta E}{\delta f}
\\
&=\,
{\sf ad}_{\beta_k^{\,\sigma}}^*
\int g_\sigma \,\widetilde\mu_{k} \,{\rm d}g
+
\int \!
\left\langle g,\left[\frac{\partial \widetilde\mu_{k+1}}{\partial g},\frac{\partial (g_a\,\beta_k^a)}{\partial g}\right]\right\rangle{\rm d}g
=\,
\frac{\,\,\partial \beta_0^{\,\sigma}}{\partial q}
\int  g_\sigma \,\widetilde\mu_{0} \,{\rm d}g
\,.
\end{align*}
At this point it is convenient to simplify the notation by defining the following dynamic quantities 
\begin{eqnarray*}
\rho&=&\int f \,{\rm d}g\,{\rm d}p
\,,
\qquad
\quad\,\,
G\,\,\,\,=\,\,\,\int g\, f\,{\rm d}g\,{\rm d}p
\,.
\end{eqnarray*}
Likewise, one defines the \emph{mobilities} as
\begin{eqnarray*}
\mu_\rho&=&\int \mu[f] \,{\rm d}g\,{\rm d}p
\,,
\qquad
\quad
\mu_G\,\,\,\,=\,\,\,\int g\, \mu[f]\,{\rm d}g\,{\rm d}p
\,.
\end{eqnarray*}
In terms of these quantities, it is possible to write the following.
\begin{framed}
\begin{theorem}\label{momeqns-thm-simplified}
The moment equations for $\rho$ and $G$ are given by
\begin{align}
\frac{\partial \rho}{\partial t}=&\,\,
\frac{\partial}{\partial q}\Bigg(\rho\,\,
\bigg(
\mu_\rho\,\, \frac{\partial}{\partial q}\frac{\delta E}{\delta \rho}
+
\bigg\langle \mu_G, \,\frac{\partial}{\partial q}\frac{\delta E}{\delta G}\bigg\rangle
\bigg)
\,\,
\Bigg)
\label{rhogen-simplified}
\\
\frac{\partial G}{\partial t}=&\,\,
\frac{\partial}{\partial q}\Bigg(G\left(\mu_\rho\,\, \frac{\partial}{\partial q}\frac{\delta E}{\delta \rho}+
\left\langle \mu_G, \,\frac{\partial}{\partial q}\frac{\delta E}{\delta G}\right\rangle\right) \Bigg)
+ {\rm ad}^*_{\left(\!{\rm ad}^*_{\frac{\delta E}{\delta G}} \mu_G\!\right)^{\!\!\sharp}}\,G
\label{Ggen-simplified}
\,.
\end{align}
\end{theorem}
\end{framed}
\begin{remark}
Equations in this family (called Geometric Order Parameter equations) were derived via a different approach in \cite{HoPu2007}.
\end{remark}

\subsection{A first property: singular solutions}\label{sinsol}
Equations (\ref{rhogen-simplified}) and (\ref{Ggen-simplified}) admit singular solutions. This means that the trajectory of a single fluid
particle is a solution of the problem and all the microscopic information
about the particles is preserved. One can prove the following.
\begin{framed}
\begin{theorem}
Equations (\ref{rhogen-simplified}) and (\ref{Ggen-simplified}) admit
solutions of the form
\begin{align}
\nonumber
\rho(q,t)&=w_\rho(t)\,\,\delta(q-Q(t))\\
G(q,t)&=w_G(t)\,\,\delta(q-Q(t))
\label{singansatz}
\end{align}
where $w_\rho$, $w_G$ and $Q$ undergo the following
dynamics
\begin{align*}
\dot{w}_\rho&=0
\\
\dot{w}_G&=\text{\rm\large ad}^*_{\left(\text{\rm ad}^*_{\delta E/\delta G}\,\mu_G\right)_{q=Q}^{\sharp}} w_G
\\
\dot{Q}&=
-\left(\mu_\rho\,\, \frac{\partial}{\partial q}\frac{\delta E}{\delta \rho}
+
\bigg\langle \mu_G, \,\frac{\partial}{\partial q}\frac{\delta E}{\delta G}\bigg\rangle\right)_{q=Q}
\rem{ 
\qquad\quad
\dot{Q}_G=
-\left(\mu_\rho\,\, \frac{\partial}{\partial q}\frac{\delta E}{\delta \rho}
+
\bigg\langle \mu_G, \,\frac{\partial}{\partial q}\frac{\delta E}{\delta G}\bigg\rangle\right)_{q=Q_G}
} 
\end{align*}
\end{theorem}
\begin{proof}
After defining the quantities
\begin{align*}
\gamma_1&:=\gamma_{1,0}=\mu_\rho\,\, \frac{\partial}{\partial q}\frac{\delta E}{\delta \rho}
+
\bigg\langle \mu_G, \,\frac{\partial}{\partial q}\frac{\delta E}{\delta G}\bigg\rangle
\\
\gamma_0&:=\gamma_{0,a}^{\sharp}\,{\bf e}_a=\left({\rm ad}^*_{\frac{\delta E}{\delta G}} \mu_G\right)^{\!\sharp}
\end{align*}
one pairs equations (\ref{rhogen-simplified}) and (\ref{Ggen-simplified}) respectively with $\phi_\rho(q)$ and $\phi_G(q)$. This yields the following results,
\begin{align*}
\int \dot\rho \,\phi_\rho\,\, {\rm d}q
&=
\int \phi_\rho\,\, \frac{\partial}{\partial q}\big(\rho\,
\gamma_1
\big)\, {\rm d}q\\
&=
-\int \frac{\partial \phi_\rho}{\partial q}\,\rho\,\gamma_1
\, {\rm d}q
\\\\
\int \left\langle\dot{G},\phi_G\right\rangle \, {\rm d}q
&=
\int \left\langle
\frac{\partial}{\partial q}\big(G\,\gamma_1 \big)
+ 
{\rm ad}^*_{\gamma_0}\,G,
\,\phi_G\right\rangle {\rm d}q
\\
&=
-\int \left\langle G,\,\gamma_1\,\frac{\partial \phi_\rho}{\partial q}\right\rangle\,{\rm d}q
+\int
\left\langle G,\Big[\gamma_0,\,\phi_\rho\Big]\right\rangle\,
\, {\rm d}q
\end{align*}
Upon substituting the singular solution ansatz (\ref{singansatz}), one
calculates
\begin{align*}
\dot{w}_\rho\,\phi_\rho(Q)\,+\,w_\rho\,\dot{Q} \,\left.\frac{\partial \phi_\rho}{\partial q}\right|_{q=Q}
\!\!\!
&=
-\,w_\rho\,\,\gamma_1(Q)\,\left.\frac{\partial \phi_\rho}{\partial q}\right|_{q=Q}
\\
\Big\langle\dot{w}_G,\,\phi_G(Q)\Big\rangle
\,+\,
\dot{Q} \left.\left\langle w_G,\,\frac{\partial \phi_G}{\partial q}\right\rangle\!\right|_{q=Q}
\!\!\!
&=
-\,\gamma_1(Q)\left.\left\langle w_G,\,\frac{\partial \phi_G}{\partial q}\right\rangle\!\right|_{q=Q}
\!\!\!
+\,\,
\left\langle \textrm{\large ad}^*_{\gamma_0}\,w_G,\phi_G(Q)\right\rangle
\end{align*}
so that identification of corresponding coefficients yields
\begin{align*}
\dot{w}_\rho&=0\,,
\qquad\qquad
\dot{w}_G=\textrm{\large ad}^*_{\gamma_0}\,w_G\,,
\qquad\qquad
\dot{Q}=
-\,\gamma_1(Q)
\end{align*}
and the thesis is proven.
\end{proof}
\end{framed}
\begin{remark}
A similar result applies for the Geometric Order Parameter (GOP) equations investigated in \cite{HoPu2007}. Indeed, the equations (\ref{rhogen-simplified})
and (\ref{Ggen-simplified}) reduce to those in \cite{HoPu2007} when one considers
a commutative  Lie algebra.
\end{remark}

\bigskip

\section{An application: rod-like particles on the line}

\subsection{Moment equations}
In this case $G={\bf m}(x)$, $\rm ad_\mathbf{v} \mathbf{w}=\mathbf{v} \times \mathbf{w}$ and $\rm ad^*_\mathbf{v} \mathbf{w}=-\mathbf{v} \times \mathbf{w}$, and the
Lie algebra pairing is represented by the dot product of vectors in $\mathbb{R}^3$. Therefore the equations are
\begin{framed}
\begin{align}
\frac{\partial \rho}{\partial t}
&=
\frac{\partial}{\partial x}
\Bigg(
\rho\,
\bigg(
\mu_\rho\, \frac{\partial}{\partial x}\frac{\delta E}{\delta \rho}
+\,
\boldsymbol\mu_{\bf m}\!\cdot\frac{\partial}{\partial x}\frac{\delta E}{\delta \bf m}
\bigg)
\,
\Bigg)
\label{rodrho}
\end{align}
and
\begin{align}
\frac{\partial {\bf m}\,}{\partial t}
&=
\frac{\partial}{\partial x}\Bigg({\bf m}
\left(\mu_\rho\frac{\partial}{\partial x}\frac{\delta E}{\delta \rho}
+
\boldsymbol\mu_{\bf m}
\cdot\frac{\partial}{\partial x}\frac{\delta E}{\delta {\bf m}}
\right)
\, \Bigg)
+
{\bf m}\times
\boldsymbol\mu_{\bf m}\times\frac{\delta E}{\delta \bf m}
\label{rodGilbert}
\end{align}
\end{framed}
Note that equations for  density $\rho$ and orientation ${\bf m}$ have conservative parts (coming from the divergence of a flux). In addition, when $\boldsymbol{\mu}_{\bf m}=a{\bf m}$ for a constant $a$, the orientation ${\bf m}$ has precisely the dissipation term ${\bf m}\times{\bf m}\times \delta E/\delta \bf m$ introduced by Gilbert \cite{Gilbert1955}. Thus, this section has derived the Gilbert dissipation term at the macroscopic level, starting from double-bracket dissipative terms in the kinetic theory description. 

\rem{ 
\begin{remark}[Directors and angular momentum]$\,$\\
Besides its applications in ferromagnetism, the equations (\ref{rodrho})
and (\ref{rodGilbert}) can be also applied to the dissipative dynamics oriented
nano-particles. In this context, the usual approach \cite{Costantin} identifies the particle orientation with a unit vector $\bf n$, called ``director''. The same approach is used in the theory of liquid crystals \cite{Ho02}. 
\end{remark}
} 

\subsection{More results: emergence and interaction of singularities}
\begin{figure}[!]\label{offsetpeaks_eps}
\center
\includegraphics[scale=0.353]{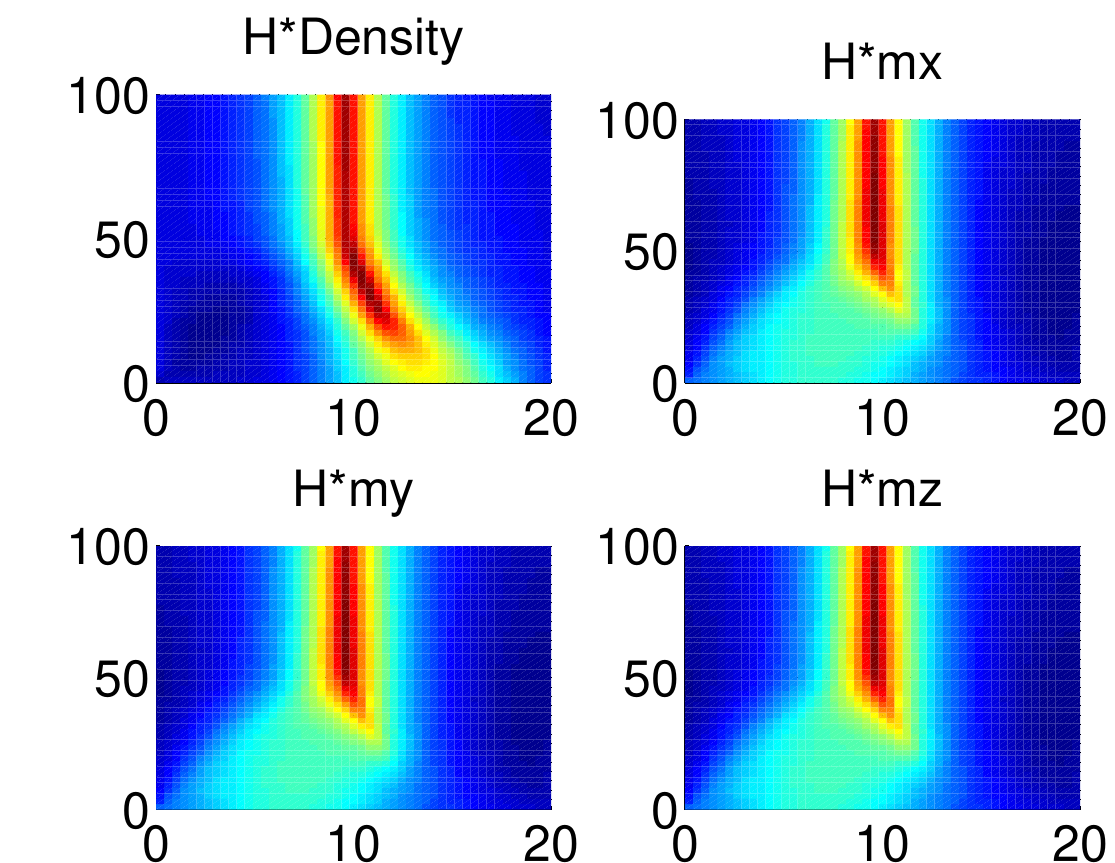}
\quad
\includegraphics[scale=0.457]{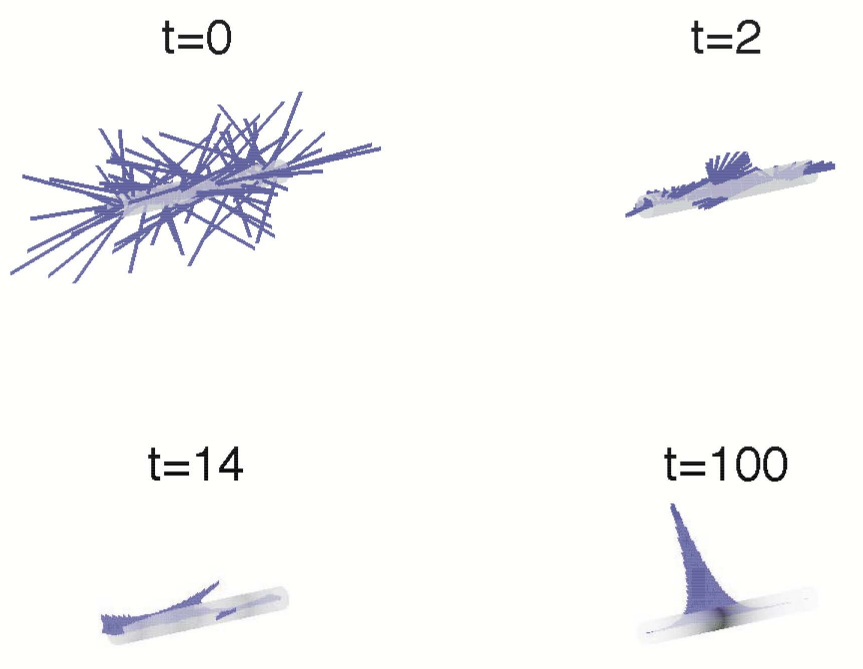}
\caption{
\emph{Left:}  Emerging singularities. Plots of the smoothed density $\bar\rho=H*\rho$
and orientation $\overline{\bf m}= H* {\bf m}$ (three components), where
the smoothing kernel is the Helmholtzian $e^{-|x|}$. The figures show how the singular solution emerges from a Gaussian initial condition
for the energy in (\ref{energy}).  Smoothed quantities are chosen to avoid the necessity to represent $\delta$-functions.
\emph{Right:} 
Orienton formation in a  $d=1$ dimensional simulation. The color-code on the cylinder denotes local \emph{averaged} density: black is maximum density while white is $\overline{\rho}=0$.  Purple lines denote the three-dimensional vector $\overline{\bf m}= H* {\bf m}$. The formation of sharp peaks in averaged quantities corresponds to the formation of $\delta$-functions. (Figures by
V. Putkaradze)
}
\end{figure}

When considering the rotation algebra $\mathfrak{so}(3)$, numerical experiments have shown \cite{HoOnTr07} that under certain conditions, the singular solutions in section~\ref{sinsol} emerge spontaneously from any confined initial distribution. This result was already known in the case of isotropic interaction for which a rigorous proof is also available \cite{HoPu2005,HoPu2006}. In that context the singular solutions
were called \emph{clumpons}. The numerics shows that the anisotropic
nature of the self interaction preserves this behavior. In particular the experiments
have shown that not only these solutions form for the density variable $\rho$,
but also for the orientation density $\bf m$. Such a situation represent
a state in which the particles are concentrated in only one point in space
and are all aligned towards only one direction (fig.~\ref{offsetpeaks_eps}).

\remfigure{ 
\begin{figure}[!]\label{orientons_eps}
\center
\includegraphics[scale=0.6]{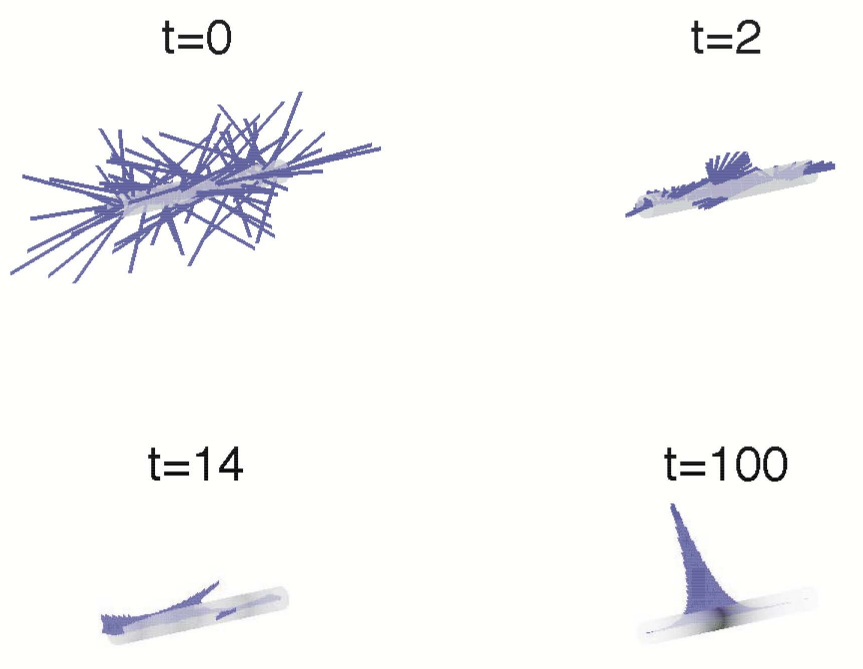}
\includegraphics[height=6cm]{densityplot.eps}
\caption{
\emph{Left:} 
Orienton formation in a  $d=1$ dimensional simulation. The color-code on the cylinder denotes local \emph{averaged} density: black is maximum density while white is $\overline{\rho}=0$.  Purple lines denote the three-dimensional vector $\overline{\bf m}= H* {\bf m}$. The formation of sharp peaks in averaged quantities corresponds to the formation of $\delta$-functions. 
\emph{Right:} The corresponding waterfall plot for evolution of averaged density $\overline{\rho}=H*\rho$. Horizontal coordinate is space.  Sharp peaks correspond to the formation of $\delta$-function singularities in the density variable $\rho$.
}
\end{figure}
} 

This section studies the interaction of two singular solutions
of the equations (\ref{rodrho}) and (\ref{rodGilbert}).  Each delta function
has the interpretation of a single particle, whose weights and positions satisfy a finite set of ordinary differential equations.  In particular one
wants to investigate the two-particle case analytically. It is possible to find conditions for which the particles tend to  merge and align.

From section~\ref{sinsol}, upon renaming the variable $w_G$ with the simpler
 notation $\boldsymbol{\lambda}$ (so that $w_G=\boldsymbol{\lambda}$), one writes
\begin{align*}
\dot{x}_i(t)=&\,V(x_i(t))
\\
\dot{\boldsymbol\lambda}_i(t)=&\,\boldsymbol{\lambda}_i(t)\times\boldsymbol\Phi(x_i(t))
\end{align*}
where $\boldsymbol{\lambda}_i$ is the orientation (or magnetic moment) of the $i$-th
particle and
\begin{align*}
V(x_i)&=-
\left(\mu_\rho\,\, \frac{\partial}{\partial x}\frac{\delta E}{\delta \rho}
+
\boldsymbol{\mu}_{\bf m}\cdot \,\frac{\partial}{\partial x}\frac{\delta E}{\delta {\bf m}}\right)_{x=x_i}
\\
\boldsymbol\Phi(x_i)&=\left(\boldsymbol{\mu}_{\bf m}\times\frac{\delta E}{\delta {\bf m}}\right)_{x=x_i}
\end{align*}
In order to specify the physical system one has to choose an energy and the
quantities $\mu_\rho$ and $\boldsymbol{\mu}_{\bf m}$. This section presents
the nonlocal purely quadratic case
\begin{align}\label{energy}
E[\rho, {\bf m}]=
\frac12\int \rho\, G_\rho*\rho\,\,{\rm d}x
+
\frac12\int {\bf m}\,G_{\bf m\!}*{\bf m}\,\,{\rm d}x
\end{align}
where * denotes convolution and $G_\rho$, $G_{\bf m}$ are the kernels of some symmetric invertible
operators (later chosen to be all equal to the Helmholtz operator). Analogously, one can
take $\mu_\rho=H_\rho*\rho$ and $\boldsymbol{\mu}_{\bf m}=H_{\bf m\!}*{\bf m}$. Under these circumstances, one writes
\begin{align*}
V(x,t)=&-H_\rho*\rho\,\,\partial_x G_\rho*\rho\,-\,H_{\mathbf{m}}*\mathbf{m}\cdot\partial_x
G_{\mathbf{m}}*\mathbf{m}
\\
\boldsymbol\Phi(x,t)=&H_{\mathbf{m}}*\mathbf{m}\times G_{\mathbf{m}}*\mathbf{m}
\,.
\end{align*}
Substituting the multi-particle solution
\begin{align*}
\rho(x,t)=&\sum_i w_i(t)\delta(x-x_i(t))
\\
{\bf m}(x,t)=&\sum_j\boldsymbol{\lambda}_j(t)\delta(x-x_j(t))
\end{align*}
yields
\begin{align*}
V(x,t)=&-
\sum_{j,k}\, w_j\,w_k\,H_\rho(x_j-x)\,\partial_x G_\rho(x_k-x)
\\
&-
\sum_{j,k}\,
\boldsymbol{\lambda}_j\cdot\boldsymbol{\lambda}_k\,
H_{\mathbf{m}}(x_j-x)\partial_x
G_{\mathbf{m}}(x_h-x)
\\
\boldsymbol\Phi(x,t)=&
\sum_{j,k}\,
H_{\mathbf{m}}(x_j-x)\,G_{\mathbf{m}}(x_k-x)\,
\boldsymbol{\lambda_j}\times\boldsymbol{\lambda}_k
\,.
\end{align*}
where all kernels are now assumed to be Helmholtz kernels (so that $H(0)=K(0)=1$).
One now considers a system of two identical clumpons ($j,k=1,2$ and $w_1=w_2=1$)
and evaluate 
\begin{align*}
V({x_1},t)=&\,-
\big(1+H_\rho(x_2-x_1)\big)\,\partial_{x_1} G_\rho(x_2-x_1)
\\
&\,-
\big(\boldsymbol{\lambda}_1+\boldsymbol{\lambda}_2\,H_{\mathbf{m}}(x_2-x_1)\big)\cdot\boldsymbol{\lambda}_2
\,\partial_{x_1} G_{\mathbf{m}}(x_2-x_1)
\\
\boldsymbol\Phi(x_1,t)=&
\big(G_{\mathbf{m}}(x_2-x_1)-H_{\mathbf{m}}(x_2-x_1)\big)\,
\boldsymbol{\lambda}_1\times\boldsymbol{\lambda}_2
\,.
\end{align*}
and analogously
\begin{align*}
V(\mathbf{x_2},t)=&\,-
\big(1+H_\rho(x_2-x_1)\big)\,\partial_{x_2} G_\rho(x_2-x_1)
\\
&\,-
\big(\boldsymbol{\lambda}_1+\boldsymbol{\lambda}_2\,H_{\mathbf{m}}(x_2-x_1)\big)\cdot\boldsymbol{\lambda}_2\,
\partial_{x_2} G_{\mathbf{m}}(x_2-x_1)
\\
\boldsymbol\Phi(x_2,t)=&\,-\,\boldsymbol\Phi(x_1,t)
\,.
\end{align*}
The equations of motion are then
\begin{align*}
\dot{x}_1=&\,
\big(1+H_\rho(x_2-x_1)\big)\,\partial_{x_1} G_\rho(x_2-x_1)\\
&+
\big(\boldsymbol{\lambda}_1+\boldsymbol{\lambda}_2\,H_{\mathbf{m}}(x_2-x_1)\big)\cdot\boldsymbol{\lambda}_2
\,\partial_{x_1} G_{\mathbf{m}}(x_2-x_1)
\\
\dot{x}_2=&\,
\big(1+H_\rho(x_2-x_1)\big)\,\partial_{x_2} G_\rho(x_2-x_1)\\
&+
\big(\boldsymbol{\lambda}_2+\boldsymbol{\lambda}_1\,H_{\mathbf{m}}(x_2-x_1)\big)\cdot\boldsymbol{\lambda}_1
\partial_{x_2} G_{\mathbf{m}}(x_2-x_1)
\\
\dot{\boldsymbol\lambda}_1=&\,\big(G_{\mathbf{m}}(x_2-x_1)-H_{\mathbf{m}}(x_2-x_1)\big)\,
\boldsymbol{\lambda}_1\times\boldsymbol{\lambda}_1\times\boldsymbol{\lambda}_2
\\
\dot{\boldsymbol\lambda}_2=&\,\big(G_{\mathbf{m}}(x_2-x_1)-H_{\mathbf{m}}(x_2-x_1)\big)\,
\boldsymbol{\lambda}_2\times\boldsymbol{\lambda}_2\times\boldsymbol{\lambda}_1
\end{align*}
Now calculate
\begin{align}
\frac{d}{dt}\left|x_1-x_2\right|=&\,{\rm sign}\,(x_2-x_1)\,\frac{d}{dt}\left(x_2-x_1\right)
\nonumber
\\
=&\,-\,{\rm sign}^{2\,}(x_2-x_1)\left\{\frac{2}{\alpha_\rho}\,\Big(1+H_\rho({x}_2-{x}_1)\Big)\,G_\rho(x_2-x_1)\right.
\nonumber
\\
&\,\,\,+
\left.\frac{1}{\alpha_{\mathbf{m}}}
\bigg[\,2\,\boldsymbol{\lambda}_2\cdot\boldsymbol{\lambda}_1+\left(\left\|\boldsymbol{\lambda}_2\right\|^2+\left\|\boldsymbol{\lambda}_1\right\|^2\right)\,H_{\mathbf{m}}({x}_2-{x}_1)\,\bigg]\,
G_{\mathbf{m}}(x_2-x_1)\right\}
\label{diff-x}
\end{align}
where one uses the fact that $\partial_{x_1} G_\rho(x_2-x_1)=-\partial_{x_2} G_\rho(x_2-x_1)$. It is easy to notice that the two particles move together after merging.

To investigate the asymptotic dynamics of alignment of $\boldsymbol{\lambda}_1$ and $\boldsymbol{\lambda}_2$, one calculates
\begin{align}\nonumber
\frac{d}{dt}\left(\boldsymbol{\lambda}_1\cdot\boldsymbol{\lambda}_2\right)&=
\dot{\boldsymbol{\lambda}}_1\cdot\boldsymbol{\lambda}_2+\boldsymbol{\lambda}_1\cdot\dot{\boldsymbol{\lambda}}_2
\\
&=
2\,\Big(H_{\mathbf{m}}(x_2-x_1)-G_{\mathbf{m}}({x}_2-{x}_1)\Big)\left\|\boldsymbol{\lambda}_1\times\boldsymbol{\lambda}_2\right\|^2
\label{lambdas}
\end{align}
where the equations for $\dot{\boldsymbol{\lambda}}_1$ and $\dot{\boldsymbol{\lambda}}_2$
have been substitute in the second step.
One notices that the dynamics of $\boldsymbol{\lambda}_1\cdot\boldsymbol{\lambda}_2$
 is nontrivial only if the particles have not clumped yet. Indeed, after the  particles merge, the angle between the $\boldsymbol{\lambda}$'s remains constant in time. The following discussion considers the case when
the time before merging is sufficiently long for the $\boldsymbol{\lambda}$'s
to reach their asymptotic equilibrium state.

Already at this stage one can conclude from equation (\ref{lambdas}) that
\begin{framed}
\begin{theorem}
The two clumpons always tend to a final state, which is either alignment
or anti-alignment. If $H_{\mathbf{m}}<G_{\mathbf{m}}$ ($H$ narrower than $G$), then $\boldsymbol{\lambda}_1\cdot\boldsymbol{\lambda}_2$
tends to its minimum value $\boldsymbol{\lambda}_1\cdot\boldsymbol{\lambda}_2\rightarrow-\left\|\boldsymbol{\lambda}_1\right\|\left\|\boldsymbol{\lambda}_2\right\|$,
so that clumpons tend to {\sl anti-align}. Alternatively, if $H_{\mathbf{m}}>G_{\mathbf{m}}$ (for $G$ is narrower than $H$), then $\boldsymbol{\lambda}_1\cdot\boldsymbol{\lambda}_2\rightarrow\left\|\boldsymbol{\lambda}_1\right\|\left\|\boldsymbol{\lambda}_2\right\|$
and the clumpons tend to {\sl align}.

This alignment process lasts as long as the two clumpons are separated by a nonzero distance.
\end{theorem}
\begin{proof}
One notices that the expression in (\ref{lambdas}) has a definite sign, positive
or negative depending on wether $H_{\mathbf{m}}>G_{\mathbf{m}}$ or $H_{\mathbf{m}}<G_{\mathbf{m}}$
respectively. Thus the product $\boldsymbol{\lambda}_1\cdot\boldsymbol{\lambda}_2$
tend to grow or decay in the two different cases. However one has $\max\{\left|\boldsymbol{\lambda}_1
\cdot\boldsymbol{\lambda}_2\right|\}=\left\|\boldsymbol{\lambda}_1\right\|\left\|\boldsymbol{\lambda}_2\right\|$,
which means that
\[
\lim_{t\to+\infty} \boldsymbol{\lambda}_1\cdot\boldsymbol{\lambda}_2
=\pm\left\|\boldsymbol{\lambda}_1\right\|\left\|\boldsymbol{\lambda}_2\right\|
\]
On the other hand, when $H_{\mathbf{m}}-G_{\mathbf{m}}=0$, then $\boldsymbol{\lambda}_1\cdot\boldsymbol{\lambda}_2$
remains constant. In particular $x_2-x_1=0\,\Rightarrow\, H_{\mathbf{m}}=G_{\mathbf{m}}=0$
 and this proves the last part of the theorem.
\end{proof}
\end{framed}
\noindent
Thus, the competition between the length scales of the smoothing functions $G$ and $H$ determines the alignment in the asymptotic state.

Two more relevant results are the following
\begin{framed}
\begin{corollary}
In the particular case $H_{\mathbf{m}}>G_{\mathbf{m}}$ ($G$ narrower than $H$) and $\boldsymbol{\lambda}_1\cdot\boldsymbol{\lambda}_2>0$
at $t=0$, then the particles will align and clump asymptotically in time. \end{corollary}
\begin{proof}
Upon using the vector identity
\[
\left\|\boldsymbol{\lambda}_1\times\boldsymbol{\lambda}_2\right\|^2=\left\|\boldsymbol{\lambda}_1\right\|^2\left\|\boldsymbol{\lambda}_2\right\|^2-\left(\boldsymbol{\lambda}_1\cdot\boldsymbol{\lambda}_2\right)^2
\]
one can write
\begin{align*}
\frac{d}{dt}\left(\boldsymbol{\lambda}_1\cdot\boldsymbol{\lambda}_2\right)=
2\,\Big(H_{\mathbf{m}}(x_2-x_1)-G_{\mathbf{m}}({x}_2-{x}_1)\Big)
\left(\left\|\boldsymbol{\lambda}_1\right\|^2\left\|\boldsymbol{\lambda}_2\right\|^2-\left(\boldsymbol{\lambda}_1\cdot\boldsymbol{\lambda}_2\right)^2\right)
\end{align*}
where $\left\|\boldsymbol{\lambda}_1\right\|^2$ and $\left\|\boldsymbol{\lambda}_2\right\|^2$ are constants. So the evolution equation is
\[
\frac{{\rm d}\left(\boldsymbol{\lambda}_1\cdot\boldsymbol{\lambda}_2\right)}
{\left(\boldsymbol{\lambda}_1\cdot\boldsymbol{\lambda}_2\right)^2-
\left\|\boldsymbol{\lambda}_1\right\|^2\left\|\boldsymbol{\lambda}_2\right\|^2}
=-
2\,\Big(H_{\mathbf{m}}(x_2-x_1)-G_{\mathbf{m}}({x}_2-{x}_1)\Big)\,{\rm d}t
\]
Upon assuming that initially $\boldsymbol{\lambda}_1(0)\cdot\boldsymbol{\lambda}_2(0)=0$,
one writes the following solution

\[
\boldsymbol{\lambda}_1\cdot\boldsymbol{\lambda}_2
=
\left\|\boldsymbol{\lambda}_1\right\|\left\|\boldsymbol{\lambda}_2\right\|\,\tanh\!
\left(
2 \left\|\boldsymbol{\lambda}_1\right\|\left\|\boldsymbol{\lambda}_2\right\|
\int_0^t\!
\Big(H_{\mathbf{m}}(x_2-x_1)-G_{\mathbf{m}}({x}_2-{x}_1)\Big)\,{\rm d}t'
\right)
\]
so that $\boldsymbol{\lambda}_1\cdot\boldsymbol{\lambda}_2$ has a definite
positive sign (if $\boldsymbol{\lambda}_1\cdot\boldsymbol{\lambda}_2>0$
at $t=0$). Consequently the expression (\ref{diff-x})  has also a positive
definite sign and thus if $G$ and $H$ are such that $H_{\mathbf{m}}>G_{\mathbf{m}}$, then the two particles indefinitely approach each other and align.
\end{proof}
\end{framed}

\begin{framed}
\begin{corollary}
When $H_{\mathbf{m}}<G_{\mathbf{m}}$, in the particular case $G_{\mathbf{m}}=G_\rho$
and $\left\|\boldsymbol{\lambda}_1\right\|\left\|\boldsymbol{\lambda}_2\right\|\leq1$,
then the clumpons tend to clump and anti-align  asymptotically in time. 
\end{corollary}
\begin{proof}
One has
\begin{multline*}
2\,\Big(1+H_\rho({x}_2-{x}_1)\Big)\,+\,2\,
\boldsymbol{\lambda}_2\cdot\boldsymbol{\lambda}_1+\left(\left\|\boldsymbol{\lambda}_2\right\|^2+\left\|\boldsymbol{\lambda}_1\right\|^2\right)\,H_{\mathbf{m}}({x}_2-{x}_1)>
\\
2\left(1+\boldsymbol{\lambda}_2\cdot\boldsymbol{\lambda}_1\right)\geq2\left(1-\left\|\boldsymbol{\lambda}_1\right\|\left\|\boldsymbol{\lambda}_2\right\|\right)
\end{multline*}
which is positive by hypothesis. By comparing with equation (\ref{diff-x}),
one finds that the expression in (\ref{diff-x}) is negative definite in sign,
so that $\left\|\boldsymbol{\lambda}_1\right\|\left\|\boldsymbol{\lambda}_2\right\|\leq1$
becomes a sufficient condition for clumping and the thesis is proven.
\end{proof}
\end{framed}

\begin{figure}[!]\label{lennon_png}
\center
\includegraphics[scale=0.195]{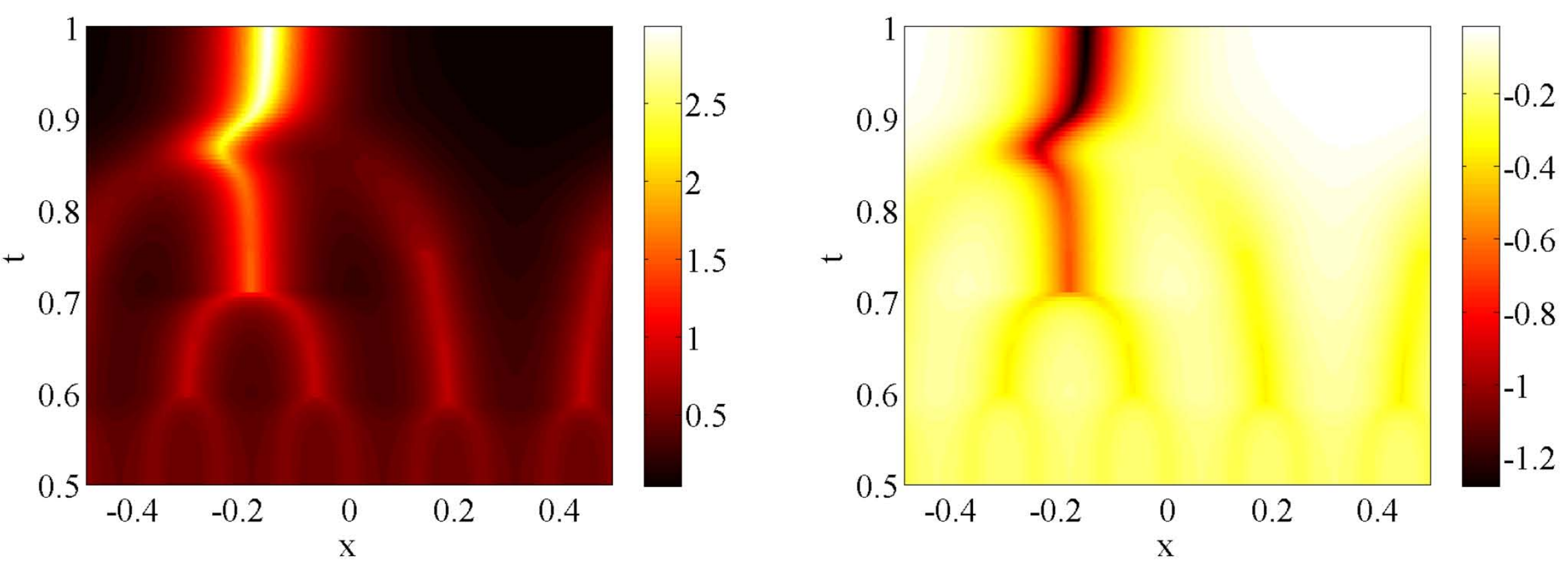}
\caption{Evolution of a flat magnetization field and a sinusoidally-varying
density. Subfigure~(a) shows the evolution of $\bar\rho=H*\rho$ for $t\in\left[0.5,1\right]$;
 (b) shows the evolution of $\bar{\bf m}_x$.  The profiles of $\bar{\bf m}_y$ and $\bar{\bf m}_z$  are similar.  At $t=0.5$, the initial data have formed eight equally spaced,  identical clumpons, corresponding to the eight density maxima in the initial configuration.  By impulsively shifting the clumpon
 at $x=0$ by a small amount, the equilibrium is disrupted and the clumpons
 merge repeatedly until only one clumpon remains. (Figures by L. \'O N\'araigh)
}
\end{figure}

\subsection{Higher dimensional treatment}
\begin{figure}[!]\label{orientation2D}
\center
\includegraphics[scale=0.7]{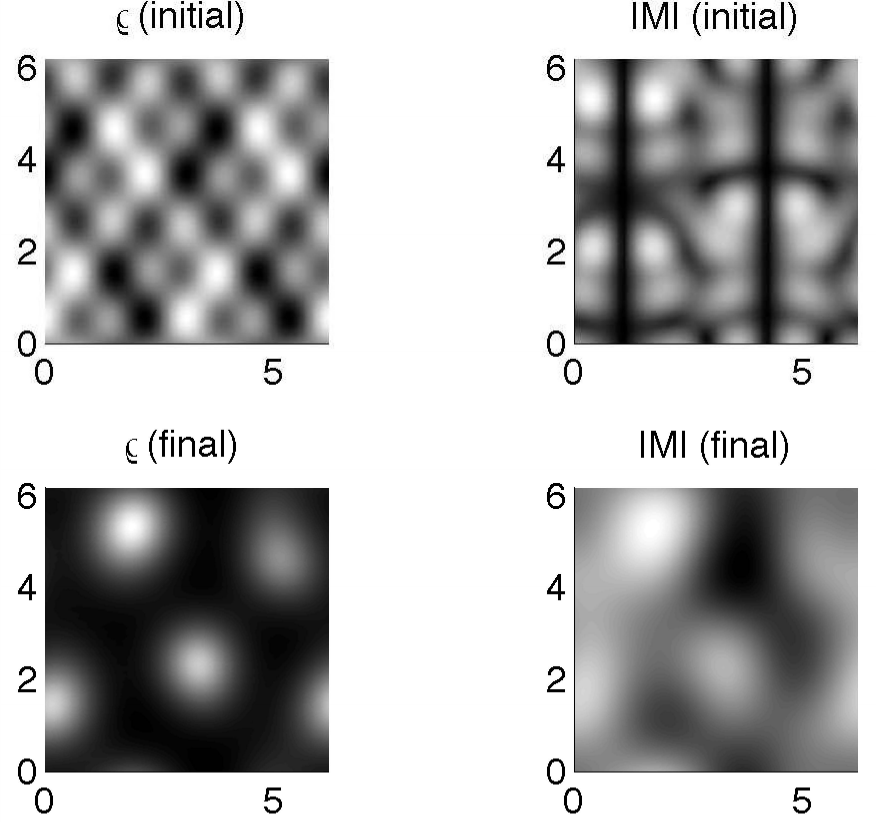}
\caption{Spontaneous emergence of clumped states in two dimensions. Random initial conditions break up into dots. The expression for the energy 
interaction is $E=1/2 \int H({\bf x-y}) \left( \rho({\bf x}) \rho({\bf y})+ \mathbf{m}({\bf x}) \cdot
\mathbf{m}({\bf y})\right)$, where $H({\bf x-y})=e^{-\left|{\bf x-y}\right|}$. The left plot shows the smoothed density $\bar\rho$,
while right plot shows the modulus $\left|\overline{\mathbf{m}}\right|$ (Figure by V. Putkaradze).}
\end{figure}

\begin{figure}[!]\label{filaments}
\center
\includegraphics[scale=0.85]{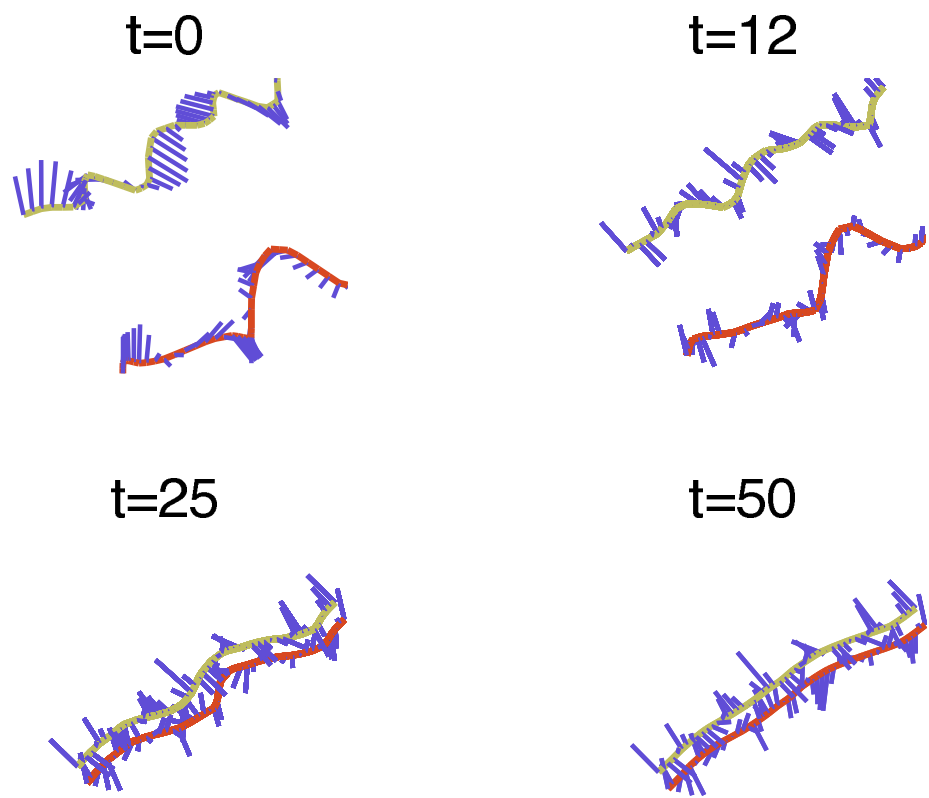}
\caption{An example of two oriented filaments (red and green) attracting each other and unwinding at the same time.  The  blue vectors illustrate the vector ${\bf m}$ at each point on the curve. Time scale is arbitrary.
(Figure by V. Putkaradze)}
\end{figure}

Although this chapter formulates a one dimensional treatment, the possibility
of going to higher dimensions is pretty clear by looking at the moment equations,
by remembering that ${\sf ad}^*_{\gamma_{1,0}}=\pounds_{\gamma_{1,0}}$. 
If one
specializes to the case of $\mathfrak{so}(3)$, then it is possible to write
the moment equations as
\begin{align*}
\frac{\partial \rho}{\partial t}
&=
\text{\rm\large div}
\Bigg(
\rho\,
\bigg(
\mu_\rho\, \text{\large$\nabla$}\frac{\delta E}{\delta \rho}
+\,
\boldsymbol\mu_{\bf m}\cdot\text{\large$\nabla$}\frac{\delta E}{\delta \bf m}
\bigg)
\,
\Bigg)\\
\frac{\partial {\bf m}\,}{\partial t}
&=
\text{\rm\large div}\Bigg({\bf m}\,\text{\large$\otimes$}
\left(\mu_\rho\text{\large$\nabla$}\frac{\delta E}{\delta \rho}
+
\boldsymbol\mu_{\bf m}
\cdot\text{\large$\nabla$}\frac{\delta E}{\delta {\bf m}}
\right)
 \Bigg)\!
+
{\bf m}\times
\boldsymbol{\mu}_{\bf m}\times\frac{\delta E}{\delta \bf m}
\end{align*}
These equations also possess singular solutions of the form \cite{HoPuTr08}
\begin{align*}
\rho(\boldsymbol{q},t)&=\int\delta(\boldsymbol{q}-{\bf Q}(s,t))\,{\rm d}s\\
{\bf m}(\boldsymbol{q},t)&=\int{\boldsymbol w}_{_\text{\!\tiny$\bf m$}}(s,t)\,\,\delta(\boldsymbol{q}-{\bf Q}(s,t))\,{\rm d}s
\end{align*}
where $s$ is a coordinate on a submanifold of $\mathbb{R}^3$. If $s$ is a
one-dimensional curvilinear coordinate, then this solution represents an {\it orientation filament} (see fig.~\ref{filaments}), while for $s$ belonging to a two-dimensional surface one obtains an {\it orientation sheet}. The
simple case of particle-like solutions (\ref{singansatz}) is still possible in higher dimensions and fig.~\ref{orientation2D} shows their spontaneous emergence in
two dimensions.

\paragraph{The ${p}\bg$-moment bracket in more dimensions.} Upon following the tensorial
interpretation for the moments established in Chapter~\ref{momLPdyn}, it is
possible to calculate the higher dimensional version for the moment bracket.
In addition the tensorial interpretation also provides the hint to calculate
moments of the general type $A_{m,k}=\int {\bf p}^m\,\boldsymbol{g}^k\,f({\bf
q,p},\bg,t)\, {\rm d}^N{\bf q}\,{\rm d}^N{\bf p}\,{\rm d}^N\bg$, where powers
have to be intended as tensor powers, so that
\begin{equation}\label{GHKmom}
A_{n,k}({\bf q}, t)=\int_\mathcal{D} \otimes^{n}{\bf p}\,\otimes^{k\!\!}{\bg}\,\, f({\bf q,
p},\bg, t)\,{\rm d}^3{\bf q}\wedge{\rm d}^3{\bf p}\wedge{\rm d}^3{\bf \bg}
\end{equation}
where $\mathcal{D}=T^*_q Q\times\mathfrak{g}$.
This can be written in terms of the basis as
\begin{align*}
A_{n,k}({\bf q}, t)&=\int_\mathcal{D} \left(p_i\,{\rm d}q^i\right)^n\,\left(g_a\,{\bf e}^a\right)^{k\,} f({\bf q,
p},\bg, t)\,{\rm d}^3{\bf q}\wedge{\rm d}^3{\bf p}\wedge{\rm d}^3{\bf \bg}
\\
&=\int_\mathcal{D} p_{\,i_1}\!\dots p_{\,i_n} \,{\rm d}q^{i_1}\!\dots{\rm d}q^{i_n}\,
g_{a_1\!}\dots g_{a_k} \,{\bf e}^{a_1\!}\dots {\bf e}^{a_k}\,
f({\bf q,
p},\bg, t)\,{\rm d}^3{\bf q}\wedge{\rm d}^3{\bf p}\wedge{\rm d}^3{\bf \bg}
\\
&=
\left(A_{n,k}({\bf q}, t)\right)_{i_1,\dots, i_n,\,a_1,\dots, a_k}
\,{\rm d}q^{i_1}\dots{\rm d}q^{i_n}\,{\bf e}^{a_1\!}\dots {\bf e}^{a_k}\,{\rm d}^3{\bf q}
\end{align*}
In order to find the higher dimensional moment Lie-Poisson bracket one can follow the same steps as in Chapter~\ref{momLPdyn}
\begin{align*}
\{\,G\,,\,H\,\}
=&
\int\hspace{-3mm}\int\hspace{-3mm}\int \!f \,
\Big\{\alpha_{m,h}({\bf q})\contract{\bf p}^m\otimes{\bg}^{h},\,\,\beta_{n,k}({\bf
q})\contract{\bf p}^n\otimes{\bg}^{k}\Big\}_1
\,{\rm d}^3{\bf q}\wedge{\rm d}^3{\bf p}\wedge{\rm d}^3{\bg}
\\
=&
\int\hspace{-3mm}\int\hspace{-3mm}\int\! f\, \bg^{k+h\!} \,
\bigg(\,p_{i_1}\dots p_{i_m}\frac{\pa \left(\alpha_m\right)^{i_1,\dots,i_m}}{\pa q^k}\frac{\pa\, p_{j_1\!}\dots p_{j_n}}{\pa p_k}\,\left(\beta_n\right)^{j_1,\dots,j_n}
\\
&\hspace{2.37cm}-
p_{j_1}\dots p_{j_n}\frac{\pa \left(\beta_n\right)^{j_1,\dots,j_n}}{\pa q^h}\frac{\pa\, p_{i_1\!}\dots p_{i_m}}{\pa p_{h}}\,\left(\alpha_m\right)^{i_1,\dots,i_m}
\bigg)
\,{\rm d}^3{\bf q}\wedge{\rm d}^3{\bf p}\wedge{\rm d}^3{\bg}
\\
&+\!
\int\hspace{-3mm}\int\hspace{-3mm}\int\!
f\left\langle\bg,\left[\frac{\pa}{\pa\bg}\Big(\alpha_{m,h}({\bf q})\contract{\bf p}^m\otimes{\bg}^{h}\Big),\,\frac{\pa}{\pa\bg}\Big(\beta_{n,k}({\bf q})\contract{\bf p}^n\otimes{\bg}^{k}\Big)\right]\right\rangle{\rm d}^3{\bf q}\wedge{\rm d}^3{\bf p}\wedge{\rm d}^3{\bg}
\\
=&\,
\int\!
A_{m+n-1,k+h\,}\,
\Big[\alpha_{m,h},\,\beta_{n,k}\Big]
{\rm\, d}^3{\bf q}
\\
&\hspace{0.63cm}+
\int\hspace{-3mm}\int\hspace{-3mm}\int\!
f\,{\bf p}^{m+n}\alpha_{m,h}({\bf q})\,\beta_{n,k}({\bf q})
\left(g_d\,C^d_{bc}\,\frac{\pa g_{a_1}\dots g_{a_h}}{\pa g_b}\,\frac{\pa g_{l_1}\dots g_{l_k}}{\pa
g_c}\right)
{\rm d}^3{\bf q}\wedge{\rm d}^3{\bf p}\wedge{\rm d}^3{\bg}
\\
=&
\int\!
A_{m+n-1,k+h\,}\,
\Big[\alpha_{m,h},\,\beta_{n,k}\Big]
{\rm\, d}^3{\bf q}
\\
&\hspace{0.63cm}+
\int\hspace{-3mm}\int\hspace{-3mm}\int\!
f\,{\bf p}^{m+n}\left(\alpha_{m,h}\right)^{a_1,\dots,a_{h-1},b}\,
\left(\beta_{n,k}\right)^{l_1,\dots,l_{k-1},c}
\\
&\hspace{4.8cm}
\Big(hk\,g_d\,\,C^d_{bc}\,\,g_{a_1}\dots g_{a_{h-1}}\,g_{l_1}\dots g_{l_{k-1}}\Big)
{\rm d}^3{\bf q}\wedge{\rm d}^3{\bf p}\wedge{\rm d}^3{\bg}
\\
=&
\int\!
A_{m+n-1,k+h\,}\,
\Big[\alpha_{m,h},\,\beta_{n,k}\Big]
{\rm\, d}^3{\bf q}
\\
&\hspace{0.63cm}+h\,k
\int\hspace{-3mm}\int\hspace{-3mm}\int\!
f\,{\bf p}^{m+n}
\Big(g_{a_1}\dots g_{a_{h+k-1}}\Big)
\\&
\hspace{3.87cm}
\,\left(\alpha_{m,h}\right)^{a_1,\dots,a_{h}}\,C^{^\text{\scriptsize\,$a_{h+k-1}$}}_{a_h\,\,a_{h+k}}\,
\left(\beta_{n,k}\right)^{a_{h+1},\dots,a_{h+k}}\,
{\rm d}^3{\bf q}\wedge{\rm d}^3{\bf p}\wedge{\rm d}^3{\bg}
\\
=&\,
\bigg\langle
A_{m+n-1,k+h\,},
\Big[\alpha_{m,h},\,\beta_{n,k}\Big]
\bigg\rangle
+h\,k\,\Big\langle
A_{\,m+n,\,h+k-1}\,,\,
C\contract \,\alpha_{m,h}\otimes\beta_{n,k}
\Big\rangle
\\
=&\!\!:\!
\bigg\langle
A_{m+n-1,k+h\,},
\Big[\alpha_{m,h},\,\beta_{n,k}\Big]
\bigg\rangle
+
\Big\langle
A_{\,m+n,\,h+k-1}\,,\,
\Big[\alpha_{m,h},\,\beta_{n,k}\Big]_\mathfrak{\!g\,}
\Big\rangle
\end{align*}

\smallskip
In conclusion, one summarizes in the following
\begin{framed}
\begin{proposition}
The moments defined in equation (\ref{GHKmom}) are symmetric contravariant
$\,\makebox{$n+k$}\,$-tensors defined on $\otimes^{n\,} T_q Q\otimes^{k\!}\mathfrak{g}$.
These quantities undergo Lie-Poisson dynamics, whose Poisson bracket is given
by the following expression
\begin{multline*}
\left\{F,G\right\}=
\left\langle
A_{m+n-1,h+k\,},\left[ n
\left(\frac{\delta E}{\delta A_{n,k}}\contract\,\nabla\right)\frac{\delta F}{\delta A_{m,h}}
-
m
\left(\frac{\delta F}{\delta A_{m,h}}\contract\,\nabla\right)\frac{\delta E}{\delta A_{n,k}}\right]
\right\rangle
\\
+
\left\langle
A_{\,m+n,\,h+k-1}\,,\,
C\contract\left(h\,\frac{\delta F}{\delta A_{m,h}}\otimes k\,\frac{\delta E}{\delta A_{n,k}}\right)
\right\rangle
\end{multline*}
where $C$ is the structure tensor of $\mathfrak{g}$ and summation over all
indexes is intended.
\end{proposition}
\end{framed}

It is easy to see that if one considers only $(m,h),(n,k)\in\{(0,0),(0,1),(1,0)\}$,
then one recovers the GHK bracket for chromohydrodynamics \cite{GiHoKu1982,GiHoKu1983}.


\section{A higher order of approximation}
\label{higher}
This section extends the previous moment equations to include also higher
order moments, which represent higher accuracy in the model. In this treatment
the equations do not close exactly and one needs to formulate a suitable
closure, such as the cold-plasma closure. Remarkably, this closure uniquely
determines the moment dynamics and does not require any other hypothesis
on the model. This introduction of a momentum dynamical variable arises in a natural way and its equation can be written in terms of the other moments
without introducing further higher terms.
\subsection{Moment dynamics}
The starting equation is (\ref{moments1})
\begin{align}
\frac{\partial A_{\,0,\lambda}}{\partial t}=
&\,
\textsf{\large ad}_{\,\gamma_{1,\nu}^\sharp}^* \,g_\nu \, g_\lambda \,A_0 \,{\rm d}g
+
\int g_\lambda \left\langle g,\left(\,
\left[\frac{\partial A_{1}}{\partial g},\frac{\partial (g_\sigma\,{\gamma}_{1,\sigma}^{\,\,\sharp})}{\partial g}\right]+\left[\frac{\partial A_{0}}{\partial g},\frac{\partial (g_\sigma\,{\gamma}_{0,\sigma}^{\,\,\sharp})}{\partial g}\right]\,\right)\right\rangle{\rm d}g
\nonumber
\\
=&\,
\frac{\partial}{\partial q}\left({\gamma_{1,\nu}^\sharp}\int \!g_\nu \, g_\lambda \,A_0 \,{\rm d}g\right)
\!+\!
\int g_\lambda \left\langle g,\left(\,
\left[\frac{\partial A_{1}}{\partial g},\frac{\partial (g_a\,{\gamma}_{1,a}^{\,\,\sharp})}{\partial g}\right]+\left[\frac{\partial A_{0}}{\partial g},\frac{\partial (g_a\,{\gamma}_{0,a}^{\,\,\sharp})}{\partial g}\right]\,\right)\right\rangle{\rm d}g \, , 
\label{moments}
\end{align}
where one defines the analogues of Darcy's velocities: 
\begin{align*}
\gamma_{0,\nu}:=\mu[f]\,\,\text{\Large$\star$}_{0,\nu}\,\,\frac{\delta E}{\delta f}
&=
\int \!g_\nu
\left\langle g,\left[\frac{\partial \widetilde\mu_{k}}{\partial g},\frac{\partial (g_a\,\beta_k^a)}{\partial g}\right]\right\rangle{\rm d}g
\\
&=\int \!g_\nu
\left\langle g,\left[\frac{\partial \widetilde\mu_{0}}{\partial g},\frac{\partial (g_a\,\beta_0^a)}{\partial g}\right]\right\rangle{\rm d}g
\end{align*}
and
\begin{align*}
\gamma_{1,\nu}:=\mu[f]\,\,\text{\Large$\star$}_{1,\nu}\,\,\frac{\delta E}{\delta f}
&=\,
{\sf ad}_{\beta_k^{\,\sigma}}^*
\int g_\nu \,g_\sigma \,\widetilde\mu_{k} \,{\rm d}g
+
\int \!g_\nu
\left\langle g,\left[\frac{\partial \widetilde\mu_{k+1}}{\partial g},\frac{\partial (g_a\,\beta_k^a)}{\partial g}\right]\right\rangle{\rm d}g
\\
&=\,
\frac{\,\,\partial \beta_0^{\,\sigma}}{\partial q}
\int \!g_\nu \, g_\sigma \,\widetilde\mu_{0} \,{\rm d}g
+
\int g_\nu
\left\langle g,\left[\frac{\partial \widetilde\mu_{1}}{\partial g},\frac{\partial (g_a\,\beta_0^a)}{\partial g}\right]\right\rangle{\rm d}g
\,.
\end{align*}
Here the only assumption is that the energy functional $E$ depends only on $A_{0,\lambda}$
(recall that $\beta_n^\lambda:=\delta E/\delta A_{n,\lambda}$), so that it
is possible to fix $k=0$ in the first line of the equation above. 

\noindent
It is now convenient to introduce the following notation 
\begin{eqnarray*}
\rho&=&\int f \,{\rm d}g\,{\rm d}p\,,
\qquad
\quad\,\,
G\,\,\,\,=\,\,\,\int g\, f\,{\rm d}g\,{\rm d}p\,,
\\
J&=&\int p\,g\, f\,{\rm d}g\,{\rm d}p\,,
\qquad
\bar{T}\,\,\,\,=\,\,\,\int gg\, f\,{\rm d}g\,{\rm d}p
\,.
\end{eqnarray*}
and analogously for the \emph{mobilities}
\begin{eqnarray*}
\mu_\rho&=&\int \mu[f] \,{\rm d}g\,{\rm d}p\,,
\qquad
\quad
\mu_G\,\,\,\,=\,\,\,\int g\, \mu[f]\,{\rm d}g\,{\rm d}p\,,
\\
\mu_J&=&\int p\,g\, \mu[f]\,{\rm d}g\,{\rm d}p\,,
\qquad
\bar{K}\,\,\,\,=\,\,\,\int gg\, \mu[f]\,{\rm d}g\,{\rm d}p\,.
\end{eqnarray*}
where $\,gg:=g_a \,g_b \,{\bf e}^a\otimes{\bf e}^b$ and $\,\bar{K}:=\bar{K}_{ab}\,\,{\bf e}^a\otimes{\bf e}^b\,$. In terms of these quantities, one may write the following.
\begin{theorem}\label{momeqns-thm}
The moment equations for $\rho$ and $G$ are given by
\begin{align}
\frac{\partial \rho}{\partial t}=&\,\,
\frac{\partial}{\partial q}\Bigg(\rho\,\,
\bigg(
\mu_\rho\,\, \frac{\partial}{\partial q}\frac{\delta E}{\delta \rho}
+
\bigg\langle \mu_G, \,\frac{\partial}{\partial q}\frac{\delta E}{\delta G}\bigg\rangle
\bigg)
\,\,
\Bigg)
\nonumber 
\\
&+
\frac{\partial}{\partial q}\left\langle G,\,
\mu_G^\sharp\,\, \frac{\partial}{\partial q}\frac{\delta E}{\delta \rho}
+
\left({\rm ad}^*_\frac{\delta E}{\delta G} \,\mu_J\right)^\sharp
+
\left(\bar{K}\cdot\frac{\partial}{\partial q}\frac{\delta E}{\delta G}\right)^\sharp
\,
\right\rangle
\label{rhogen}
\end{align}
and
\begin{align}
\frac{\partial G}{\partial t}=&\,\,
\frac{\partial}{\partial q}\Bigg(G\left(\mu_\rho\,\, \frac{\partial}{\partial q}\frac{\delta E}{\delta \rho}+
\left\langle \mu_G, \,\frac{\partial}{\partial q}\frac{\delta E}{\delta G}\right\rangle\right) \Bigg)
\nonumber
\\
&+
\frac{\partial}{\partial q}\left(\bar{T}\cdot
\bigg(
\mu_G^\sharp\,\, \frac{\partial}{\partial q}\frac{\delta E}{\delta \rho}
+
\left(\bar{K}\cdot\frac{\partial}{\partial q}\frac{\delta E}{\delta G}\right)^\sharp
+
\left({\rm ad}^*_\frac{\delta E}{\delta G} \,\mu_J\right)^\sharp
\bigg)\right)
\nonumber
\\
&+\text{\rm\large ad}^*_{
\left(\!
\mu_G\, \frac{\partial}{\partial q}\!\frac{\delta E}{\delta \rho}
\,+\,
\bar{K}\cdot\frac{\partial}{\partial q}\!\frac{\delta E}{\delta G}
\,+\,{\rm ad}^*_\frac{\delta E}{\delta G} \,\mu_J\!
\right)^{\!\!\sharp}
}\,J
+ {\rm ad}^*_{\left(\!{\rm ad}^*_{\frac{\delta E}{\delta G}} \mu_G\!\right)^{\!\!\sharp}}\,G
\label{Ggen}
\end{align}
where the symbol $(\,\cdot\,)$ stands for contraction in the Lie algebra, for example $(\bar{T}\cdot\Gamma)_a:=\bar{T}_{ab}\,\Gamma^b$.
\end{theorem}
The proof of this theorem is given in Section \ref{appB2}.

\begin{remark}
In  equation (\ref{Ggen}), the tensor $\bar T$ plays the role of a \emph{Lie algebra-pressure
tensor} which generates the second advection term in the equation for $G$,
exactly as it happens for the ordinary pressure tensor in the motion of compressible fluids.
Moreover, one can see that the last term in the equation for $G$ is a dissipative
term, which involves only quantities in the (dual) Lie algebra and does
not introduce any further advection term in space. This
term generalizes the Landau-Lifschitz dissipation in $\mathfrak{so}(3)$ to
any Lie algebra $\mathfrak{g}$.
\end{remark}

These equations need a suitable closure, obtained, for example, 
by expressing the unknown quantities $\bar{T}$, $\bar{K}$, and $J$ in terms of the dynamical variables $\rho$ and $G$.  This can be done by using the {\bfi cold plasma formulation}. (Other closures would also be possible, but these are not considered here). In this way, one can easily find a closure for $\bar T$ and $\bar K$. However, this is not enough for the closure of the flux $J$, which instead will be determined by the evolution equation for the first order moments.

\subsection{Cold plasma formulation and moment closure}

The cold-plasma solution of the Vlasov equation is given by the following product of delta functions in momentum and orientation, 
\begin{equation} 
f(q,p,t)=\rho(q,t)\,\delta(p-\bar{p}(q,t))\,\delta(g-\bar{g}(q,t)) \, , 
\label{coldplasma} 
\end{equation}
so that (with indices suppressed)
\begin{eqnarray*}
G=\rho\,\bar{g}\,,
\qquad
J=G\,\bar{p}\,,
\qquad
\bar{T}=\frac1\rho\,GG\,,
\end{eqnarray*}

It remains to model the phase space mobility $\mu[f]$ appropriately. One possibility would be to take $\mu[f]=\mu_\rho(q,t)\,\delta(p-\mu_{\bar{p}}(q,t))\,\delta(g-\mu_{\bar{g}}(q,t))$
so that
\begin{eqnarray*}
\mu_G=\mu_\rho\,\mu_{\bar{g}}\,,
\qquad
\mu_J=\mu_G\,\mu_{\bar{p}}\,,
\qquad
\bar{K}=\frac1\mu_{\!\rho}\mu_G\,\mu_G\, . 
\end{eqnarray*}
In this case the equation for $\rho$ is
\begin{equation}
\frac{\partial \rho }{\partial t}=
\frac{\partial}{\partial q}\Bigg(\rho\,\bigg(1+\bigg\langle \frac{G}\rho,\,
\frac{\mu_G^\sharp}{\mu_\rho}
\bigg\rangle\bigg)\!
\bigg(
\mu_\rho\,\, \frac{\partial}{\partial q}\frac{\delta E}{\delta \rho}
+
\bigg\langle \mu_G, \,\frac{\partial}{\partial q}\frac{\delta E}{\delta G}\bigg\rangle
\bigg)
+
\left\langle G,\,
\left({\rm ad}^*_\frac{\delta E}{\delta G} \,\mu_J\right)^{\!\sharp}\,
\right\rangle
\Bigg)
\end{equation}
Similarly, one finds the following equation for the macroscopic orientation $G$: 
\begin{align}
\hspace{-0.4cm}
\frac{\partial G}{\partial t}=&\,\,
\frac{\partial}{\partial q}\left(G\,\bigg(1+\bigg\langle \frac{G}\rho,\,
\frac{\mu_G^\sharp}{\mu_\rho}
\bigg\rangle\bigg)\!\!
\left(\mu_\rho\,\, \frac{\partial}{\partial q}\frac{\delta E}{\delta \rho}+
\left\langle \mu_G, \,\frac{\partial}{\partial q}\frac{\delta E}{\delta G}\right\rangle\right)
+
\frac{G}\rho\left\langle G,
\left({\rm ad}^*_\frac{\delta E}{\delta G} \,\mu_J\right)^{\!\sharp} 
\right\rangle\,
\right)
\nonumber 
\\
&+
\left(\mu_\rho\,\frac{\partial}{\partial q}\frac{\delta E}{\delta \rho}\,
+
\left\langle \mu_G,\,\frac{\partial}{\partial q}\frac{\delta E}{\delta G}\right\rangle
\right)
\textrm{\large ad}^*_{
\frac{\mu_G^{\sharp}}{\mu_\rho}
}\,\,
J
+\,
\textrm{\large ad}^*_{
\left({\rm ad}^*_\frac{\delta E}{\delta G} \,\mu_J\right)^{\!\sharp}
}\,\,J
+ {\rm ad}^*_{\left(\!{\rm ad}^*_{\frac{\delta E}{\delta G}} \mu_G\!\right)^{\!\!\sharp}}\,G
\end{align}
These equations comprise the cold-plasma closure of the exact (but incomplete)  equations  (\ref{rhogen},\ref{Ggen}).  To complete the process, one needs to find  a closure for the Lie algebra-valued flux $J$. This closure arises very naturally from the moment equation
for $A_{1,\lambda}$.

\begin{remark}
It is worth to emphasize that the cold plasma approximation and the linearity Assumption \ref{linassump} are sufficient for complete closure of the system. No additional assumptions will be needed. 
\end{remark}

From (\ref{bracket1}) one deduces that
\begin{align*}
\frac{\partial A_{1,\lambda}}{\partial t}=&\,\,
{\sf ad}_{\gamma_{0,\nu}^{\,\sharp}}^*\int g_\nu \, g_\lambda \,A_0 \,{\rm d}g
+
{\sf ad}_{\gamma_{1,\nu}^{\,\sharp}}^*\int g_\nu \, g_\lambda \,A_1 \,{\rm d}g
\\
&+
\,\int g_\lambda \left\langle g,\left(
\left[\frac{\partial A_{2}}{\partial g},\frac{\partial (g_a\,\gamma_{1,a}^{\,\sharp})}{\partial g}\right]
+\left[\frac{\partial A_{1}}{\partial g},\frac{\partial (g_a\,{\gamma}_{0,a}^{\,\sharp})}{\partial g}\right]\right)
\right\rangle\,{\rm d}g \, . 
\end{align*}
In the particular case $\lambda=0$, the equation is written as
\begin{align*}
\frac{\partial A_{1,0}}{\partial t}&=
\,
{\sf ad}_{\gamma_{0,\nu}^{\,\sharp}}^*A_{0,\nu}
+
{\sf ad}_{\gamma_{1,\nu}^{\,\sharp}}^*A_{1,\nu}
\\&
=
\,
A_{0,\nu}\,\,\frac{\!\partial_{\,} \gamma_{0,\nu}^{\,\sharp}}{\partial q}
\,+\,
\textit{\large\pounds}_{\!\gamma_{1,\nu}^{\,\sharp}}A_{1,\nu}
=
A_{0,\nu}\,\diamond\,\gamma_{0,\nu}^{\,\sharp}
\,+\,
\textit{\large\pounds}_{\!\gamma_{1,\nu}^{\,\sharp}}A_{1,\nu}
\end{align*}
Now, from the cold plasma approximation (\ref{coldplasma}) one obtains
\[
A_{1,0}=\int \!p\, \rho\,\delta(p-\bar{p})\,\delta(g-\bar{g})\,{\rm d}p \,{\rm d}g
=\rho\,\bar{p} \, . 
\]
Physically, the quantity $A_{1,0}=\rho\,\bar{p}=:M$ is the macroscopic momentum. Since $\gamma_{0,0}=0$, then the evolution equation for $M$ is 
\begin{align*}
\frac{\partial M}{\partial t}=&\,\,
A_{0,a}\,\,\frac{\!\partial_{\,} \gamma_{0,a}^{\,\sharp}}{\partial q}
+
\text{\large\it\pounds}_{\gamma_{1,0}^{\sharp}}M
+
\text{\large\it\pounds}_{\gamma_{1,a}^\sharp}J_a
\\
=&\,
\left\langle G,\,\frac{\partial}{\partial q}\!
\left({\rm ad}^*_\frac{\delta E}{\delta G}\,\mu_G\right)^{\!\sharp}\right\rangle
+
\gamma\,\frac{\partial M}{\partial q}
+
2M\,\frac{\partial \gamma}{\partial q}
+
\left\langle
\frac{\partial J}{\partial q}
,\,\bar\gamma
\right\rangle
+
2\left\langle
J
,\,\frac{\partial \bar\gamma}{\partial q}
\right\rangle
\end{align*}
where  the following notation is introduced
\begin{align*}
\gamma:\!\!&=\gamma_{1,0}^{\sharp}=\mu_\rho\,\frac{\partial}{\partial q}\frac{\delta E}{\delta \rho}
+
\left\langle
\!\mu_G,\,\frac{\partial}{\partial q}\frac{\delta E}{\delta G}
\right\rangle
\\
\bar\gamma:\!\!&=
\frac{\mu_G^\sharp}{\mu_\rho}
\left(
\mu_\rho\,\frac{\partial}{\partial q}\frac{\delta E}{\delta \rho}
+
\left\langle
\!\mu_G,\,\frac{\partial}{\partial q}\frac{\delta E}{\delta G}
\right\rangle
\right)
+
\left(
{\rm ad}^*_\frac{\delta E}{\delta G} \,\mu_J
\right)^{\!\sharp}
\end{align*}
so that $\bar\gamma^{\,a}=\gamma_{1,a}^{\,\sharp}$.
Now, by the cold  plasma approximation (\ref{coldplasma}), the flux 
$J$ and its corresponding generalized mobility $\mu_J$ may be written as 
\begin{align*} 
J&=\frac{1}\rho\,\,G\,\otimes M
\,, \\ 
\mu_J&=\frac1{\mu_\rho}\,\,\mu_G\otimes\mu_{M} \, . 
\end{align*} 
Thus, the flux of orientation $J$ is associated with an induced mean momentum $M$. 
\begin{framed}\noindent
The final equation for $M$ can be written as
\begin{align}
\frac{\partial M}{\partial t}=&\,\,
\left(
\mu_\rho\,\frac{\partial}{\partial q}\frac{\delta E}{\delta \rho}
+
\left\langle
\!\mu_G,\,\frac{\partial}{\partial q}\frac{\delta E}{\delta G}
\right\rangle
\right)\!
\left(\,
\frac{\partial}{\partial q}
\bigg(\frac{M}\rho
\bigg(
\rho+
\bigg\langle G
,\,
\frac{\mu_G^\sharp}{\mu_\rho}
\bigg\rangle
\bigg)
\bigg)
+\frac{M}\rho\,
\bigg\langle G
,\,
\frac{\partial}{\partial q}
\frac{\mu_G^\sharp}{\mu_\rho}
\bigg\rangle\,
\nonumber 
\right)
\\
&+
2\,\frac{M}\rho\,
\bigg(\rho+
\bigg\langle G
,\,
\frac{\mu_G^\sharp}{\mu_\rho}
\bigg\rangle
\bigg)\,
\frac{\partial}{\partial q}\!
\left(
\mu_\rho\,\frac{\partial}{\partial q}\frac{\delta E}{\delta \rho}
+
\left\langle
\!\mu_G,\,\frac{\partial}{\partial q}\frac{\delta E}{\delta G}
\right\rangle
\right)
\nonumber
\\
&+
\left\langle
\frac{\partial}{\partial q}\bigg(\frac{M}\rho\, G\bigg)
+2\,
\frac{M}\rho\, G
,\,
\left({\rm ad}^*_\frac{\delta E}{\delta G} \mu_G\right)^{\!\sharp}
\right\rangle\,
+\,
\left\langle G,\,\frac{\partial}{\partial q}\!
\left({\rm ad}^*_\frac{\delta E}{\delta G}\,\mu_G\right)^{\!\sharp}\right\rangle \, . 
\label{momentumeq}
\end{align}
This equation for the fluid momentum provides the necessary closure of the system.
The corresponding equations for the density $\rho$ and orientation density  $G$ become
\begin{align}\label{cold-plasma-rho}
\frac{\partial \rho}{\partial t}=&
\frac{\partial}{\partial q}\Bigg(\rho\,\bigg(1+\bigg\langle \frac{G}\rho,\,
\frac{\mu_G^\sharp}{\mu_\rho}
\bigg\rangle\bigg)\!
\bigg(
\mu_\rho\,\, \frac{\partial}{\partial q}\frac{\delta E}{\delta \rho}
+
\bigg\langle \mu_G, \,\frac{\partial}{\partial q}\frac{\delta E}{\delta G}\bigg\rangle
\bigg)
+
\frac{\mu_M}{\mu_\rho}
\left\langle G,\,
\left({\rm ad}^*_\frac{\delta E}{\delta G} \,\mu_G\right)^{\!\sharp}\,
\right\rangle
\Bigg)\\
\label{G-eq}
\frac{\partial G}{\partial t}
=&
\,\,
\frac{\partial}{\partial q}\left(\,G\,\bigg(1+\bigg\langle \frac{G}\rho,\,
\frac{\mu_G^\sharp}{\mu_\rho}
\bigg\rangle\bigg)\!\!
\left(\mu_\rho\,\, \frac{\partial}{\partial q}\frac{\delta E}{\delta \rho}+
\left\langle \mu_G, \,\frac{\partial}{\partial q}\frac{\delta E}{\delta G}\right\rangle\right)
+
\frac{G}\rho\frac{\mu_M}{\mu_\rho}\left\langle G,\,
\left({\rm ad}^*_\frac{\delta E}{\delta G} \,\mu_G\right)^{\!\sharp} 
\right\rangle\,
\right)
\nonumber 
\\
&+
\frac{M}\rho
\left(\mu_\rho\,\frac{\partial}{\partial q}\frac{\delta E}{\delta \rho}\,
+
\left\langle \mu_G,\,\frac{\partial}{\partial q}\frac{\delta E}{\delta G}\right\rangle
\right)
\textrm{\large ad}^*_{
\frac{\mu_G^{\sharp}}{\mu_\rho}
}\,\,
G
+\,\left(1+\frac{M}\rho\frac{\mu_M}{\mu_\rho}\right)\,
\textrm{\large ad}^*_{
\left({\rm ad}^*_\frac{\delta E}{\delta G} \,\mu_G\right)^{\!\sharp}
}\,\,G
\end{align}
\end{framed}
\begin{remark}
The last term in equation (\ref{momentumeq}) is a source of momentum $M$ which must vanish for  $M=0$ to be a steady solution. 
\end{remark}

\medskip

\subsection{Proof of Theorem \ref{momeqns-thm}}
\label{appB2}
The moment equations in Theorem \ref{momeqns-thm} for particles with anisotropic interactions are derived as follows. One starts with equation (\ref{moments})
\begin{align*}
\frac{\partial A_{\,0,\lambda}}{\partial t}=
&\,\,
\frac{\partial}{\partial q}\left({\gamma_{1,\nu}^\sharp}\int \!g_\nu \, g_\lambda \,A_0 \,{\rm d}g\right)
\!+\!
\int g_\lambda \left\langle g,\left(\,
\left[\frac{\partial A_{1}}{\partial g},\frac{\partial (g_a\,{\gamma}_{1,a}^{\,\,\sharp})}{\partial g}\right]+\left[\frac{\partial A_{0}}{\partial g},\frac{\partial (g_a\,{\gamma}_{0,a}^{\,\,\sharp})}{\partial g}\right]\,\right)\right\rangle{\rm d}g
\end{align*}
with velocities given by
\begin{align*}
\gamma_{0,\nu}&=\int \!g_\nu
\left\langle g,\left[\frac{\partial \widetilde\mu_{0}}{\partial g},\frac{\partial (g_a\,\beta_0^a)}{\partial g}\right]\right\rangle{\rm d}g
\\
\gamma_{1,\nu}
&=\,
\frac{\,\,\partial \beta_0^{\,\sigma}}{\partial q}
\int \!g_\nu \, g_\sigma \,\widetilde\mu_{0} \,{\rm d}g
+
\int g_\nu
\left\langle g,\left[\frac{\partial \widetilde\mu_{1}}{\partial g},\frac{\partial (g_a\,\beta_0^a)}{\partial g}\right]\right\rangle{\rm d}g
\,.
\end{align*}
Now, fix $\lambda=0$ in equation (\ref{moments}), so that the equation for
$\rho:=A_{0,0}$ is
\begin{align*}
\frac{\partial \rho}{\partial t}=
&\,\,
\frac{\partial}{\partial q}\left({\gamma_{1,0}^\sharp}\int \! A_0 \,{\rm d}g\right)
+
\frac{\partial}{\partial q}\left({\gamma_{1,a}^\sharp}\int \!g_a \,A_0 \,{\rm d}g\right)
\\
=&
\,\,
\frac{\partial}{\partial q}\left(\gamma\,\rho\right)
+
\frac{\partial}{\partial q}\left\langle G,{\overline\gamma}\right\rangle
\end{align*}
where the other terms in (\ref{moments}) cancel by integration by parts and
one defines $\gamma:=\gamma_{1,0}$ and $\overline\gamma^{\,a}:=\gamma_{1,a}^\sharp$.
For $\gamma_{1,\nu}$ one writes
\[
\gamma_{1,\nu}=
{\sf ad}_\frac{\delta E}{\delta \rho}^*\,\int g_\nu\,\mu_0 \,{\rm d}g
+
\int g_\nu \,\, g_a \,\,{\sf ad}_\frac{\delta E}{\delta G_{\!a}}^*\,\,\mu_0 \,{\rm d}g
+
\int g_\nu
\left\langle g,\left[\frac{\partial \mu_1}{\partial g},\frac{\delta E}{\delta G}\right]\right\rangle{\rm d}g
\]
where $\mu_n\,:=\,\int p^n\,\mu[f]\,dpdg$ and one remembers that $\beta_n^\lambda:=\delta E/\delta A_{n,\lambda}$. Therefore
\begin{align*}
\gamma:=\gamma_{1,0}\,=\,
{\sf ad}_\frac{\delta E}{\delta \rho}^*\,\mu_\rho
+
{\sf ad}_\frac{\delta E}{\delta G_a}^*\int\! g_a\, \mu_{0}\,dpdg
\,=\,
\mu_\rho\,\frac{\partial}{\partial q}\frac{\delta E}{\delta \rho}
+
\left\langle\mu_G,\frac{\partial}{\partial q}\frac{\delta E}{\delta G}\right\rangle
\, . 
\end{align*}
In what follows the following Lemma will be useful.
\begin{lemma}\label{lemmaB}
Let $\mathfrak{g}$ be a finite-dimensional Lie algebra. Given 
 $\eta\in\mathfrak{g}$ and a function $f(g)$ on $\mathfrak{g}^*$, the following holds
\[
\int g \left\langle g,\left[\frac{\partial f}{\partial g},\eta\right]\right\rangle{\rm d}g
=
\text{\rm\large ad}^*_\eta\, G
\]
where $G:=\int g\,f(g)\,{\rm d}g$ and  $g\in\mathfrak{g}^*$. 
\end{lemma}

\begin{proof}
One calculates
\begin{align*}
\int g \left\langle g,\left[\frac{\partial f}{\partial g},\eta\right]\right\rangle{\rm d}g
&=
-
\int g \left\langle \textrm{\large ad}^*_\eta\, g,\frac{\partial f}{\partial g}\right\rangle{\rm d}g
\\
&=
-
\int g \,\,\frac{\partial }{\partial g}\cdot\!\Big( f\,\textrm{\large ad}^*_\eta\, g\Big)\,\,{\rm d}g
=
\int f\,\textrm{\large ad}^*_\eta\,g\,\,{\rm d}g
\end{align*}
where we have used respectively the definition of $\textrm{\large ad}$ and $\textrm{\large ad}^*$, the
Leibnitz rule and the integration by parts. The thesis follows immediately.
\end{proof}

\medskip
\noindent
By using the Lemma above one finds $\overline\gamma$
\[
\overline\gamma:=
\mu_G^\sharp\,\, \frac{\partial}{\partial q}\frac{\delta E}{\delta \rho}
+\left({\rm ad}^*_\frac{\delta E}{\delta G} \,\mu_J\right)^\sharp
+
\left(\bar{K}\cdot\frac{\partial}{\partial q}\frac{\delta E}{\delta G}\right)^\sharp
\,.
\]
Substituting these expressions into the equation for $\rho$ yields the explicit moment equation for the mass density
\begin{align*}
\frac{\partial \rho}{\partial t}=&
\,\,
\frac{\partial}{\partial q}\Bigg(\rho\,\,
\bigg(
\mu_\rho\,\, \frac{\partial}{\partial q}\frac{\delta E}{\delta \rho}
+
\bigg\langle \mu_G, \,\frac{\partial}{\partial q}\frac{\delta E}{\delta G}\bigg\rangle
\bigg)
\,\,
\Bigg)\\
&\,+
\frac{\partial}{\partial q}\left\langle G,\,
\mu_G^\sharp\,\, \frac{\partial}{\partial q}\frac{\delta E}{\delta \rho}
+
\left({\rm ad}^*_\frac{\delta E}{\delta G} \,\mu_J\right)^\sharp
+
\left(\bar{K}\cdot\frac{\partial}{\partial q}\frac{\delta E}{\delta G}\right)^\sharp
\,
\right\rangle
\,.
\end{align*}
Now let $\lambda=a$ in (\ref{moments}). The equation becomes
\begin{align*}
\frac{\partial G_a}{\partial t}
=&
\,\,
{\sf ad}_{\gamma_{1,0}^\sharp}^*\, G_a 
+
{\sf ad}_{\gamma_{1,b}^\sharp}^*\int g_b \, g_a \,A_0 \,{\rm d}g
\\&+
\int g_a \left\langle g,\left(\,
\left[\frac{\partial A_{1}}{\partial g},\frac{\partial (g_b\,{\gamma}_{1,b}^{\,\sharp})}{\partial g}\right]+\left[\frac{\partial A_{0}}{\partial g},\frac{\partial (g_b\,{\gamma}_{0,b}^{\,\sharp})}{\partial g}\right]\,\right)\right\rangle\,{\rm d}g
\end{align*}
which may be written more compactly as
\begin{align*}
\frac{\partial G}{\partial t}=
\frac{\partial}{\partial q}\left(\gamma\, G+\bar{T}\cdot\overline\gamma\right)
+
{\rm ad}^*_{\overline\gamma}\,J
+
{\rm ad}^*_{\overline\Gamma}\,G
\end{align*}
where one uses again the Lemma \ref{lemmaB} and introduces $\overline\Gamma^{\,a}:={\gamma}_{0,a}^{\,\sharp}$.
On the other hand, one has (again by Lemma \ref{lemmaB})
\[
\overline\Gamma=\left({\rm ad}^*_\frac{\delta E}{\delta G}\, \mu_G\right)^\sharp
\]
and substituting into the equation for $G$, one has
\begin{align*}
\frac{\partial G}{\partial t}=&
\,\,\frac{\partial}{\partial q}\Bigg(G\left(\mu_\rho\,\, \frac{\partial}{\partial q}\frac{\delta E}{\delta \rho}+
\left\langle \mu_G, \,\frac{\partial}{\partial q}\frac{\delta E}{\delta G}\right\rangle\right) \Bigg)
\\
&+\frac{\partial}{\partial q}\left(\bar{T}\cdot
\bigg(
\mu_G^\sharp\,\, \frac{\partial}{\partial q}\frac{\delta E}{\delta \rho}
+
\left(\bar{K}\cdot\frac{\partial}{\partial q}\frac{\delta E}{\delta G}\right)^\sharp
+
\left({\rm ad}^*_\frac{\delta E}{\delta G} \,\mu_J\right)^\sharp
\bigg)\right)
\\
&+\textrm{\large ad}^*_{
\left(\!
\mu_G\, \frac{\partial}{\partial q}\!\frac{\delta E}{\delta \rho}
\,+\,
\bar{K}\cdot\frac{\partial}{\partial q}\!\frac{\delta E}{\delta G}
\,+\,{\rm ad}^*_\frac{\delta E}{\delta G} \,\mu_J\!
\right)^{\!\!\sharp}
}\,J
+ {\rm ad}^*_{\left(\!{\rm ad}^*_{\frac{\delta E}{\delta G}} \mu_G\!\right)^{\!\!\sharp}}\,G
\end{align*}
This finishes the derivation of the moment equations in Theorem \ref{momeqns-thm} for particles with anisotropic interactions.

The present treatment has shown how the kinetic moments in Vlasov dynamics
can be extended to include anisotropic interactions. Their Lie-Poisson dynamics
has been found explicitly and different levels of approximations have been
presented for the dynamics of the first moments. The simplest model extends
Darcy's law to anisotropic interactions and allows for singular solutions in
any spatial dimension. A second level of approximation is given by a truncation
of the moment Lie algebra and determines a non trivial dynamics of the fluid
momentum variable. The next section focuses on another kind of moments, which
do not depend only on the spatial coordinate, but also on the orientation.These
moments are commonly known as ``Smoluchowski moments''.

\section{Smoluchowski approach to moment dynamics\label{sec:Smoluchowski} }
This section considers the Smoluchowski approach to the description of the interaction of anisotropic particles. Usually, these particles are assumed to be rod-like, so their orientation can be described by a point on a two-dimensional sphere $S^2$ \cite{DoEd1988}. However, this section considers particles of arbitrary shape, for which one needs the full $SO(3)$ to define their orientation. The next section presents an example of Smoluchowski approach for the corresponding Lie algebra $\mathfrak{so}(3)$, while the later sections
deal with a general finite-dimensional Lie algebra $\mathfrak{g}$.

\subsection{A new GOP-Smoluchowski equation}
A first example of a Smoluchowski approach can be given in geometric terms
as follows. An equation can be formulated for a distribution function on
the $\left(\boldsymbol{x},\,\boldsymbol{m}\right)$-space spanned by position and orientation.
The evolution equation for the distribution $\varphi({\boldsymbol{x},\boldsymbol{m}},t)$ may be written as a conservation form along the velocity $\mathbf {U}=\left(\mathbf{U_x},\mathbf{U_{\boldsymbol{m}}}\right)$  on the  $\left(\boldsymbol{x},\,\boldsymbol{m}\right)$-space. For $\bx\in
\mathbb{R}^3$ one writes
\begin{equation*}
\frac{\partial \varphi}{\partial t}=
-\text{\rm\large div}_{\left(\boldsymbol{x},\,\boldsymbol{m}\right)}
\left(\varphi\,{\bf U}\right)
=-\text{\rm\large$\nabla$}_{\boldsymbol{x}}\cdot
\left(\varphi\,{\bf U}_{\boldsymbol{x}}\right)-
\frac{\partial}{\partial\boldsymbol{m}}\cdot
\left(\varphi\,{\bf U}_{\boldsymbol{m}}\right)
\, . 
\end{equation*}
At this point one has to choose appropriate velocities in order to respect
the nature of Darcy's law (or, more mathematically, GOP theory). A possibility
is to introduce Darcy velocity
\[
{\bf U}_{\boldsymbol{x}}=\dot{\bx}
=\mu[\varphi]\text{\large$\nabla$}\frac{\delta E}{\delta \varphi}
\]
while a suitable choice for $\mathbf{U_m}$ is given by the rigid body
dynamics on $\mathfrak{so}(3)$
\[
{\bf U}_{\boldsymbol{m}}=\dot{\boldsymbol{m}}=
\boldsymbol{m}\times\frac{\partial}{\partial\boldsymbol{m}} \frac{\delta E}{\delta \varphi}
\]
so that the final equation can be written
as
\begin{framed}
\begin{align}\label{GOP-smolu}
\frac{\partial \varphi}{\partial t}
=
\text{\large\rm div}\!\left(\varphi\,\mu[\varphi]\text{\large$\nabla$}\frac{\delta E}{\delta \varphi}\right)
+
\left\{\varphi,\left\{\mu[\varphi],\frac{\delta E}{\delta \varphi}\right\}\right\}
\end{align}
\end{framed}
\noindent
where $\{\cdot,\,\cdot\}$ denotes the rigid body bracket 
$\left\{g,\,h\right\}:=\boldsymbol{m}\,\cdot\partial_{\boldsymbol{m}\,} g\times\partial_{\boldsymbol{m}}h$.
\begin{theorem}
The equation (\ref{GOP-smolu}) is a GOP equation with respect to the direct
sum Lie algebra $\mathfrak{X}(\mathbb{R}^3)\oplus\mathfrak{X}_\textnormal{can}(\mathfrak{so}^*(3))$.
\end{theorem}
\begin{proof}
Consider the action of a vector field $v\in\mathfrak{X}(\mathbb{R}^3)$ on the
density variable $\varphi\in{\rm Den}(\mathbb{R}^3)$:
\[
\boldsymbol{v}\cdot\varphi=\pounds_{\boldsymbol{v}}\varphi=
\text{\rm div}\left(\boldsymbol{v}\varphi\right)
\]
and consider the action of the Hamiltonian function $h\in\mathfrak{X}_\text{can}(\mathfrak{so}^*(3))$ on $\varphi$
\[
h\cdot\varphi={\rm ad}^*_{{\partial h}/{\partial\boldsymbol{m}}}\,\boldsymbol{m}\cdot\frac{\partial \varphi}{\partial
\boldsymbol{m}}=
\left\{h,\,\varphi\right\}
\]
Now consider the action of the direct sum on the densities on the ($\boldsymbol{x},\boldsymbol{m}$)-space
\[
(\boldsymbol{v}\oplus h)\cdot\varphi={\rm div}(\varphi\,\boldsymbol{v})+ {\rm ad}^*_{{\partial h}/{\partial
\boldsymbol{m}}}\,\boldsymbol{m}\cdot\frac{\partial \varphi}{\partial
\boldsymbol{m}}
\]
This is the action of the Lie algebra $\mathfrak{X}(\mathbb{R}^3)\oplus\mathfrak{X}_\text{can}(\mathfrak{so}^*(3))$.
Define the dual action
\[
\langle\varphi\diamond k,\boldsymbol{v}\oplus h\rangle:=\langle k,(\boldsymbol{v}\oplus h)\cdot\varphi\rangle
\]
The GOP equation is defined as
\[
\dot\varphi=\left(\mu[\varphi]\diamond\frac{\delta E}{\delta \varphi}\right)^{\!\!\sharp}\cdot\varphi
\]
By integration by parts, it is easy to see that
\[
\left(\varphi\diamond k\right)^{\sharp}=\varphi\nabla k\cdot\frac{\partial}{\partial \boldsymbol{x}}
+
{\rm ad}^*_{{\partial\left\{\varphi,\,k\right\}}/{\partial \boldsymbol{m}}
}\,\boldsymbol{m}\cdot\frac{\partial}{\partial \boldsymbol{m}}
\]
so that the GOP equation is
\[
\frac{\partial \varphi}{\partial t}
=
\text{\large\rm div}\!\left(\varphi\,\mu[\varphi]\text{\large$\nabla$}\frac{\delta E}{\delta \varphi}\right)
+
\left\{\varphi,\left\{\mu[\varphi],\frac{\delta E}{\delta \varphi}\right\}\right\}
\]
and the thesis is proven.
\end{proof}

Consequently this equation expresses a geometric dissipative flow in the context of GOP-double bracket theory. By the usual arguments, one shows that
when $\delta E/\delta\varphi$ is sufficiently smooth, this equation allows
for singular solutions of the form
\[
\varphi(\boldsymbol{x},{\bf m},t)=\sum_i w_i\int\delta(\boldsymbol{x}-\boldsymbol{Q}_i(s,t))\,\delta({\bf m}-\boldsymbol{\Lambda}_i(s,t))\,\,{\rm d}s
\]
where $w_i$ denotes the weight of the $i$-th particle and $s$ is a coordinate
of a submanifold of $\mathbb{R}^3\times\mathfrak{so}(3)\simeq\mathbb{R}^6$.

The next sections show how a more complete Somulchowski approach can be derived from the dissipative Vlasov equation, which is preferable to the present formulation obtained by ad hoc arguments.

\rem{ 
\paragraph{Singular solutions.\\}
\comment{Write the dynamics of the singular solutions.}
} 

\subsection{Systematic derivation of moment equations}
In the Smoluchowski approach, moments are  defined  as
\[
A_n(q,g):=\!\int p^n\,f(q,p,g)\,dp \, . 
\]
From the general theory, these moments are dual to $\beta_n(q,g)$, which are introduced by expanding the Hamiltonian function
$h(q,p,g)$ as $h(q,p,g)=p^n\,\beta_n(q,g)$. The quantities
$\beta_n$ have a {\bfi Lie algebra bracket} given by 
\[
\left[\!\left[\beta_n,\alpha_m\right]\!\right]_1=
\left[\!\left[\beta_n,\alpha_m\right]\!\right]+\left\langle g,\left[\beta_n^{\,\prime},\alpha_m^{\,\prime}\right]\right\rangle
\,,
\]
where prime denotes partial derivative with respect to $g$ and $[[\cdot,\cdot]]$
denotes the moment Lie bracket. The {\bfi Lie algebra action} is given by
\[
\beta_n\cdot  f=\textit{\large\pounds}_{\!\widehat{X}_{p^n\!\beta_n}}f
\]
where the vector field $\widehat{X}_h$ was defined in Section \ref{sec:Anisotropic}. The {\bfi dual action} is given by
\begin{align*}
\Big\langle f\,\text{\Large$\star$}_{n}\, k,\,\beta_n\Big\rangle:=
\Big\langle f,\, \beta_n\, k\Big\rangle &=
\Big\langle f\!\star k\,,\, p^n \beta_n(q,g) \Big\rangle =
 \left\langle \int p^n\,\{f, k\}_{1}\,dp\,,\,\beta_n \right\rangle
 \,,
\end{align*}
and the {\bfi star operator} is defined  explicitly for $k=p^m\alpha_m$ as 
\begin{align*}
f\,\,\text{\Large$\star$}_{n}\,k
\,=&\,
{\sf ad}_{\alpha_m}^*A_{m+n-1}
+
\left\langle g,\left[\frac{\partial A_{m+n}}{\partial g},\frac{\partial \alpha_m}{\partial g}\right]\right\rangle \, . 
\end{align*}
The {\bfi coadjoint action operator} {\sf ad}$^*$ is the Kupershmidt-Manin
operator defined section \ref{Kup-Man}. One  introduces the {\bfi dissipative bracket} by 
\begin{equation} 
\dot{F}=\{\{F,E\}\}=-\left\langle \mu[f]\,\text{\large$\star$}_{n}\,\frac{\delta E}{\partial f}\,,\,f\,\text{\large$\star$}_{n}\,\frac{\delta F}{\partial f} \right\rangle \, . 
\label{dissbracketSmol}
\end{equation} 
By using this evolution equation for an arbitrary functional $F$, the rate of change for zero-th moment  $A_0$ is found to be
\[
\frac{\partial A_0}{\partial t}={\sf ad}^*_{\gamma_n}A_{n-1}+
\left\{A_n,\,\gamma_n\right\}  
\]
where $\{\cdot,\,\cdot\}$ stands for the Lie-Poisson bracket on $\mathfrak{g}$
\[
\left\{A_n,\,\gamma_n\right\}:=
\left\langle g,\left[\frac{\partial A_n}{\partial g},\frac{\partial \gamma_n}{\partial g}\right]\right\rangle\, .
\]
As usual, summation over repeated indices is assumed, $n\geq 0$.%
One truncates this sum, by taking $n \leq 1$ so that
\begin{align}
\frac{\partial A_0}{\partial t}=\frac{\partial}{\partial
q}\left({\gamma_1}A_{0}\right)+
\left\{
A_0,\,\gamma_0
\right\}
+
\left\{
A_1,\,\gamma_1
\right\}
\label{A0-smolu}
\end{align}
where the Darcy velocities are given by
\[
\gamma_n:=\int p^n \left\{\mu[f],\frac{\delta E}{\delta f}\right\}_1\, {\rm d}p
\,.
\]
In particular, one finds
\begin{align*}
\gamma_0&= \left\{\mu_0,\,\beta_0\right\}
\,,
\\
\gamma_1&={\sf ad}^*_{\beta_0} \,\mu_0+\left\{\mu_1,\,\beta_0\right\}
\,.
\end{align*}

\subsection{A cold plasma-like closure}
To close the system for $A_0$, it is necessary to find an evolution equation for the first moment $A_1$. 
Again, one uses the dissipative bracket (\ref{dissbracketSmol}), and truncates the sum in the bracket to include  $A_0$, $A_1$ and $A_2$ terms. Continuing this procedure to write an equation for $A_k$, would require including 
$A_0,A_1,\dots,A_{k+m}$. Such extensions are possible, but they lead to very cumbersome calculations and there is no clear physical way of justifying the closure. The equation for $A_1$ is the following:  
\begin{align*}
\frac{\partial A_1}{\partial t}&=
{\sf ad}^*_{\gamma_0}A_0+{\sf ad}^*_{\gamma_1}A_1
+
\left\{ A_1,\,\gamma_0\right\}
+
\left\{ A_2,\,\gamma_1\right\}
\\
&=
A_0\frac{\partial \gamma_0}{\partial q}+\pounds_{\gamma_1}A_1
+
\left\{ A_1,\,\gamma_0\right\}
+
\left\{ A_2,\,\gamma_1\right\}
\end{align*}
where  $A_1$ is a one-form density in the position space (from the moment theory), and the Lie derivative has to be computed accordingly.  
One introduces the cold-plasma approximation (cf. equation (\ref{coldplasma})) 
\[
f(q,p,g)=A_0(q,g)\,\,\delta\!\left(p-\frac{A_1(q,g)}{A_0(q,g)}\right)
\]
so that
\[
A_2=\frac{A_1^{\,2}}{A_0}
\]
and the equation for $A_1$ closes to become
\begin{align*}
\frac{\partial A_1}{\partial t}&=
A_0\frac{\partial \gamma_0}{\partial q}+\textit{\large\pounds}_{\gamma_1}A_1
+
\Big\{ A_1,\,\gamma_0\Big\}
+
\left\{ \frac{A_1^2}{A_0},\,\gamma_1\right\}
\, . 
\end{align*}

\begin{framed}
\noindent
 The final bracket form of the moment equations is thus 
\begin{align}
\hspace{-2.65mm}
\frac{\partial A_0}{\partial t}=\,&\frac{\partial}{\partial
q}\left(\!A_{0}\left(\mu_0\,\frac{\partial \beta_0}{\partial q}+
\Big\{\mu_1,\,\beta_0\Big\}
%
\right)\right)
+
\Big\{A_0,\,
\Big\{\mu_0,\,\beta_0\Big\}
\Big\}
+
\left\{A_1,\,\left(
\mu_0\frac{\partial \beta_0}{\partial q} 
+ 
\Big\{\mu_1,\,\beta_0\Big\}
\right)
\right\}
\label{A0-brkt}
\end{align}
and
\begin{align}
\frac{\partial A_1}{\partial t}=&\,\,
A_0\,\frac{\partial}{\partial q}
\Big\{\mu_0,\,\beta_0\Big\}
+
\left(
\mu_0\frac{\partial \beta_0}{\partial q} 
+ 
\Big\{\mu_1,\,\beta_0\Big\}
\right)
\frac{\partial A_1}{\partial q}
+
2A_1\,\frac{\partial}{\partial q}\!
\left(
\mu_0\frac{\partial \beta_0}{\partial q}
+
\Big\{\mu_1,\,\beta_0\Big\}
\right)
\nonumber\\
&+
\left\{A_1,\,\Big\{\mu_0,\,\beta_0\Big\}\right\}
+
\left\{\frac{A_1^{\,2}}{A_0},\,
\left(
\mu_0\frac{\partial \beta_0}{\partial q} 
+ 
\Big\{\mu_1,\,\beta_0\Big\}
\right)
\right\}
\label{A1-brkt}
\end{align}
\end{framed}
\noindent
These equations contain spatial gradients combined with both single and double Poisson brackets. By defining a flux 
\begin{equation}
\mathcal{F}_{01}
=
\mu_0\frac{\partial \beta_0}{\partial q} 
+ 
\Big\{\mu_1,\,\beta_0\Big\}
\label{flux01}
\end{equation}
the previous equations may be written compactly as
\begin{equation}
\frac{\partial A_0}{\partial t}
=
\frac{\partial}{\partial q}
\Big(\!A_0
\mathcal{F}_{01}
\Big)
+
\Big\{A_0,\,
\Big\{\mu_0,\,\beta_0\Big\}
\Big\}+
\Big\{A_1,\,\mathcal{F}_{01}\Big\}
\label{A0-brkt-short}
\end{equation}
and
\begin{align}
\frac{\partial A_1}{\partial t}
=\,&
\frac{\partial}{\partial q}
\Big(\!A_1
\mathcal{F}_{01}
\Big)
+ 
A_0\,\frac{\partial}{\partial q}
\Big\{\mu_1,\,\beta_0\Big\}
+
A_1\frac{\partial \mathcal{F}_{01}}{\partial q}
+
\left\{A_1,\,\Big\{\mu_0,\,\beta_0\Big\}\right\}+
\left\{\frac{A_1^{\,2}}{A_0},\,
\mathcal{F}_{01}
\right\}
\label{A1-brkt-short}
\end{align}


\subsection{Some results on specializations and truncations}
An interesting feature of the Smoluchowski moment equations is that they
recover both the well know Landau-Lifshitz equation and the GOP-Smoluchowski
equation (\ref{GOP-smolu}) as particular cases. First, one sees that upon
considering only $\gamma_0$ in the equation (\ref{A0-smolu}) this equation
becomes
\[
\frac{\pa A_0}{\pa t}+\left\{\left\{\mu_0,\,\frac{\delta E}{\delta A_0}\right\},\,A_0\right\}=0\,.
\]
which is an equation in double bracket form. Now if one considers the linear
moment $G(q,t)=\int g\,A_0(q,g,t)\,{\rm d}g$ and repeats the same treatment as in the previous sections for the moment equations, then it is possible to express the equation for $G$ as
\[
\frac{\pa G}{\pa t}=\textrm{\large ad}^*_{\left(\textrm{\small ad}_\frac{\delta E}{\delta G}\, \mu_G\right)^{\!\sharp}}\,G
\,.
\]
Specializing to the case $G=\bm\in\mathfrak{so}(3)$ yields the purely dissipative
Landau-Lifshitz equation
\[
\frac{\pa \bm}{\pa t}=\bm\times\boldsymbol{\mu}_{\bm} \times\frac{\delta E}{\delta \bm}
\,.
\]
Another specialization is to neglect the first-order moments $A_1$, $\mu_1$
in equation (\ref{A0-brkt}). It is easy to see that this yields 
\begin{align}\nonumber
\frac{\partial A_0}{\partial t}=\,&\frac{\partial}{\partial
q}\left(\!A_{0}\left(\mu_0\,\frac{\partial \beta_0}{\partial q}
\right)\right)
+
\Big\{A_0,\,
\Big\{\mu_0,\,\beta_0\Big\}
\Big\}
\end{align}
which is exactly the equation (\ref{GOP-smolu}) for $\beta_0=\delta
E/\delta A_0$ and $A_0=\varphi$.

Thus different specializations in the Smoluchowski moment equations
yield different order of approximations. Indeed one can see, that the difference
between the dissipative Landau-Lifshitz equation and the GOP Smoluchowski
equation (\ref{GOP-smolu}) differ in that the latter allows for particle
motion with a velocity which is proportional to the collective force (Darcy's
velocity), while the first takes into account only magnetization effects
without considering particle displacement.

\subsection{A divergence form for the moment equations}
At this point it is convenient to introduce the following
\begin{lemma}
Given any two functions $h$ and $f$ on the Lie algebra $\mathfrak{g}$, the following relation holds
\[
\Big\{h,\,f\Big\}:=
\left\langle g,\left[\frac{\partial h}{\partial g},\frac{\partial f}{\partial g}\right]\right\rangle
\,=\,-\,
\frac{\partial}{\partial g}\cdot\left(h\,\,\,\frac{\partial}{\partial g}\cdot\left(f\,\widehat{g}\,\right)\right)
\quad\text{ with}\quad
g\in\mathfrak{g}
\]
where the antisymmetric tensor $\widehat{g}$ is defined in terms of the structure constants $C^a_{bc}$ as
\[
\widehat{g}_{\,bc}:=g_a\, C^a_{bc}
\]
\end{lemma}
\begin{proof}
By the Leibnitz rule one has
\begin{align*}
\left\langle g,\left[\frac{\partial h}{\partial g},\frac{\partial f}{\partial g}\right]\right\rangle
\,=\,-\,
\frac{\partial}{\partial g}\cdot\left(h\,\,\textrm{\large
ad}^*_{\frac{\partial f}{\partial g}}\, g\right)
\,+\,
h\,\left(\frac{\partial}{\partial g}\cdot\,\textrm{\large
ad}^*_{\frac{\partial f}{\partial g}}\, g\right)
\,.
\end{align*}
Also, one calculates, by the Leibnitz rule again and the antisymmetry of the structure constants that
\begin{align*}
\textrm{\large
ad}^*_{\frac{\partial f}{\partial g}}\, g
&=
g_a\, C^a_{bc}\,\frac{\partial f}{\partial g_b}\,{\bf e}^c
=
\frac{\partial}{\partial g_b}\big(f\,g_a\, C^a_{bc}\big)\,{\bf e}^c
=
\frac{\partial}{\partial g}\!\cdot\!\big(f\,\widehat{g}\,\big)
\\
\frac{\partial}{\partial g}\cdot\,\textrm{\large
ad}^*_{\frac{\partial f}{\partial g}}\, g
&=
\frac{\partial}{\partial g_c}\left(g_a\,C^a_{bc}\,\frac{\partial f}{\partial g_b}\right)
=
g_a\,C^a_{bc}\,\frac{\partial^2 f}{\partial g_c \,\partial g_b}
=\,
\widehat{g}\,:\frac{\partial}{\partial g}\!\otimes\!\frac{\partial}{\partial g}\,f=0
\end{align*}
where the symbol : stands for contraction of all indices. The result in the second line is justified by symmetry, as it involves a contraction of an antisymmetric tensor $\widehat{g}\,$ with the symmetric tensor $\partial_g \otimes
\partial_g$. This completes the proof.
\end{proof}

\medskip

\noindent

By making use of this Lemma, one can rearrange equations (\ref{A0-brkt}-\ref{A1-brkt}) into the following form
\begin{align*}
\frac{\partial A_0}{\partial t}=&\,\frac{\partial}{\partial
q}\left(A_{0}\,\left(\mu_0\,\frac{\partial \beta_0}{\partial q}
\,-\,
\frac{\partial}{\partial g}\cdot\left(\mu_1\,\frac{\partial}{\partial g}\cdot\left(\beta_0 \,\widehat{g}\,\right)\right)\,
\right)\,\right)
\\
&+
\frac{\partial}{\partial g}\cdot\left(A_0\,\,\frac{\partial}{\partial g}\cdot\left(\,\widehat{g}\,\,\,
\frac{\partial}{\partial g}\cdot\left(\mu_0\,\frac{\partial}{\partial g}\cdot\left(\beta_0 \,\widehat{g}\,\right)\right)\,
\right)\,\right)
\\
&+
\frac{\partial}{\partial g}\cdot\left(A_1\,\,\frac{\partial}{\partial g}\cdot\left(\,\widehat{g}\,\,\,
\frac{\partial}{\partial g}\cdot\left(\mu_1\,\frac{\partial}{\partial g}\cdot\left(\beta_0 \,\widehat{g}\,\right)\right)\,
\right)\,\right)
\\
&-
\frac{\partial}{\partial g}\cdot\left(A_1\,\,\frac{\partial}{\partial g}\cdot\left(\,\widehat{g}\,\,
\mu_0\,\frac{\partial \beta_0}{\partial q}\right)\right)
\end{align*}
and
\begin{align*}
\frac{\partial A_1}{\partial t}\,=&\,
-A_0\,\,\frac{\partial}{\partial q}\left(
\frac{\partial}{\partial g}\cdot\left(\mu_0\,\frac{\partial}{\partial g}\cdot\left(\beta_0 \,\widehat{g}\,\right)\right)\right)
\\
&+
\left(
\mu_0\frac{\partial \beta_0}{\partial q}-
\frac{\partial}{\partial g}\cdot\left(\mu_1\,\frac{\partial}{\partial g}\cdot\left(\beta_0 \,\widehat{g}\,\right)\right)
\right)\frac{\partial A_1}{\partial q}
\\
&+
2A_1\,\frac{\partial}{\partial q}\!
\left(\mu_0\frac{\partial \beta_0}{\partial q}-
\frac{\partial}{\partial g}\cdot\left(\mu_1\,\frac{\partial}{\partial g}\cdot\left(\beta_0 \,\widehat{g}\,\right)\right)
\right)
\\
&+
\frac{\partial}{\partial g}\cdot\left(A_1\,\frac{\partial}{\partial g}\cdot\left(\,\widehat{g}\,\,\,
\frac{\partial}{\partial g}\cdot\left(\mu_0\,\frac{\partial}{\partial g}\cdot\left(\beta_0 \,\widehat{g}\,\right)\right)\,
\right)\,\right)
\\
&+
\frac{\partial}{\partial g}\cdot\left(\frac{A_1^{\,2}}{A_0}\,\frac{\partial}{\partial g}\cdot\left(\,\widehat{g}\,\,\,
\frac{\partial}{\partial g}\cdot\left(\mu_1\,\frac{\partial}{\partial g}\cdot\left(\beta_0 \,\widehat{g}\,\right)\right)\,
\right)\,\right)
\\
&-
\frac{\partial}{\partial g}\cdot\left(\frac{A_1^{\,2}}{A_0}\,\frac{\partial}{\partial g}\cdot\left(\,\widehat{g}\,\,\,
\mu_0\,\frac{\partial \beta_0}{\partial q}\right)\right)
\end{align*}

\medskip
\noindent
If one inserts the notation
\begin{equation}
\lambda_0(q,g)
=
\frac{\partial}{\partial g}\cdot\left(\mu_0\,\frac{\partial}{\partial g}\cdot\left(\beta_0 \,\widehat{g}\,\right)\right)
=
\frac{\partial}{\partial g}\cdot\left(\mu_0\,\,\textrm{\large ad}^*_{\frac{\partial \beta_0}{\partial g}}\,g\right)=-\,\Big\{\mu_0,\,\beta_0\Big\}
\end{equation}
and similarly, $\lambda_1(q,g)=-\,\{\mu_1,\,\beta_0\}$,
then it is possible to can write the $(A_0,A_1)$ dynamics more compactly as
\begin{eqnarray}
\frac{\partial A_0}{\partial t}
&=&
\,\frac{\partial}{\partial q}\Big(A_0\mathcal{F}_{01} \Big)
+
\frac{\partial}{\partial g}\cdot\left(A_0\,\,
\textrm{\large ad}^*_\frac{\partial
\lambda_0}{\partial g}\,g
\,-\,
A_1\,\,
\textrm{\large ad}^*_\frac{\partial
\mathcal{F}_{01}}{\partial g}\,g\right)
\end{eqnarray}
and
\begin{align}\nonumber
\frac{\partial A_1}{\partial t}
=\,&
\,\frac{\partial}{\partial q}\Big(A_1\mathcal{F}_{01} \Big)
-
A_0\,\,\frac{\partial \lambda_1}{\partial q}
\,+\,
A_1\,\frac{\partial}{\partial q}\mathcal{F}_{01}
+
\frac{\partial}{\partial g}\cdot\left(A_1\,\textrm{\large ad}^*_\frac{\partial
\lambda_0}{\partial g}\,g
\,-\,
\frac{A_1^{\,2}}{A_0}\,\textrm{\large ad}^*_\frac{\partial
\mathcal{F}_{01}}{\partial g}\,g\right)
\,.
\end{align}
These equations may also be written in slightly more familiar form by writing the ${\rm ad}^*$ operations explicitly in terms of  derivatives on the Lie algebra,
\begin{framed}
\begin{align}
\frac{\partial A_0}{\partial t}
=\,&
\,\frac{\partial}{\partial q}\Big(A_0\mathcal{F}_{01} \Big)
+\,
\frac{\partial}{\partial g}\cdot\left(A_0\,\frac{\partial}{\partial g}\cdot\left(\,\widehat{g}\,\,
\lambda_0
\right)
\,-\,
A_1\,\,\frac{\partial}{\partial g}\cdot
\,\widehat{g}\,\mathcal{F}_{01}\right)
\label{A0dyn}
\\
\frac{\partial A_1}{\partial t}
=\,&
\,\frac{\partial}{\partial q}\Big(A_1\mathcal{F}_{01} \Big)
-
A_0\,\frac{\partial \lambda_1}{\partial q}
+
A_1\,\frac{\partial}{\partial q}\mathcal{F}_{01}
+
\frac{\partial}{\partial g}\cdot\left(A_1\,\frac{\partial}{\partial g}\cdot\left(\,\widehat{g}\,\,
\lambda_0\,
\right)
\,-\,
\frac{A_1^{\,2}}{A_0}\,\frac{\partial}{\partial g}\cdot\,
(\widehat{g}\,\mathcal{F}_{01})\right)
\label{A1dyn}
\end{align}
\end{framed}

\begin{remark}[Relation to Smoluchowski equations]
A connection may exist with the nonlinear ``diffusion'' term ${\rm div}_{\!\rm
g}\,(\mathbf{G} f)$ in equation (6) in \cite{Co2005}, where subscript $\rm g$ denotes the metric on $S^2$ and $\mathbf{G}=\nabla_{\!\rm g}\, U+\mathbf{W}$ for some scalar $U$ and a vector field  $\mathbf{W}$ on $S^2$.  In the present formulation, $g$ is an element of Lie algebra $\mathfrak{g}$, \emph{not} of the Lie group, the divergence terms are of the type ${\rm div}_{\!g}\!\left( A_0\, {\rm div}_{\!g} \,\bar{F}\right)$, where $\bar{F}$ is a $(0,2)$ \emph{antisymmetric} tensor over the Lie algebra $\mathfrak{g}$. It is not possible for this tensor to be diagonal. In particular, if one considers the case of the GOP Smoluchowski equation in the divergence form
\begin{align*}
\frac{\partial A_0}{\partial t}=&\,\frac{\partial}{\partial
q}\left(A_{0}\,\left(\mu_0\,\frac{\partial}{\partial q}\frac{\delta E}{\delta A_0}
\right)\,\right)
+
\frac{\partial}{\partial g}\cdot\left(A_0\,\,\frac{\partial}{\partial g}\cdot
\big(\,\widehat{g}\,\lambda_0\,
\big)\,\right)
\end{align*}
the possibility of a connection appears more explicitly.

\medskip

In addition, classical Smoluchowski equations in \cite{Co2005} do not have the $A_1$ contribution of the inherent particle momentum. Instead, they couple the evolution of $A_0$ to the ambient fluid motion described by a variant of the Navier-Stokes equations. In the present approach, no ambient fluid motion is imposed, rather the continuum flow is induced by the dynamics of orientation, leading to the induced momentum $A_1$. The presence of $A_1$ is another difference between the physical interpretation of the present approach and the classical Smoluchowski treatment. The meaning of these differences between the results obtained here and the Smoluchowski approach \cite{Co2005} will be pursued further in future work. 
\end{remark}

\section{Summary and outlook}

The double-bracket Vlasov moment dynamics discussed here provides an alternative to both the variational-geometric approach of \cite{HoPu2007} and the Smoluchowski treatment reviewed in \cite{Co2005}. These are early days in this study of the benefits afforded by the double-bracket approach to Vlasov moment dynamics. However, the derivations of Darcy's law in (\ref{Darcy-rho}) and the Gilbert dissipation term in (\ref{rodGilbert}) by this approach lends hope that this direction will provide the systematic derivations needed for modern technology of macroscopic models for microscopic processes involving interactions of particles that depend on their relative orientations. Although some of these formulas may look daunting, they possess an internal consistency and systematic derivation that might be worth pursuing further. Possible next steps will be the following:
\begin{itemize}
\item Extend the theory of straight filament consistent of rod-like particles to deformable media, 
\item Perform the analysis of the mobility functionals in kinetic space $\mu[f]$ as well as the mobilities for each particular geometric quantity $\mu_\rho$, $\mu_G$ \emph{etc.} 
\item Study the conditions for the emergence of weak solutions (singularities) in the macroscopic (averaged) equations. 

\item Add more physics to the moment approach. For example, it could be worthwhile to investigate the behavior of singularities in a relativistic version of the nonlocal Darcy's law (\ref{Darcy-rho}). This might provide some insight into galaxy clustering in the Universe, especially if the spontaneous emergence of singularities persists in the relativistic approach.
\end{itemize}



\chapter{Conclusions and perspectives}
This thesis has developed a geometric basis for modeling continuum dynamics using double brackets. It has established the geometric approach, proven its effectiveness and used it to reveal new perspectives for modeling dissipative structures. It has developed new types of integral-PDE systems that are available in this approach with a special focus on emergent singular solutions.  

The result is a framework and vista for possible applications for the new science of geometric moment equations. These equations address physical and technological very promising phenomena whose modeling description lies at the boundary between continuum mechanics and kinetic theory. 

The particles that aggregate and form patterns are allowed to be anisotropic. The internal degrees of freedom of such particles (including, for example, the nano-rods developed recently for exploring shape and orientation effects in nanotechnology) influence their aggregation into patterns. The derivation of a wide variety of these new models shows the richness of the modeling approach developed here. Future investigations will seek the appropriate applications of this new geometric approach for deriving moment equations that possess singular solutions.

This chapter summarizes the results obtained and outlines a plan for future research.

\section{The Schouten concomitant and moment dynamics}

This work has used the geometric formulation of moment dynamics to
obtain macroscopic continuum description from the microscopic kinetic theory.
The key idea is that the operation of taking the moments is a Poisson map
leading to the Kupershmidt-Manin structure. 

The first result is a geometric interpretation of the moments in terms of {\bfi symmetric tensors} on the configuration space \cite{GiHoTr2008}. This idea provides the identification of the moment Lie bracket with the {\bfi symmetric Schouten bracket}
(or ``concomitant'') \cite{BlAs79,Ki82,DuMi95}, which is different from the Lie bracket presented by Kupershmidt \cite{Ku1987,Ku2005} in terms of multi-indexes. This fact relates moment dynamics with the theory of invariant differential operators \cite{Ki82}. In formulas, the Schouten form of the moment bracket is given by \cite{GiHoTr2008}
\[
\left\{F,G\right\}=
\sum_{n,m=0}^\infty
\left\langle
A_{m+n-1},\left[ n
\left(\frac{\delta E}{\delta A_n}\contract\,\nabla\right)\frac{\delta F}{\delta A_m}
-
m
\left(\frac{\delta F}{\delta A_m}\contract\,\nabla\right)\frac{\delta E}{\delta A_n}\right]
\right\rangle
\]
where $\beta_n\contract\,\nabla:=\beta_n^{\,\,i_1,\,\dots\,,\,i_n}\,\partial_{\,i_n}$
denotes the usual tensor contraction of indexes.

The symmetry property of the moments relates their geometry
with the symmetric group $\mathcal{S}_n$ involving permutations of the $n$ components of the $n$-th moment $A_n$. After all, this is not surprising,
since the symmetric group $\mathcal{S}_n$ is already involved in other kinds
of moments in kinetic theory, i.e. the statistical Vlasov moments \cite{HoLySc1990} and the BBGKY moments \cite{MaMoWe1984} of the Liouville equation for the
phase space distribution of a discrete number of particles. For the kinetic
moments treated here, the role of the symmetric group is not clear since
it is related with the nature of coadjoint motion, which is not known yet. Questions concerning the nature of the coadjoint motion for the moments provide
an interesting topic for future research.

After showing how diffeomorphisms act on the moments yielding the equations
of fluid dynamics, this work has reviewed some of the physical applications
where moments play a central role. In particular a new result concerns {\bfi
beam dynamics in particle accelerators} \cite{GiHoTr2007}: the dynamics of coasting beams \cite{Venturini} is governed
by the integrable Benney equation \cite{Be1973,Gi1981} and this explains the observation of nonlinear
coherent structures \cite{ScFe2000} in several experiments \cite{KoHaLi2001,CoDaHoMa04,BlBrGlRaRy03,MoBaJaLeNgShTa}. So far these nonlinear excitations have been explained in terms of soliton behavior, while the fact that the
Benney equation is dispersionless suggests that solitons are unlikely
to appear in this context. Rather these are coherent structures
that cannot be studied through simple perturbative approaches.

\section{Geodesic moment equations and EPSymp}
The main objective of the first part of this work (chapters~\ref{momLPdyn}
and~\ref{EPSymp}) is the study of \emph{geodesic}
motion on the moments. This investigation has provided \cite{GiHoTr05,GiHoTr2007} a clear explanation
of this dynamics in terms of {\bfi geodesic flow on the symplectomorphisms} Symp$(T^*Q)$
of the
cotangent bundle, which is the natural extension of the geodesic flow on the diffeomorphisms Diff$(Q)$ of the configuration space, known as EPDiff
\cite{HoMa2004}.
(By analogy the geodesic flow on the symplectic group has been called EPSymp.)
Surprising similarities of this system have been shown with the integrable
Bloch-Iserles system \cite{BlIs,BlIsMaRa05}, which is again a geodesic motion
on the linear symplectomorphisms Sp$(T^*Q)$, i.e. the group of symplectic
matrices. This direction provides an interesting topic to be pursued
in the next future. For example, one wonders what relation holds between
the Bloch-Iserles system and EPSymp. Do integrability issues arise for the
latter?

Also, {\bfi singular solutions} have been analyzed for the geodesic moment equations
and they coincide with the single particle trajectory \cite{GiHoTr05,GiHoTr2007}. The fact that a tensor
power appears in the singular solution
\[
A_n({\bf q},t)=\int \otimes^n {\bf P}(s)\,\delta({\bf q-Q}(s,t))\,{\rm d}s
\]
is not only justified by the single particle nature, but also by the fact
that the power is the only function that always restricts these contravariant
tensors to be fully symmetric. The last observation provides an interpretation
of these solutions in terms of momentum map \cite{MaRa99} defined on the cotangent bundle
of the embeddings ${\bf Q}:s\mapsto{\bf x}\in\mathbb{R}^3$. The evaluation
of the momentum map at the point $(\bf Q,P)$ always yields a sequence of contravariant symmetric tensors (the symmetry is guaranteed by the power $\otimes^n{\bf P}$), that is a
\emph{kinetic moment}. By following the same treatment in \cite{HoMa2004},
 one writes the momentum map as
\[
J\,:\,({\bf Q,P})\mapsto\int \otimes^n {\bf P}(s)\,\delta({\bf q-Q}(s,t))\,{\rm d}s
\,.
\]

The geodesic moment equations have been shown to possess {\bfi remarkable
specializations}, whose first example is the integrable Camassa-Holm equation
\cite{CaHo1993,HoMa2004}
(obtained for Hamiltonians depending only on $A_1$). When considering both
moments $A_0$ and $A_1$, the geodesic moment equations yield the two component
Camassa-Holm equation \cite{ChLiZh2005,Falqui06,Ku2007}, which is again an integrable system. 

The geodesic moment equations have also been extended \cite{GiHoTr2007} to include {\bfi aniso-tropic interaction} by following the treatment in \cite{GiHoKu1983}. Singular solutions have been analyzed as well as their mutual interaction, yielding the problem of the interaction of two rigid bodies.

\section{Geometric dissipation}
The second part of this work (starting with chapter~\ref{GOP}) presented a form of geometric flow for geometric
order parameters (GOP equations). This flow arose in the work of Holm and
Putkaradze \cite{HoPu2007} during their efforts to
establish a geometric interpretation of Darcy's law \cite{HoPu2005,HoPu2006}.
Darcy's law (\ref{DLaw}) is also known as the ``porous media equation'' and is used to model
self-aggregation phenomena in physical applications. The fact that these phenomena can be modelled
by Darcy's law makes this equation an interesting opportunity for its mathematical
inspection. Indeed Darcy's law turns out to have a geometric structure that
suggests its applicability to any quantity
belonging to any vector space $V$ acted on by a Lie algebra $\mathfrak{g}$ (i.e. a $\mathfrak{g}$-module $V$). Nevertheless, the geometric structure of Darcy's law
presents an ambiguity (cf. chapter~\ref{GOP}), which makes it not sufficient for the extension to generic order parameters.
Chapter~\ref{GOP} has shown how the requirement of singular solutions uniquely determines
a general geometric structure, thereby generating what has been called {\bfi GOP equation} \cite{HoPu2007,HoPuTr2007}. This is is a dissipative flow \cite{HoPu2007}, which
is generated by the Lie group $G$ corresponding to the Lie algebra $\mathfrak{g}=T_e
G$ and is completely justified by thermodynamic arguments \cite{HoPuTr2007}. In formulas, when
$\mathfrak{g}=T_e {\rm Diff}$, the
GOP equation for an order parameter $\kappa\in V$ is given by
\[
\frac{\pa\kappa}{\pa t}+\textit{\large$\pounds$}
_\text{\!\small$\left(\mu[\kappa]\left.\diamond\right.\frac{\delta E}{\delta \kappa}\right)^\sharp$}
\,\,\kappa=0
\] 
The mathematical geometric structure of GOP equations can be interpreted in terms of an
invariant Riemannian metric defined on $V^*$ \cite{HoPu2007}. The symmetric nature of the metric is the mathematical reason for dissipation, in agreement with the work of Kaufmann and Morrison \cite{Ka1984,Mo1984,Mo1986}.

The main result in this work concerning the GOP equations is the {\bfi existence of singular solutions},
which is made possible by the appropriate introduction of a ``mobility functional'',
that is a smoothed version of the dynamical variable itself. The smoothing
process yields an equation which is nonlocal. In the case when the dynamical
variable is acted on by diffeomorphisms (Lie derivative), the GOP equation
is a characteristic equation along a smooth vector field, which includes
the nonlocal effects. 

Applications of this flow have been proposed for {\bfi differential forms}, which are cases of interest in physical applications (e.g. the magnetic field in magnetized plasmas \cite{HoMaRa}). In the case of exact forms,
it has been shown that singular solutions are allowed for both the forms
themselves and their potentials and these solutions are different in the
two cases \cite{HoPuTr2007}.

\section{Dissipative equation for fluid vorticity}
A special case of dissipative dynamics is provided by the vorticity exact two--form in fluid dynamics \cite{MaWe83}. Indeed, it has been shown
how the GOP equation for the vorticity in section~\ref{sec:euler} yields a {\bfi double bracket dissipation
for perfect incompressible fluids}, thereby recovering the results in 
\cite{BlKrMaRa1996} previously introduced in \cite{VaCaYo1989}. The dissipative
equation for the vorticity
\[
\frac{\partial\boldsymbol\omega}{\pa t}
+\text{curl}\left\{\boldsymbol\omega
\times\text{curl}\left[\frac{\delta H}{\delta\boldsymbol\omega}
- \text{curl}\left(\boldsymbol\mu[\boldsymbol\omega]
\times
\text{curl}\,\frac{\delta E}{\delta\boldsymbol\omega}\right)
\,\right]\right\}=0
\]
 has been shown to preserve many properties of
the ideal case, such as Ertel's theorem, Kelvin circulation theorem and the
conservation of helicity \cite{HoPuTr2007}. The main difference from the vorticity equation
in the ideal case is the presence of a {\bfi modified velocity}, such that the characteristic
velocity of the equation is given by the sum of the ideal velocity and (minus)
the ``Darcy velocity'', which takes into account the dissipation and ``slows
down'' the fluid particles while preserving the coadjoint orbits as in the
theory of double bracket dissipation. The two--dimensional case has also
been presented to possess the same structure of the ideal case, but with
a velocity suitably decreased in time by the dissipative effects. The {\bfi
point
vortex solution} has been analyzed \cite{HoPuTr2007}.

Another application has been presented to the case of {\bfi one form--densities}, involving the Camassa-Holm equation \cite{CaHo1993}. In this case the peakon solutions undergo dissipative dynamics \cite{HoPuTr2007} and the equations of the peakon lattice have been presented.

GOP equations have been shown to reduce to double bracket equations when applied to variables whose Hamiltonian dynamics is given in Lie-Poisson form. The cases of the vorticity equation and the Camassa--Holm equation are clear
examples of this situation. This fact constitutes one
of the mathematical motivations for the remaining discussions in this work.
\rem{ 
 In the double bracket form, the GOP Riemannian
structure assumes a central role as a ``normal metric'', that is a metric
which is preserved on coadjoint orbits.
} 

\section{Geometric dissipation in kinetic theory}

The dissipative flow for geometric order parameters provides a basis for
deriving a dissipative kinetic flow in terms of Vlasov equation. The fact
that kinetic {\bfi moments are a Poisson map} \cite{Gi1981} yields the corresponding {\bfi
dissipative flow for the moments}.

The idea of a GOP equation for the Vlasov distribution directly involves the action of symplectomorphisms on the phase space densities. At the Lie
algebra level, the Lie derivative is written as Poisson bracket and the fact
that the GOP theory reduces to double bracket determines the dissipative
Vlasov equation \cite{HoPuTr2007-CR}
\[
\frac{\pa f}{\pa t}+\left\{f,\frac{\delta H}{\delta f}\right\}=
\left\{f,\left\{\mu[f],\frac{\delta E}{\delta f}\right\}\right\}
\]
where $E$ is a suitable energy functional, which is possibly different from
the Hamiltonian $H$ and it is usually chosen to be the collective potential.
The case $E=H$ and $\mu[f]\propto f$ reduces to the equation presented in
\cite{BlKrMaRa1996} and the choice $E=J$ (always with $\mu[f]\propto f$)
is presented in the work of Kandrup \cite{Ka1991}, who first introduced this
form of dissipative Vlasov equation for applications in astrophysics.

The first consequence of this equation is that the evolution of $f$ occurs
in the form of coadjoint motion and thus it allows all the Casimirs of the
Hamiltonian case and more importantly the entropy functional $S=\int f \log
f$ is also conserved \cite{HoPuTr2007-CR}. The conservation of entropy can be physically interpreted
in terms of {\bfi reversibility of the dynamics}. Indeed, being a form of coadjoint motion, the evolution is given by the action of canonical transformations
that are always invertible, by definition of Lie group. Thus the evolution
of $f$ can always be inverted without any loss of information and this fact
is the key to understand the preservation of entropy. This case differs from
the conventional Fokker--Planck approach, which is based on the hypothesis
of brownian motion through the Langevin stochastic equation. However, in
principle it is possible to recover the increasing entropy by adding a diffusion
term to the equation. This process would {\it still} be different from the Langevin
approach that involves a linear dissipation in the microscopic equation,
but would recover the {\it brownian} property and thus would increase entropy.
The combination of stochastic effects with the deterministic effects discussed here is a subject for future research.

The {\bfi single particle solution} has been shown to be consistent with the two--dimensional
vorticity equation. The existence of the single particle solution is an important
property, which is {\it not\,} shared with any other dissipative kinetic equation.
Besides its absence in the Fokker-Planck theory, it is worth mentioning that even the equations presented by Kandrup \cite{Ka1991} and Bloch et al.
\cite{BlKrMaRa1996} {\it do not} possess the single particle solution. Moreover
it is also important to notice that the existence of this solution has nothing to do with the preservation of entropy, which is instead {\it shared} with the theory of Kandrup \cite{Ka1991} and Bloch et al. \cite{BlKrMaRa1996}.

\section{Double bracket equations for the moments}

Once the dissipative Vlasov equation has been established, the present work
has investigated the corresponding moment dynamics. Making use of the double
bracket theory, one can find the dissipative {\bfi double bracket form of moment
dynamics} \cite{HoPuTr2007-CR}, whose full expression (\ref{mom-diss-dyn}) is rather complicated. However it has been
shown how it is possible to construct different closures of this hierarchy
by considering truncations at the zero--th or first order \cite{HoPuTr2007-CR}.

The simplest example is Darcy's law: it has been shown how the {\bfi simplest truncation}
of the hierarchy involving only the zero--th order moment coincides with
Darcy's law. The importance of this result is that Darcy's law can now be provided
with a complete {\bfi justification in terms of kinetic theory} and this is the
first time this result has been accomplished. There have been important results
concerning this point involving the Fokker-Planck treatment \cite{Chavanis04};
however they require $\mu[\rho]=const$ for the mobility functional and the
diffusion cannot be neglected as done in the present treatment.

Another simplification of the moment hierarchy is what has been called ``Darcy
fluid'', i.e. the closure of the hierarchy given by considering only the
zero-th and the first moments. The resulting equations
\begin{align*}
\frac{\pa \rho}{\pa t}+\textit{\large$\pounds$}_\text{\!\!\small$\left(\mu_\rho\left.\diamond\right.\frac{\delta E}{\delta \rho}
\left.+\right.
\mu_m\diamond\frac{\delta E}{\delta m}\right)^{\!\sharp}$}\,\,\rho&=0
\\
\frac{\pa m}{\pa t}+\textit{\large$\pounds$}_\text{\!\!\small$\left(\mu_\rho\left.\diamond\right.\frac{\delta E}{\delta \rho}
\left.+\right.
\mu_m\diamond\frac{\delta E}{\delta m}\right)^{\!\sharp}$}\,\,m&=-\,
\rho\,\diamond
\Big(\textit{\large$\pounds$}_\text{\!\small$\frac{\delta E}{\delta m}$}\,\mu_\rho \Big)^\sharp
\end{align*}
 are rather complicated,
although each single term can be identified both physically and mathematically,
thereby showing clearly the underlying geometric structure and its physical
meaning. The importance of this example is that it shows once again how the
 moment bracket is an extremely powerful tool to obtain macroscopic fluid
 models starting from microscopic kinetic equations. The {\bfi Darcy fluid equations}
 still allow for the single particle solution (in the purely dissipative
 case) and for suitable choices of the energy may model the dissipative version
 of the two--component Camassa-Holm equation \cite{ChLiZh2005,Falqui06,Ku2007}. This point can be an interesting
 opportunity for pursuing this direction further in terms of dissipative
 dynamics on semidirect product group of the type $G\circledS H$, where $H$
 is an appropriate $G$-module (in fluid dynamics this is Diff$\circledS$Den,
 where Den denotes density variables). For example, one may wonder what kind
 of peakon dynamics corresponds to this kind of flow and the peakon-peakon
 interaction would be a possible case of study. 
 
Further moment equations have been shown to possess interesting behavior.
For example, the dissipative moment bracket has been used to formulate a
{\bfi double bracket form of the $b$-equation} (\ref{DB-beq}), which embodies to the dissipative
Camassa-Holm equation \cite{HoPuTr2007}
as a special case. Also the {\bfi GOP equation for the moments} (\ref{GOPmom}) has been formulated,
recovering both Darcy's law and the dissipative Camassa-Holm equation as special
cases. The behavior of singular solutions in these cases is also a possible
direction to be pursued further.

\section{Anisotropic interactions}

The previous efforts to geometrize Darcy's law have yielded its microscopic justification in terms of kinetic theory. This can provide important insight
into self-aggregation phenomena, especially at nano-scales, which is a wide
open area of physical research. However, many of the collective interactions
of interest in this area are anisotropic and most of the time the interactions
between two particles depend on their mutual orientation. The {\bfi orientation
of a nano-particle} can be interpreted in terms of rigid-body dynamics, so
that each single particle is not a point particle, but rather it carries
a moment of inertia and thus it has a non-zero spatial length.

A possible approach for such systems has been formulated by Gibbons, Holm
and Kupershmidt (GHK) \cite{GiHoKu1982,GiHoKu1983}. Thus the {\bfi extension of Darcy's law to anisotropic interactions} \cite{HoPuTr2007-Poisson} has been shown to arise from the moment equations of the double bracket form of the Vlasov-GHK equation. Indeed,
since the Vlasov-GHK equation is in Lie-Poisson form, all the double bracket theory can be transferred to this case. In this treatment the Vlasov distribution $f$ depends on position, momentum {\it and} orientation \cite{GiHoKu1982,GiHoKu1983}
\[
f=f({\bf q,p},\bg,t)
\,.
\]
Once the double bracket equation \cite{HoPuTr2007-Poisson}
\[
\frac{\partial f}{\partial t}=\left\{f,\left\{\mu[f],\frac{\delta E}{\delta f}\right\}_{\!1}\,\right\}_{\!1}
\]
is established, kinetic moments are introduced in the form
\[
A_{n,k}({\bf q},t)=\int {\bf p}^n\, \bg^k\, f({\bf q,p},\bg,t)\,{\rm d}{\bf p}\,{\rm d}\bg
\]
and the moment theory can be transferred to these $p\bg$-moments. The moment
equations are again rather complicated but the main result of this work concerns
the truncation of the hierarchy to consider the special case $n=0,\,k=0,1$.
The resulting equations are \cite{HoPuTr08}
\begin{align*}
\frac{\partial \rho}{\partial t}
&=
\text{\rm\large div}
\Bigg(
\rho\,
\bigg(
\mu_\rho\, \text{\large$\nabla$}\frac{\delta E}{\delta \rho}
+\,
\boldsymbol\mu_{\bf m}\cdot\text{\large$\nabla$}\frac{\delta E}{\delta \bf m}
\bigg)
\,
\Bigg)\\
\frac{\partial {\bf m}\,}{\partial t}
&=
\text{\rm\large div}\Bigg({\bf m}\,\text{\large$\otimes$}
\left(\mu_\rho\text{\large$\nabla$}\frac{\delta E}{\delta \rho}
+
\boldsymbol\mu_{\bf m}
\cdot\text{\large$\nabla$}\frac{\delta E}{\delta {\bf m}}
\right)
 \Bigg)\!
+
{\bf m}\times
\boldsymbol{\mu}_{\bf m}\times\frac{\delta E}{\delta \bf m}
\end{align*}
where $\rho:=A_{0,0}\,$, ${\bf m}:=A_{0,1}$ and $\mu_\rho$, $\boldsymbol{\mu}_{\bf m}$ are
filtered versions of $\rho$ and $\bf m$ respectively. As one can easily see,
the right hand side of the second equations {\bfi recovers the Landau-Lifshitz-Gilbert
dissipative dynamics} for the magnetization in ferromagnetic media and this
constitutes one of the main results of this work: the dissipative magnetization
dynamics of Landau, Lifshitz and Gilbert has been recovered from microscopic arguments in kinetic theory, by following a double bracket approach for the
Vlasov equation. To the author's knowledge this is the first time that the Landau-Lifshitz-Gilbert dynamics is {\bfi derived from a microscopic kinetic
treatment}. This term is recovered at all levels of approximation, since the double bracket preserves the geometric structure of the dynamics, as explained in chapter~\ref{DBVlasov}.

The {\bfi singular solutions} allowed by this model have been extensively analyzed
in the present work in the one-dimensional case. However, important questions
concern their behavior in three dimensions, when the two variables are supported
on submanifolds of the Euclidean space (filaments and sheets), each following
its own dynamics. For example, in one dimension the singularities have been
shown to emerge spontaneously, but does this feature persist in more dimensions?
How do the orientation filaments interact? All these questions need to be
answered in future research. Possible applications are suggested in protein
folding and other issues in nano-sciences.

As a further step, a higher order of approximation has been introduced in
the truncation of the moment equations, which takes into account the evolution
of the fluid momentum $A_{1,0}$ as well as of the polarization flux $A_{1,1}$.
However these equations are complicated and they do not allow for singular
solutions.

\section{The Smoluchowski approach}

The last part of this work has presented what is known as Smoluchowski kinetic approach. In this context, the moments are still $p$-moments and they depend
on both position and orientation. Two possibilities have been presented.
The first is simpler in construction and it leads to the {\bfi GOP-Smoluchowski equation} \cite{HoPuTr08}
\begin{align}
\frac{\partial \varphi}{\partial t}
=
\text{\large\rm div}\!\left(\varphi\,\mu[\varphi]\text{\large$\nabla$}\frac{\delta E}{\delta \varphi}\right)
+
\left\{\varphi,\left\{\mu[\varphi],\frac{\delta E}{\delta \varphi}\right\}\right\}
\end{align}
where $\{\cdot,\,\cdot\}$ denotes here the rigid body bracket 
$\left\{g,\,h\right\}:=\boldsymbol{m}\,\cdot\partial_{\boldsymbol{m}\,} g\times\partial_{\boldsymbol{m}}h$.
The interesting feature of this equation is that it leads naturally to the
Landau-Lifshitz equation for the magnetization ${\bf m}=\int \bm\,\varphi(\bq,\bm,t)\,{\rm d}\bm$, when $\delta E/\delta \varphi$ is constant in $\bq$ (otherwise it also leads to the previous equations for $\rho$ and $\bf m$). This equation is {\it not}
rigorously derived from the dissipative Vlasov equation; rather it is established
as a GOP continuity equation in the $({\bf q}, \bm)$-space. 

The second approach follows the process of {\bfi taking the $p$-moments} of the
dissipative Vlasov equation. The resulting equations are complicated and
the truncation to the first moment requires a cold plasma-like closure that
does not allow for singular solutions. However it has been shown how two
particular truncations are possible, whose simplest one is identical
to the Landau-Lifshitz-Gilbert equation. The GOP equation for $\varphi$ is
also obtained as the second specialization.

Possible roads for future research involve the analysis of this hierarchy
and in particular it is not clear how the appearance of the GOP equation
can be rigorously justified by considering the geometric structure of the
whole hierarchy. Also, the singular solutions allowed by the GOP equation
may deserve further study.

\section{Future objectives in geometric moment dynamics}

The study presented in this thesis raises new open questions,
which are sketched in this section. The following scheme presents a plan
of objectives that is divided in two main topics, i.e. Hamiltonian and dissipative moment flows. The final part is devoted to the question of coadjoint moment
dynamics.

\subsubsection{Geodesic motion on Vlasov moments}

\rem{ 
The Vlasov equation for the collisionless evolution of the single-particle
probability distribution function (PDF) is a well-known example of 
coadjoint motion, which comes from its Lie-Poisson dynamics. Remarkably, the Lie-Poisson dynamics survives the
process of taking moments. That is, the evolution of the moments of the Vlasov
PDF is also regulated by a Lie-Poisson structure. The resulting dynamics of the Vlasov moments is found to be integrable is several cases. For
example, I have identified the dynamics for coasting beams in particle accelerators with the integrable dynamics of the Benney shallow-water equation. 

\smallskip
The moment Hamiltonian always include quadratic terms and this represents
the first inspiration for my joint work with J. Gibbons and D. Holm on purely quadratic Hamiltonians, which regulate geodesic motion on the moments. The second and more important observation
is that this dynamics is actually a geodesic motion on the group of canonical transformations (EPSymp). 
} 

\paragraph{\bfi Singular solutions.} 
Study of singular solutions for geodesic moment equations and their closures.
Analysis of their spontaneous emergence. Study of the geometric properties of the fluid closure (dual pairs and plasma-to-fluid
momentum map). Analysis of filaments and sheets in higher dimensions.

\rem{ 
\smallskip
In one dimension, the singular solutions are the single particle solution of the corresponding Vlasov description. Thus their importance is immediately
related with the particle information, which is lost when the singular solutions
are not allowed. (An example in which this happens is provided by the integrable
Vlasov-Benney equation.) However in higher dimensions these solutions are concentrated on embedded subspaces of physical space. When the embedded subspace has dimension 1,
then one calls it a ``filament'', while the word ``sheet'' is used in the
2D case. The motion and interactions of these embedded subspaces are governed by canonical Hamiltonian equations for their geodesic evolution. So another
question concerns the interaction of higher dimensional singularities in
the geodesic moment equations.
}  

\paragraph{\bfi Extension to oriented nano-particles.}
Study of singular solutions in the anisotropic case for nano-particles. Analysis of their interaction in higher dimensions, ``orientation filaments'' and sheets.

\paragraph{\bfi Connections to integrable PDE's.} 
Development of further connections with integrable systems, in particular the Bloch-Iserles equation, which has the same geometric nature as the geodesic moment equations.


\subsubsection{Double-bracket dissipation for moment dynamics}

\paragraph{\bfi Singular solutions.} 

Study of singular solutions of the dissipative moment equations, which have
a very different behavior from the Hamiltonian geodesic case. Analysis of the dissipative fluid closure both in terms of singular solutions and its
geometric properties (e.g. dual pair analogues for the dissipative EPDiff equation). Study of singular solutions in higher dimensions.

\paragraph{\bfi Anisotropic interactions.} 
Study of singular solutions in the anisotropic case (oriented nano-particles),
especially in higher dimensions (some results on the interaction of two oriented filaments have just been published \cite{HoPuTr08}). An important question
is whether these filaments emerge spontaneously in two or three dimensions.

\paragraph{\bfi Relations with complex fluids.}  Study of connections between
the anisotropic moment equations and the Lie algebraic treatment of complex fluids (Lie algebra cocyles). Development of the double bracket theory in
this context.

\paragraph{\bfi The Smoluchowski approach.} Further development of the geometric
background of the Smoluchowski moment equations. Analysis of their closures
and study of singular solutions.

\subsubsection{Other mathematical issues}

\paragraph{\bfi Moments and momentum maps.} Further development of the geometry
underlying the moment hierarchy. Interpretation of moments as momentum maps under the action of the symplectic group, by the properties of the symmetric
Schouten bracket.





\end{document}